\documentclass[12pt]{elsarticle}
\usepackage{nomencl}
\usepackage{ifthen}
\renewcommand{\nomgroup}[1]{%
\ifthenelse{\equal{#1}{S}}{\item[\textbf{Subscripts}]}{%
\ifthenelse{\equal{#1}{V}}{\item[\textbf{Variables}]}{%
\ifthenelse{\equal{#1}{S}}{\item[\textbf{sets}]}{}}}
}
\makenomenclature

\usepackage{amssymb}

\usepackage[a4paper]{geometry}  
\usepackage{times} 
\usepackage{xurl} 
\usepackage[italian,english]{babel}
\usepackage{latexsym} 
\usepackage{xcolor}
\usepackage{multicol}

\makeatletter
\newcommand{\@BIBLABEL}{\@emptybiblabel}
\newcommand{\@emptybiblabel}[1]{}
\usepackage{graphicx}
\usepackage{latexsym}
\usepackage{amssymb}
\usepackage{bm}
\usepackage{amsmath}
\usepackage{epsfig}
\usepackage{caption}
\usepackage{subcaption}
\usepackage{graphicx}
\usepackage{caption}
\usepackage{subcaption}
\usepackage{mwe} 
\usepackage{morefloats}
\usepackage{epstopdf}
\begin{document}
\begin{frontmatter}


\title{Effect of Uniform and Non-uniform wall heating on Three-Dimensional Magneto-Hydrodynamics Natural Convection and Entropy Generation: A computational study using New Higher Order Super Compact Scheme}

\makeatletter
\def\ps@pprintTitle{%
  \let\@oddhead\@empty
  \let\@evenhead\@empty
  \let\@oddfoot\@empty
  \let\@evenfoot\@oddfoot
}
\makeatother
\author{\textbf{Ashwani Punia$^{1}$, Rajendra K. Ray$^{2^*}$} \\
  1,2. School of Mathematical and Statistical Sciences, Indian Institute of Technology Mandi,\\
  Mandi, Himachal Pradesh, 175005, India \\
 {\tt mr.punia11@gmail.com}, {\tt rajendra@iitmandi.ac.in}}

\begin{abstract}
Current research work deals with the effect of uniform and non-uniform wall heating on magnetohydrodynamic (MHD) natural convection within a three-dimensional (3D) cavity filled with molten lithium. A new Higher-Order Super Compact (HOSC) finite difference scheme is used to analyze the thermal behavior under both heating scenarios. After the quantitative and qualitative validations, the computed results are analyzed for a range of Hartman number ($Ha = 25, 50, 100, 150$) and Rayleigh number ($Ra = 10^3, 10^4, 10^5$) with fixed $Pr=0.065$ (molten lithium). Three distinct heating scenarios, i.e., uniform heating ($T_h = 1$), $y$-dependent non-uniform heating ($T_h = sin(\pi y$)), and a combination of $y$ and $z$-dependent non-uniform heating ($T_h = sin(\pi y)sin(\pi z)$) are investigated on the left wall ($x=0$) of the cubic cavity. It is found that variations in the $Ha$ and $Ra$, along with distinct thermal boundary conditions, exert significant effects on both the temperature distribution and flow field inside the 3D cubical cavity. Specifically, an increase in $Ra$ corresponds to enhanced heat transfer, highlighting the dominance of convection. Conversely, an increase in $Ha$ leads to a reduction in heat transfer due to the deceleration of fluid velocity. The scenario in which walls are uniformly heated exhibits the most significant total entropy generation. It is observed that with an increase in the $Ra$, the Bejan number ($Be$) decreases, which ultimately leads to an increase in total entropy generation. The implementation of the new HOSC scheme in this analysis showcases its effectiveness in capturing the complexities of 3D MHD-driven natural convection and entropy generation. This study offers significant information that might help improve the optimization and design of relevant engineering systems. Thus, our work stands out as genuinely novel and pioneering in its approach.

\end{abstract}
\begin{keyword}
HOSC Scheme \sep MHD Natural Convection \sep Heat Transfer \sep Bejan Number \sep Entropy Generation \sep Three-dimensional Cavity 
\end{keyword}
\end{frontmatter}

\newcommand{\nomunit}[1]{%
\renewcommand{\nomentryend}{\hspace*{\fill}#1}}
\renewcommand{\nompreamble}{\begin{multicols}{2}}
\renewcommand{\nompostamble}{\end{multicols}}

\nomenclature{$Be$}{Bejan number} 
\nomenclature{$\alpha$}{Thermal diffusivity $\left(\mathrm{m}^2 \mathrm{~s}^{-1}\right)$}
\nomenclature{$\rho$}{Density $\left(\mathrm{kg} \mathrm{~m}^{-3}\right)$}
\nomenclature{$g$}{Acceleration due to gravity $\left(\mathrm{m\ } \mathrm{s}^{-2}\right)$}
\nomenclature{$\theta$}{Dimensionless temperature}
\nomenclature{$U, V, W$}{Velocity components in $X, Y, Z$ directions $\left(\mathrm{m\ } \mathrm{s}^{-1}\right)$}
\nomenclature{$X, Y, Z$}{Cartesian coordinates ($m$)}
\nomenclature{$pr$}{Dimensionless pressure}
\nomenclature{$Ra$}{Rayleigh number}
\nomenclature{MHD}{Magnetohydrodynamic}
\nomenclature{$Ha$}{Hartmann number}
\nomenclature{$T_h$}{Temperature at hot wall}
\nomenclature{$T_c$}{Temprature at cold wall}
\nomenclature{HOSC}{Higher-order super compact}
\nomenclature{$Pr$}{Prandtl number}
\nomenclature{$\tau$}{Dimensionless time}
\nomenclature{$L$}{Cavity length ($m$)}
\nomenclature{$t$}{Time ($s$)}
\nomenclature{$T$}{Temperature $(\mathrm{K})$}
\nomenclature{$x, y, z$}{Dimensionless Cartesian coordinates}
\nomenclature{$u, v, w$}{Dimensionless velocity components in $x, y, z$ directions}
\nomenclature{$Nu$}{Nusselt number}
\nomenclature{$T_0$}{Bulk temperature $(\mathrm{K})$}
\nomenclature{$\Delta T$}{Temperature difference}
\nomenclature{$P$}{Pressure $(\mathrm{Pa})$}
\nomenclature{$\varphi^{*}$}{Irreversibility distribution ratio}
\nomenclature{$v^{*}$}{Kinematic viscosity $\left(\mathrm{m}^2 \mathrm{~s}^{-1}\right)$}

\nomenclature{$E$}{Entropy}
\nomenclature{$\beta$}{Coefficient of thermal expansion $\left(\mathrm{K}^{-1}\right)$}
\nomenclature{$\mu$ }{Dynamic viscosity $\left(\mathrm{kg\ } \mathrm{m}^{-1} \mathrm{~s}^{-1}\right)$}
\nomenclature[Sp]{$L$}{Local}
\nomenclature[Sp]{A}{Average}
\nomenclature[Sp]{$LE$}{Local Entropy}
\nomenclature[Sp]{$LE,F$}{Local entropy by fluid friction}
\nomenclature[Sp]{$LE,H$}{Local Entropy by heat transfer}
\nomenclature[Sp]{$LE,M$}{Local Entropy by magnetic field}
\nomenclature[Sp]{$TE$}{Total Entropy}

\printnomenclature

\section{Introduction}
\label{sec:Introduction}
In recent decades, there has been a notable focus on investigating laminar natural convection in electrically conducting fluids, subjected to a magnetic field, because of its broad engineering applications \cite{Vives_1987,Utech_1996,Sarris_2006}. 
Magnetohydrodynamic (MHD) natural convection is vital for understanding the transfer of heat in various engineering applications, ranging from geophysics to advanced technological systems such as astrophysics, aeronautics, nuclear fusion devices, MHD accelerators, and aerospace. Due to the interconnection between the governing equations for velocity and temperature resulting from buoyancy forces, and the magnetic field that induces the Lorentz force, the exploration of MHD natural convection becomes intricately complex.

Numerous researchers have explored MHD natural convection through a combination of analytical, experimental, and numerical methodologies. Ozoe and Maruo \cite{Ozoe_1987} conducted a 2D numerical investigation of natural convection in presence of a magnetic field, with a specific focus on an electrically conducting fluid with $Pr=0.054$. Ozoe and Okada \cite{Ozoe_1989} used a finite difference method based on a 3D vector potential-vorticity formulation to examine the influence of magnetic field orientation. Okada and Ozoe \cite{Okada_1992} conducted an experimental study of natural convection involving molten gallium ($Pr$ = 0.024). The setup involved cooling from one side wall and heating from the opposite wall, while all other walls were kept insulated. The study explored the impact of various magnetic field directions. Their findings revealed that a vertical external magnetic field demonstrated superior effectiveness when contrasted with a magnetic field aligned parallel to the hot vertical wall. In modeling the impact of a transverse magnetic field on natural convection within a two-dimensional (2D) cavity, they utilized an analytical solution proposed by Garandet et al. \cite{Garandet_1992}. Vasseur et al. \cite{Vasseur_1995} derived both numerical and analytical solutions for a tall inclined cavity subjected to a transverse magnetic field in presence of a porous medium.  Rudraiah et al. \cite{Rudraiah_1995} designed a finite difference technique using the modified ADI and Successive Line Over Relaxation (SLOR) methods. This hybrid technique is used to address the 2D streamfunction-vorticity form of the governing equations. The numerical study, conducted for $Pr=0.733$, revealed a decreasing trend in the $Nu_{\text {A}}$ with an increase in $Ha$. Subsequently, Hadid et al. \cite{Hadid_1997_P1, Hadid_1997} employed an ADI approach that incorporates the central differentiation of order 2 and the compact Hermitian technique of order 4 for both 2D and 3D scenarios. Furthermore, they introduced a projection method for solving three-dimensional scenarios. Al-Najem et al. \cite{Al‐Najem_1998} utilized an ADI-based control volume method to study the 2D natural convection within an inclined enclosure, in presence of a magnetic field. Pirmohammadi et al. \cite{Pirmohammadi_2008} study the laminar 2D MHD natural convection of fluid flows employing the SIMPLER algorithm at a constant $Pr=0.733$. They further expanded the idea to analyze an inclined 2D square enclosure. Lo et al. \cite{Lo_2010} introduced a high-resolution approach using the differential quadrature method, relying on the 2D velocity-vorticity formulation of the Navier-Stokes equations. This approach is applied to investigate the influence of a magnetic field on buoyancy-driven MHD flow within a confined space. The results suggested that the heat transfer rate reaches its maximum at higher $Pr$ and without the presence of MHD effects. Many researchers have expanded the scope of the conventional problem to explore various boundary conditions. For instance, Venkatachalappa et al. \cite{Venkatachalappa_1993} provided 2D numerical findings for a fluid with $Pr=0.733$ confined in a rectangular cavity. The study considered the impact of a vertical magnetic field with constant heat flow from the adjacent wall. Sivasankaran et al. \cite{Sivasankaran_2011} solve a two-dimensional mixed convection problem within a cavity. They incorporated the sinusoidal boundary conditions for the temperature and investigated the impact of the Richardson number, phase deviation, amplitude ratio, and $Ha$ on the heat transfer process. They observed that the Nusselt number experiences an increase with an increase in amplitude ratio. Oztop et al. \cite{Oztop_2011} studied mixed convection of laminar flow under the influence of a magnetic field in a 2D lid-driven cavity with heating facilitated by a corner. The study explored various $Ha$ and Grashof numbers, employing the finite volume method. The results show that the heat transfer rate diminishes as the $Ha$ increases.
Bhuvaneswari et al. \cite{Bhuvaneswari_2011} investigated convective heat transfer within a 2D cavity featuring sinusoidal thermal boundary conditions on both side walls under the influence of a uniform magnetic field. Using the finite volume approach, they discovered that increasing the phase deviation causes the heat transfer rate to initially increase and subsequently drop. Sajjadi et al. \cite{Sajjadi_2012} used Lattice Boltzmann Method (LBM) to study how a magnetic field affects mixed convection flow in a two-sided lid-driven cavity. Zhang and Che \cite{Zhang_2016} employed LBM to address a 2D MHD flow in an inclined cavity featuring four heat sources. They utilized a double Multiple Relaxation Time (MRT) model to solve the governing equations to study the effects of the $Ha$ on heat and mass transfer. Their findings revealed that the $Nu_{\text {A}}$ decreases with an increasing the $Ha$ for all $Ra$. Sajjadi et al. \cite{Sajjadi_2019} conducted a 3D simulation of MHD natural convection in a cubic cavity using a novel approach based on the Lattice Boltzmann method with a double Multi-Relaxation-Time (MRT) model. Their findings revealed that an increase in the $Ha$ significantly reduces heat transfer. Ahmad et al. \cite{Ahmad_2024} investigated the heat and mass transfer in MHD natural convection flow within a 2D cavity considering the impact of source and sink. The study employed the finite element method, and the results indicate that the incorporation of sources and sinks enhances the effectiveness of heat transfer systems and influences flow patterns.
Numerous researchers have explored MHD natural convection \cite{Okada_1992,Venkatachalappa_1993,Bhuvaneswari_2011,Sajjadi_2012,Zhang_2016,Sajjadi_2019,Mirzaei_2023, Moderres_2023} by employing experimental, analytical, and numerical approaches.

In the literature, it's observed that the majority of prior studies focused on two-dimensional flows primarily due to the complexities associated with simulating three-dimensional flows. To capture a more realistic representation of the flow, numerical simulations involving three-dimensional flows are necessary. 
Unraveling the complexities of 3D MHD natural convection is of paramount importance due to its relevance in phenomena such as astrophysical flows, magnetic confinement in fusion reactors, and the design of innovative cooling systems. While extensive studies have been conducted on natural convection in conventional fluids, the integration of magnetic fields introduces unique challenges and opportunities. Motivated by the necessity to understand the collective influence of the $Ha$ and $Ra$ on heat transfer, we systematically explore a diverse range of scenarios. The study involves three cases: one with a uniformly heated wall and two with non-uniform sinusoidal heated wall conditions. Sinusoidal thermal boundary conditions prove applicable in various natural convection scenarios, especially when dealing with periodic or oscillatory temperature variations. This approach facilitates a comprehensive understanding of the system's response to changing magnetic and thermal conditions. Real-life applications of sinusoidal thermal boundary conditions include electronic device cooling, where the cyclic operation of devices can be mimicked; solar radiation influence on components in space systems; thermal management in LED arrays responding to cyclic loads or environmental changes, and battery charging and discharging cycles. In these scenarios, sinusoidal thermal boundary conditions serve as valuable tools for simulating and optimizing heat dissipation strategies. Also, it's important to highlight that there is a notable gap in the literature regarding 3D MHD natural convection studies utilizing finite difference methods. Additionally, the widespread use and user-friendly nature of finite difference methods underscore the necessity of developing higher-order finite difference schemes for 3D MHD natural convection flows. 
Kalita \cite{Kalita_2014} initially introduced the super-compact higher-order approach and solved the fundamental lid-driven cavity problem without considering heat or energy transfer. However, more recently, Punia and Ray \cite{Punia_2024} developed the Higher-Order Super Compact (HOSC) scheme to address 3D natural convection problem and found that the HOSC scheme can capture the fluid flow and heat transfer phenomena very accurately. The intricate complexity of exploring magnetohydrodynamic (MHD) natural convection arises from the interconnection between the governing equations for velocity and temperature, driven by buoyancy forces, and the involvement of the Lorentz force induced by the magnetic field. This research extends the HOSC finite difference scheme that can effectively simulate three-dimensional MHD natural convection, providing a more comprehensive understanding of the complex fluid dynamics involving buoyancy and Lorentz forces. Here, we consider a 3D cavity filled with molten lithium, a scenario relevant to various industrial and energy-related processes. The choice of molten lithium as the working fluid adds a layer of complexity to the analysis due to its unique thermophysical properties. Lithium is known for its high thermal conductivity and low viscosity, making it an interesting medium for applications where efficient heat transfer is essential subjected to MHD natural convection. The computational approach employed leverages the capabilities of the HOSC scheme, notable for its second-order accuracy in time and fourth-order accuracy in space.\\
 The current study not only contributes to the fundamental understanding of complex fluid dynamics but also paves the way for practical applications in various fields such as energy transport, magnetic fluid dynamics, and advanced thermal management systems. However, to the best of the authors' knowledge, no existing study has employed the HOSC finite difference method to address 3D MHD convection problems.

\section{Description of the Problem and Discretization of Governing Equations}
\label{sec:Problem Description and Discretization of Governing Equations}
\begin{figure}
    \centering
    \includegraphics[width=\textwidth]{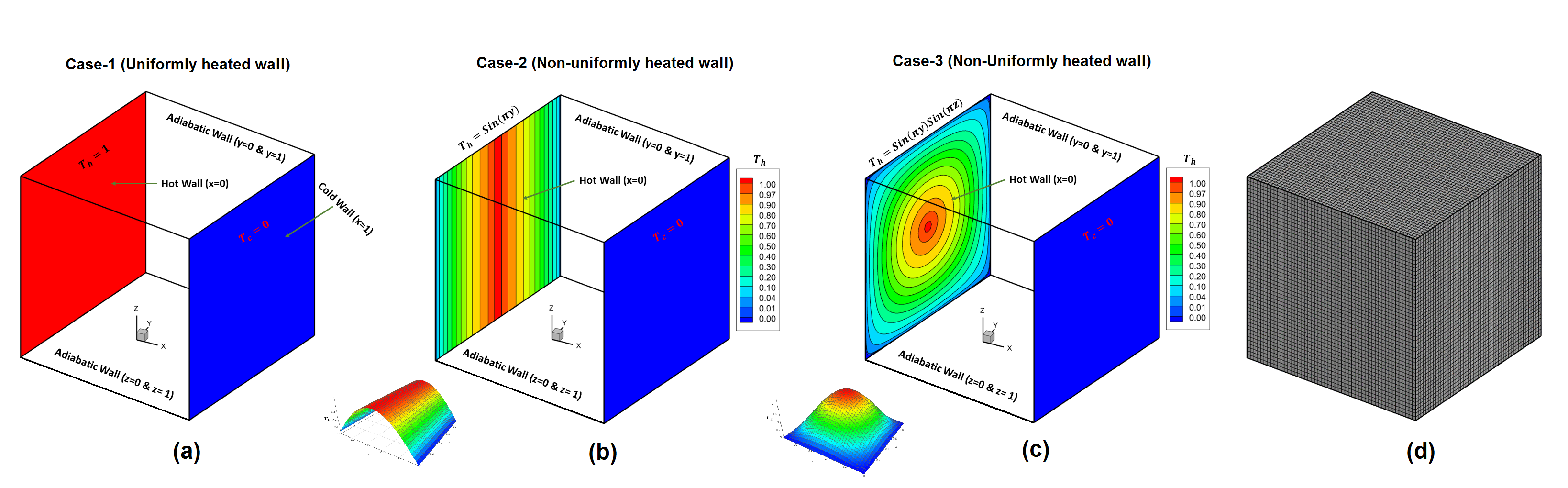}
    \caption{(a-c) Schematic view of the problem for three different cases (d) Visualization of the grids with $51\times51\times51$ grid size.}
    \label{fig:Sche_diag}
\end{figure}

This study focuses on the effect of uniformly heated and non-uniformly heated wall conditions on the transient flow of MHD natural convective fluid in a closed 3D cubic cavity filled with molten lithium ($Pr=0.065$). Three different cases, based on distinct thermal boundary conditions, one with a uniformly heated wall and two with non-uniform sinusoidal heated wall, are considered. The schematic diagram of these three cases considered in this paper is shown in Figure \ref{fig:Sche_diag}. In Case 1, the left wall has a constant temperature $(T_h=1)$. In Case 2, the left wall is subjected to a non-uniform sinusoidal boundary condition $(T_h=Sin(\pi y))$. The surface plot of the temperature function, shown in the subfigure of Figure \ref{fig:Sche_diag}(b), is like a smoothly curved sheet, with its peak in the middle line $(y=0.5, 0 \leq z \leq 1 )$. Remarkably, the temperature contours on the left heated wall appear as straight lines in this case. Transitioning to Case 3, where the left wall is heated with a more intricate boundary condition $((T_h=Sin(\pi y)Sin(\pi z))$, the surface plot of the temperature function resembles a convex bowl-like shape with its peak at the midpoint $(y=0.5, z=0.5)$, as illustrated in subfigure of Figure \ref{fig:Sche_diag}(c). Intriguingly, the temperature contours on the left wall form a circular pattern, showing the interaction of the sinusoidal heating profiles in two spatial dimensions. This discrepancy in temperature distribution and contour shapes clearly highlights the various thermal behaviors that result from different boundary conditions, offering significant insights into the system's thermal dynamics. The investigation involves the unsteady laminar fluid flow model and gravity acts perpendicular to the $xz$ plane. The fluid, characterized by constant density and thermophysical properties, is treated as incompressible. In this context, the Boussinesq approximation is incorporated to account for the body force, detailing the density fluctuation as a function of isotherms. This coupling mechanism links the isothermal and flow fields. Additionally, a uniform magnetic field is applied externally to the 3D cubical cavities, oriented in the negative $y$ direction. Under these assumptions, the transport equations for mass, momentum, and energy conservation in the non-dimensional primitive variable formulation can be described as \cite{Purusothaman_2016, Zhang_2017, Sajjadi_2018}:
\begin{equation}\label{Main_governing_eq_1}
\frac{\partial u}{\partial \tau}+v \frac{\partial u}{\partial y}+w \frac{\partial u}{\partial z}+u \frac{\partial u}{\partial x}=Pr \left[\frac{\partial^2 u}{\partial y^2}+\frac{\partial^2 u}{\partial z^2}+\frac{\partial^2 u}{\partial x^2}\right] -\frac{\partial pr}{\partial x} 
\end{equation}
\begin{equation}\label{Main_governing_eq_2}
\frac{\partial v}{\partial \tau}+v \frac{\partial v}{\partial y}+w \frac{\partial v}{\partial z}+u \frac{\partial v}{\partial x}=Pr \left[\frac{\partial^2 v}{\partial y^2}+\frac{\partial^2 v}{\partial z^2}+\frac{\partial^2 v}{\partial x^2}\right]-\frac{\partial pr}{\partial y} - Ha^2 \ Pr \ v+ R a \ P r \ \theta 
\end{equation}

\begin{equation}\label{Main_governing_eq_3}
\frac{\partial w}{\partial \tau}+v \frac{\partial w}{\partial y}+w \frac{\partial w}{\partial z}+u \frac{\partial w}{\partial x}=Pr \left[\frac{\partial^2 w}{\partial y^2}+\frac{\partial^2 w}{\partial z^2}+\frac{\partial^2 w}{\partial x^2}\right]-\frac{\partial pr}{\partial z}
\end{equation}

\begin{equation}\label{Main_governing_eq_4}
\frac{\partial \theta}{\partial \tau}+v \frac{\partial \theta}{\partial y}+ w \frac{\partial \theta}{\partial z}+u \frac{\partial \theta}{\partial x}=\frac{\partial^2 \theta}{\partial y^2}+\frac{\partial^2 \theta}{\partial z^2}+\frac{\partial^2 \theta}{\partial x^2}
\end{equation}

\begin{equation} \label{Main_governing_eq_5} 
\frac{\partial v}{\partial y}+\frac{\partial w}{\partial z}+\frac{\partial u}{\partial x}=0 
\end{equation}
The process of non-dimensionalizing physical quantities involves the following parameter transformations: 
$$
\begin{aligned}
&  y=\frac{Y}{L},  v=\frac{V L}{\alpha}, z=\frac{Z}{L},  w=\frac{W L}{\alpha}, x=\frac{X}{L}, u=\frac{U L}{\alpha}, P r=\frac{v^{*}}{\alpha}, \\
&  \theta=\frac{T-T_c}{T_h-T_c}, \tau=\frac{t \alpha}{L^2}, pr=\frac{P L^2}{\rho \alpha^2}, Ha^2=\frac{B_0^2L^2\sigma}{\mu}, R a=\frac{L^3\left(T_h-T_c\right)\beta g }{{\alpha v^{*} }} 
\end{aligned}
$$
where $U,V,W$ are the dimensional fluid velocity component in $X,Y$ and $Z$ direction, respectively, $L$ is the length of the cavity, $\alpha$ is the thermal diffusivity, $v^{*}$ is kinematic viscosity, $P$ is pressure, $B_0$ is the external magnetic field, $\sigma$ is electric conductivity and $g$ is acceleration due to gravity. The dimensionless expressions for the initial and boundary conditions are as follows:
$$
\begin{array}{lll}
\tau=0: & w=v=u=0 ; \quad \theta=0 ; \quad 0 \leqslant z, y, x \leqslant 1 \\
\tau>0:  & w=v=u=0 ; \quad \frac{\partial \theta}{\partial z}=0 ; \quad z=0,1\\
  & w=v=u=0 ; \quad \frac{\partial \theta}{\partial y}=0 ; \quad y=0,1 \\
  & w=v=u=0 ; \quad \theta=1 \text{ (for Case 1) or } sin(\pi y) \text{ (for Case 2) or } \\
 &   \quad \quad  \quad \quad \quad \quad \quad \quad \quad \quad sin(\pi y)sin(\pi z)  \text{ (for Case 3)} ; \text{   } x=0 \\
   & w=v=u=0 ; \quad \theta=0 ; \quad x=1\\
\end{array}
$$

\subsection{Discretization of Governing Equations}
\begin{figure}
    \centering
    \includegraphics[width=0.8\textwidth]{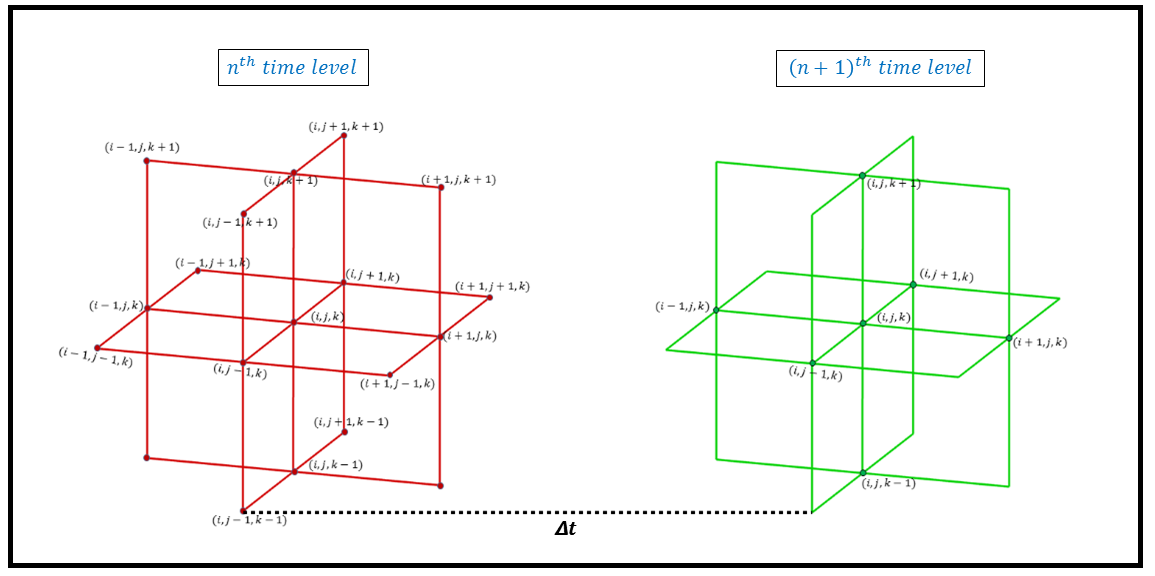}
    \caption{The super-compact unsteady stencil}
    \label{fig:stencil}
\end{figure}

In this section, we discretize the nonlinear coupled transport equations (\ref{Main_governing_eq_1})-(\ref{Main_governing_eq_4}) using newly developed HOSC scheme on a uniform grid. Here we showcase the HOSC discretization procedure on an unsteady three-dimensional (3D) convection-diffusion-reaction process that encompasses a transport variable `$\varphi$' over the continuous domain, which can be directly extended to discretize equations (\ref{Main_governing_eq_1})-(\ref{Main_governing_eq_4}). This phenomenon, governed by variable coefficients, is captured by the following equation:
\begin{equation}
\begin{aligned}
& L\frac{\partial \varphi}{\partial\tau}+M^{*}(z, y, x,\tau) \frac{\partial \varphi}{\partial x}+N^{*}(z, y, x,\tau) \frac{\partial \varphi}{\partial y}+O^{*}(z, y, x,\tau) \frac{\partial \varphi}{\partial z}+K^{*}(z, y, x,\tau) \varphi \\
& \quad=\nabla^2 \varphi+F^{*}(z, y, x,\tau),
\end{aligned}\label{eq1}
\end{equation}
Within this setting, the equation incorporates a constant identified as ``$L$" and various coefficients, including ``$M^{*}$", ``$N^{*}$", and ``$O^{*}$", which signify convection, along with ``$K^{*}$" represents the reaction term, and ``$F^{*}$ indicates the presence of body force. The above equation (\ref{eq1}) describes the convection-diffusion process for a wide range of fluid variables, such as heat, velocity, streamfunction, energy, vorticity, etc., inside the specified continuous domain.
This equation can accurately represent the properties of the momentum and heat equations by making suitable selections for the parameters '$L$', '$M^{*}$', '$N^{*}$', '$O^{*}$', '$K^{*}$', and '$F^{*}$'. 
Hence, this equation provides a complete foundation for incorporating several fluid dynamics processes into a single mathematical structure.
It is noteworthy that in the case of a simple natural convection problem \cite{Punia_2024}, the coefficient '$K^{*}$' is typically zero for all momentum and heat equations. However, the presence of a magnetic field introduces heterogeneity, resulting in non-zero values for '$K^{*}$' in the y-momentum equation, which makes the system more complex. Establishing proper domain boundary conditions is very important to guarantee a well-defined and physically applicable problem formulation. To discretize the cubical problem domain, we use uniform mesh with increments $h$, $k$, and $l$ for the $x$, $y$, and $z$-direction, respectively. In this process, we employ the Forward-Time Centered-Space (FTCS) scheme to discretized the equation (\ref{eq1}). Using the FTCS approximation, we can approximate Eq. (\ref{eq1}) at the general node $(i, j, k)$ as follows:
\begin{equation} \begin{aligned}
\left(L \delta_\tau^{+}+O^{*}\delta_z+N^{*}\delta_y+M^{*} \delta_x-\delta_z^2-\delta_y^2-\delta_x^2 + K^{*}\right) \varphi_{i j k}-\xi_{i j k}=F^{*}_{i j k},
\end{aligned} \label{eq2_} \end{equation} 
In the above equation, $\varphi_{ijk}$ represents the functional value of transport variable $\varphi$ at a three-dimensional grid point $(x_i, y_j, z_k)$. The operator $\delta_x, \delta_x^2, \delta_y, \delta_y^2, \delta_z$ and $\delta_z^2$ are associated with first and second-order central differences along the space variables $x, y, z,$ respectively, and $\delta_\tau^{+}$ is the first order forward difference along the time variable. Truncation error $\xi_{i j k}$ associated with the present numerical method, employing a uniform time step $\Delta\tau$, serves as a measure of the introduced error due to the discretization process, as described by:
\begin{equation}
\begin{aligned}
\xi_{i j k}= & {\left[L \frac{\Delta\tau}{2} \frac{\partial^2 \varphi}{\partial\tau^2}-\frac{k^2}{12}\left(\frac{\partial^4 \varphi}{\partial y^4} - 2 N^{*} \frac{\partial^3 \varphi}{\partial y^3}\right)-\frac{h^2}{12}\left(\frac{\partial^4 \varphi}{\partial x^4}-2 M^{*} \frac{\partial^3 \varphi}{\partial x^3}\right)\right.} \\
& \left.-\frac{l^2}{12}\left(\frac{\partial^4 \varphi}{\partial z^4}-2 O^{*} \frac{\partial^3 \varphi}{\partial z^3}\right)\right]_{i j k}+O\left(\Delta\tau^2, h^4, k^4, l^4\right).
\end{aligned}\label{eq3}
\end{equation}
To achieve higher temporal precision (second-order accuracy) and spatial precision (fourth-order accuracy) for equation (\ref{eq1}), a compact approximation for the derivatives of the leading term in equation (\ref{eq3}) is employed, as discussed in \cite{MacKinnon_1991, Spotz_1995}. This approach results in a formulation characterized by reduced truncation error.
To fulfill this objective, Eq. (\ref{eq1}) is treated as an auxiliary relationship, allowing the derivation of higher-order derivatives, which means higher derivatives are computed from Eq. (\ref{eq1}). For instance, the backward temporal difference method 
is applied to the variables $L, A^{}, N^{*}, O^{*}, M^{*},$ and $K^{*}$ and forward difference method is employed for the transport variable $\varphi$ \cite{Kalita_2014}. 
This enables the representation of derivatives in the initial right-hand sided term of Eq. (\ref{eq3}) in the following manner:
\begin{equation}
\begin{aligned}
\left.L \frac{\partial^2 \varphi}{\partial t^2}\right|_{i j k}= & \left(\delta_z^2+\delta_y^2+\delta_x^2-K^{*}_{i j k}-M^{*}_{i j k} \delta_x-N^{*}_{i j k} \delta_y-O^{*}_{i j k} \delta_z\right) \delta_\tau^{+} \varphi_{i j k} \\
& -\left(\delta_\tau^{-} M^{*}_{i j k} \delta_x+\delta_\tau^{-} N^{*}_{i j k} \delta_y+\delta_\tau^{-} K^{*}_{i j k}+\delta_\tau^{-} O^{*}_{i j k} \delta_z \right)\varphi_{i j k} +\delta_\tau^{-} F^{*}_{i j k}\\
&+O\left(\Delta\tau, h^2, k^2, l^2\right),
\end{aligned} \label{eq4}
\end{equation}
The operator $\delta_\tau^{+}$ and $\delta_\tau^{-}$ signifies a first-order forward and backward difference with respect to the time, respectively. The operators $\delta_x, \delta_x^2, \delta_y, \delta_y^2, \delta_z$ and $\delta_z^2$ are associated with first and second-order central differences along the spatial variables. As a result, by replacing the derivatives in Eq. (\ref{eq3}) with the provided approximations in Eq. (\ref{eq4}) and subsequently substituting $\xi_{i j k}$ in the Eq. (\ref{eq2_}), we derive the following estimation, achieving an order of accuracy $O\left(\Delta \tau^2, h^4, k^4, l^4\right)$ for the primary governing equation (\ref{eq1}).
$$
\begin{aligned}
L[1 & +\left(\frac{k^2}{12}-\frac{\Delta\tau}{2 L}\right)\left(\delta_y^2-N^{*}_{i j k} \delta_y\right) +\left(\frac{h^2}{12}-\frac{\Delta\tau}{2 L}\right)\left(\delta_x^2-M^{*}_{i j k} \delta_x\right) \\
& \left.+\left(\frac{l^2}{12}-\frac{\Delta\tau}{2 L}\right)\left(\delta_z^2-O^{*}_{i j k} \delta_z\right)+\frac{\Delta\tau}{2 L} K^{*}_{i j k}\right] \delta_\tau^{+} \varphi_{i j k} \\
& +\left(-\alpha_{i j k} \delta_x^2-\beta_{i j k} \delta_y^2-\gamma_{i j k} \delta_z^2+M1_{i j k} \delta_x+M2_{i j k} \delta_y+M3_{i j k} \delta_z+M4_{i j k}\right) \varphi_{i j k} \\
& -\frac{h^2+k^2}{12}\left(\delta_x^2 \delta_y^2-M^{*}_{i j k} \delta_x \delta_y^2-N^{*}_{i j k} \delta_x^2 \delta_y-p1_{i j k} \delta_x \delta_y\right) \varphi_{i j k}
\end{aligned}
$$
\begin{equation}
\begin{aligned}
& -\frac{k^2+l^2}{12}\left(\delta_y^2 \delta_z^2-N^{*}_{i j k} \delta_y \delta_z^2-O^{*}_{i j k} \delta_y^2 \delta_z-q1_{i j k} \delta_y \delta_z\right) \varphi_{i j k} \\
& -\frac{l^2+h^2}{12}\left(\delta_z^2 \delta_x^2-O^{*}_{i j k} \delta_z \delta_x^2-M^{*}_{i j k} \delta_z^2 \delta_x-r1_{i j k} \delta_z \delta_x\right) \varphi_{i j k} = R_{i j k}\\
\end{aligned}\label{eq5}
\end{equation}
The coefficients $\alpha_{i j k}, \beta_{i j k}, \gamma_{i j k}, M1_{i j k}, M2_{i j k}, M3_{i j k}, M4_{i j k}, R_{i j k}, p1_{i j k}, q1_{i j k}$ and $r1_{i j k}$ are as follows:
$$
\begin{aligned}
& \alpha_{i j k}=\frac{h^2}{12}\left({M^{*}}_{i j k}^2-{K^{*}}_{i j k}-2 \delta_x {M^{*}}_{i j k}\right)+1 \text {, } \\
& \beta_{i j k}=\frac{k^2}{12}\left({N^{*}}_{i j k}^2-{K^{*}}_{i j k}-2 \delta_y {N^{*}}_{i j k}\right)+1 \text {, } \\
& \gamma_{i j k}=\frac{l^2}{12}\left({O^{*}}_{i j k}^2-{K^{*}}_{i j k}-2 \delta_z {O^{*}}_{i j k}\right)+1\text {, } \\
& M1_{i j k}=\left[\frac{h^2}{12}\left(\delta_x^2-{M^{*}}_{i j k} \delta_x\right)+\frac{k^2}{12}\left(\delta_y^2-{N^{*}}_{i j k} \delta_y\right)+\frac{l^2}{12}\left(\delta_z^2-{O^{*}}_{i j k} \delta_z\right)+\frac{\Delta\tau}{2} \delta_\tau^{-}+1\right] {M^{*}}_{i j k} \\
& -\frac{h^2}{12}\left({M^{*}}_{i j k}-2 \delta_x\right) {K^{*}}_{i j k}, \\
& M2_{i j k}=\left[\frac{h^2}{12}\left(\delta_x^2-{M^{*}}_{i j k} \delta_x\right)+\frac{k^2}{12}\left(\delta_y^2-{N^{*}}_{i j k} \delta_y\right)+\frac{l^2}{12}\left(\delta_z^2-{O^{*}}_{i j k} \delta_z\right)+\frac{\Delta\tau}{2} \delta_\tau^{-}+1\right] {N^{*}}_{i j k} \\
& -\frac{k^2}{12}\left({N^{*}}_{i j k}-2 \delta_y\right) {K^{*}}_{i j k} \text {, } \\
& M3_{i j k}=\left[\frac{h^2}{12}\left(\delta_x^2-{M^{*}}_{i j k} \delta_x\right)+\frac{k^2}{12}\left(\delta_y^2-{N^{*}}_{i j k} \delta_y\right)+\frac{l^2}{12}\left(\delta_z^2-{O^{*}}_{i j k} \delta_z\right)+\frac{\Delta\tau}{2} \delta_\tau^{-}+1\right] {O^{*}}_{i j k} \\
& -\frac{l^2}{12}\left({O^{*}}_{i j k}-2 \delta_z\right) {K^{*}}_{i j k} \text {, } \\
& M4_{i j k}=\left[\frac{h^2}{12}\left(\delta_x^2-{M^{*}}_{i j k} \delta_x\right)+\frac{k^2}{12}\left(\delta_y^2-{N^{*}}_{i j k} \delta_y\right)+\frac{l^2}{12}\left(\delta_z^2-{O^{*}}_{i j k} \delta_z\right)+\frac{\Delta\tau}{2} \delta_\tau^{-}+1\right] {K^{*}}_{i j k}, \\
& R_{i j k}=\left[\frac{h^2}{12}\left(\delta_x^2-{M^{*}}_{i j k} \delta_x\right)+\frac{k^2}{12}\left(\delta_y^2-{N^{*}}_{i j k} \delta_y\right)+\frac{l^2}{12}\left(\delta_z^2-{O^{*}}_{i j k} \delta_z\right)+\frac{\Delta\tau}{2} \delta_\tau^{-}+1\right] {F^{*}}_{i j k}, \\
& p1_{i j k}=-{M^{*}}_{i j k} {N^{*}}_{i j k}+\frac{2}{h^2+k^2}\left(k^2 \delta_y {M^{*}}_{i j k}+h^2 \delta_x {N^{*}}_{i j k}\right) \text {, } \\
& q1_{i j k}=-{N^{*}}_{i j k} {O^{*}}_{i j k}+\frac{2}{k^2+l^2}\left(l^2 \delta_z {N^{*}}_{i j k}+k^2 \delta_y {O^{*}}_{i j k}\right) \text {, } \\
& r1_{i j k}=-{O^{*}}_{i j k} {M^{*}}_{i j k}+\frac{2}{l^2+h^2}\left(h^2 \delta_x {O^{*}}_{i j k}+l^2 \delta_z {M^{*}}_{i j k}\right) . \\
&
\end{aligned}
$$
By employing Eq. (\ref{eq5}), an implicit finite difference method is developed, demonstrating second-order temporal accuracy \(O(\tau^2)\) and fourth-order accuracy \(O(h^4, k^4, l^4)\) in spatial grid spacing. This is achieved through the utilization of a $(19,7)$ stencil as shown in Figure \ref{fig:Sche_diag}(d). This strategy yields a compact seven-point stencil at the $(n + 1)^{th}$ time step, leading to a notable reduction in computational complexity. It is worth mentioning that numerous high-order compact methods designed even for two-dimensional convection-diffusion equations, as seen in \cite{Gupta_1991, Spotz_1995, Kalita_2002, Ray_2017}, typically require a nine-point stencil at the $(n + 1)^{th}$ time level. However, in the current approach, even for the three-dimensional case, the requirement is reduced to only a seven-point stencil at the $(n + 1)^{th}$ level of time. The HOSC scheme provides dual benefits. Firstly, it facilitates an easier seven-point stencil, necessitating only the point $(i, j, k)$ $^{th}$ and its six adjacent points (as depicted in Figure \ref{fig:stencil}) at the $(n + 1)^{th}$ time level. Secondly, it effectively removes the need for extensive corner points and considerably decreases the quantity of points necessary for the approximation, thereby improving the computational efficiency.
When the HOSC scheme is applied to Equation (\ref{eq5}), at all grid points, it produces a system of algebraic equations, where the coefficient matrix is an asymmetric sparse matrix and lacks diagonal dominance. Consequently, conventional iterative methods like, Gauss-Seidel and SOR are ineffective in this context. Therefore, a Hybrid Biconjugate Gradient Stabilized method is employed for solving the convection-diffusion equation, without the use of any preconditioning \cite{Kelley_1995, Spotz_1995, Saad_2003, Kalita_2014}. When this HOSC approach is applied to the momentum and energy equations (Eqs. (\ref{Main_governing_eq_1}) - (\ref{Main_governing_eq_4})), it also produces a system of algebraic equations that must be solved by the Hybrid Biconjugate Stabilized method for the required solutions. \\
Table \ref{Coefficient_Table} presents the actual coefficients need to be discretized for the momentum and energy equations (Eqs. (\ref{Main_governing_eq_1}) - (\ref{Main_governing_eq_4})). Moreover, It's crucial to acknowledge that due to the lack of known analytical expressions for pressure, we resort to numerical approximations for pressure gradients. To achieve this, central difference approximations are used at the inner grid points of the domain, while standard one-sided approximations are used at the boundary nodes. After solving the equations ((\ref{Main_governing_eq_1}) - (\ref{Main_governing_eq_3})) with constant initial pressure, the next step involves determining the pressure $pr$ for the subsequent time step. We chose to employ the modified compressibility technique proposed by Cortes and Miller \cite{Cortes_1994}. This approach is selected for its efficiency, simplicity, and straightforward implementation.
In the context of this approach, the modified continuity equation is expressed as:
$$
\lambda \nabla \cdot \mathbf{v}+pr=0 .
$$
At every time step, once pressure gradients are computed and momentum equations are solved, we calculate the dilation parameter $D = v_y + u_x + w_z$. If the greatest absolute value of $D$, indicated as $|D|_{\max}$, comes under a specified tolerance level, it is inferred that the pressure value has attained the required level of accuracy.
The whole computation subsequently advances to solving the energy equation (Eq. (\ref{Main_governing_eq_4})) and eventually to the next time level. This process will continue until the steady state is reached. However, if the greatest value of $|D|$ surpasses the designated tolerance threshold, we initiate a pressure correction step to enhance the accuracy of the pressure value:
$$
pr^{\text {new }}=pr^{\text {old }}-\lambda \nabla \cdot \mathbf{v} .
$$
Here, $pr^{\text{new}}$ denotes the updated pressure, $p_r^{\text{old}}$ represents the obtained pressure value in the preceding iteration, and $\lambda$ signifies a relaxation parameter. This iterative procedure continues until the greatest absolute value of $|D|_{\max}$, satisfies the specified tolerance limit.\\

{\small\begin{table}[htbp]
\caption{\small Coefficient for different equations}\label{Coefficient_Table}
\centering
 \begin{tabular}{cccccccc}  \hline \hline
Equation & $\varphi$ & $L$   &    $M^{*}$   &  $N^{*}$  &  $O^{*}$ & $K^{*}$ & $F^{*}$   \\
$x$-momentum &$u$ & ($ \frac{1}{Pr}$)   &    $(\frac{u}{Pr})$   &  $(\frac{v}{Pr})$  &  $(\frac{w}{Pr})$ & 0 & $ \frac{-1}{Pr}\cdot \frac{\partial pr}{\partial x}$ \\

$y$-momentum  &$v$& ($ \frac{1}{Pr}$)   &    $(\frac{u}{Pr})$   &  $(\frac{v}{Pr})$  &  $(\frac{w}{Pr})$ & $Ha^2$ & $\frac{-1}{Pr} \cdot \frac{\partial pr}{\partial y} + Ra \cdot \theta $ \\

$z$-momentum  &$w$& ($ \frac{1}{Pr}$)   &    $(\frac{u}{Pr})$   &  $(\frac{v}{Pr})$  &  $(\frac{w}{Pr})$ & 0 & $ \frac{-1}{Pr} \cdot \frac{\partial pr}{\partial z}$ \\

Energy & $\theta$ & 1   &    $u$   &  $v$  &  $w$ & 0 & $0$ \\

\hline \\
\hline
 \end{tabular}
\end{table}
}
\section{Sensitivity Test and Scheme Validation}
\label{sec:SENSITIVITY TESTS AND SCHEME VALIDATION}
\subsection{Grid Independence Test}
To ensure the grid independence of our results and optimize computational efficiency, we conducted a grid-independence test in which we varied the mesh size but allowed all other parameters constant. We examined three distinct grid resolutions:  $91\times91\times91$, $51\times51\times51$, and $11\times11\times11$ with constant values for $Pr$, $Ha$, $\Delta \tau$ and $Ra$ set as $ 0.065, 50, 0.02$ and $10^5$, respectively. Table \ref{grid_independent_test} provides the $u, v,$ and $w$ values at two different specified monitoring locations ($(0.65, 0.65, 0.65)$ and $(0.4, 0.4, 0.4)$) towards the center of the cavity for each grid size. Our analysis indicates that the maximum relative velocity error between the $91\times91\times91$ and $51\times51\times51$ grids is merely $1.2\%$. This result indicates that enlarging the grid size does not have a substantial affect on the solution. Therefore, in our quest to balance computational efficiency and solution accuracy, we have opted to utilize the $51\times51\times51$ grid size for our study. The grid arrangement at a resolution of $51\times51\times51$ is visualized in Figure \ref{fig:Sche_diag}(d). 
{\small\begin{table}[htbp]
\caption{\small $u, v$ and $w$ values at two distinct location ($(0.65, 0.65, 0.65)$ and $(0.4, 0.4, 0.4)$) near the center area of the cavity with fixed time $=50$, $Ra = 10^5$, $Ha = 50$ and $\Delta \tau$ = 0.02 at three different grid sizes}\label{grid_independent_test}
\centering
 \begin{tabular}{cccccc}  \hline \hline
Monitoring point & M$\times$N$\times$O    &    $u$   &  $v$  &  $w$ & Max. relative error ($\%$)    \\ \hline
 (0.65, 0.65, 0.65) &(11 $\times$ 11 $\times$ 11)           &  -2.122 &   2.710    &   -0.179  & --- \\
&(51 $\times$ 51 $\times$ 51)         &  -2.613  &  3.121  &   -0.205 & 18.7\\
&(91 $\times$ 91 $\times$ 91)              &  -2.617 &  3.119  &   -0.203  & 0.9 
  \\ \hline
 (0.40, 0.40, 0.40) &(11 $\times$ 11 $\times$ 11)           &  -4.825 &   4.137    &   0.079  & --- \\
&(51 $\times$ 51 $\times$ 51)         &  -5.534  &  4.744  &   0.084 & 14.6\\
&(91 $\times$ 91 $\times$ 91)              &  -5.498 &  4.801  &   0.085  & 1.2  \\
\hline
 \end{tabular}
\end{table}
}
\subsection{Validation of the HOSC finite difference Scheme}
{\small\begin{table}[htbp]
\caption{\small Comparison of the highest $Nu_{\text {A}}$ value at the uniformly heated wall obtained using a grid resolution of $51\times51\times51$ with existing findings \cite{Fusegi_1991, Wang_2017, Benhamou_2022}}\label{Average_Nusselt_Number_Comparison}
\centering
 \begin{tabular}{cccccccccc}  \hline \hline
& &   $Ra$=$10^3$  & $\delta_e$(\%)  &  $Ra$=$10^4$  & $\delta_e$(\%)  &  $Ra$=$10^5$  & $\delta_e$(\%)   \\ \hline 
& Present             &  1.098 & -- &  2.256 & -- &  4.638 & -- &   \\
& Benhamou et al. \cite{Benhamou_2022}     &  1.101 & 0.27 & 2.281  & 1.10 &    4.612 & 0.56\\
& Wang et al. \cite{Wang_2017}          &  1.088 & 0.91 &    2.247  & 0.39 &    4.599 & 0.84 \\
& Fusegi et al. \cite{Fusegi_1991}       &  1.105 & 0.63  & 2.320  & 2.83 &    4.646 & 0.17\\

\hline\hline
 \end{tabular}
\end{table}
}

\begin{figure}
    \centering
    \includegraphics[width=\textwidth]{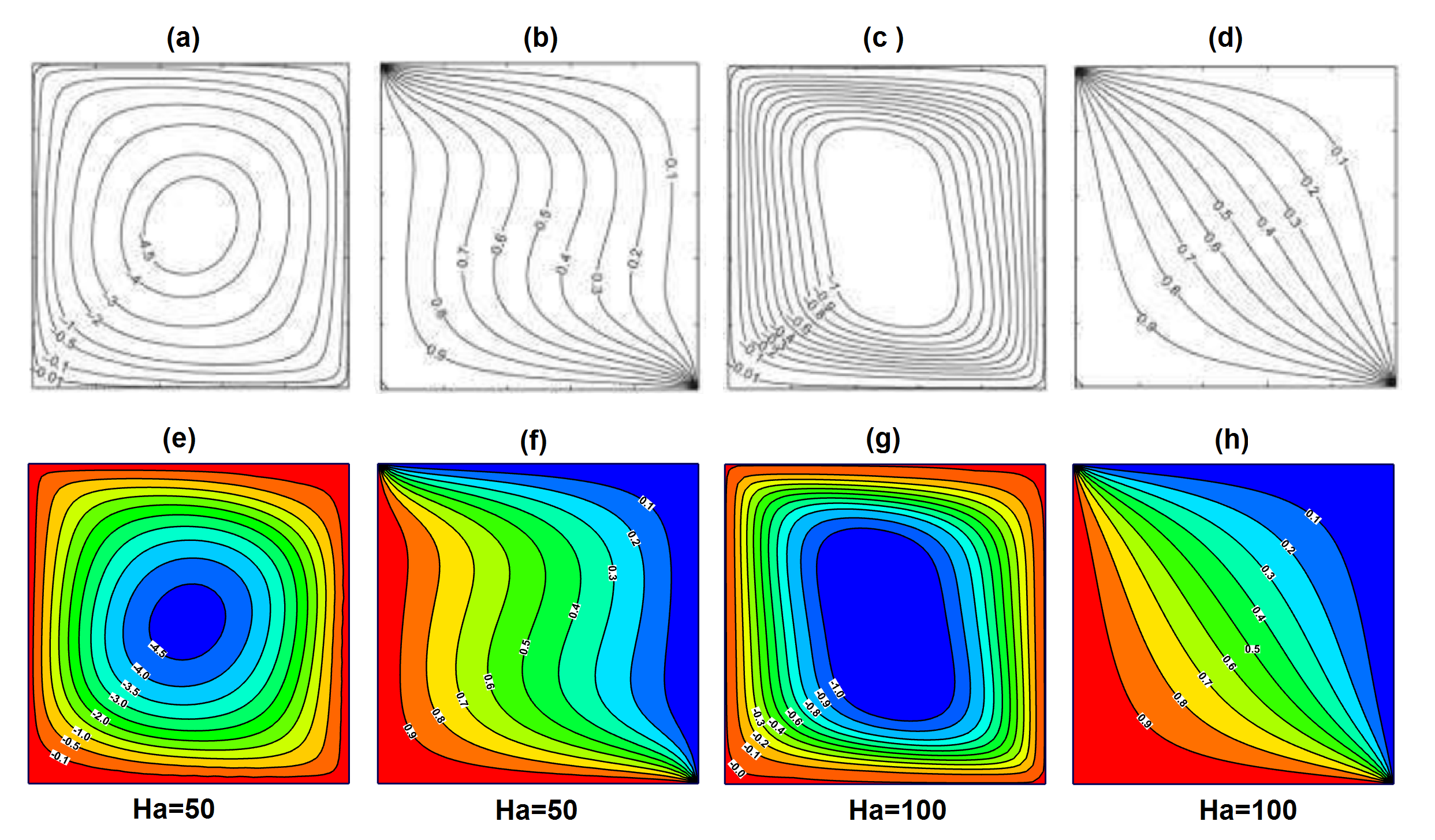}
    \caption{Comparison of stream‐function and isotherm contours (at $Ra = 10^5$) between (a-d) existing \cite{Chamkha_2012} and (e–h) present computed results}
    \label{fig:3d_isotherm_streamlines_comparison}
\end{figure}

\begin{figure}
    \centering
    \includegraphics[width=\textwidth]{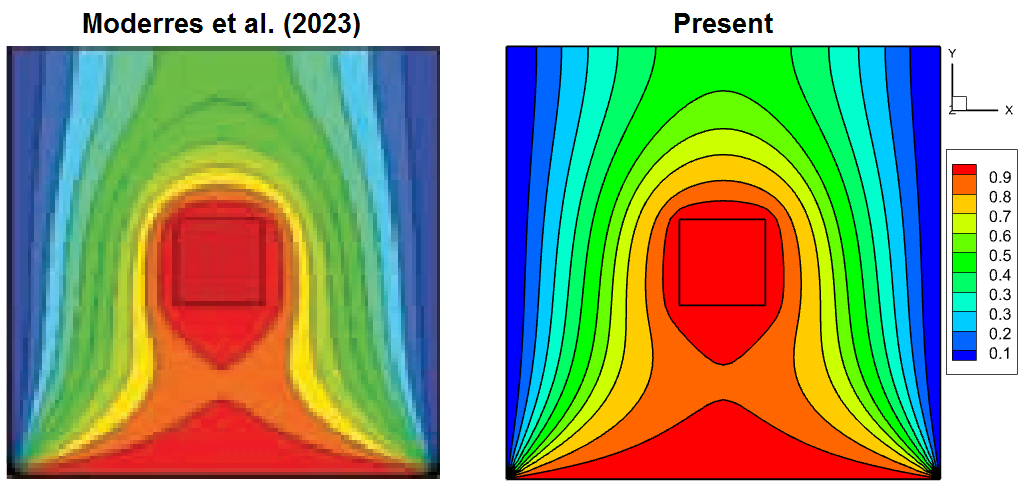}
    \caption{Comparison of Isotherms (at $Ra=10^3$, $Ha=90$) for natural convection on symmetry plane ($z=0.5$): Moderres et al. (2023) \cite{Moderres_2023} vs. present result}
    \label{fig:3d_isotherm_comparison_1}
\end{figure}

We performed a thorough validation to confirm the accuracy of our in-house code, as well as the validity and robustness of the HOSC scheme. This validation encompassed both quantitative and qualitative comparisons with the existing results for MHD natural convective flows inside a cavity, spanning a variety of $Ha$ and $Ra$. The problems identified in references \cite{Fusegi_1991, Chamkha_2012, Wang_2017} have been addressed and compared by using the same code used in this study. In Table \ref{Average_Nusselt_Number_Comparison}, the highest $Nu_{\text {A}}$ value at the uniformly heated wall obtained with a grid resolution of $51\times51\times51$ for various 
$Ra$ and fixed $Ha=0.0$ are compared with the results from already published work.
Furthermore, we analyze the greatest relative error between our computed results and the existing results. When compared to Fusegi et al.'s (1991) results \cite{Fusegi_1991}, the maximum relative error is $2.83\%$. However, this error decreases to just $1.10\%$ when compared to the latest findings by Wang et al. (2017) \cite{Wang_2017} and Benhamou et al. (2022) \cite{Benhamou_2022}. This indicates the great effectiveness of the current numerical technique in accurately replicating heat transport phenomena. Figure \ref{fig:3d_isotherm_streamlines_comparison} shows a comparative analysis between our computed results and those published by Chamkha et al. \cite{Chamkha_2012}, using a two-dimensional model, considering various magnetic field strengths ($Ha$). Figure \ref{fig:3d_isotherm_comparison_1} depicts the comparison of isotherms at the $z=0.5$ plane for $Ra=10^3$ and $Ha=90$ with the result of Moderres et al. \cite{Moderres_2023}. It is worth mentioning that our computed results are in great agreement with the available standard results, underscoring the high level of accuracy and reliability inherent in our method for accurately capturing the complicated 3D heat transfer phenomena.
\section{Results and Discussion}
\label{sec:Results and Discussion}
\subsection{Study of Streamlines and Isotherm Patterns}
In this section, we discuss the distribution of streamlines and isotherms patterns, systematically exploring their variations across different Hartmann numbers ($Ha$) and Rayleigh numbers ($Ra$) for all three cases. The presentation of this study follows a methodical structure, where we first discuss the impact of $Ha$ and $Ra$ on streamlines and isotherms for Case 1. Subsequently, we extend our analysis for Case 2, followed by Case 3. Furthermore, we undertake a comparative assessment of the results from all three cases, shedding light on the distinctive effects introduced by different heated wall configurations.
\subsubsection{Case 1}
\begin{figure}[htbp]
 \centering
 \vspace*{0pt}%
 \hspace*{\fill}%
\begin{subfigure}{0.33\textwidth}     
    \centering
    \includegraphics[width=\textwidth]{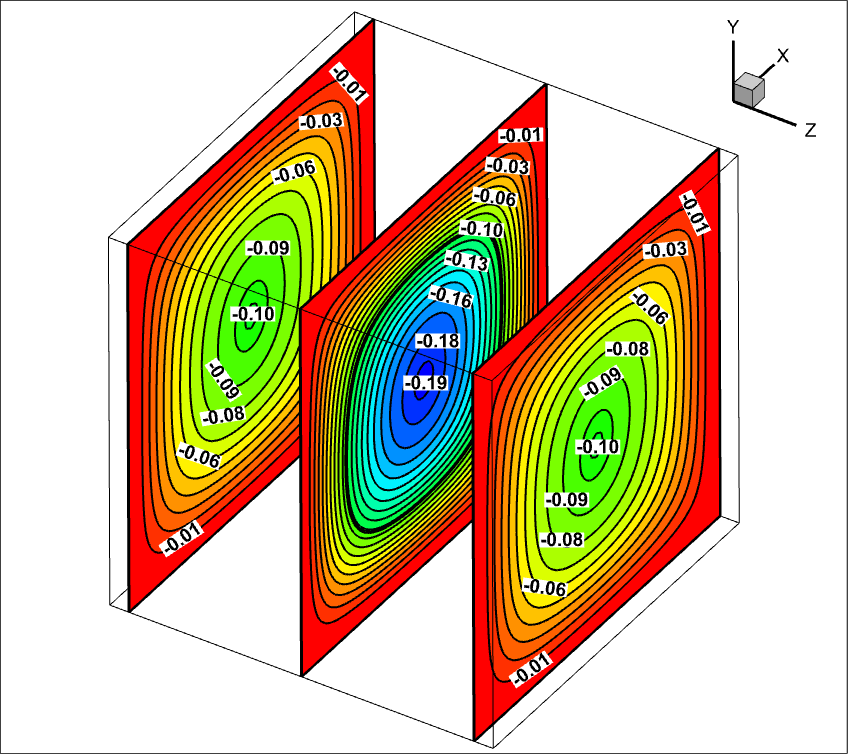}%
    \captionsetup{skip=2pt}%
    \caption{(a) $Ha=25, Ra = 10^3$ }
    \label{fig:Ra_10^3_Ha_25_P1_Streamlines}
  \end{subfigure}%
 \begin{subfigure}{0.33\textwidth}        
   \centering
    \includegraphics[width=\textwidth]{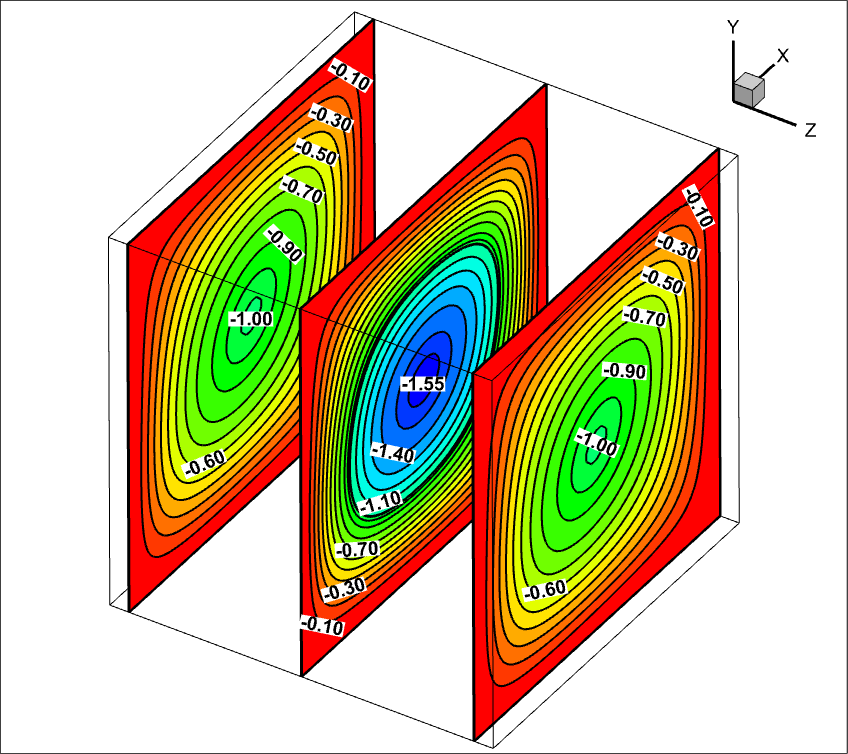}%
    \captionsetup{skip=2pt}%
    \caption{(b) $Ha=25, Ra = 10^4$}
    \label{fig:Ra_10^4_Ha_25_P1_Streamlines}
  \end{subfigure}
   \begin{subfigure}{0.33\textwidth}        
   \centering
    \includegraphics[width=\textwidth]{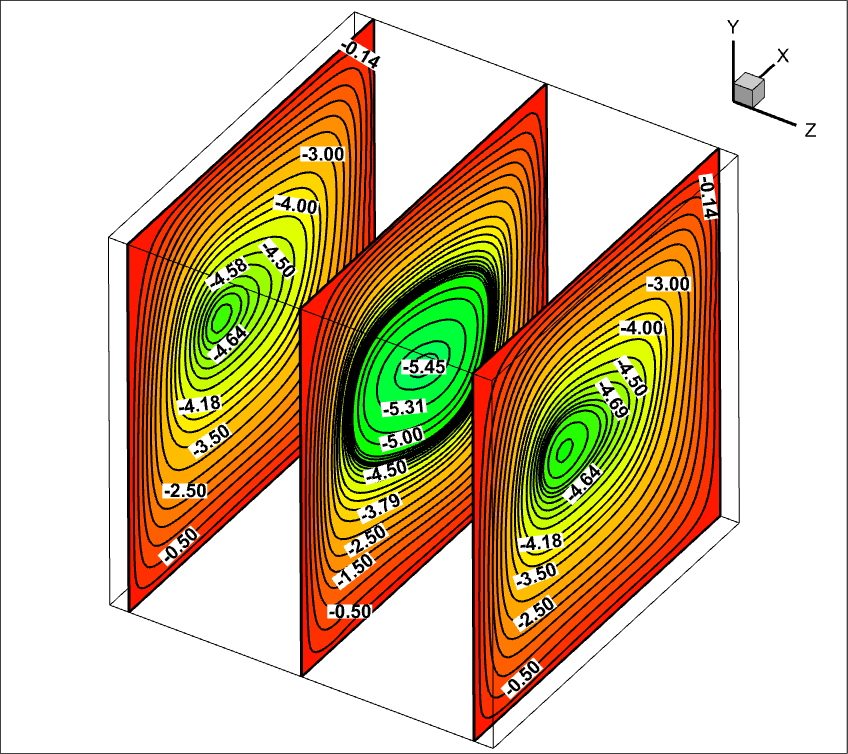}%
    \captionsetup{skip=2pt}%
    \caption{(c) $Ha=25, Ra = 10^5$}
    \label{fig:Ra_10^5_Ha_25_P1_Streamlines.png}
  \end{subfigure}%
  \hspace*{\fill}

  \vspace*{8pt}%
  \hspace*{\fill}%
  \begin{subfigure}{0.33\textwidth}     
    \centering
    \includegraphics[width=\textwidth]{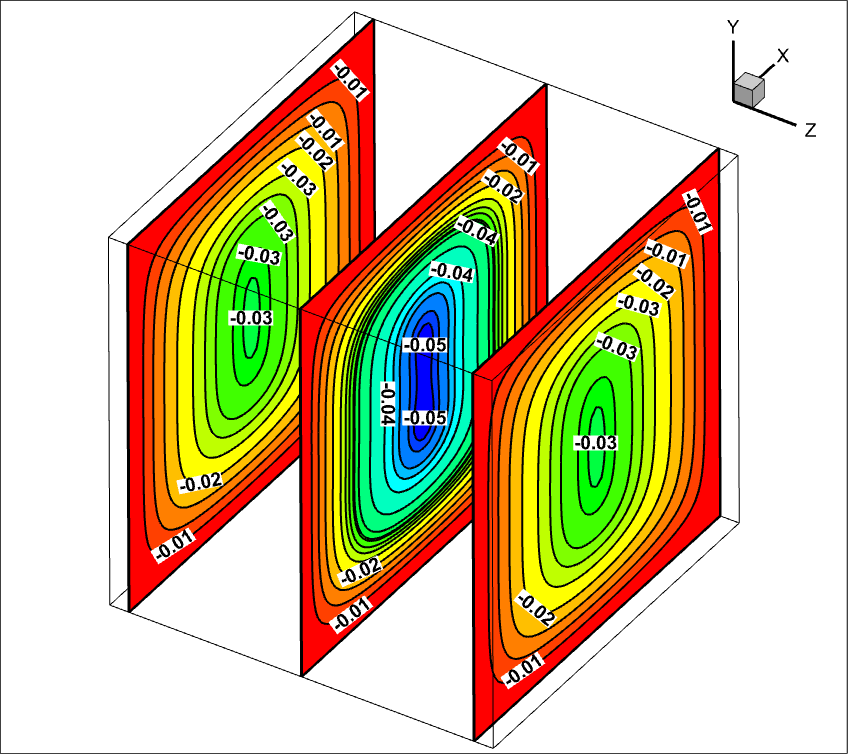}%
    \captionsetup{skip=2pt}%
    \caption{(d) $Ha=50, Ra = 10^3$}
    \label{fig:Ra_10^3_Ha_50_P1_Streamlines}
  \end{subfigure}%
 \begin{subfigure}{0.33\textwidth}        
   \centering
    \includegraphics[width=\textwidth]{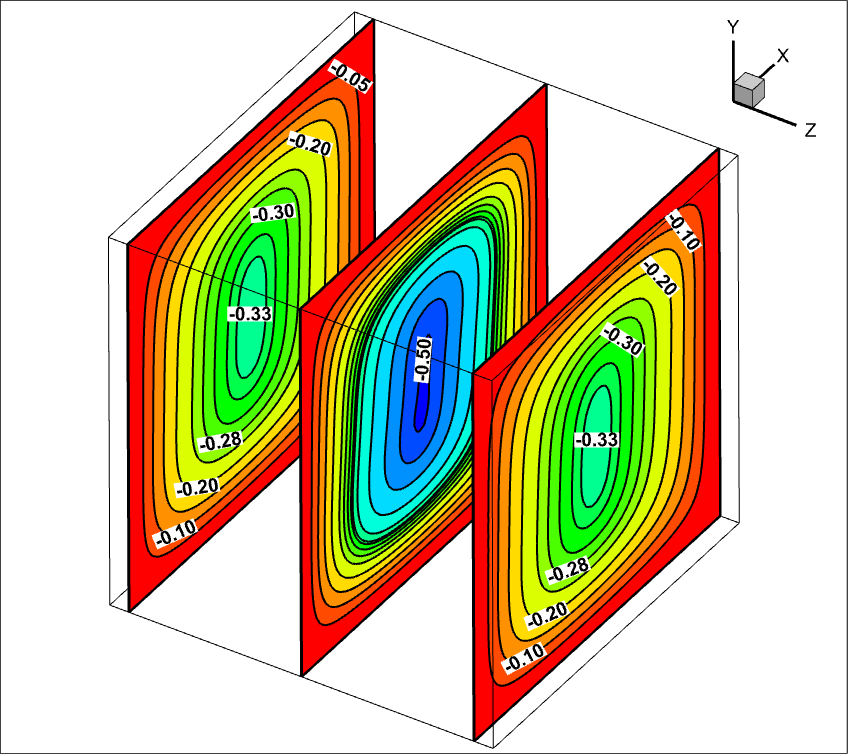}%
    \captionsetup{skip=2pt}%
    \caption{(e) $Ha=50, Ra = 10^4$}
    \label{fig:Ra_10^4_Ha_50_P1_Streamlines}
  \end{subfigure}
   \begin{subfigure}{0.33\textwidth}        
   \centering
    \includegraphics[width=\textwidth]{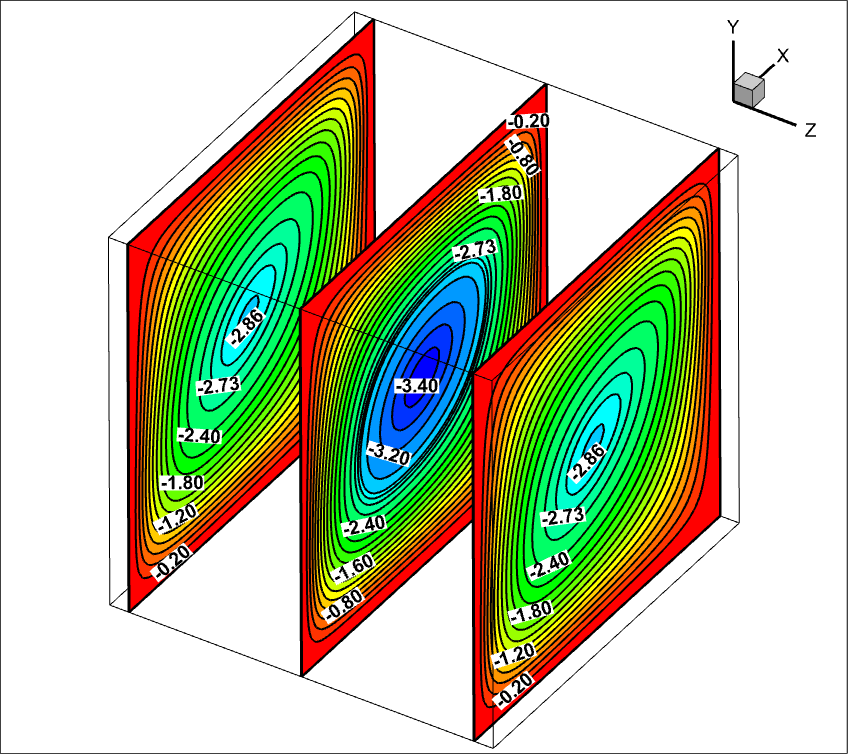}%
    \captionsetup{skip=2pt}%
    \caption{(f) $Ha=50, Ra = 10^5$}
    \label{fig:Ra_10^5_Ha_50_P1_Streamlines3}
  \end{subfigure}%
  \hspace*{\fill}

  \vspace*{8pt}%
  \hspace*{\fill}%
  \begin{subfigure}{0.33\textwidth}     
    \centering
    \includegraphics[width=\textwidth]{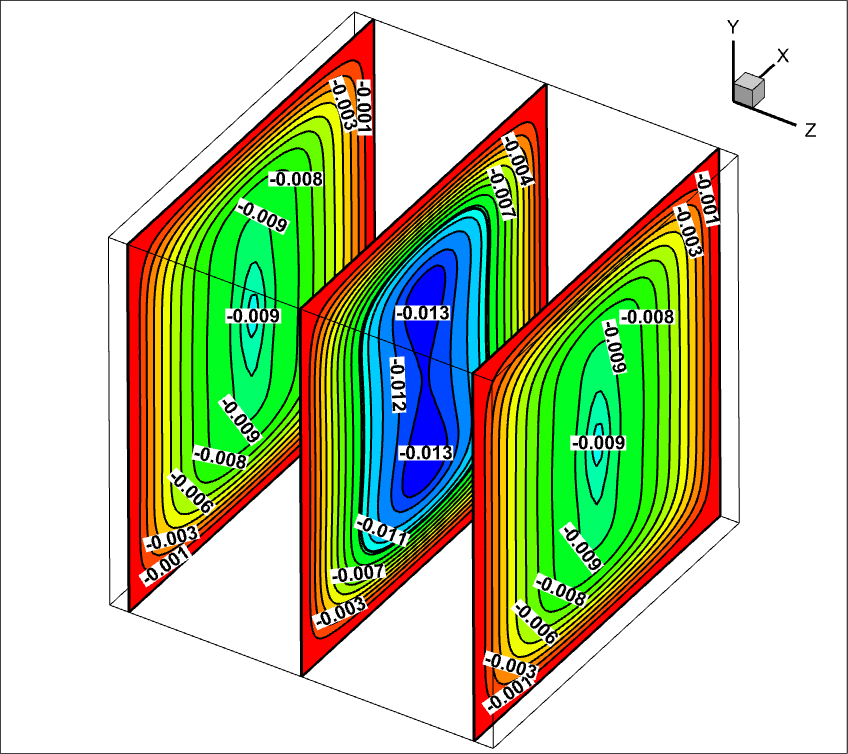}%
    \captionsetup{skip=2pt}%
    \caption{(g) $Ha=100, Ra = 10^3$}
    \label{fig:Ra_10^3_Ha_100_P1_Streamlines}
  \end{subfigure}%
 \begin{subfigure}{0.33\textwidth}        
   \centering
    \includegraphics[width=\textwidth]{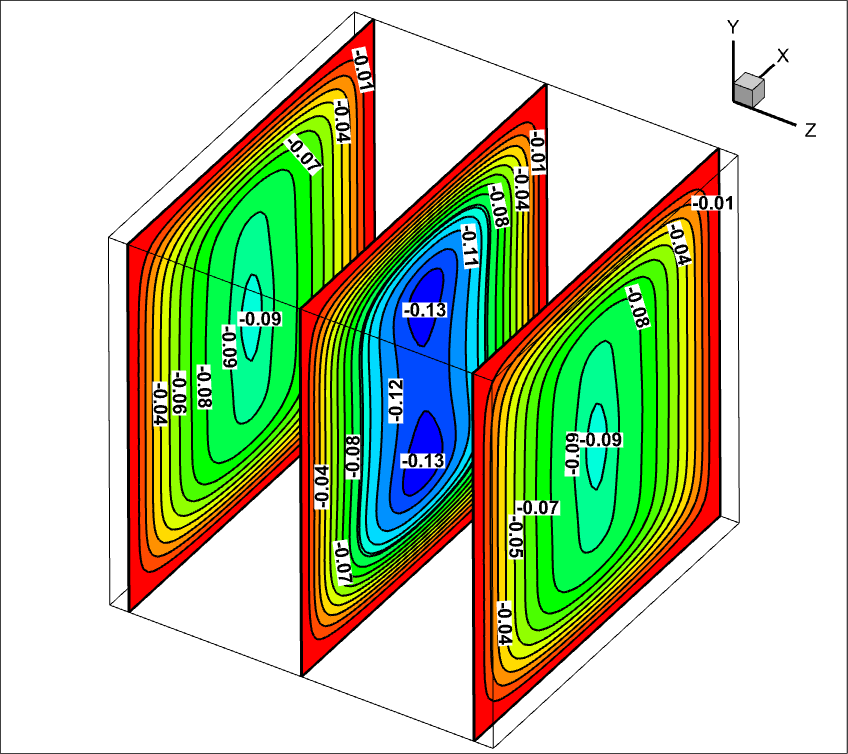}%
    \captionsetup{skip=2pt}%
    \caption{(h) $Ha=100, Ra = 10^4$}
    \label{fig:Ra_10^4_Ha_100_P1_Streamlines}
  \end{subfigure}
   \begin{subfigure}{0.33\textwidth}        
   \centering
    \includegraphics[width=\textwidth]{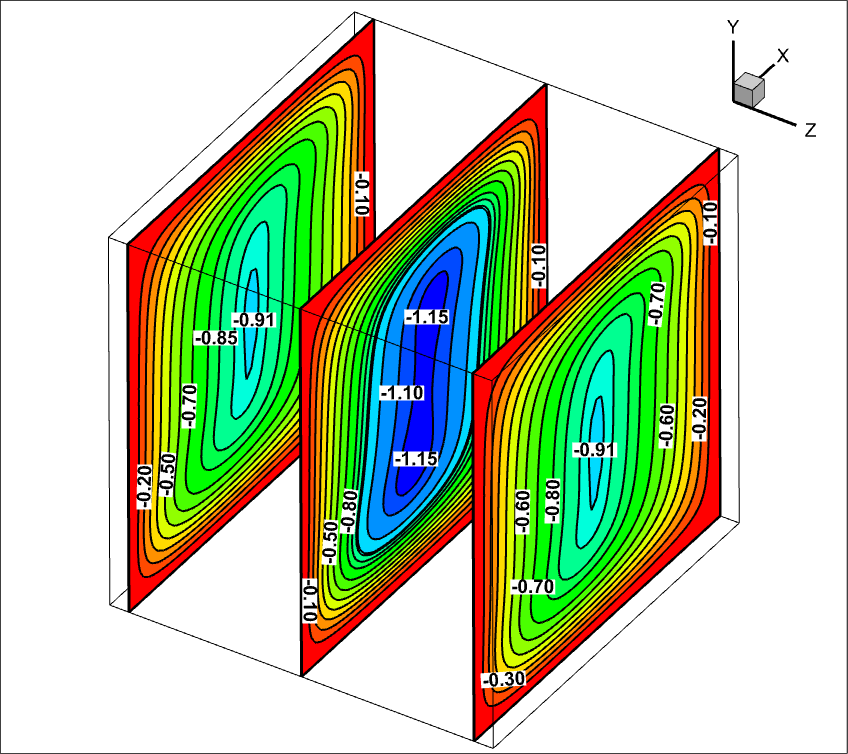}%
    \captionsetup{skip=2pt}%
    \caption{(i) $Ha=100, Ra = 10^5,$}
    \label{fig:Ra_10^5_Ha_100_P1_Streamlines}
  \end{subfigure}%
  \hspace*{\fill}

  \vspace*{8pt}%
  \hspace*{\fill}%
  \begin{subfigure}{0.33\textwidth}     
    \centering
    \includegraphics[width=\textwidth]{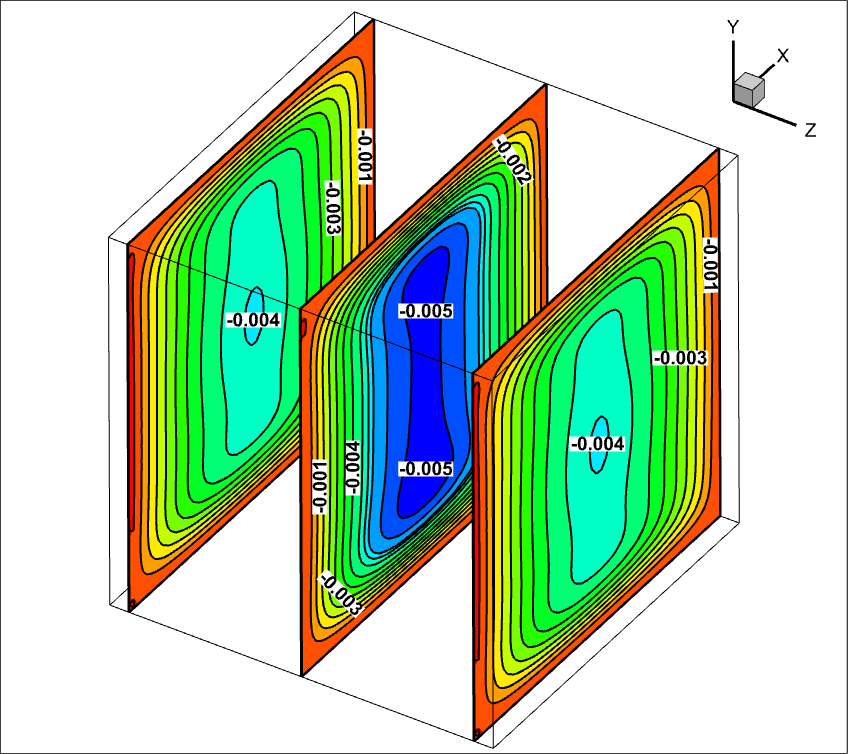}%
    \captionsetup{skip=2pt}%
    \caption{(j) $Ha=150, Ra = 10^3$}
    \label{fig:Ra_10^3_Ha_150_P1_Streamlines}
  \end{subfigure}%
 \begin{subfigure}{0.33\textwidth}        
   \centering
    \includegraphics[width=\textwidth]{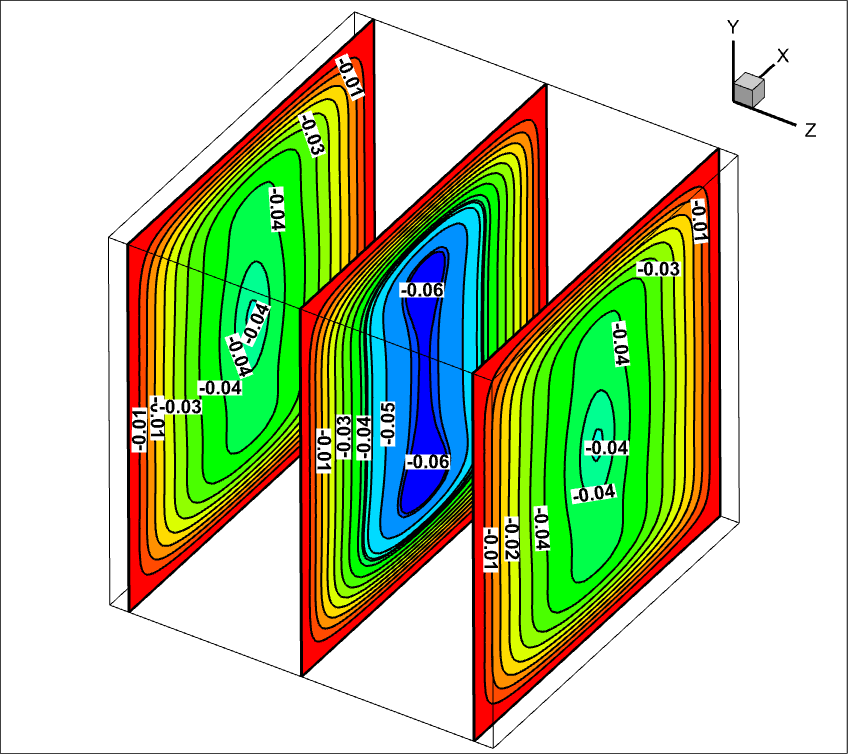}%
    \captionsetup{skip=2pt}%
    \caption{(k) $Ha=150, Ra = 10^4$}
    \label{fig:Ra_10^4_Ha_150_P1_Streamlines}
  \end{subfigure}
   \begin{subfigure}{0.33\textwidth}        
   \centering
    \includegraphics[width=\textwidth]{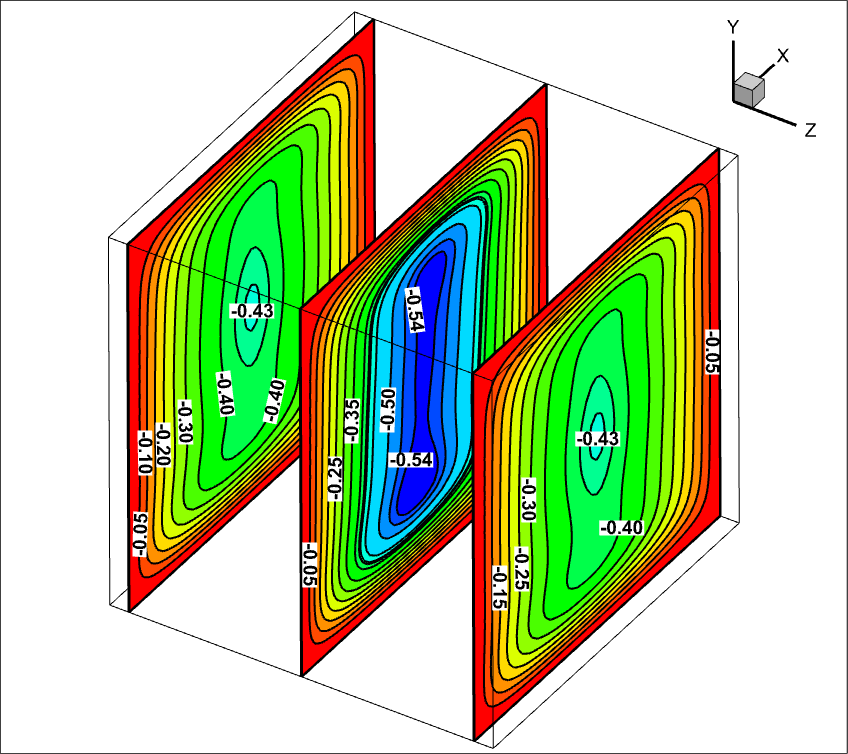}%
    \captionsetup{skip=2pt}%
    \caption{(l) $Ha=150, Ra = 10^5$}
    \label{fig:Ra_10^5_Ha_150_P1_Streamlines}
  \end{subfigure}%
  \vspace*{1pt}%
  \hspace*{\fill}%
  \caption{Case 1. Effect of different $Ha$ (Column-wise) and $Ra$ (Row-wise) on streamlines pattern at different $xy$ planes $(z=0.05, z=0.5, z=0.95)$ with fixed $Pr = 0.065$ }
  \label{fig:case-1_Streamlines_Contours}
\end{figure}

\begin{figure}[htbp]
 \centering
 \vspace*{0pt}%
 \hspace*{\fill}%
\begin{subfigure}{0.33\textwidth}     
    \centering
    \includegraphics[width=\textwidth]{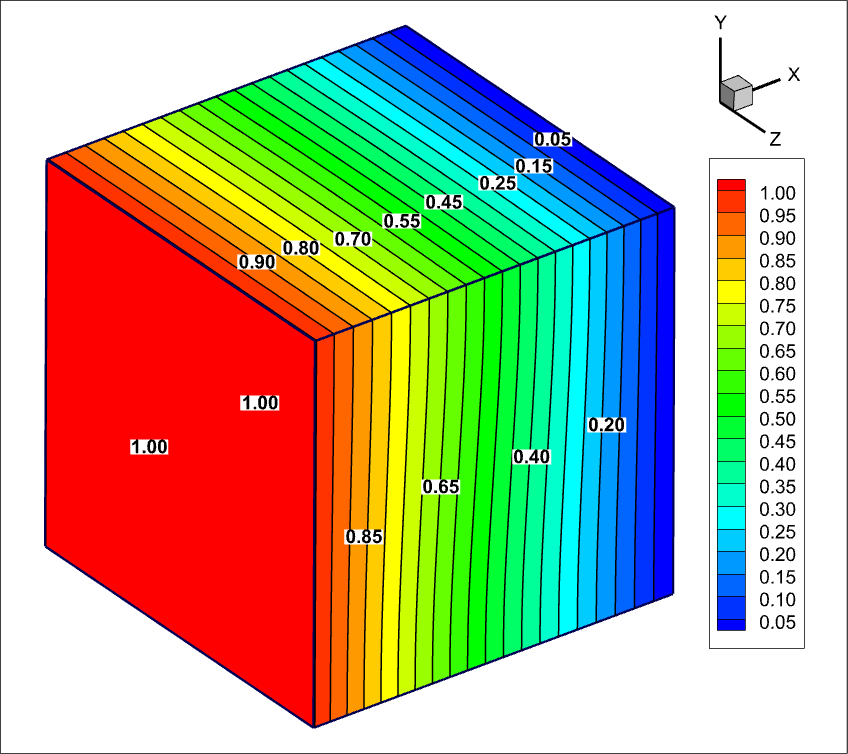}%
    \captionsetup{skip=2pt}%
    \caption{(a) $Ha=25, Ra=10^3$}
    \label{fig:Ra_10^3_Ha_25_P1_3D}
  \end{subfigure}%
 \begin{subfigure}{0.33\textwidth}        
   \centering
    \includegraphics[width=\textwidth]{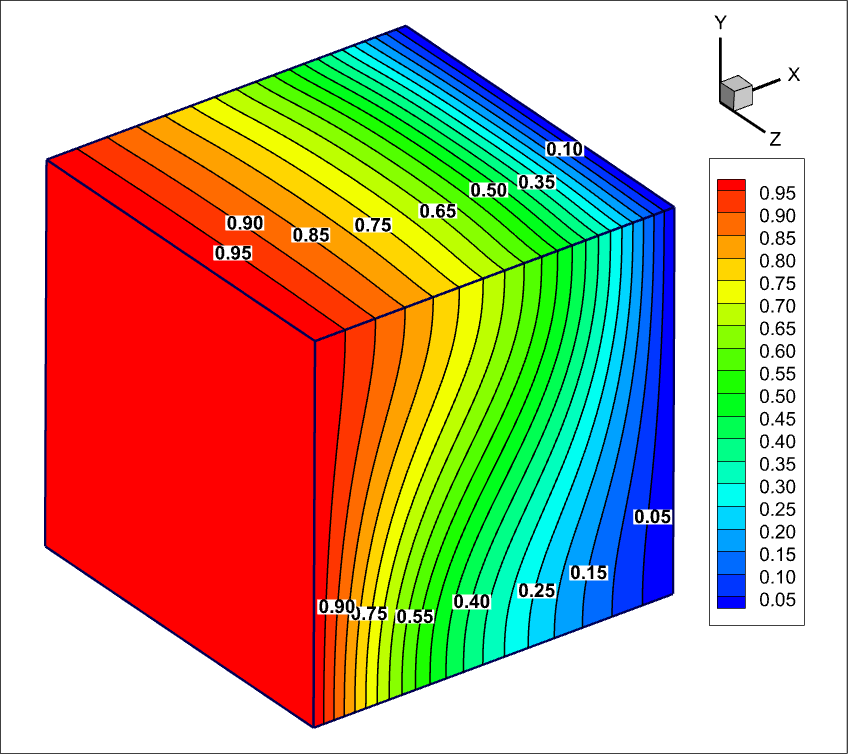}%
    \captionsetup{skip=2pt}%
    \caption{(b) $ Ha=25, Ra=10^4$}
    \label{fig:Ra_10^4_Ha_25_P1_3D}
  \end{subfigure}
   \begin{subfigure}{0.33\textwidth}        
   \centering
    \includegraphics[width=\textwidth]{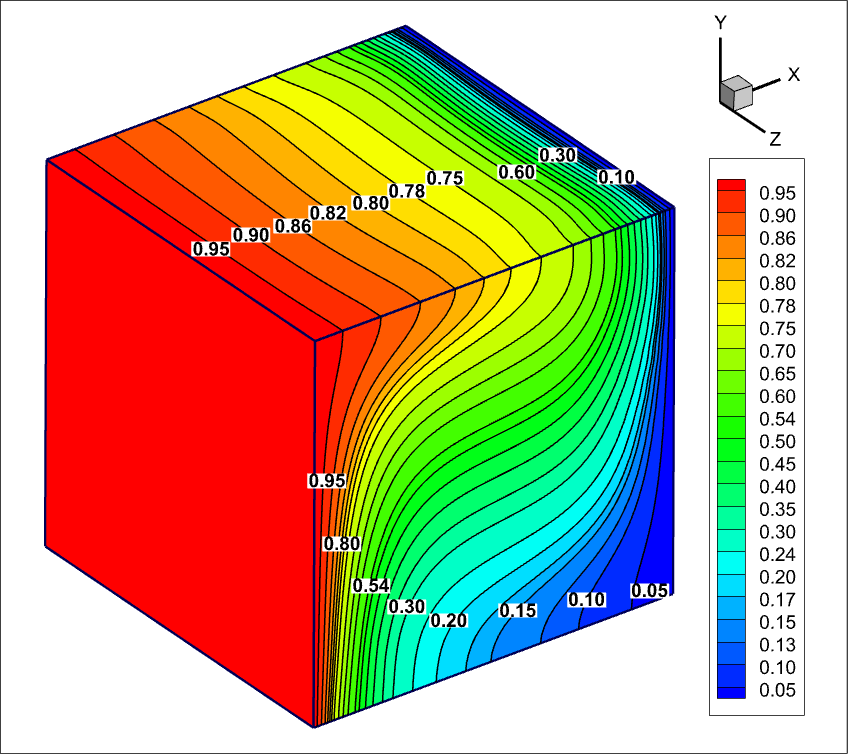}%
    \captionsetup{skip=2pt}%
    \caption{(c) $Ha=25, Ra=10^5$}
    \label{fig:Ra_10^5_Ha_25_P1_3D}
  \end{subfigure}%
  \hspace*{\fill}

  \vspace*{8pt}%
  \hspace*{\fill}%
  \begin{subfigure}{0.33\textwidth}     
    \centering
    \includegraphics[width=\textwidth]{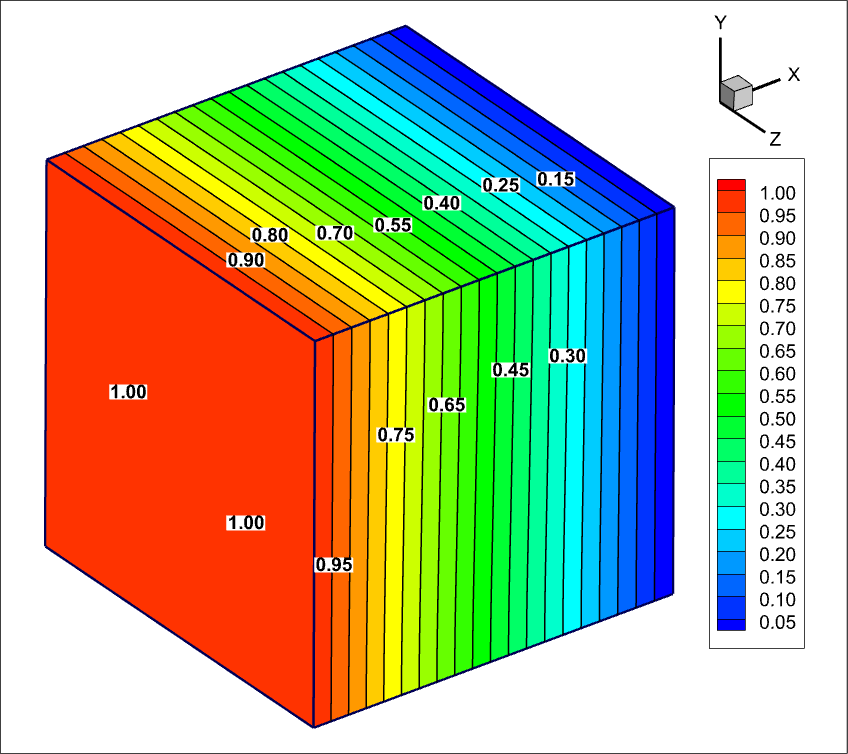}%
    \captionsetup{skip=2pt}%
    \caption{(d) $Ha=50, Ra=10^3$}
    \label{fig:Ra_10^3_Ha_50_P1_3D}
  \end{subfigure}%
 \begin{subfigure}{0.33\textwidth}        
   \centering
    \includegraphics[width=\textwidth]{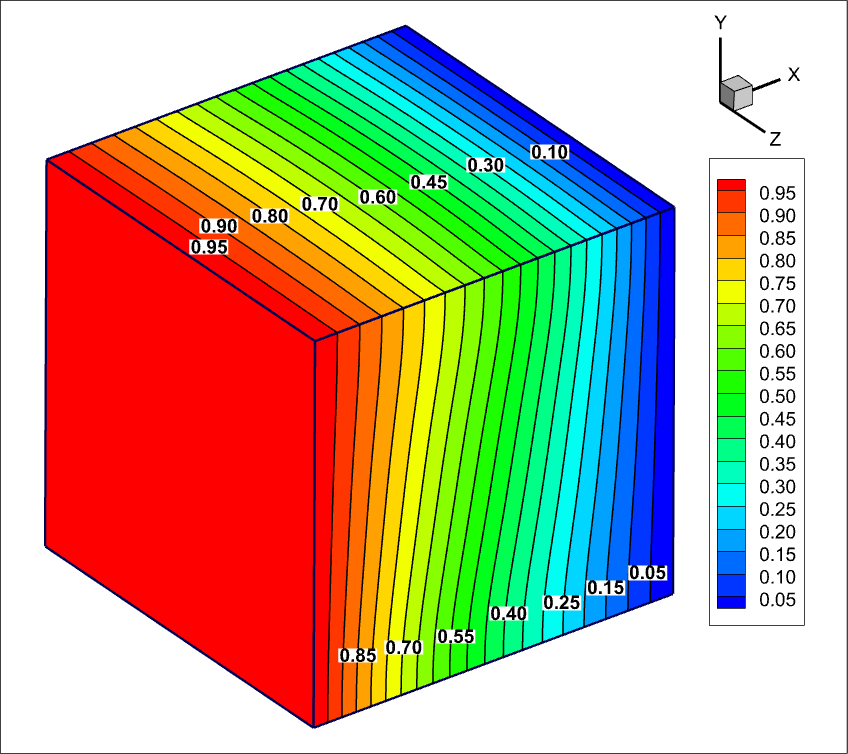}%
    \captionsetup{skip=2pt}%
    \caption{(e) $Ha=50, Ra=10^4$}
    \label{fig:Ra_10^4_Ha_50_P1_3D}
  \end{subfigure}
   \begin{subfigure}{0.33\textwidth}        
   \centering
    \includegraphics[width=\textwidth]{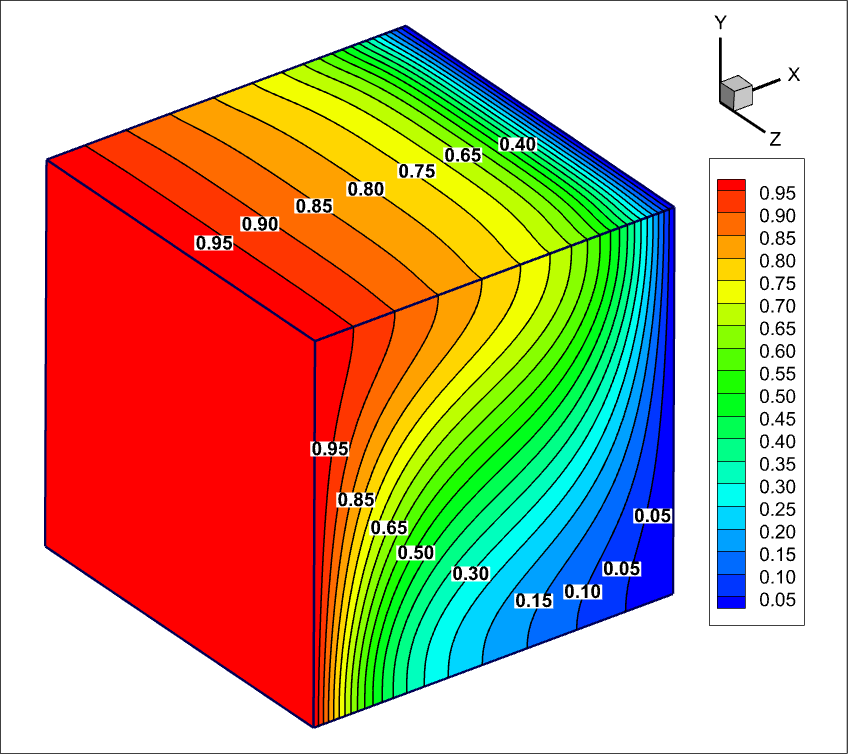}%
    \captionsetup{skip=2pt}%
    \caption{(f) $Ha=50, Ra=10^5$}
    \label{fig:Ra_10^5_Ha_50_P1_3D.png}
  \end{subfigure}%
  \hspace*{\fill}

  \vspace*{8pt}%
  \hspace*{\fill}%
  \begin{subfigure}{0.33\textwidth}     
    \centering
    \includegraphics[width=\textwidth]{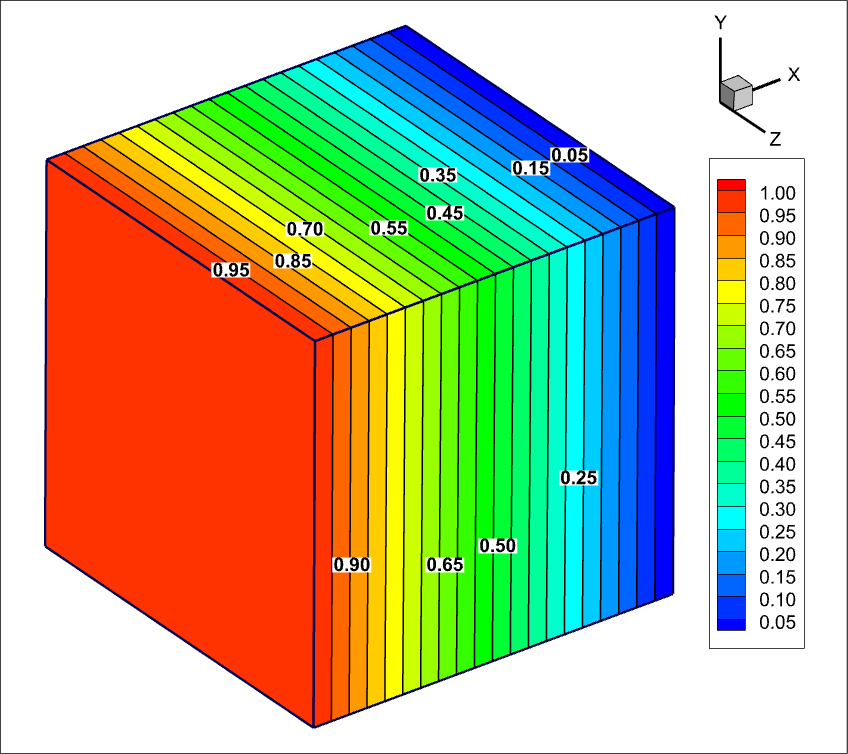}%
    \captionsetup{skip=2pt}%
    \caption{(g) $Ha=100, Ra=10^3$}
    \label{fig:Ra_10^3_Ha_100_P1_3D.png}
  \end{subfigure}%
 \begin{subfigure}{0.33\textwidth}        
   \centering
    \includegraphics[width=\textwidth]{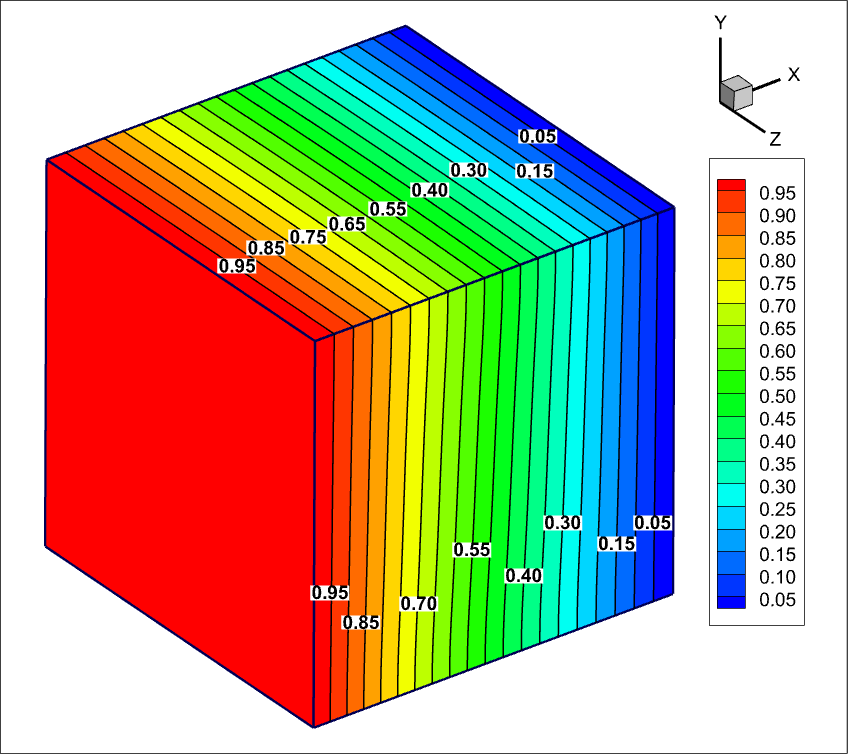}%
    \captionsetup{skip=2pt}%
    \caption{(h) $Ha=100, Ra=10^4$}
    \label{fig:Ra_10^4_Ha_100_P1_3D.png}
  \end{subfigure}
   \begin{subfigure}{0.33\textwidth}        
   \centering
    \includegraphics[width=\textwidth]{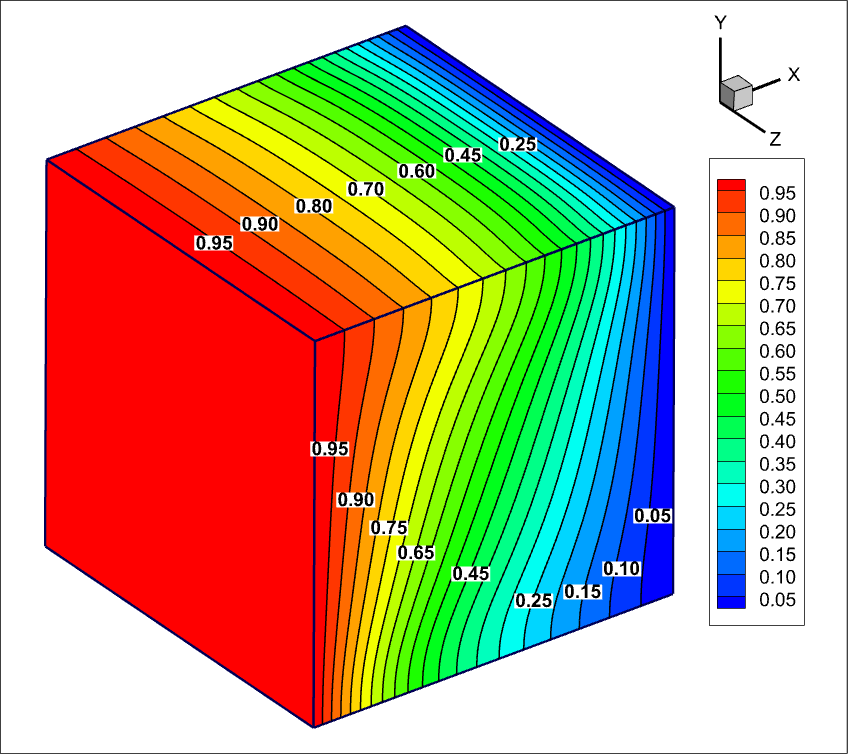}%
    \captionsetup{skip=2pt}%
    \caption{(i) $ Ha=100, Ra=10^5$}
    \label{fig:Ra_10^5_Ha_100_P1_3D.png}
  \end{subfigure}%
  \hspace*{\fill}

  \vspace*{8pt}%
  \hspace*{\fill}%
  \begin{subfigure}{0.33\textwidth}     
    \centering
    \includegraphics[width=\textwidth]{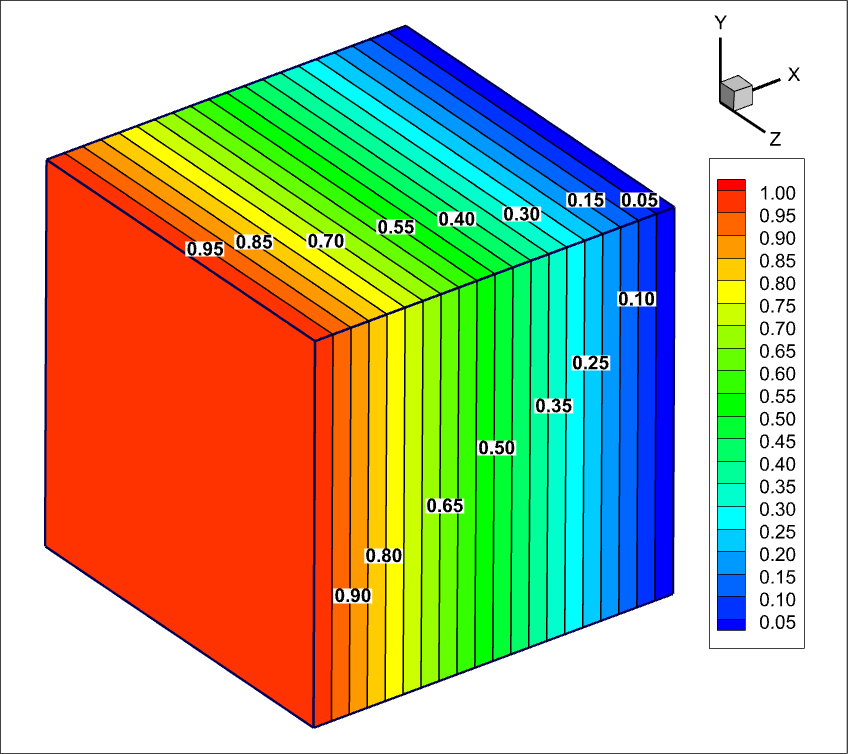}%
    \captionsetup{skip=2pt}%
    \caption{(j) $ Ha=150, Ra=10^3$}
    \label{fig:Ra_10^3_Ha_150_P1_3D.png}
  \end{subfigure}%
 \begin{subfigure}{0.33\textwidth}        
   \centering
    \includegraphics[width=\textwidth]{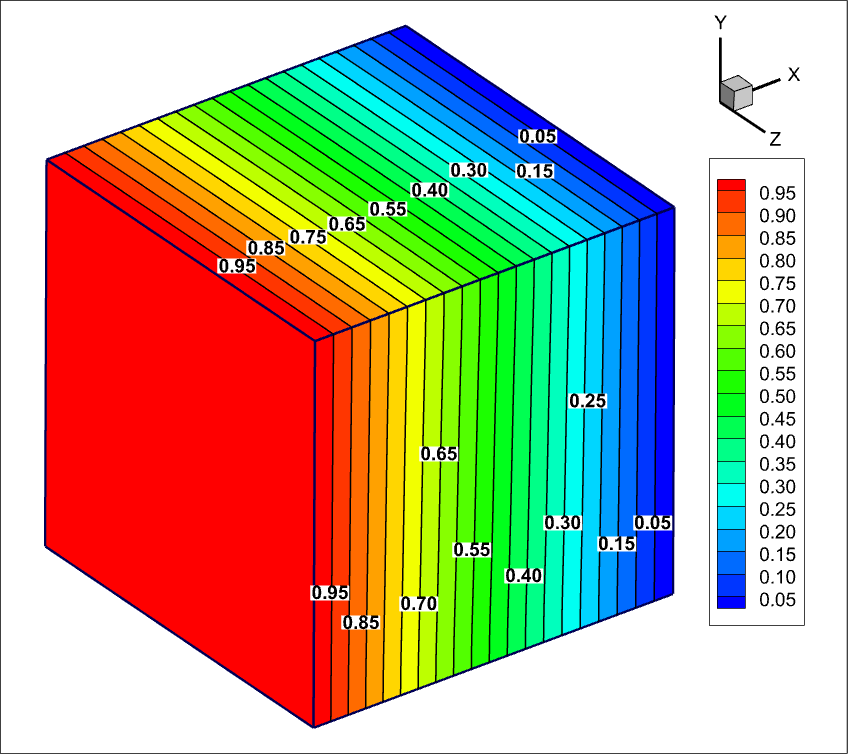}%
    \captionsetup{skip=2pt}%
    \caption{(k) $ Ha=150, Ra=10^4$}
    \label{fig:Ra_10^4_Ha_150_P1_3D.png}
  \end{subfigure}
   \begin{subfigure}{0.33\textwidth}        
   \centering
    \includegraphics[width=\textwidth]{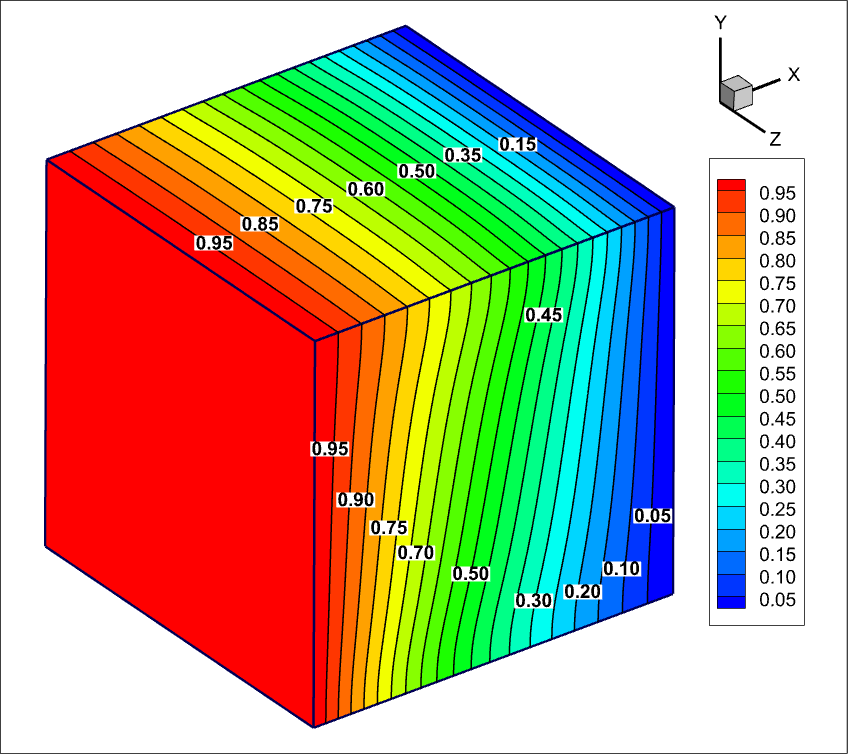}%
    \captionsetup{skip=2pt}%
    \caption{(l) $ Ha=150, Ra=10^5$}
    \label{fig:Ra_10^5_Ha_150_P1_3D.png}
  \end{subfigure}%
  \hspace*{\fill}
  \vspace*{1pt}%
  \hspace*{\fill}%
  \caption{Case 1. Effect of different $Ha$ (Column-wise) and $Ra$ (Row-wise) on isotherm contours at a fixed $Pr = 0.065$}
  \label{fig:case-1_Isotherm_Contours_3D}
\end{figure}

\begin{figure}[htbp]
 \centering
 \vspace*{2pt}%
 \hspace*{\fill}%
\begin{subfigure}{0.33\textwidth}     
    \centering
    \includegraphics[width=\textwidth]{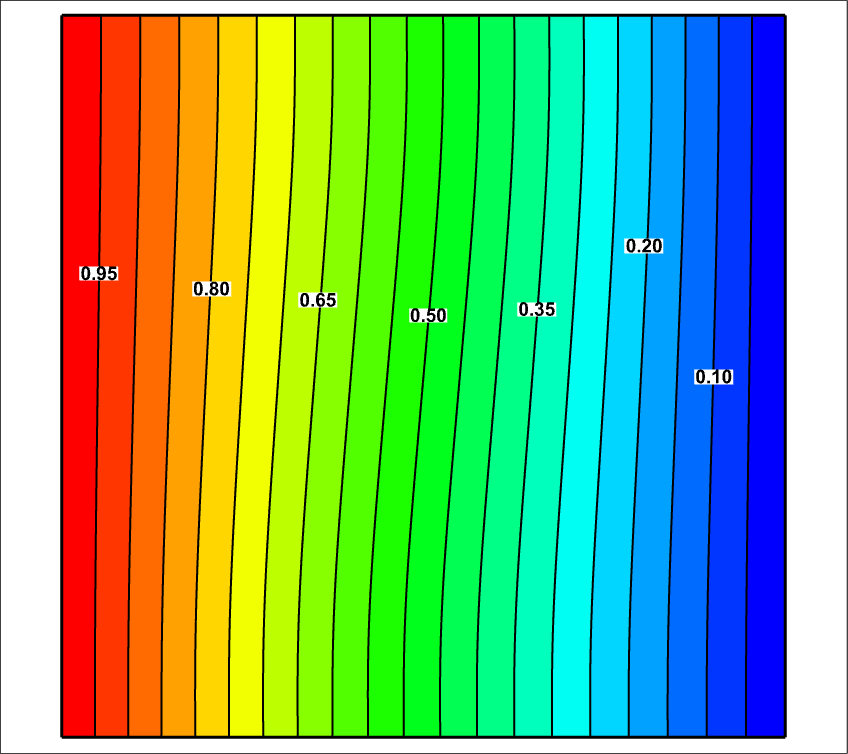}%
    \captionsetup{skip=2pt}%
    \caption{(a) $ Ha=25, Ra=10^3$}
    \label{fig:Ra_10^3_Ha_25_P1}
  \end{subfigure}%
 \begin{subfigure}{0.33\textwidth}        
   \centering
    \includegraphics[width=\textwidth]{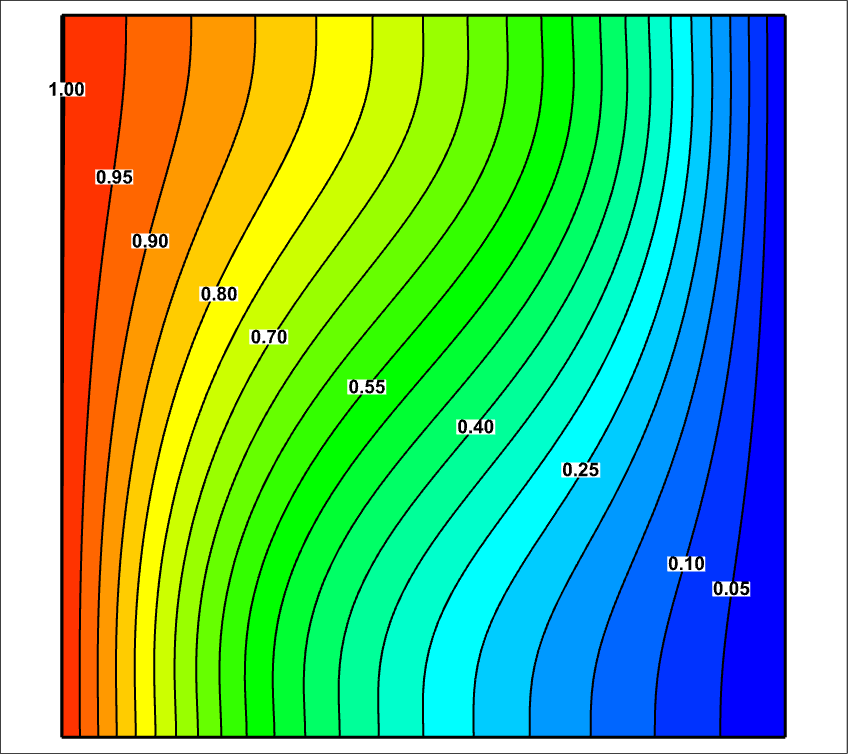}%
    \captionsetup{skip=2pt}%
    \caption{(b) $ Ha=25, Ra=10^4$}
    \label{fig:Ra_10^4_Ha_25_P1}
  \end{subfigure}
   \begin{subfigure}{0.33\textwidth}        
   \centering
    \includegraphics[width=\textwidth]{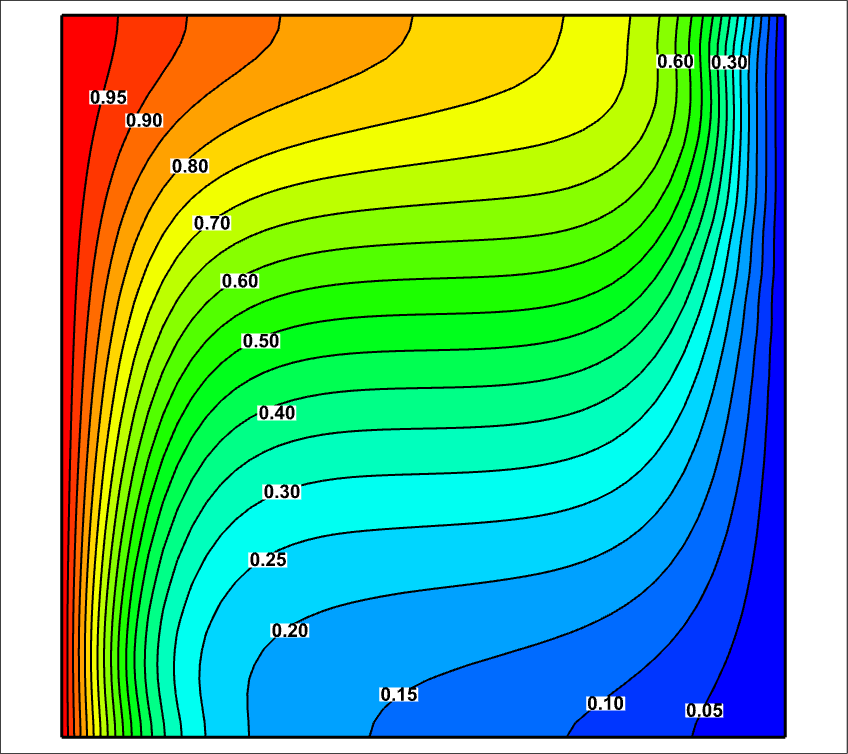}%
    \captionsetup{skip=2pt}%
    \caption{(c) $ Ha=25, Ra=10^5$}
    \label{fig:Ra_10^5_Ha_25_P1}
  \end{subfigure}%
  \hspace*{\fill}

  \vspace*{8pt}%
  \hspace*{\fill}%
  \begin{subfigure}{0.33\textwidth}     
    \centering
    \includegraphics[width=\textwidth]{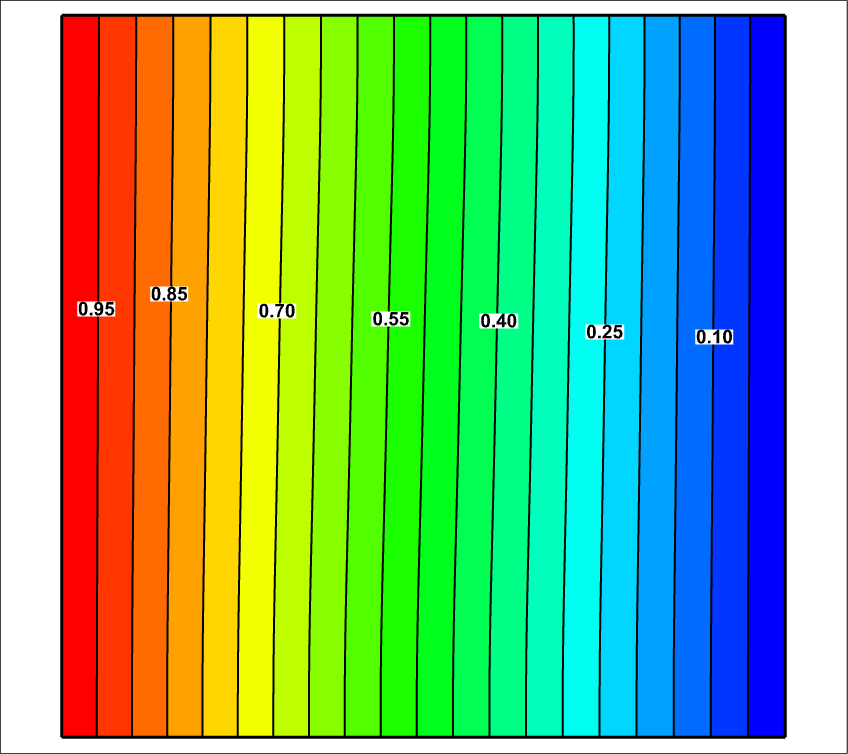}%
    \captionsetup{skip=2pt}%
    \caption{(d) $ Ha=50, Ra=10^3$}
    \label{fig:Ra_10^3_Ha_50_P1}
  \end{subfigure}%
 \begin{subfigure}{0.33\textwidth}        
   \centering
    \includegraphics[width=\textwidth]{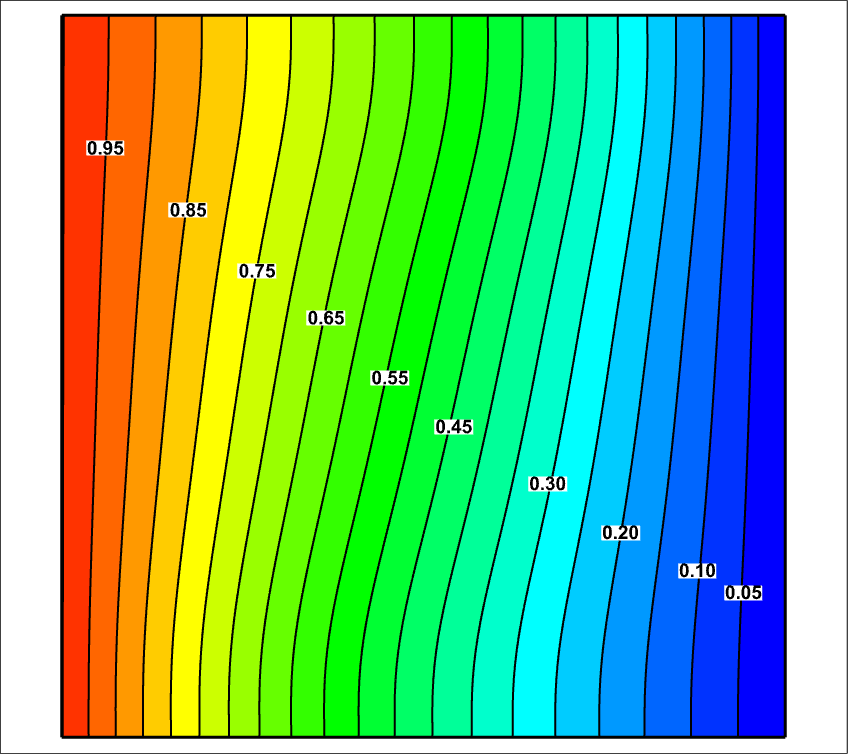}%
    \captionsetup{skip=2pt}%
    \caption{(e) $ Ha=50, Ra=10^4$}
    \label{fig:Ra_10^4_Ha_50_P1}
  \end{subfigure}
   \begin{subfigure}{0.33\textwidth}        
   \centering
    \includegraphics[width=\textwidth]{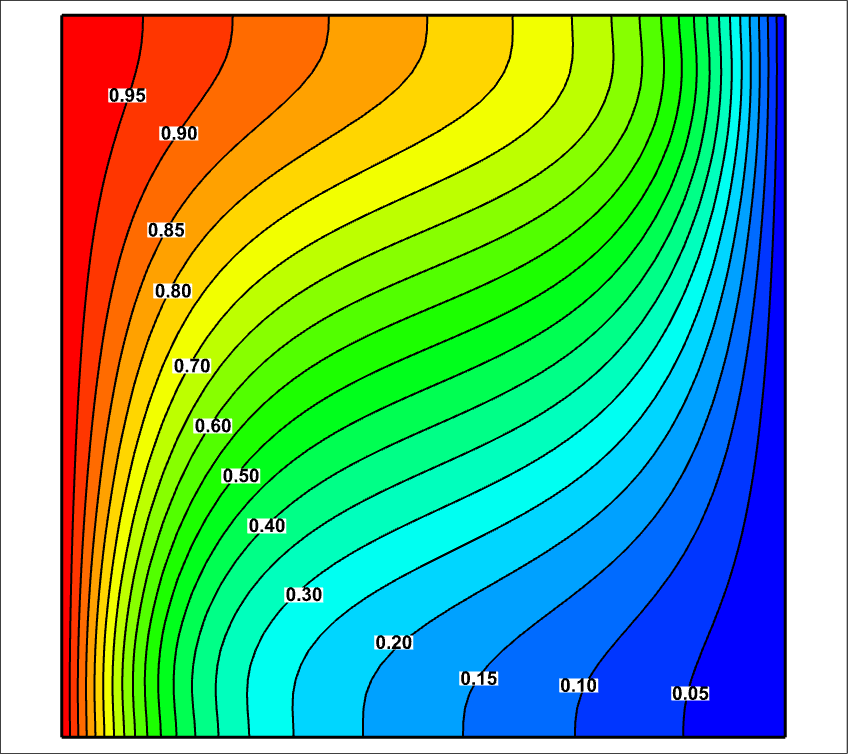}%
    \captionsetup{skip=2pt}%
    \caption{(f) $ Ha=50, Ra=10^5$}
    \label{fig:Ra_10^5_Ha_50_P1.png}
  \end{subfigure}%
  \hspace*{\fill}

  \vspace*{8pt}%
  \hspace*{\fill}%
  \begin{subfigure}{0.33\textwidth}     
    \centering
    \includegraphics[width=\textwidth]{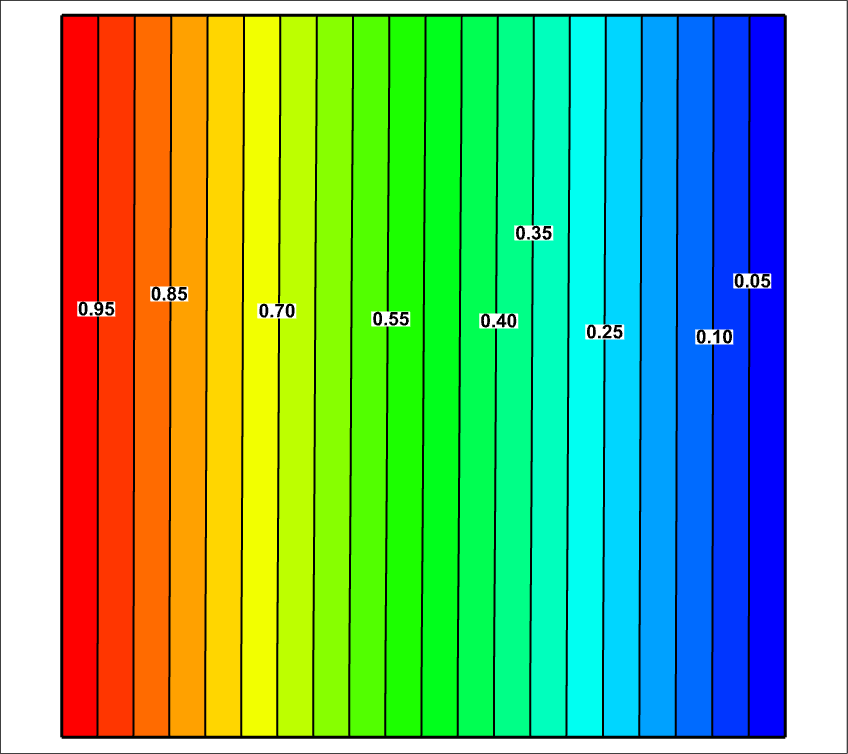}%
    \captionsetup{skip=2pt}%
    \caption{(g) $ Ha=100, Ra=10^3$}
    \label{fig:Ra_10^3_Ha_100_P1.png}
  \end{subfigure}%
 \begin{subfigure}{0.33\textwidth}        
   \centering
    \includegraphics[width=\textwidth]{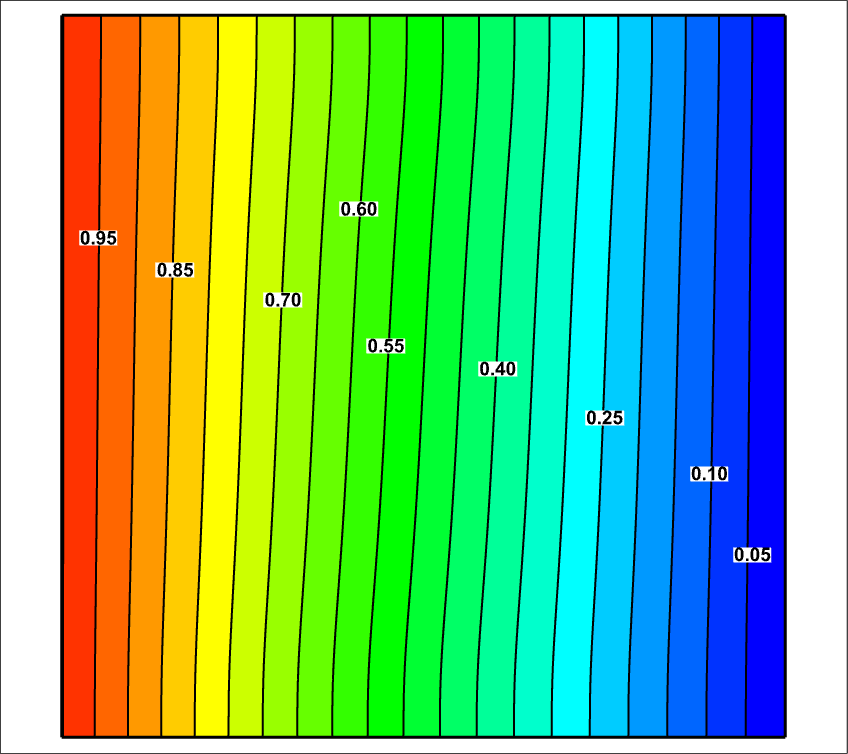}%
    \captionsetup{skip=2pt}%
    \caption{(h) $ Ha=100, Ra=10^4$}
    \label{fig:Ra_10^4_Ha_100_P1.png}
  \end{subfigure}
   \begin{subfigure}{0.33\textwidth}        
   \centering
    \includegraphics[width=\textwidth]{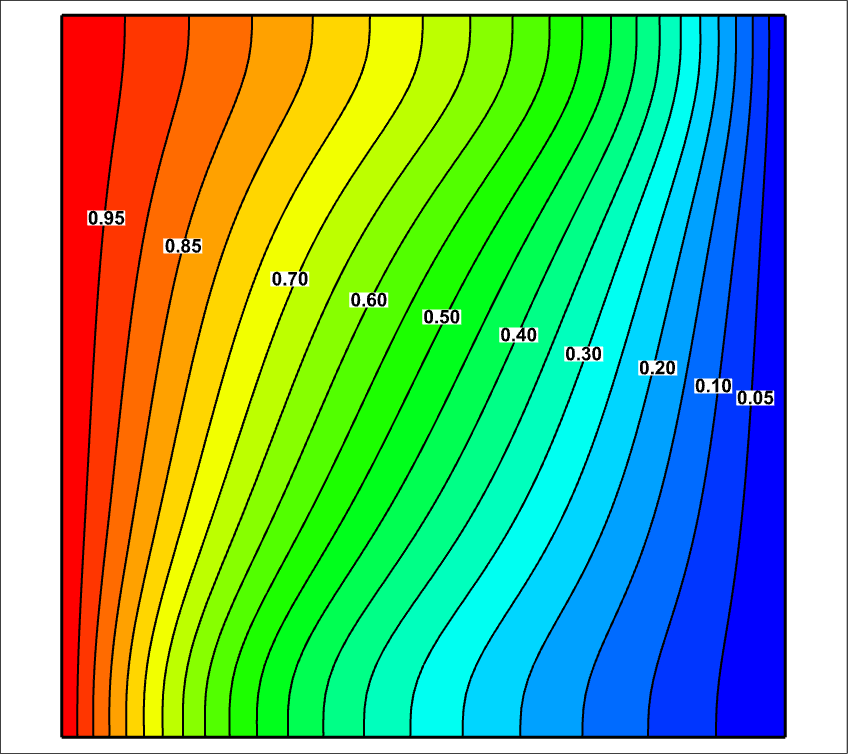}%
    \captionsetup{skip=2pt}%
    \caption{(i) $ Ha=100, Ra=10^5$}
    \label{fig:Ra_10^5_Ha_100_P1.png}
  \end{subfigure}%
  \hspace*{\fill}

  \vspace*{8pt}%
  \hspace*{\fill}%
  \begin{subfigure}{0.33\textwidth}     
    \centering
    \includegraphics[width=\textwidth]{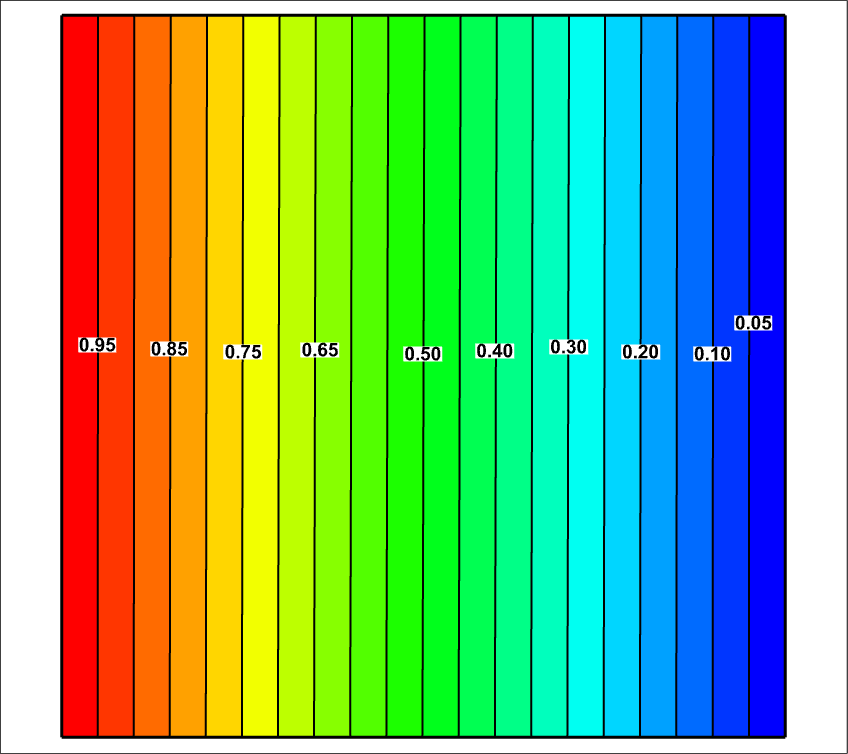}%
    \captionsetup{skip=2pt}%
    \caption{(j) $ Ha=150, Ra=10^3$}
    \label{fig:Ra_10^3_Ha_150_P1.png}
  \end{subfigure}%
 \begin{subfigure}{0.33\textwidth}        
   \centering
    \includegraphics[width=\textwidth]{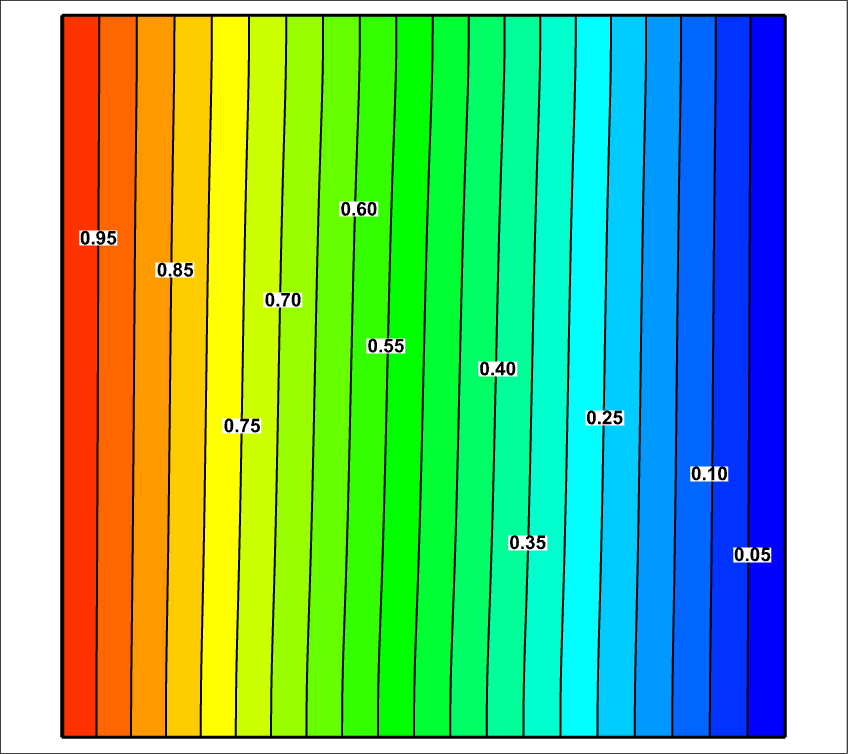}%
    \captionsetup{skip=2pt}%
    \caption{(k) $ Ha=150, Ra=10^4$}
    \label{fig:Ra_10^4_Ha_150_P1.png}
  \end{subfigure}
   \begin{subfigure}{0.33\textwidth}        
   \centering
    \includegraphics[width=\textwidth]{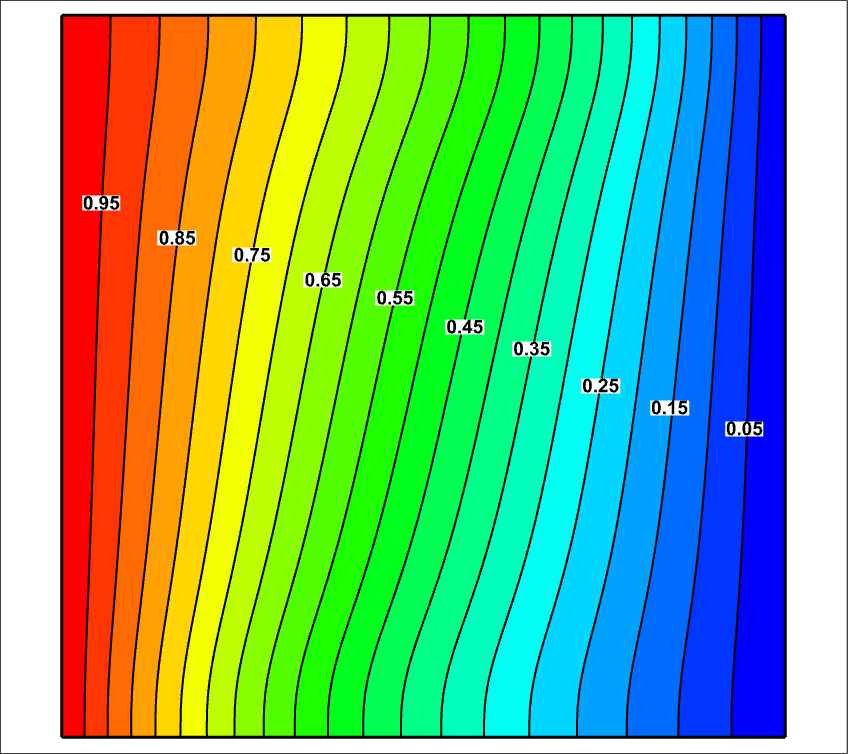}%
    \captionsetup{skip=2pt}%
    \caption{(l) $ Ha=150, Ra=10^5$}
    \label{fig:Ra_10^5_Ha_150_P1.png}
  \end{subfigure}%
  \hspace*{\fill}
  \vspace*{1pt}%
  \hspace*{\fill}%
  \caption{Case 1. Effect of different $Ha$ (Column-wise) and $Ra$ (Row-wise) on isotherm contours (at $z=0.5$ plane) at a fixed $Pr = 0.065$}
  \label{fig:case-1_Isotherm_Contours}
\end{figure}

In Figure \ref{fig:Sche_diag}(a), we present the schematic representation of Case 1, where the left wall ($x=0$) of the cavity undergoes uniform heating ($T_h=1$), the right wall ($x=1$) remains cold, and the other walls are thermally insulated or adiabatic. The flow field and isotherms for this case are depicted in Figures \ref{fig:case-1_Streamlines_Contours}-\ref{fig:case-1_Isotherm_Contours}, showcasing the impact of $Ha$ within the range of $25-150$ and $Ra$ within $10^3-10^5$. The visualization of streamlines is presented at three distinct planes ($z=0.05, z=0.5, z=0.95$). The presence of negative stream function values signifies a clockwise rotation of the fluid, consistently observed across all considered parameters. Notably, there is discernible modulation in the streamfunction magnitude; as $Ha$ increases, the magnitude decreases for a specific $Ra$, while an increase in $Ra$ leads to an elevation in streamfunction magnitude for a given $Ha$. This trend implies a correlation between flow velocity within the cavity, $Ra$, and $Ha$, suggesting that higher $Ra$ values contribute to increased flow velocity, while higher $Ha$ values result in reduced flow velocity. The streamline patterns showcase a noticeable transformation as $Ha$ increases from $25$ to $50$, transforming the primary vortices from round shape to stretched shape in the $y$-direction for $Ra=10^3$ and $Ra=10^4$. In the case of $Ra=10^5$, the primary vortices stretch diagonally. This trend is consistent across all three considered planes ($z=0.05, z=0.5, z=0.95$). Further increase in $Ha$ to $100$ results in the breakup of primary vortices into a bicellular structure at the $z=0.5$ plane, while other planes ($z=0.05, z=0.95$) exhibit single primary vortices for all $Ra$ values. Notably, for low Rayleigh numbers ($Ra=10^3, 10^4$) and $Ha\ge 50$ the vortex undergoes vertical stretching, whereas, for high $Ra$, diagonal stretching is observed. 
As $Ra$ rises, isotherms (Figure \ref{fig:case-1_Isotherm_Contours_3D}, \ref{fig:case-1_Isotherm_Contours}) exhibit increasing curvature for any fixed $Ha$, reflecting the heightened influence of convective heat transfer. Conversely, with an increase in $Ha$, the isothermal contours transition from curved to straight lines, particularly evident at $Ra=10^4, 10^5$. This shift indicates a decrease in convective heat transfer rates because of the declining fluid flow velocity associated with a higher magnetic field ($Ha$). Hence, we can say that the interplay of $Ha$ and $Ra$ significantly influences both the flow patterns and temperature distribution within the cavity.
\subsubsection{Case 2}
\begin{figure}[htbp]
 \centering
 \vspace*{0pt}%
 \hspace*{\fill}%
\begin{subfigure}{0.33\textwidth}     
    \centering
    \includegraphics[width=\textwidth]{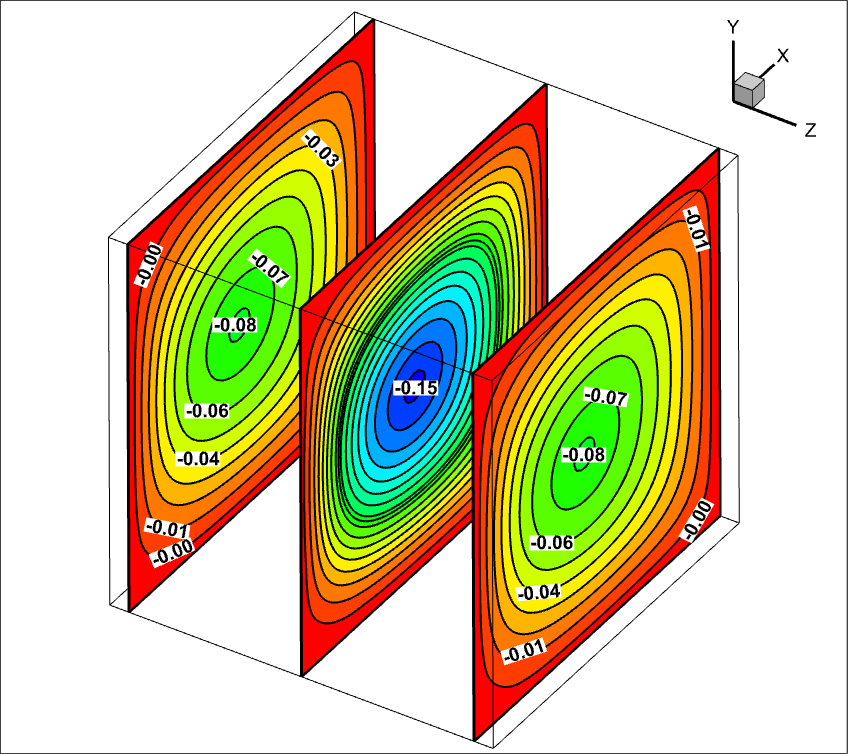}%
    \captionsetup{skip=2pt}%
    \caption{(a) $Ha=25, Ra=10^3$}
    \label{fig:Ra_10^3_Ha_25_P2_Streamlines}
  \end{subfigure}%
 \begin{subfigure}{0.33\textwidth}        
   \centering
    \includegraphics[width=\textwidth]{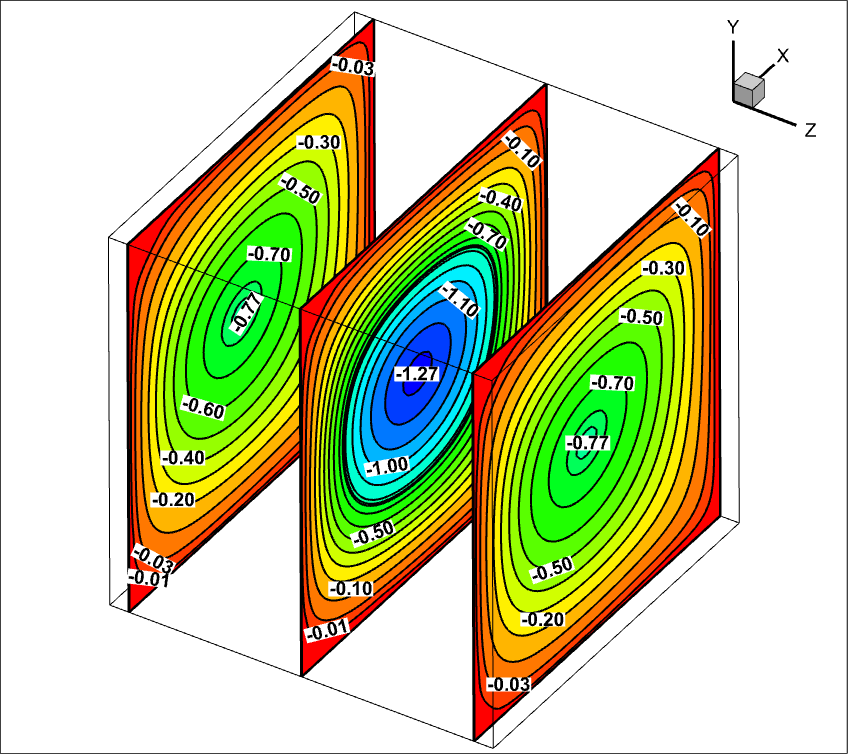}%
    \captionsetup{skip=2pt}%
    \caption{(b) $Ha=25, Ra=10^4$}
    \label{fig:Ra_10^4_Ha_25_P2_Streamlines}
  \end{subfigure}
   \begin{subfigure}{0.33\textwidth}        
   \centering
    \includegraphics[width=\textwidth]{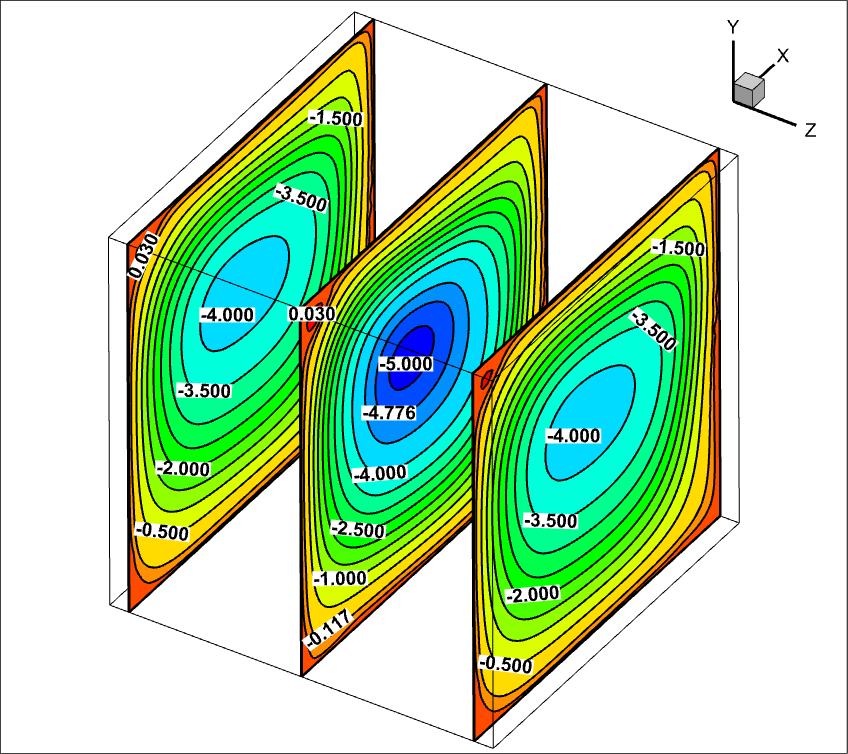}%
    \captionsetup{skip=2pt}%
    \caption{(c) $Ha=25, Ra=10^5$}
    \label{fig:Ra_10^5_Ha_25_P2_Streamlines.png}
  \end{subfigure}%
  \hspace*{\fill}

  \vspace*{8pt}%
  \hspace*{\fill}%
  \begin{subfigure}{0.33\textwidth}     
    \centering
    \includegraphics[width=\textwidth]{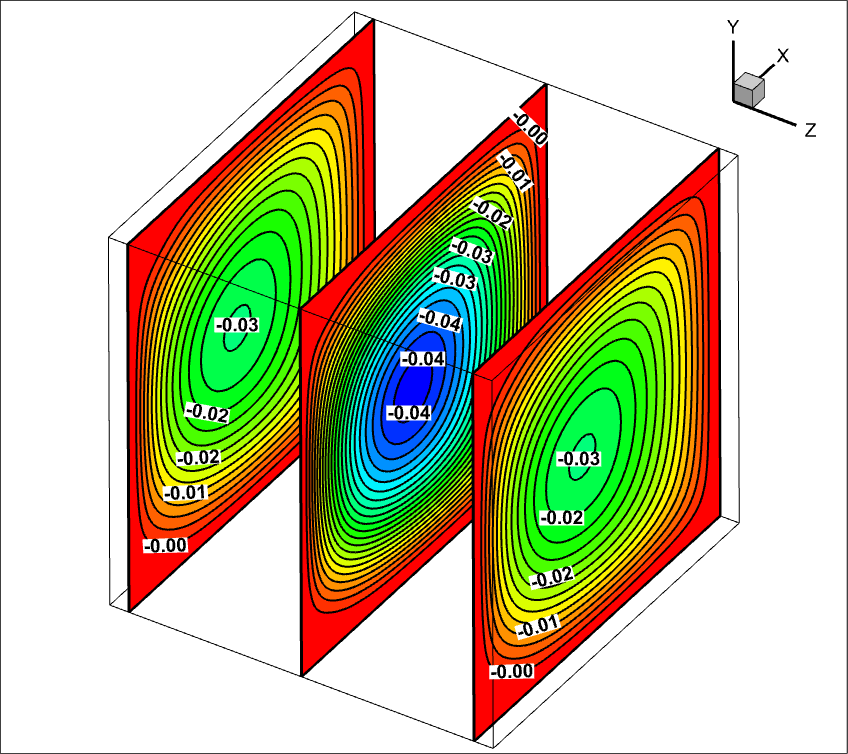}%
    \captionsetup{skip=2pt}%
    \caption{(d) $Ha=50, Ra=10^3$}
    \label{fig:Ra_10^3_Ha_50_P2_Streamlines}
  \end{subfigure}%
 \begin{subfigure}{0.33\textwidth}        
   \centering
    \includegraphics[width=\textwidth]{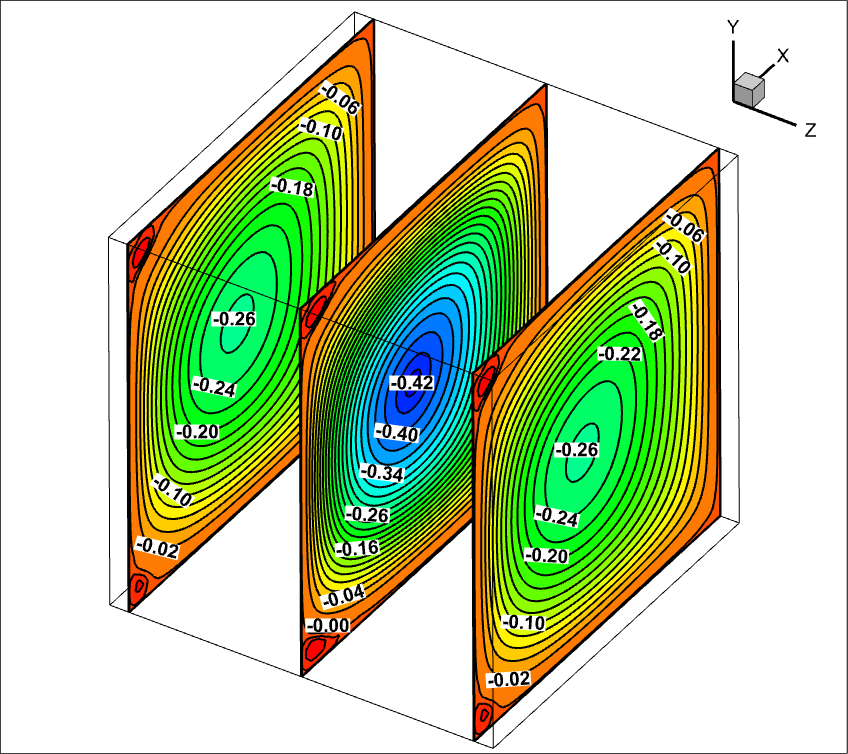}%
    \captionsetup{skip=2pt}%
    \caption{(e) $Ha=50, Ra=10^4$}
    \label{fig:Ra_10^4_Ha_50_P2_Streamlines}
  \end{subfigure}
   \begin{subfigure}{0.33\textwidth}        
   \centering
    \includegraphics[width=\textwidth]{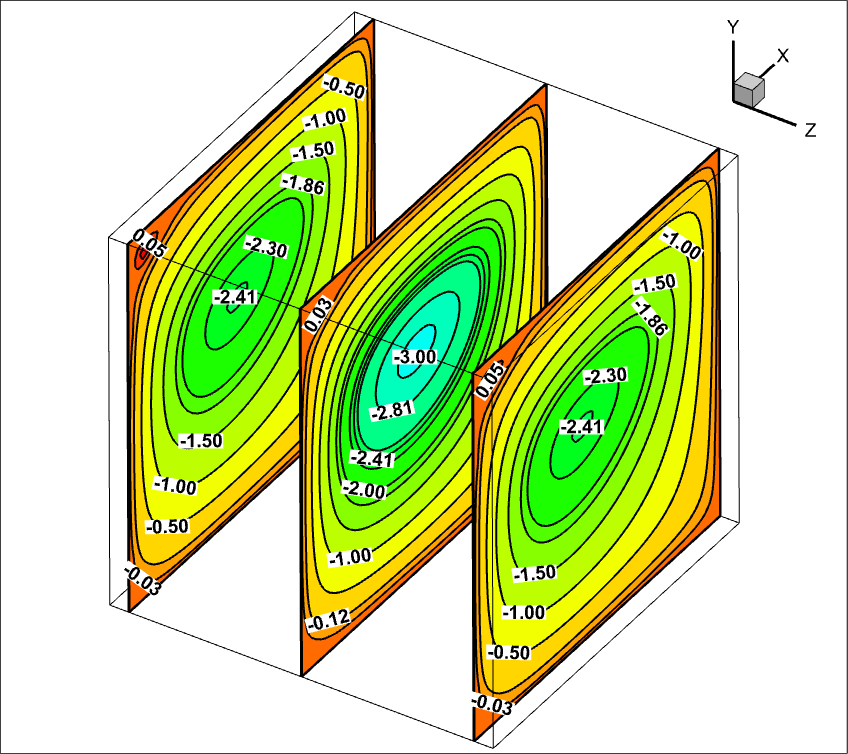}%
    \captionsetup{skip=2pt}%
    \caption{(f) $Ha=50, Ra=10^5$}
    \label{fig:Ra_10^5_Ha_50_P2_Streamlines3}
  \end{subfigure}%
  \hspace*{\fill}

  \vspace*{8pt}%
  \hspace*{\fill}%
  \begin{subfigure}{0.33\textwidth}     
    \centering
    \includegraphics[width=\textwidth]{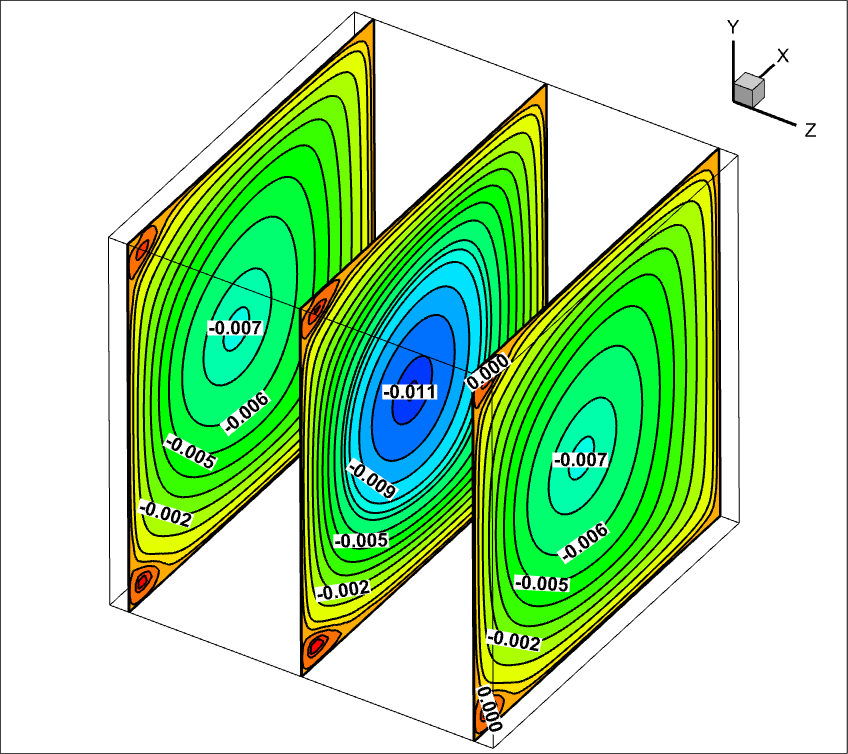}%
    \captionsetup{skip=2pt}%
    \caption{(g) $Ha=100, Ra=10^3$}
    \label{fig:Ra_10^3_Ha_100_P2_Streamlines}
  \end{subfigure}%
 \begin{subfigure}{0.33\textwidth}        
   \centering
    \includegraphics[width=\textwidth]{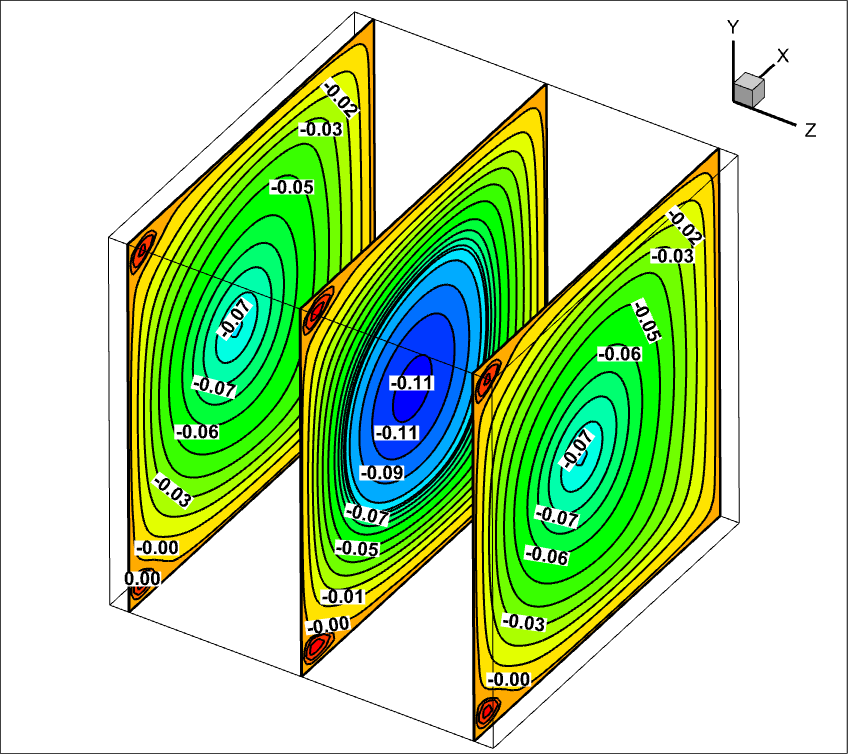}%
    \captionsetup{skip=2pt}%
    \caption{(h) $Ha=100, Ra=10^4$}
    \label{fig:Ra_10^4_Ha_100_P2_Streamlines}
  \end{subfigure}
   \begin{subfigure}{0.33\textwidth}        
   \centering
    \includegraphics[width=\textwidth]{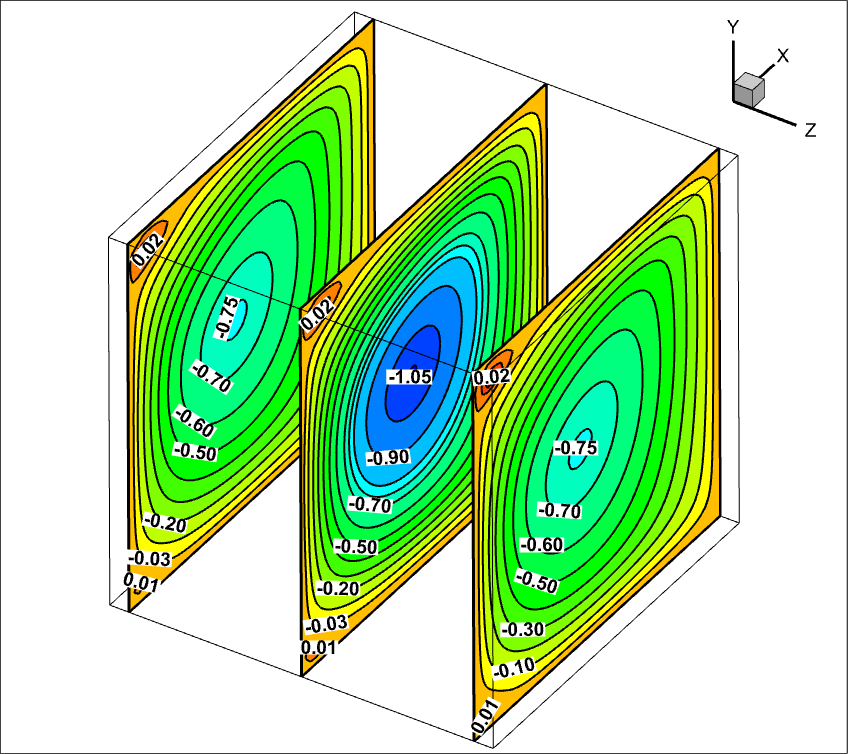}%
    \captionsetup{skip=2pt}%
    \caption{(i) $Ha=100, Ra=10^5$}
    \label{fig:Ra_10^5_Ha_100_P2_Streamlines}
  \end{subfigure}%
  \hspace*{\fill}

  \vspace*{8pt}%
  \hspace*{\fill}%
  \begin{subfigure}{0.33\textwidth}     
    \centering
    \includegraphics[width=\textwidth]{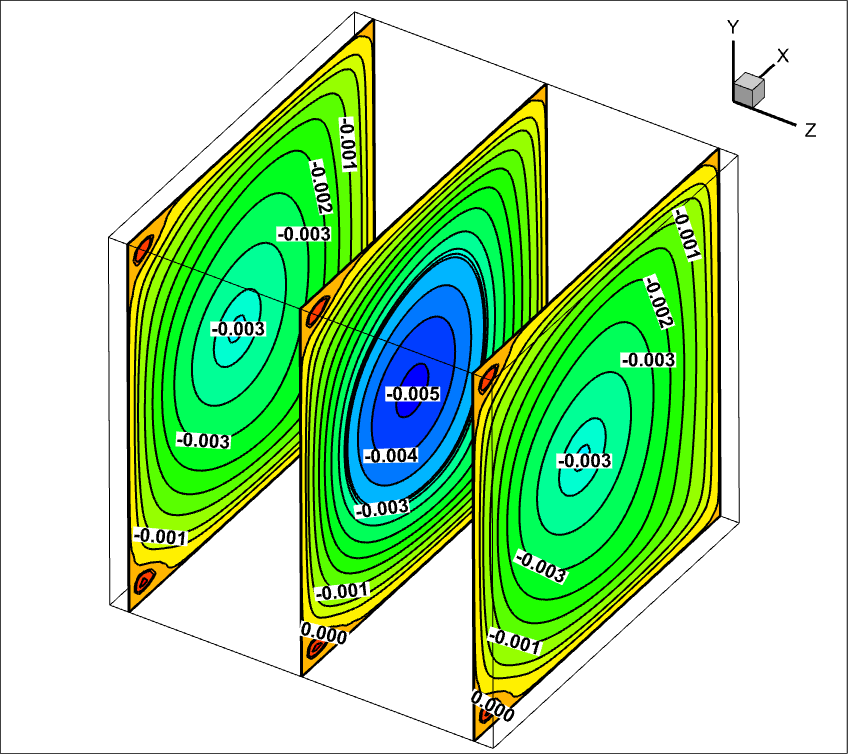}%
    \captionsetup{skip=2pt}%
    \caption{(j) $Ha=150, Ra=10^3$}
    \label{fig:Ra_10^3_Ha_150_P2_Streamlines}
  \end{subfigure}%
 \begin{subfigure}{0.33\textwidth}        
   \centering
    \includegraphics[width=\textwidth]{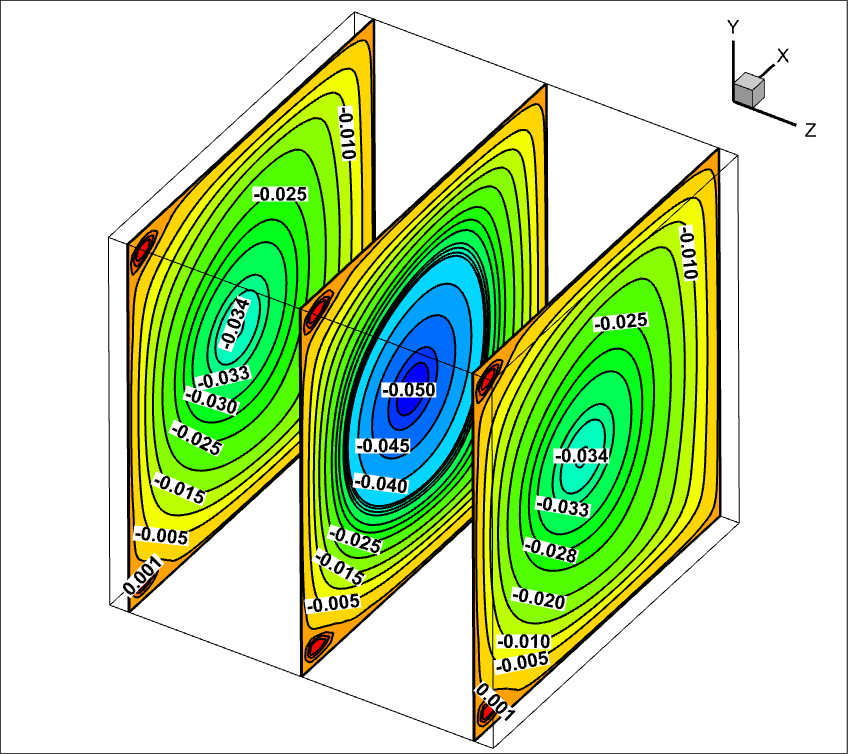}%
    \captionsetup{skip=2pt}%
    \caption{(k) $Ha=150, Ra=10^4$}
    \label{fig:Ra_10^4_Ha_150_P2_Streamlines}
  \end{subfigure}
   \begin{subfigure}{0.33\textwidth}        
   \centering
    \includegraphics[width=\textwidth]{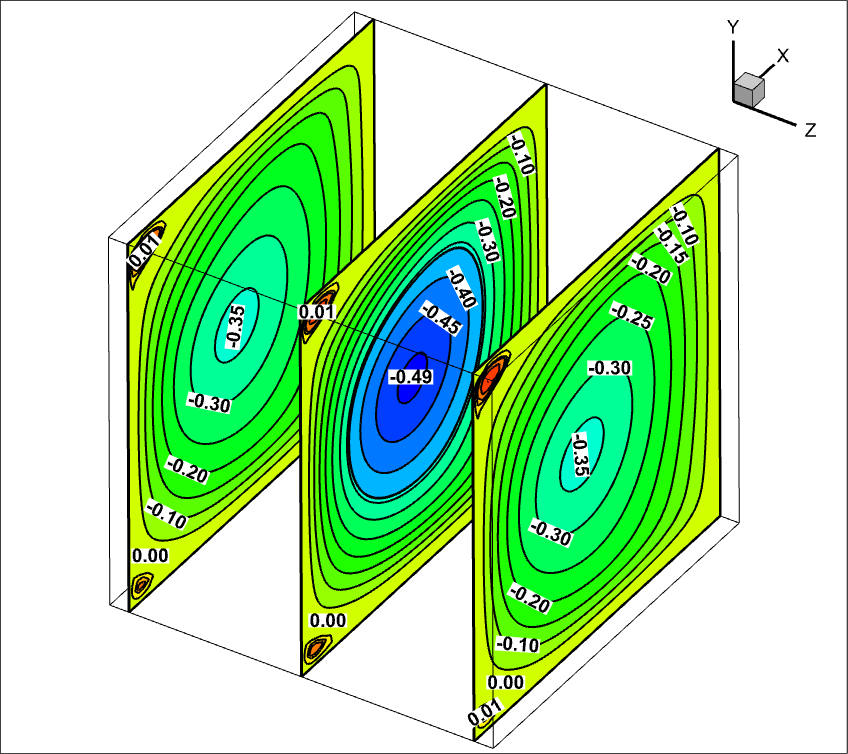}%
    \captionsetup{skip=2pt}%
    \caption{(l) $Ha=150, Ra=10^5$}
    \label{fig:Ra_10^5_Ha_150_P2_Streamlines}
  \end{subfigure}%
  \hspace*{\fill}
  \vspace*{1pt}%
  \hspace*{\fill}%
  \caption{Case 2. Effect of different $Ha$ (Column-wise) and $Ra$ (Row-wise) on streamlines pattern at different $xy$ planes $(z=0.05, z=0.5, z=0.95)$ with fixed $Pr = 0.065$}
  \label{fig:case-2_Streamlines_Contours}
\end{figure}

\begin{figure}[htbp]
 \centering
 \vspace*{0pt}%
 \hspace*{\fill}%
\begin{subfigure}{0.33\textwidth}     
    \centering
    \includegraphics[width=\textwidth]{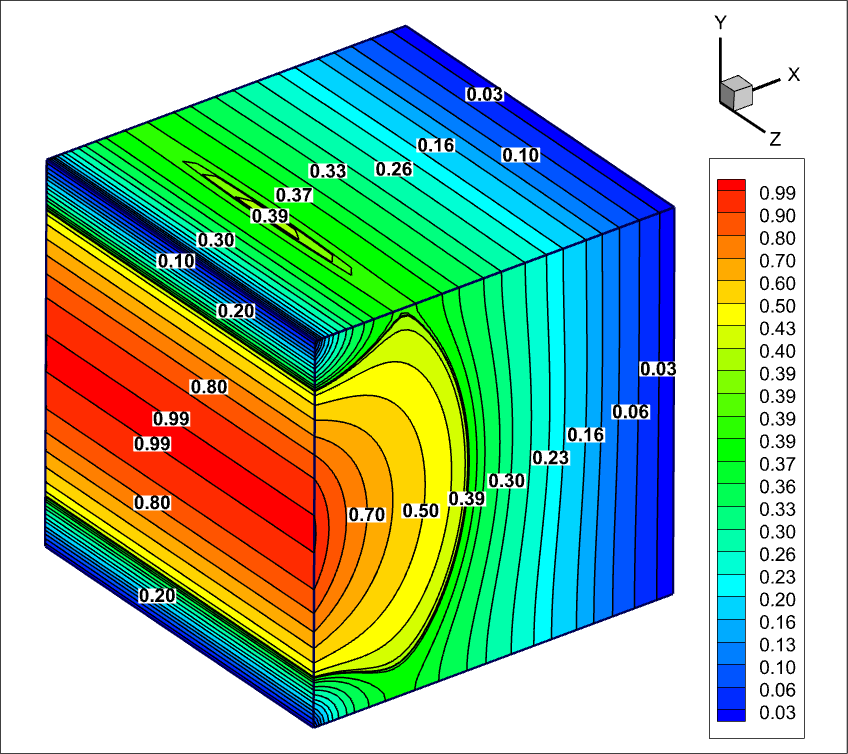}%
    \captionsetup{skip=2pt}%
    \caption{(a) $Ha=25,Ra=10^3$}
    \label{fig:Ra_10^3_Ha_25_P2_3D}
  \end{subfigure}%
 \begin{subfigure}{0.33\textwidth}        
   \centering
    \includegraphics[width=\textwidth]{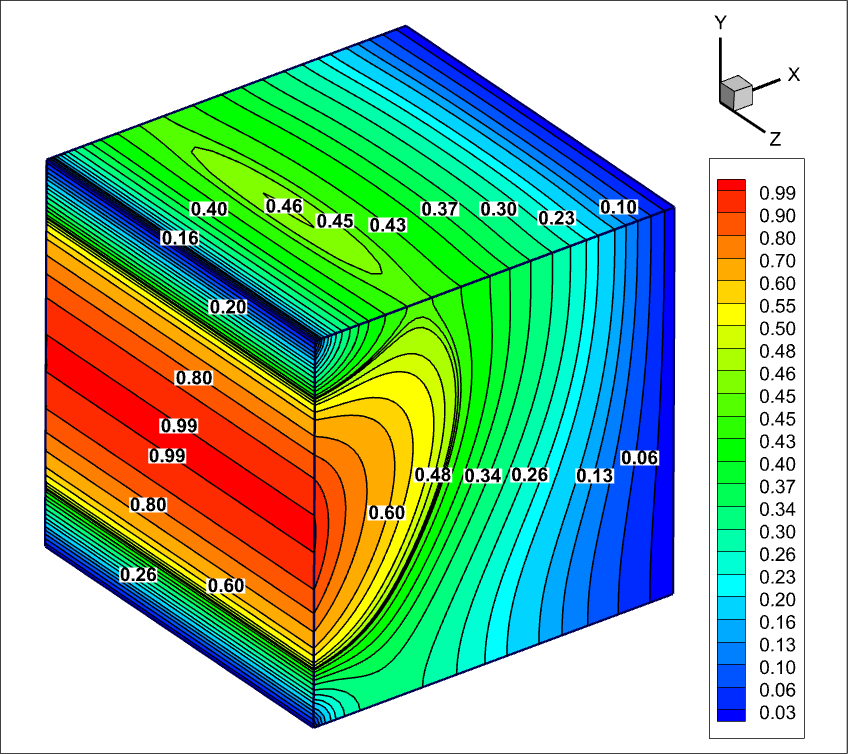}%
    \captionsetup{skip=2pt}%
    \caption{(b) $ Ha=25, Ra=10^4$}
    \label{fig:Ra_10^4_Ha_25_P2_3D}
  \end{subfigure}
   \begin{subfigure}{0.33\textwidth}        
   \centering
    \includegraphics[width=\textwidth]{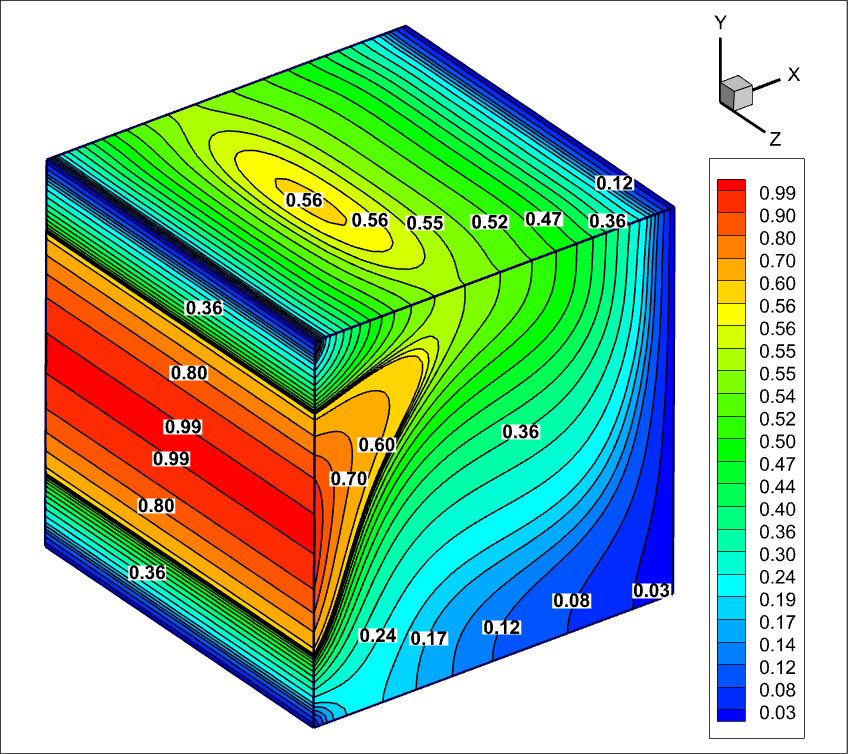}%
    \captionsetup{skip=2pt}%
    \caption{(c) $ Ha=25, Ra=10^5$}
    \label{fig:Ra_10^5_Ha_25_P2_3D}
  \end{subfigure}%
  \hspace*{\fill}

  \vspace*{8pt}%
  \hspace*{\fill}%
  \begin{subfigure}{0.33\textwidth}     
    \centering
    \includegraphics[width=\textwidth]{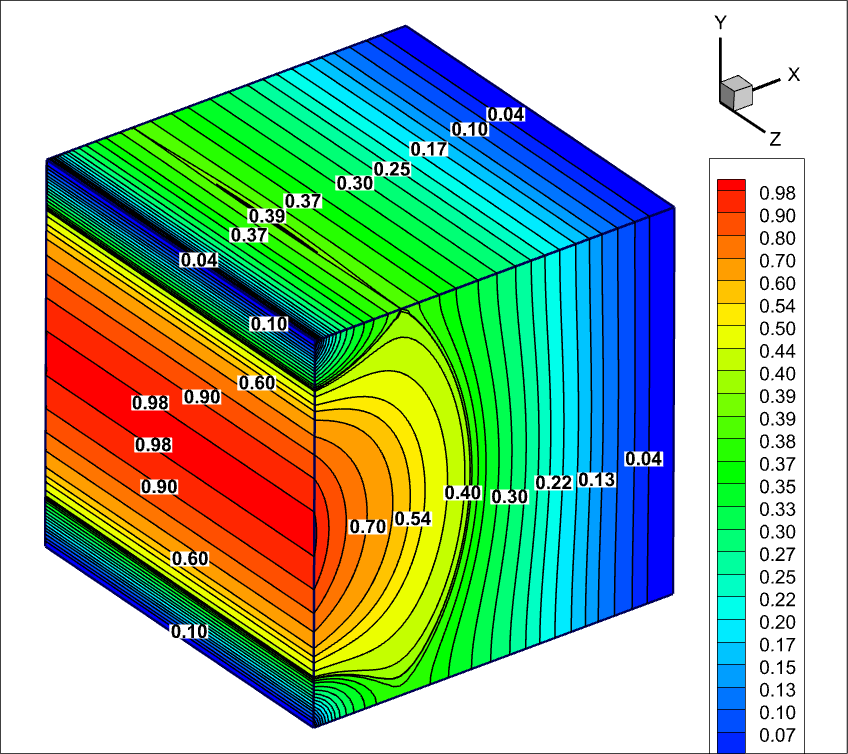}%
    \captionsetup{skip=2pt}%
    \caption{(d) $ Ha=50, Ra=10^3$}
    \label{fig:Ra_10^3_Ha_50_P2_3D}
  \end{subfigure}%
 \begin{subfigure}{0.33\textwidth}        
   \centering
    \includegraphics[width=\textwidth]{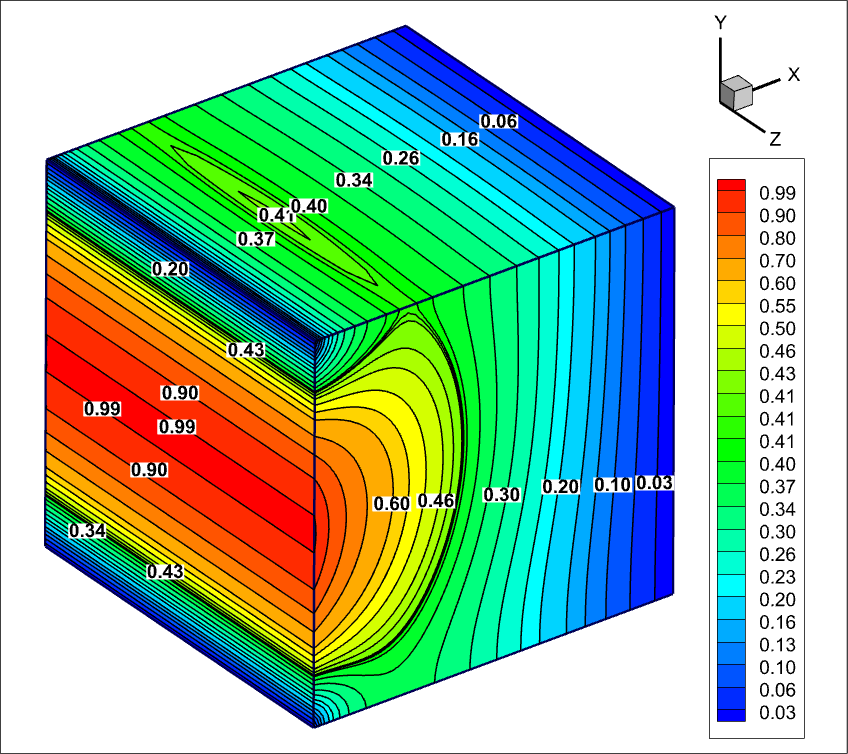}%
    \captionsetup{skip=2pt}%
    \caption{(e) $Ha=50, Ra=10^4$}
    \label{fig:Ra_10^4_Ha_50_P2_3D}
  \end{subfigure}
   \begin{subfigure}{0.33\textwidth}        
   \centering
    \includegraphics[width=\textwidth]{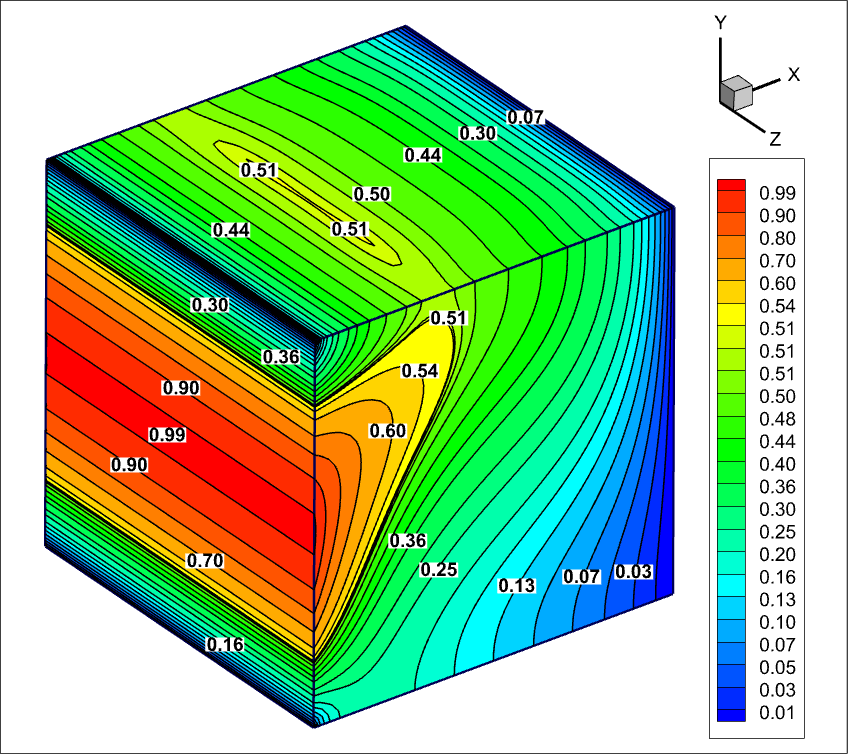}%
    \captionsetup{skip=2pt}%
    \caption{(f) $ Ha=50, Ra=10^5$}
    \label{fig:Ra_10^5_Ha_50_P2_3D.png}
  \end{subfigure}%
  \hspace*{\fill}

  \vspace*{8pt}%
  \hspace*{\fill}%
  \begin{subfigure}{0.33\textwidth}     
    \centering
    \includegraphics[width=\textwidth]{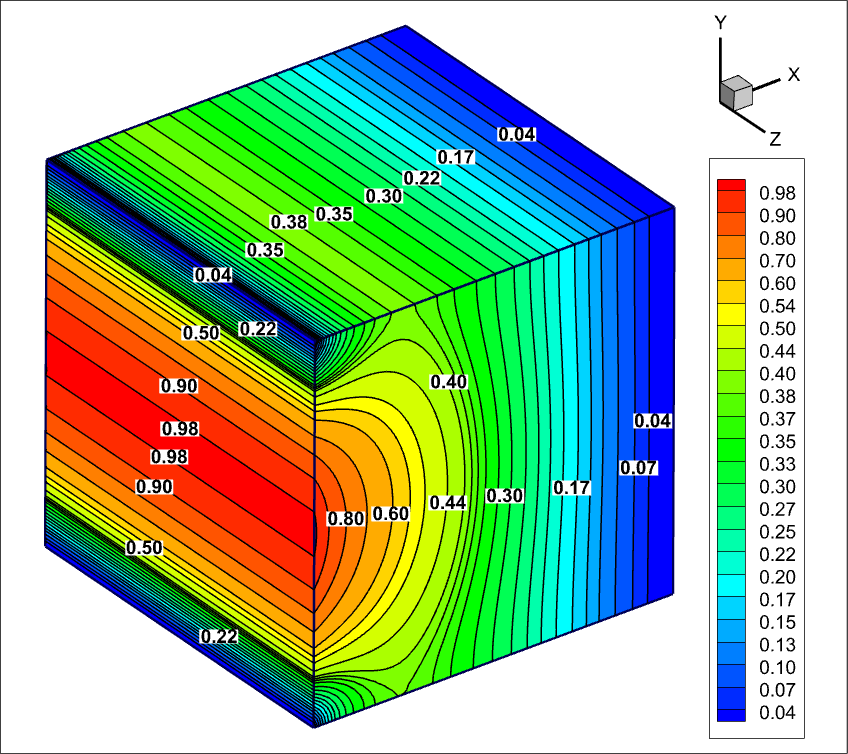}%
    \captionsetup{skip=2pt}%
    \caption{(g) $ Ha=100, Ra=10^3$}
    \label{fig:Ra_10^3_Ha_100_P2_3D.png}
  \end{subfigure}%
 \begin{subfigure}{0.33\textwidth}        
   \centering
    \includegraphics[width=\textwidth]{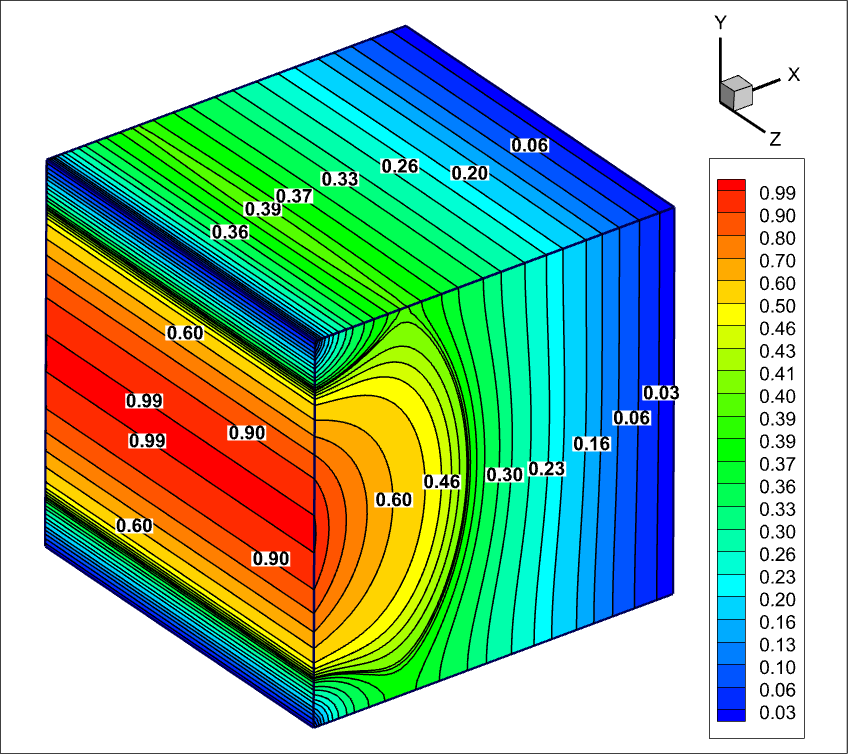}%
    \captionsetup{skip=2pt}%
    \caption{(h) $Ha=100, Ra=10^4$}
    \label{fig:Ra_10^4_Ha_100_P2_3D.png}
  \end{subfigure}
   \begin{subfigure}{0.33\textwidth}        
   \centering
    \includegraphics[width=\textwidth]{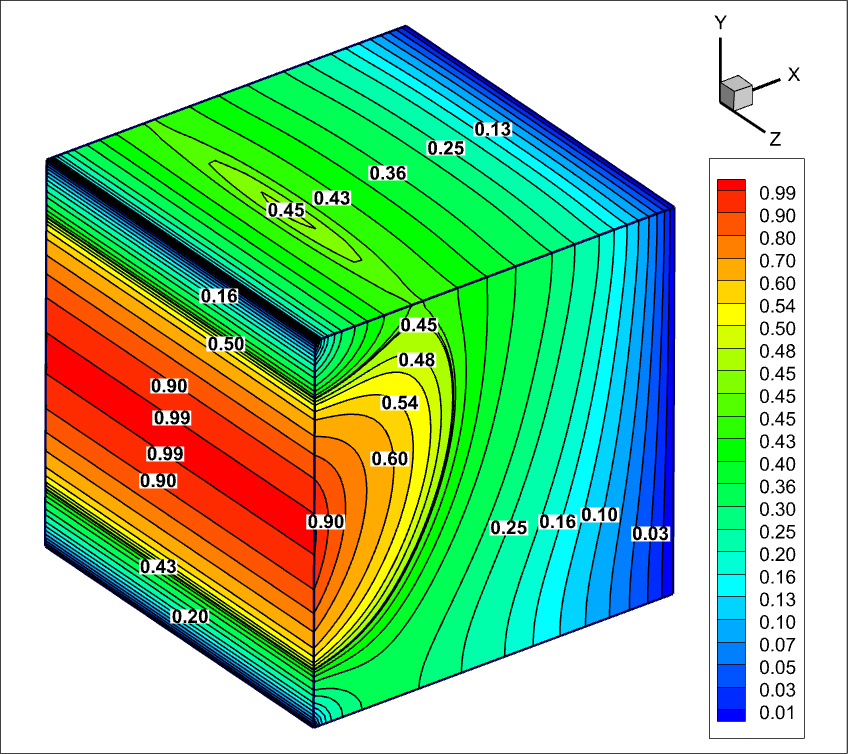}%
    \captionsetup{skip=2pt}%
    \caption{(i) $ Ha=100, Ra=10^5$}
    \label{fig:Ra_10^5_Ha_100_P2_3D.png}
  \end{subfigure}%
  \hspace*{\fill}

  \vspace*{8pt}%
  \hspace*{\fill}%
  \begin{subfigure}{0.33\textwidth}     
    \centering
    \includegraphics[width=\textwidth]{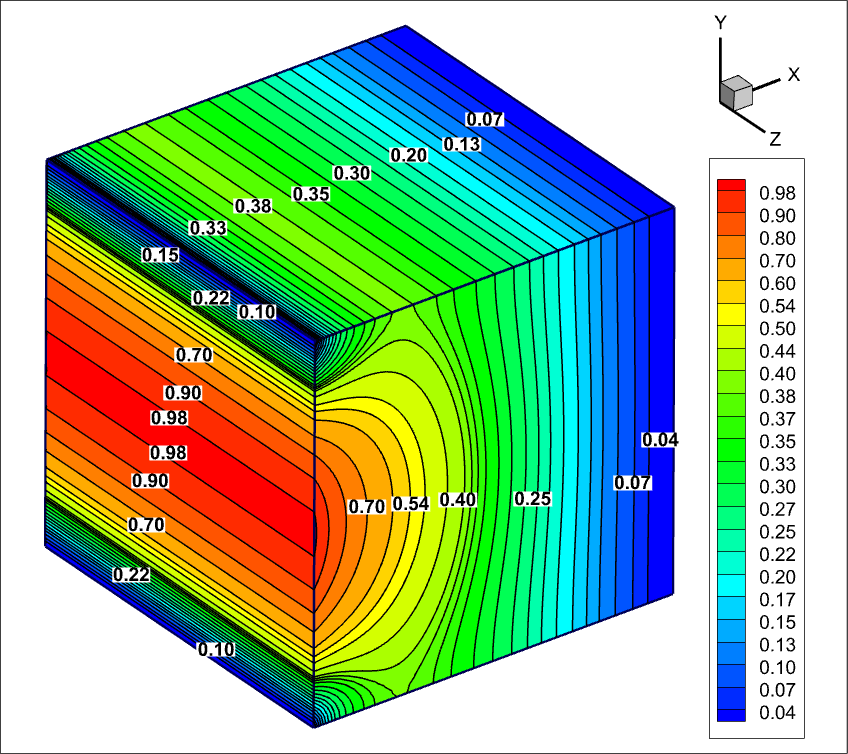}%
    \captionsetup{skip=2pt}%
    \caption{(j) $ Ha=150, Ra=10^3$}
    \label{fig:Ra_10^3_Ha_150_P2_3D.png}
  \end{subfigure}%
 \begin{subfigure}{0.33\textwidth}        
   \centering
    \includegraphics[width=\textwidth]{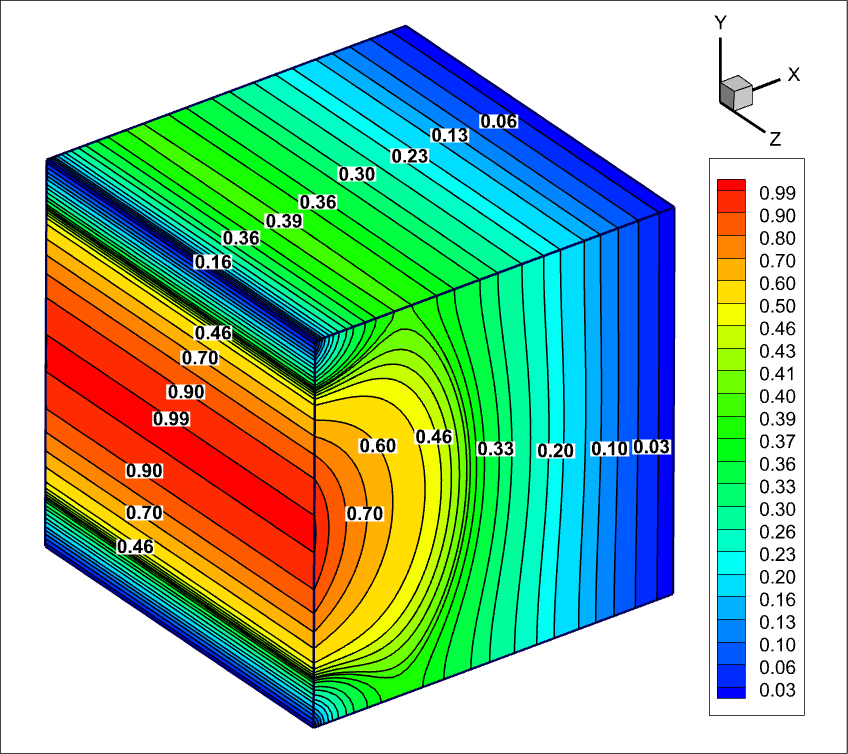}%
    \captionsetup{skip=2pt}%
    \caption{(k) $Ha=150, Ra=10^4$}
    \label{fig:Ra_10^4_Ha_150_P2_3D.png}
  \end{subfigure}
   \begin{subfigure}{0.33\textwidth}        
   \centering
    \includegraphics[width=\textwidth]{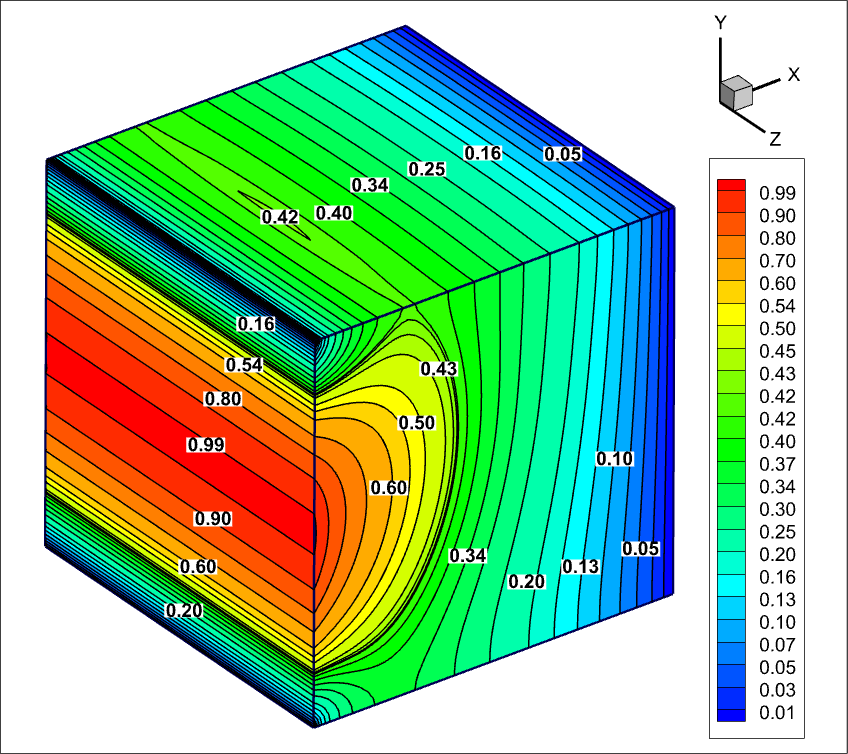}%
    \captionsetup{skip=2pt}%
    \caption{(l) $Ha=150, Ra=10^5$}
    \label{fig:Ra_10^5_Ha_150_P2_3D.png}
  \end{subfigure}%
  \hspace*{\fill}
  \vspace*{1pt}%
  \hspace*{\fill}%
  \caption{Case 2. Effect of different $Ha$ (Column-wise) and $Ra$ (Row-wise) on isotherm contours at a fixed $Pr = 0.065$}
  \label{fig:case-2_Isotherm_Contours_3D}
\end{figure}

\begin{figure}[htbp]
 \centering
 \vspace*{0pt}%
 \hspace*{\fill}%
\begin{subfigure}{0.33\textwidth}     
    \centering
    \includegraphics[width=\textwidth]{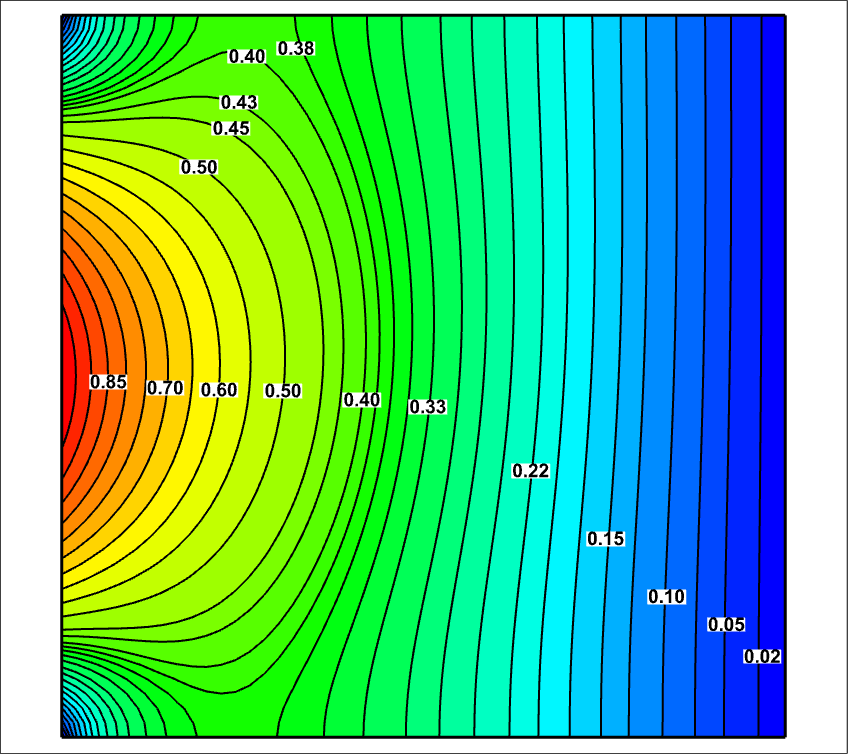}%
    \captionsetup{skip=2pt}%
    \caption{(a) $ Ha=25, Ra=10^3$}
    \label{fig:Ra_10^3_Ha_25_P2}
  \end{subfigure}%
 \begin{subfigure}{0.33\textwidth}        
   \centering
    \includegraphics[width=\textwidth]{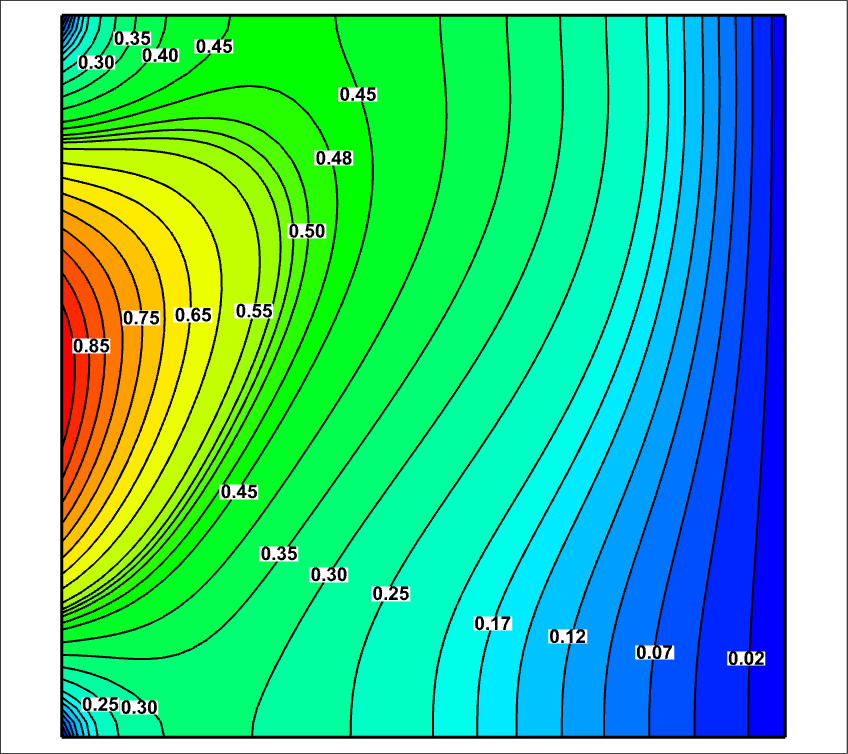}%
    \captionsetup{skip=2pt}%
    \caption{(b) $ Ha=25, Ra=10^4$}
    \label{fig:Ra_10^4_Ha_25_P2}
  \end{subfigure}
   \begin{subfigure}{0.33\textwidth}        
   \centering
    \includegraphics[width=\textwidth]{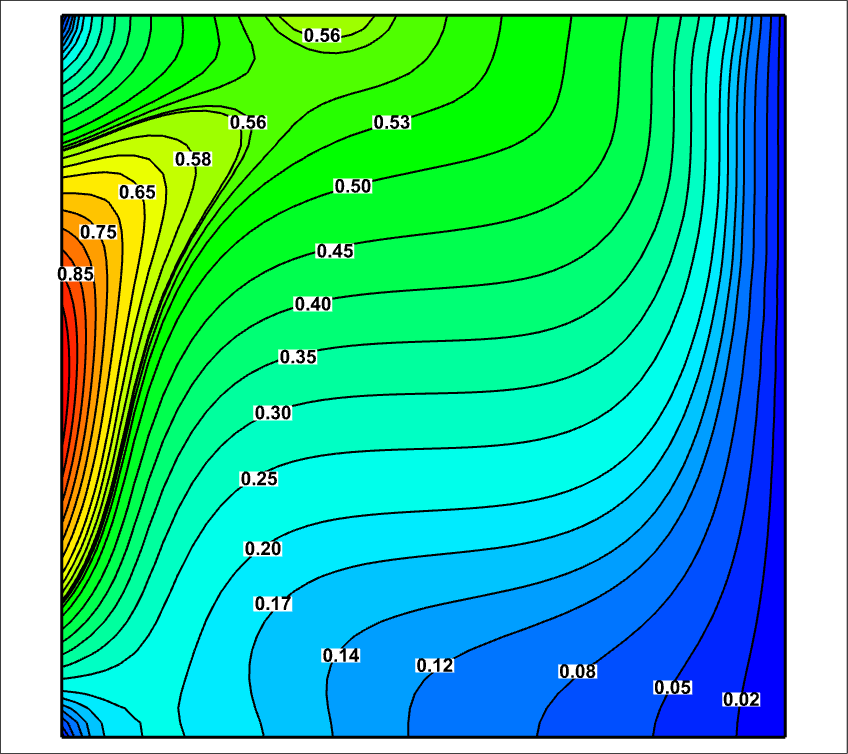}%
    \captionsetup{skip=2pt}%
    \caption{(c) $ Ha=25, Ra=10^5$}
    \label{fig:Ra_10^5_Ha_25_P2}
  \end{subfigure}%
  \hspace*{\fill}

  \vspace*{8pt}%
  \hspace*{\fill}%
  \begin{subfigure}{0.33\textwidth}     
    \centering
    \includegraphics[width=\textwidth]{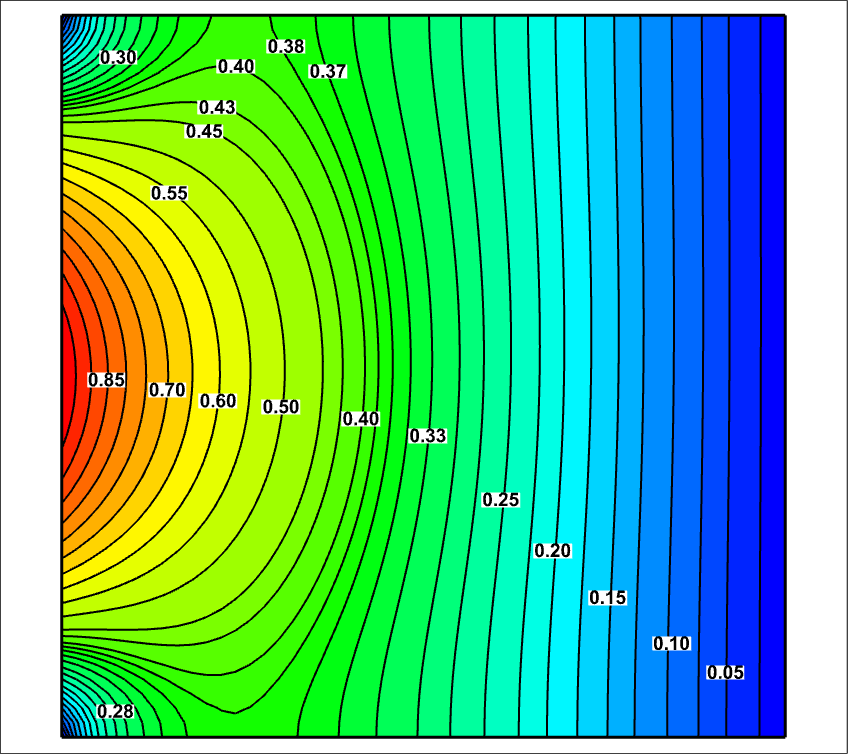}%
    \captionsetup{skip=2pt}%
    \caption{(d) $Ha=50, Ra=10^3$}
    \label{fig:Ra_10^3_Ha_50_P2}
  \end{subfigure}%
 \begin{subfigure}{0.33\textwidth}        
   \centering
    \includegraphics[width=\textwidth]{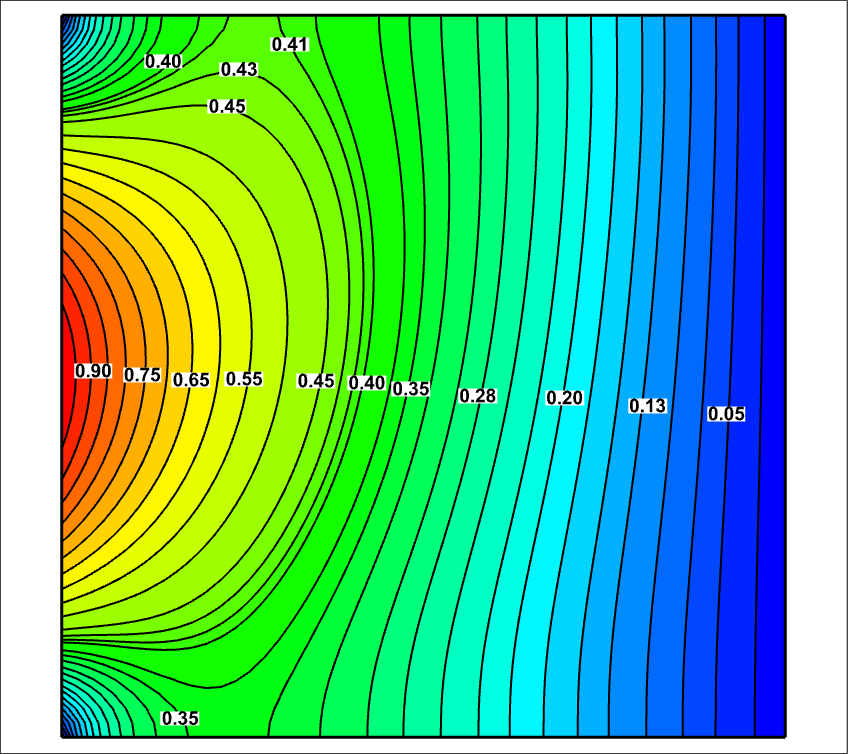}%
    \captionsetup{skip=2pt}%
    \caption{(e) $ Ha=50, Ra=10^4$}
    \label{fig:Ra_10^4_Ha_50_P2}
  \end{subfigure}
   \begin{subfigure}{0.33\textwidth}        
   \centering
    \includegraphics[width=\textwidth]{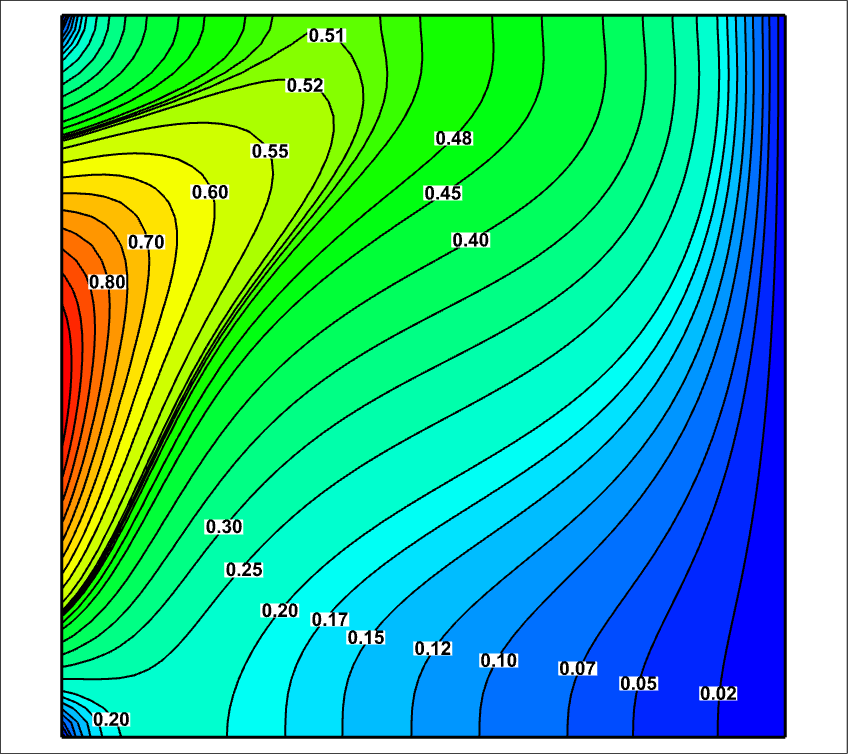}%
    \captionsetup{skip=2pt}%
    \caption{(f) $ Ha=50, Ra=10^5$}
    \label{fig:Ra_10^5_Ha_50_P2.png}
  \end{subfigure}%
  \hspace*{\fill}

  \vspace*{8pt}%
  \hspace*{\fill}%
  \begin{subfigure}{0.33\textwidth}     
    \centering
    \includegraphics[width=\textwidth]{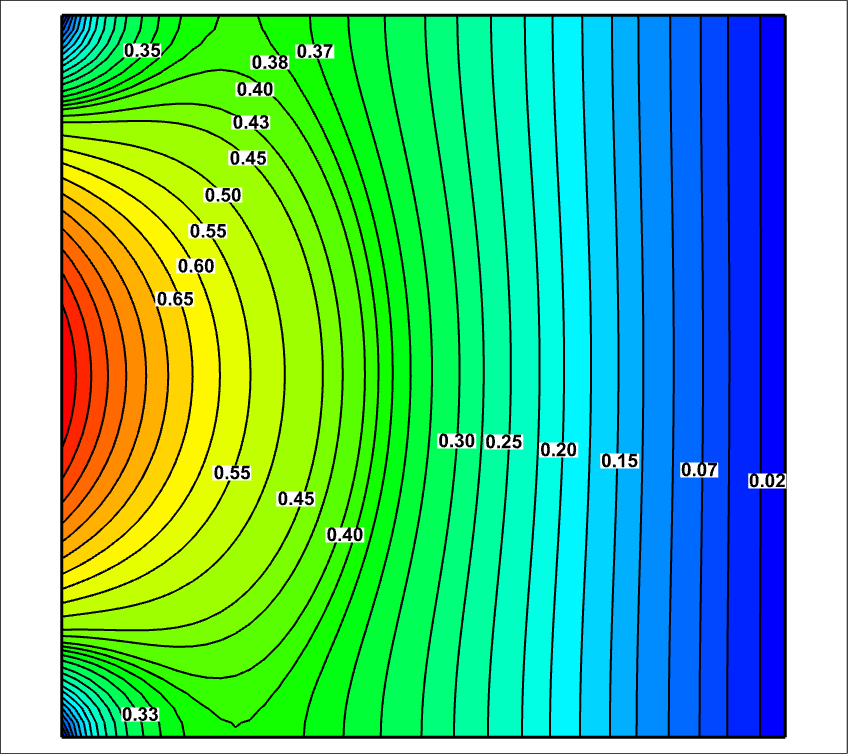}%
    \captionsetup{skip=2pt}%
    \caption{(g) $Ha=100, Ra=10^3$}
    \label{fig:Ra_10^3_Ha_100_P2.png}
  \end{subfigure}%
 \begin{subfigure}{0.33\textwidth}        
   \centering
    \includegraphics[width=\textwidth]{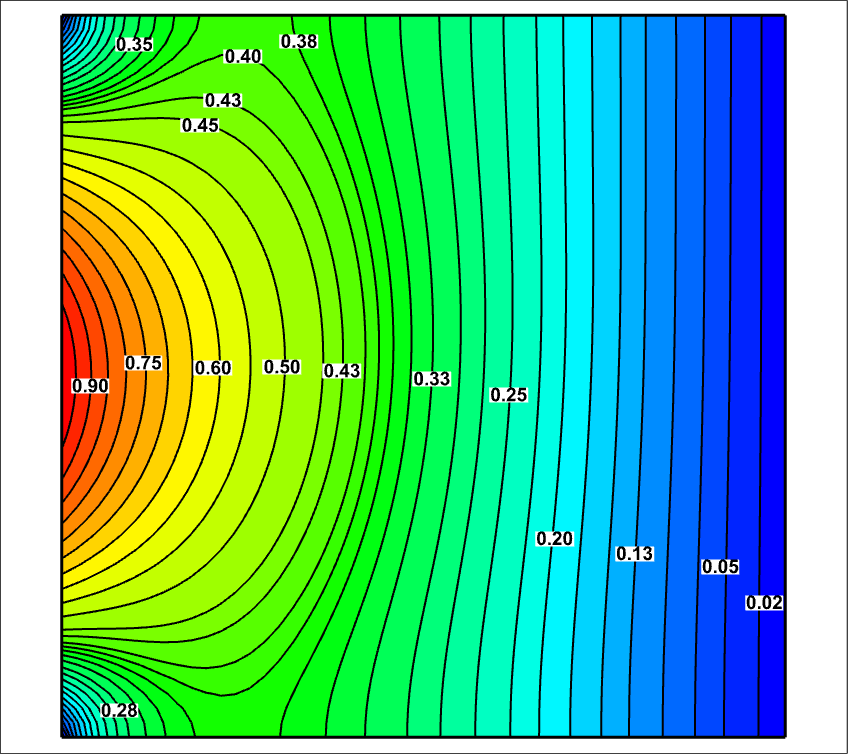}%
    \captionsetup{skip=2pt}%
    \caption{(h) $ Ha=100, Ra=10^4$}
    \label{fig:Ra_10^4_Ha_100_P2.png}
  \end{subfigure}
   \begin{subfigure}{0.33\textwidth}        
   \centering
    \includegraphics[width=\textwidth]{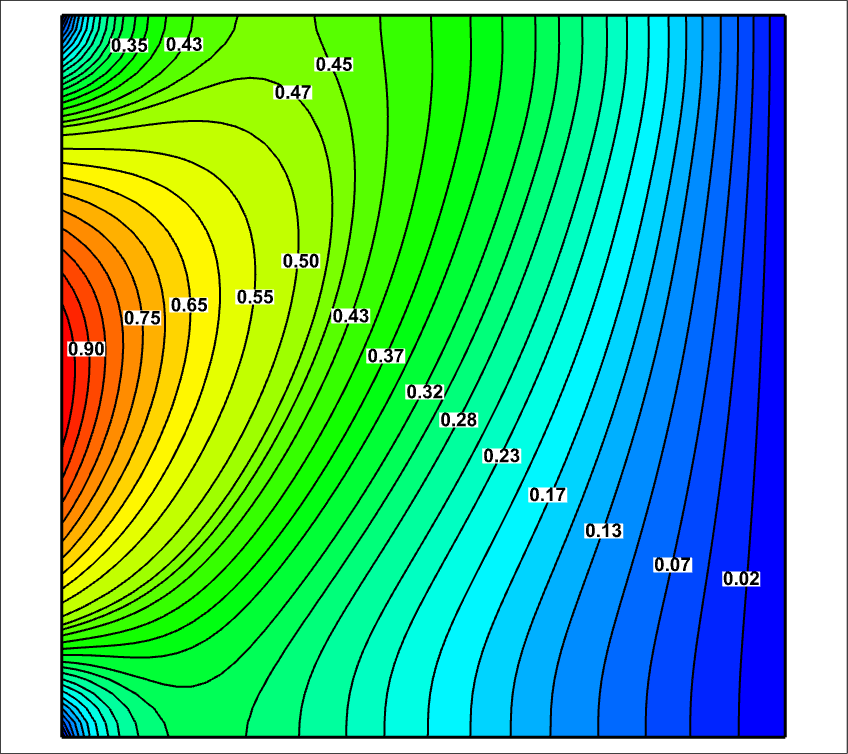}%
    \captionsetup{skip=2pt}%
    \caption{(i) $ Ha=100, Ra=10^5$}
    \label{fig:Ra_10^5_Ha_100_P2.png}
  \end{subfigure}%
  \hspace*{\fill}

  \vspace*{8pt}%
  \hspace*{\fill}%
  \begin{subfigure}{0.33\textwidth}     
    \centering
    \includegraphics[width=\textwidth]{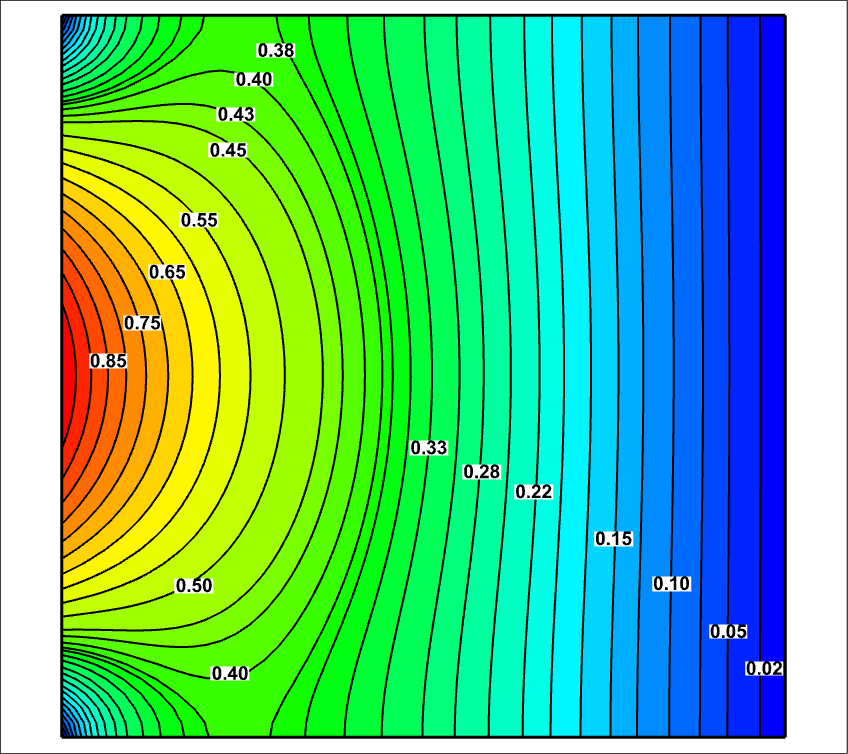}%
    \captionsetup{skip=2pt}%
    \caption{(j) $Ha=150, Ra=10^3$}
    \label{fig:Ra_10^3_Ha_150_P2.png}
  \end{subfigure}%
 \begin{subfigure}{0.33\textwidth}        
   \centering
    \includegraphics[width=\textwidth]{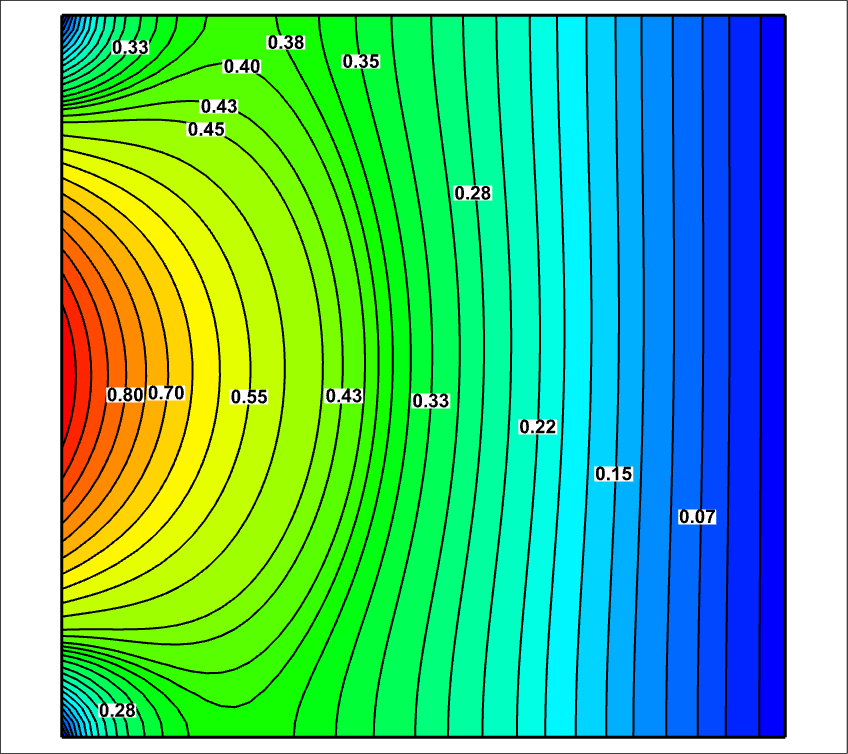}%
    \captionsetup{skip=2pt}%
    \caption{(k) $Ha=150, Ra=10^4$}
    \label{fig:Ra_10^4_Ha_150_P2.png}
  \end{subfigure}
   \begin{subfigure}{0.33\textwidth}        
   \centering
    \includegraphics[width=\textwidth]{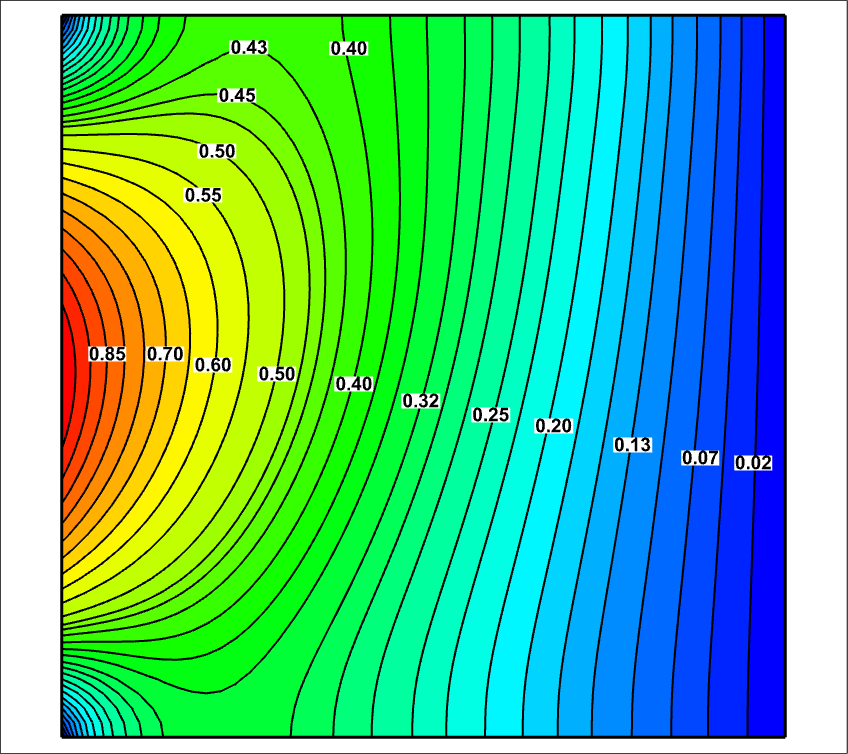}%
    \captionsetup{skip=2pt}%
    \caption{(l) $ Ha=150, Ra=10^5$}
    \label{fig:Ra_10^5_Ha_150_P2.png}
  \end{subfigure}%
  \hspace*{\fill}
  \vspace*{1pt}%
  \hspace*{\fill}%
  \caption{Case 2. Effect of different $Ha$ (Column-wise) and $Ra$ (Row-wise) on isotherm contours (at $z=0.5$ plane) at a fixed $Pr = 0.065$}
  \label{fig:case-2_Isotherm_Contours}
\end{figure}

The schematic representation of Case 2 is depicted in Figure \ref{fig:Sche_diag}(b). In this scenario, the left wall ($x=0$) experiences non-uniform heating characterized by $T_h=\sin(\pi y)$, while the right wall remains cold, and the remaining walls are adiabatic. Due to the non-uniform heating, the temperature distribution on the left wall is not constant. The isotherms are vertical and maximum temperature occurs in the middle of the left wall. The corresponding surface plot of the non-uniform heating condition is presented as a small sub-figure in Figure \ref{fig:Sche_diag}(b). Figure \ref{fig:case-2_Streamlines_Contours} illustrates streamline contours at three distinct planes ($z=0.05, z=0.5, z=0.95)$. The first column examines the impact of $Ha$ while maintaining a constant $Ra=10^3$, the second column explores the effect of $Ha$ with a fixed $Ra=10^4$, and the third column delves into the influence of $Ha$ at a fixed $Ra=10^5$. The impact of $Ra$ can be observed in a row-wise manner. It is evident from the results that the streamfunction values are consistently negative, indicating a clockwise direction of the flow. Additionally, the magnitude of the streamfunction values exhibits an increase with $Ra$ and a decrease with $Ha$, signifying an augmentation in velocity with $Ra$ and a reduction with $Ha$. Notably, two small secondary vortices are observable near the corners of the heated wall for high $Ha$, characterized by a slightly positive value, implying a counterclockwise flow in the left corners.
These small secondary vortices are not observed when the left wall is uniformly heated (Case 1). The shape of the streamline pattern is also influenced by both $Ha$ and $Ra$. When $Ha=25$, a noticeable circular pattern becomes evident at $Ra=10^3, 10^4$. However, at $Ra=10^5$, the primary vortices shift upward due to the prevailing influence of buoyancy forces over viscous forces. Increasing $Ha$ to $50$ exhibits a similar trend across all examined values of $Ra$, displaying a slightly altered and more stretched shape. Further elevating $Ha$ to 100 and 150 results in a noticeable effect, with the main vortex center shifting toward the cavity's center. This shift is attributed to the reduction in fluid velocity accompanying an increase in $Ha$. The streamlines are very similar to Case 1 for $Ha=25$. However, with an increase in $Ha$, the patterns diverge significantly. In Case 1, the pattern becomes highly stretched, whereas in Case 2, for $Ha \ge 50$ and $10^3 \leq Ra \leq 10^5$, the circular-shaped pattern emerges. This striking difference illustrates the effect of non-uniform heating wall condition.
Figure \ref{fig:case-2_Isotherm_Contours_3D} presents isothermal contours in the 3D cavity, while Figure \ref{fig:case-2_Isotherm_Contours} displays 2D isothermal patterns in the middle of the $z$-plane ($z=0.5$). Within the rows of Figures \ref{fig:case-2_Isotherm_Contours_3D} and \ref{fig:case-2_Isotherm_Contours}, the influence of the magnetic field on the temperature field can be observed.
In Figure \ref{fig:case-2_Isotherm_Contours_3D}, the non-uniformly heated wall of the 3D cavity is evident, with maximum heat ($T_h=1$) observed at the middle of the $x=0, 0 \leq z \leq 1$ plane. 
The examination reveals significant alterations in isotherms surfaces and a reduction in the gradient of the boundary layer near the cold wall with an increase in the $Ha$, particularly evident for high $Ra=10^4$ and $10^5$. This indicates that the heat transfer rate, influenced by the temperature gradient, decreases as the magnetic field strength rises. This emphasizes the interplay between magnetic field and convective heat transfer, revealing that higher $Ha$ values introduce a stabilizing effect, influencing fluid motion and altering the shape of isothermal contours. The isotherm patterns remain nearly identical at low $Ra=10^3$ for all $Ha$ values, indicating a minimal influence of the magnetic field at this lower $Ra$ value. The isotherm patterns show significant differences for the case of uniform and non-uniform heating conditions. At $z=0.5$, the temperature contours exhibit diagonal symmetry for the uniform heating condition (Case 1) across all parameters. Whereas, for the non-uniform heating condition (Case 2), an asymmetrical isotherm pattern emerges. 

Along the row of the Figures \ref{fig:case-2_Isotherm_Contours_3D} and \ref{fig:case-2_Isotherm_Contours} (i.e., for fixed $Ha$ with varying $Ra$), we can see the effect of the $Ra$ on the temperature field. An increase in $Ra$ increases the heat transfer rate, leading to an elevated temperature difference between the hot fluid near the left wall and the cold fluid near the right wall within the cavity. At the fixed $Ha=25$, We can see the most pronounced impact of the $Ra$. Here, the isotherm contours exhibit varying shapes, patterns, and values relative to the $Ra$. The isotherms display a twisted configuration, indicating a departure from convective effects and the prevalence of conduction effects within the enclosure. At a lower $Ra$ value ($Ra=10^3$), the isotherm lines run almost parallel near the cold walls, resulting in a reduced temperature gradient near the right wall. These parallel isotherms are also present in Case 1 at $Ra=10^3$, spanning the entire cavity instead of being confined solely near the cold wall due to the uniformly heated left wall.
As $Ra$ values increase, the isothermal lines for Case-2 display greater curvature for any specific $Ha$, leading to an increase in the heat transfer rate. However, this effect decreases as $Ha$ increases. This indicates that conductive heat transfer predominates in the enclosure at high $Ra$.


\subsubsection{Case 3}
\begin{figure}[htbp]
 \centering
 \vspace*{1pt}%
 \hspace*{\fill}%
\begin{subfigure}{0.33\textwidth}     
    \centering
    \includegraphics[width=\textwidth]{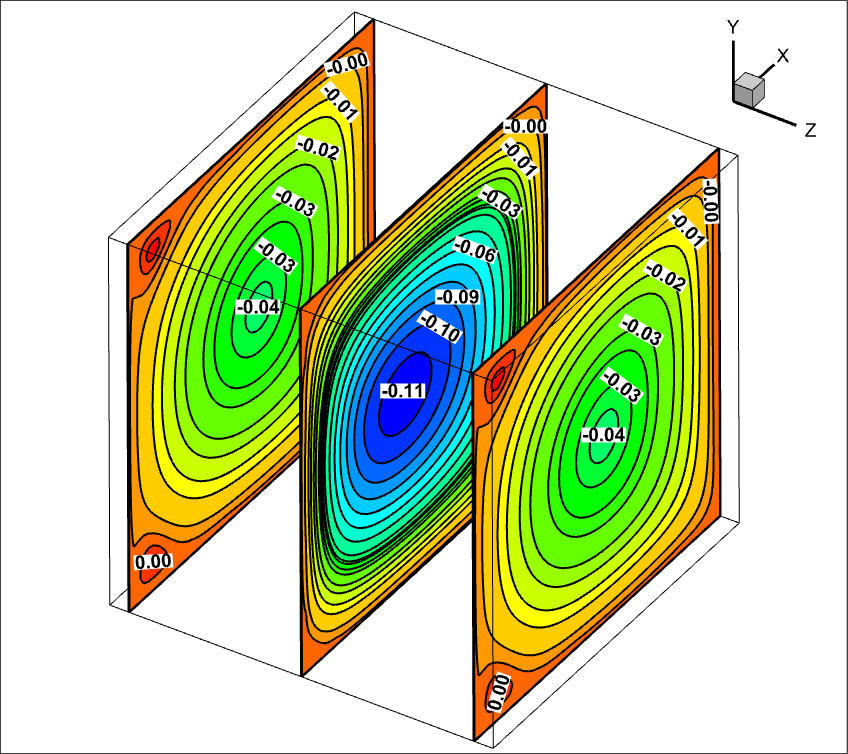}%
    \captionsetup{skip=2pt}%
    \caption{(a) $ Ha=25, Ra=10^3$}
    \label{fig:Ra_10^3_Ha_25_P3_Streamlines}
  \end{subfigure}%
 \begin{subfigure}{0.33\textwidth}        
   \centering
    \includegraphics[width=\textwidth]{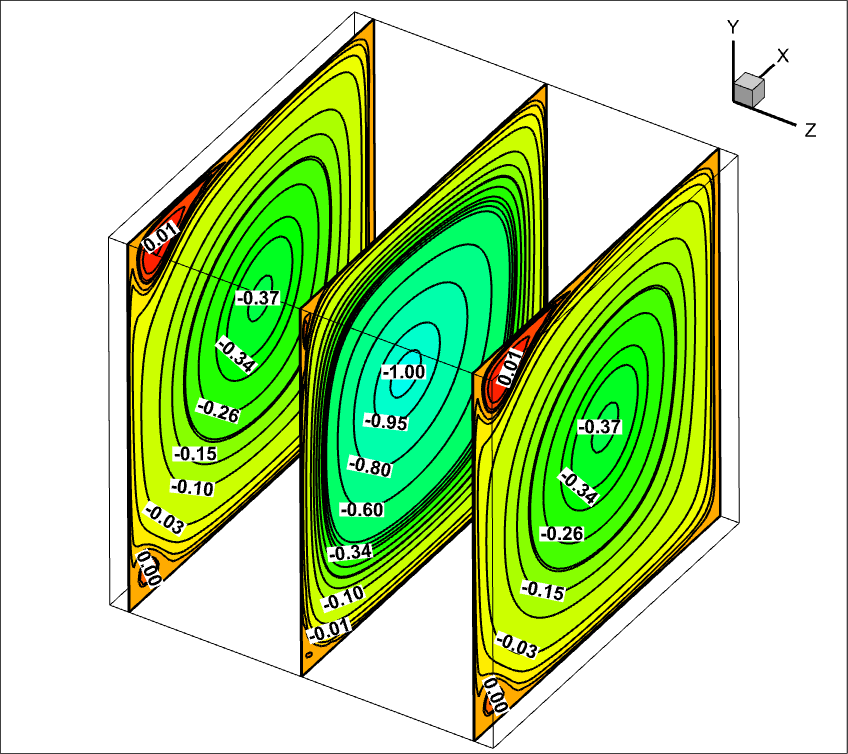}%
    \captionsetup{skip=2pt}%
    \caption{(b) $ Ha=25, Ra=10^4$}
    \label{fig:Ra_10^4_Ha_25_P3_Streamlines}
  \end{subfigure}
   \begin{subfigure}{0.33\textwidth}        
   \centering
    \includegraphics[width=\textwidth]{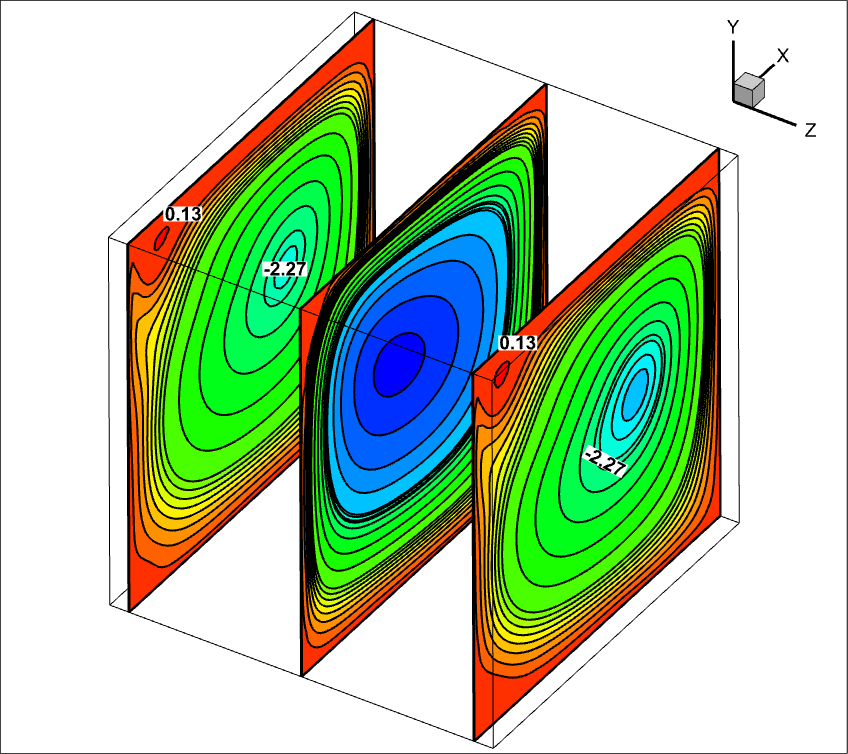}%
    \captionsetup{skip=2pt}%
    \caption{(c) $ Ha=25, Ra=10^5$}
    \label{fig:Ra_10^5_Ha_25_P3_Streamlines.png}
  \end{subfigure}%
  \hspace*{\fill}

  \vspace*{8pt}%
  \hspace*{\fill}%
  \begin{subfigure}{0.33\textwidth}     
    \centering
    \includegraphics[width=\textwidth]{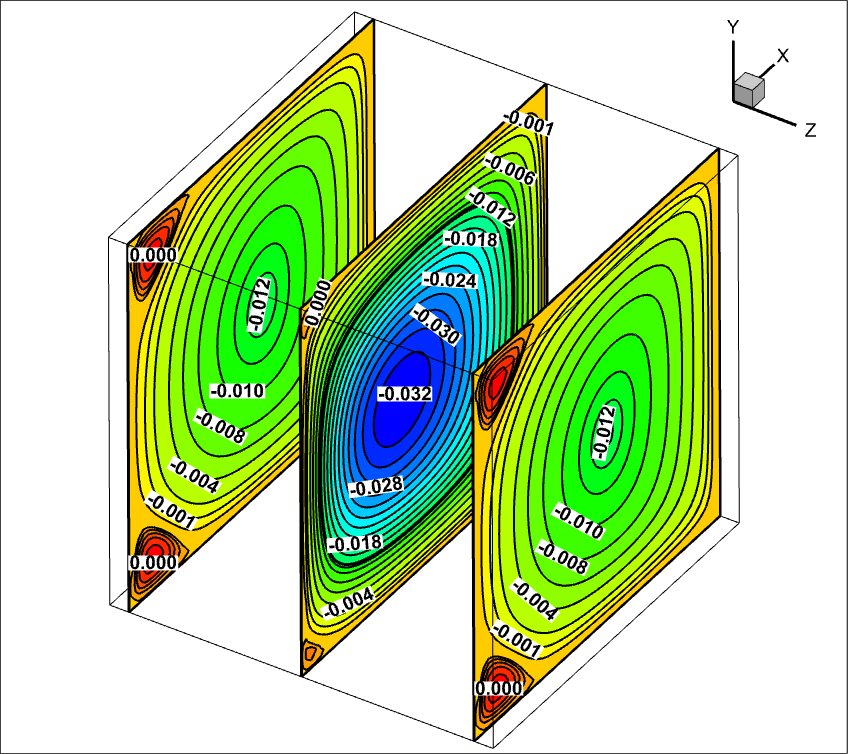}%
    \captionsetup{skip=2pt}%
    \caption{(d) $ Ha=50, Ra=10^3$}
    \label{fig:Ra_10^3_Ha_50_P3_Streamlines}
  \end{subfigure}%
 \begin{subfigure}{0.33\textwidth}        
   \centering
    \includegraphics[width=\textwidth]{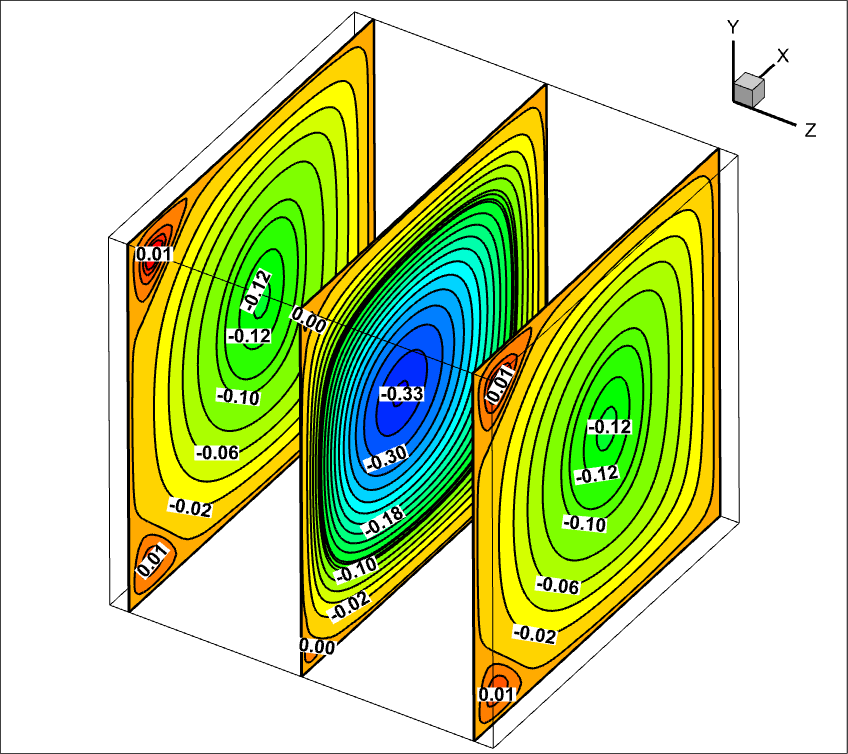}%
    \captionsetup{skip=2pt}%
    \caption{(e) $Ha=50, Ra=10^4$}
    \label{fig:Ra_10^4_Ha_50_P3_Streamlines}
  \end{subfigure}
   \begin{subfigure}{0.33\textwidth}        
   \centering
    \includegraphics[width=\textwidth]{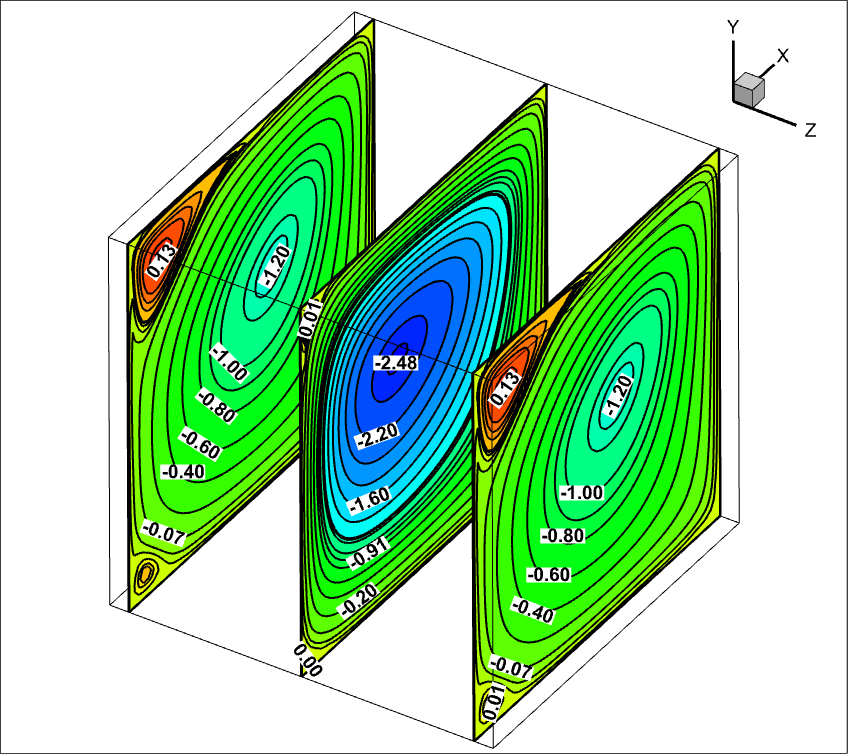}%
    \captionsetup{skip=2pt}%
    \caption{(f) $ Ha=50, Ra=10^5$}
    \label{fig:Ra_10^5_Ha_50_P3_Streamlines3}
  \end{subfigure}%
  \hspace*{\fill}

  \vspace*{8pt}%
  \hspace*{\fill}%
  \begin{subfigure}{0.33\textwidth}     
    \centering
    \includegraphics[width=\textwidth]{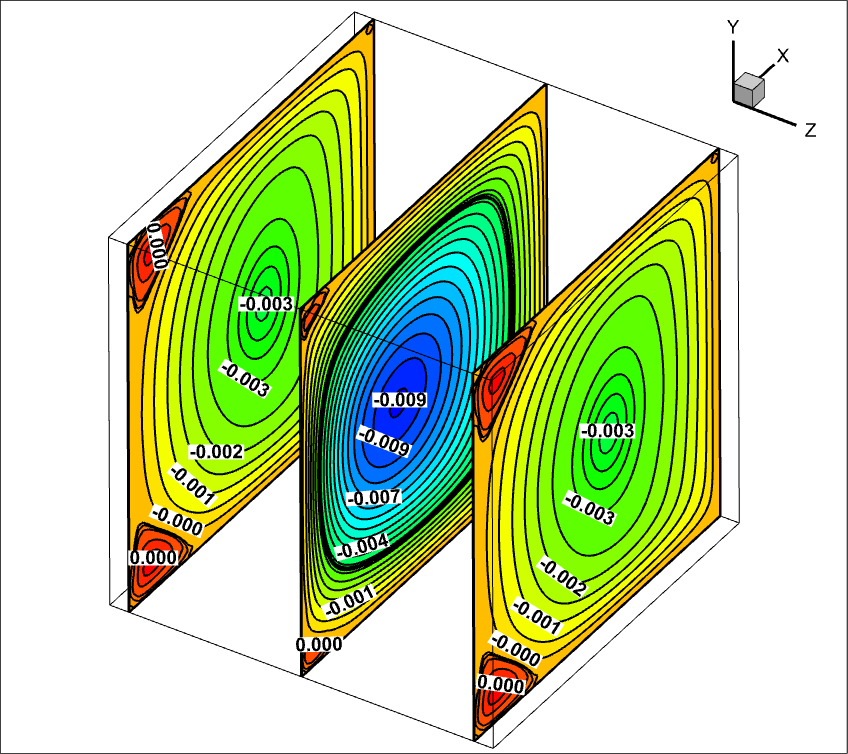}%
    \captionsetup{skip=2pt}%
    \caption{(g) $ Ha=100, Ra=10^3$}
    \label{fig:Ra_10^3_Ha_100_P3_Streamlines}
  \end{subfigure}%
 \begin{subfigure}{0.33\textwidth}        
   \centering
    \includegraphics[width=\textwidth]{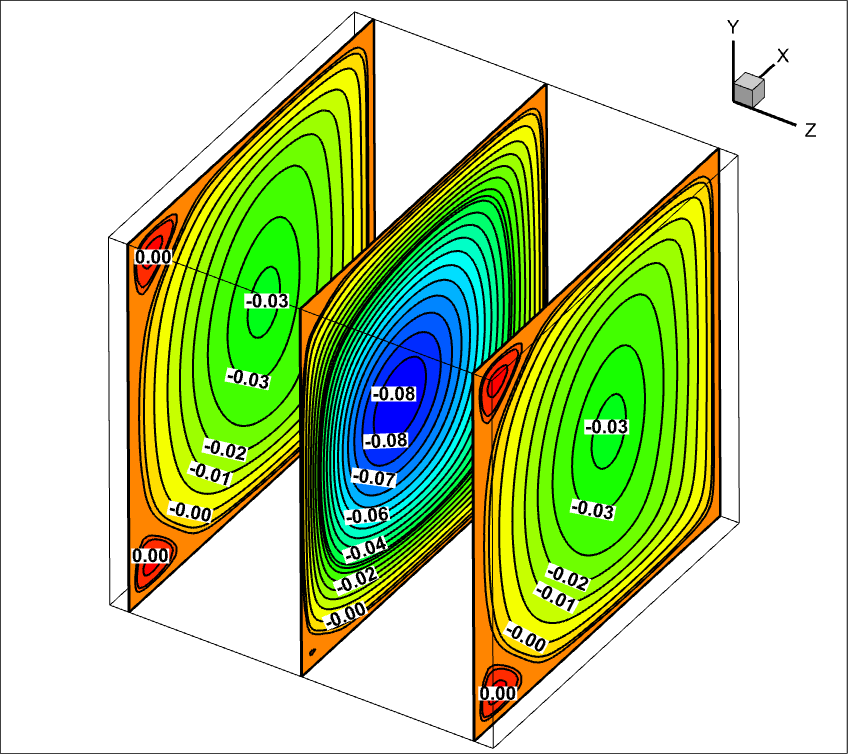}%
    \captionsetup{skip=2pt}%
    \caption{(h) $ Ha=100, Ra=10^4$}
    \label{fig:Ra_10^4_Ha_100_P3_Streamlines}
  \end{subfigure}
   \begin{subfigure}{0.33\textwidth}        
   \centering
    \includegraphics[width=\textwidth]{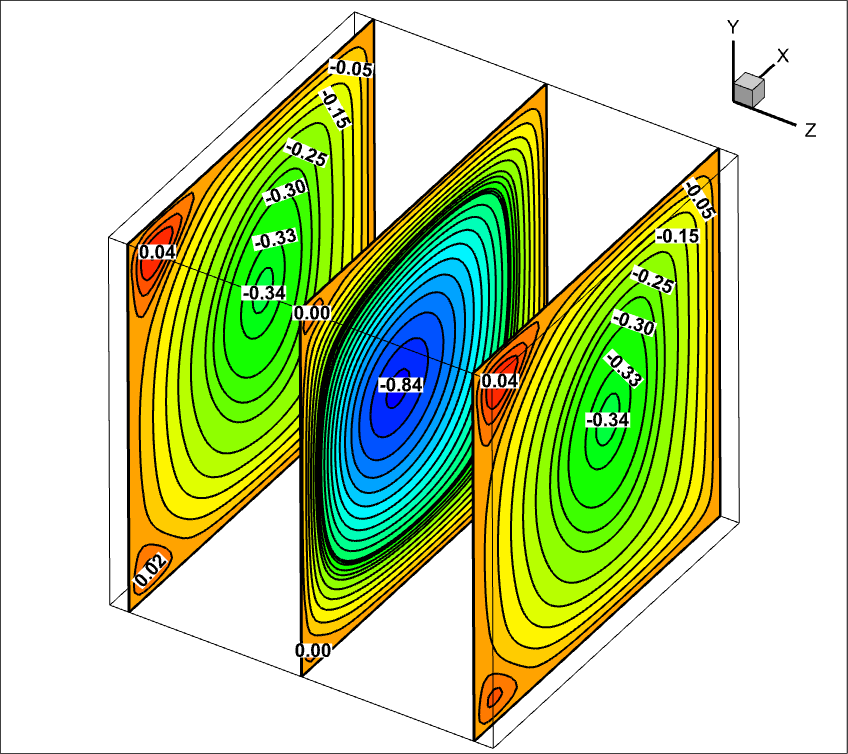}%
    \captionsetup{skip=2pt}%
    \caption{(i) $ Ha=100, Ra=10^5$}
    \label{fig:Ra_10^5_Ha_100_P3_Streamlines}
  \end{subfigure}%
  \hspace*{\fill}

  \vspace*{8pt}%
  \hspace*{\fill}%
  \begin{subfigure}{0.33\textwidth}     
    \centering
    \includegraphics[width=\textwidth]{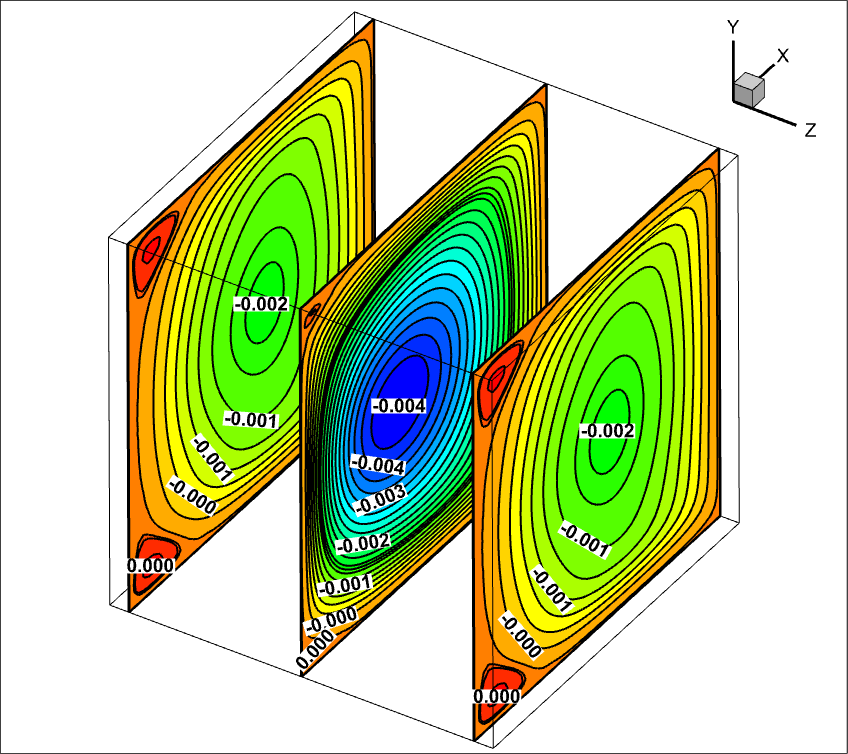}%
    \captionsetup{skip=2pt}%
    \caption{(j) $ Ha=150, Ra=10^3$}
    \label{fig:Ra_10^3_Ha_150_P3_Streamlines}
  \end{subfigure}%
 \begin{subfigure}{0.33\textwidth}        
   \centering
    \includegraphics[width=\textwidth]{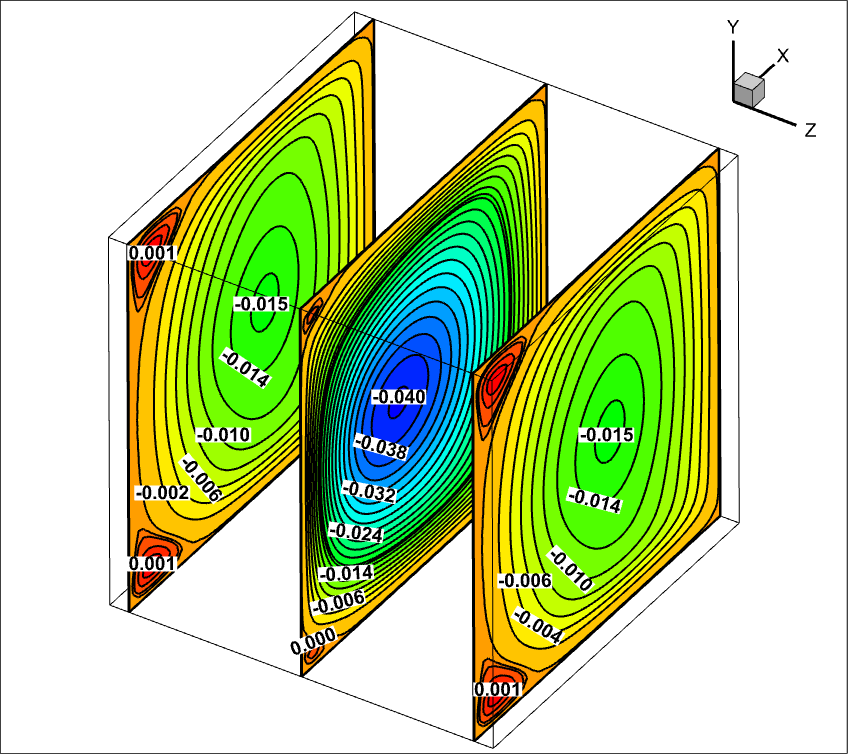}%
    \captionsetup{skip=2pt}%
    \caption{(k) $ Ha=150, Ra=10^4$}
    \label{fig:Ra_10^4_Ha_150_P3_Streamlines}
  \end{subfigure}
   \begin{subfigure}{0.33\textwidth}        
   \centering
    \includegraphics[width=\textwidth]{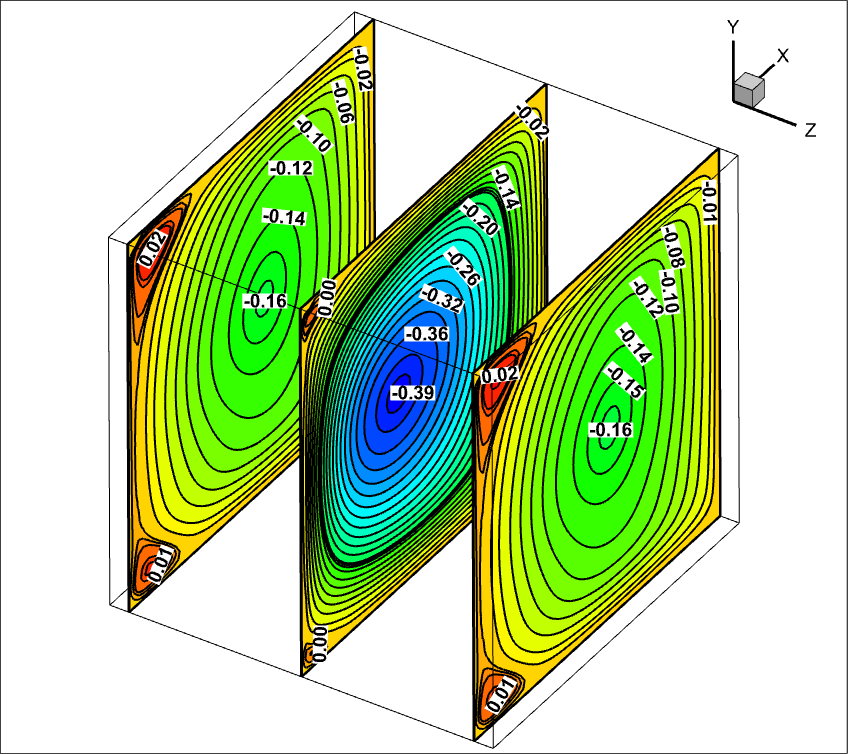}%
    \captionsetup{skip=2pt}%
    \caption{(l) $ Ha=150, Ra=10^5$}
    \label{fig:Ra_10^5_Ha_150_P3_Streamlines}
  \end{subfigure}%
  \hspace*{\fill}
  \vspace*{2pt}%
  \hspace*{\fill}%
  \caption{Case 3. Effect of different $Ha$ (Column-wise) and $Ra$ (Row-wise) on streamlines pattern at different $xy$ planes $(z=0.05, z=0.5, z=0.95)$ with fixed $Pr = 0.065$}
  \label{fig:case-3_Streamlines_Contours}
\end{figure}

\begin{figure}[htbp]
 \centering
 \vspace*{0pt}%
 \hspace*{\fill}%
\begin{subfigure}{0.33\textwidth}     
    \centering
    \includegraphics[width=\textwidth]{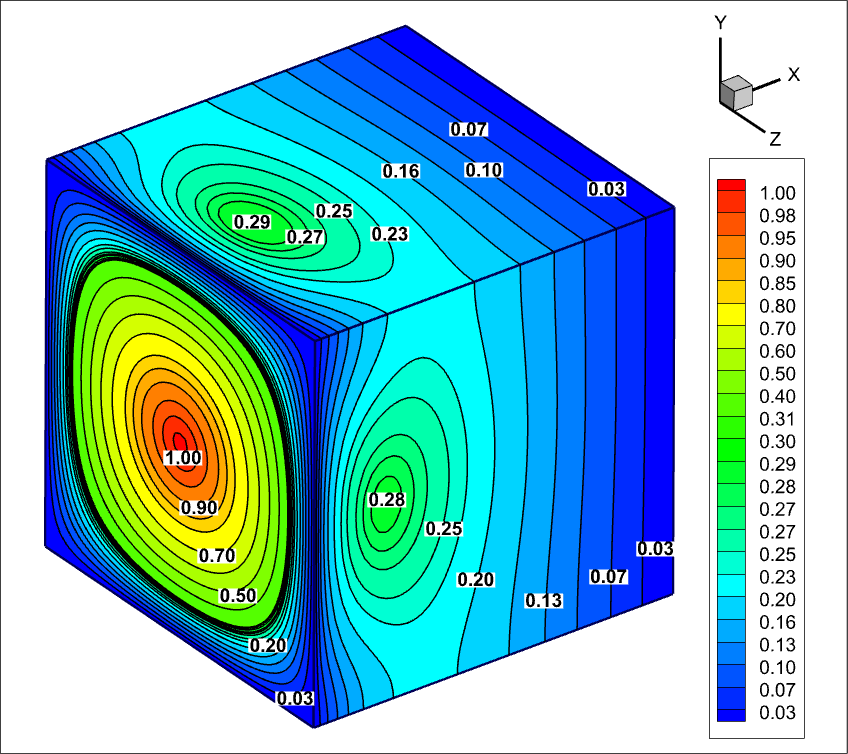}%
    \captionsetup{skip=2pt}%
    \caption{(a) $ Ha=25, Ra=10^3$}
    \label{fig:Ra_10^3_Ha_25_P3}
  \end{subfigure}%
 \begin{subfigure}{0.33\textwidth}        
   \centering
    \includegraphics[width=\textwidth]{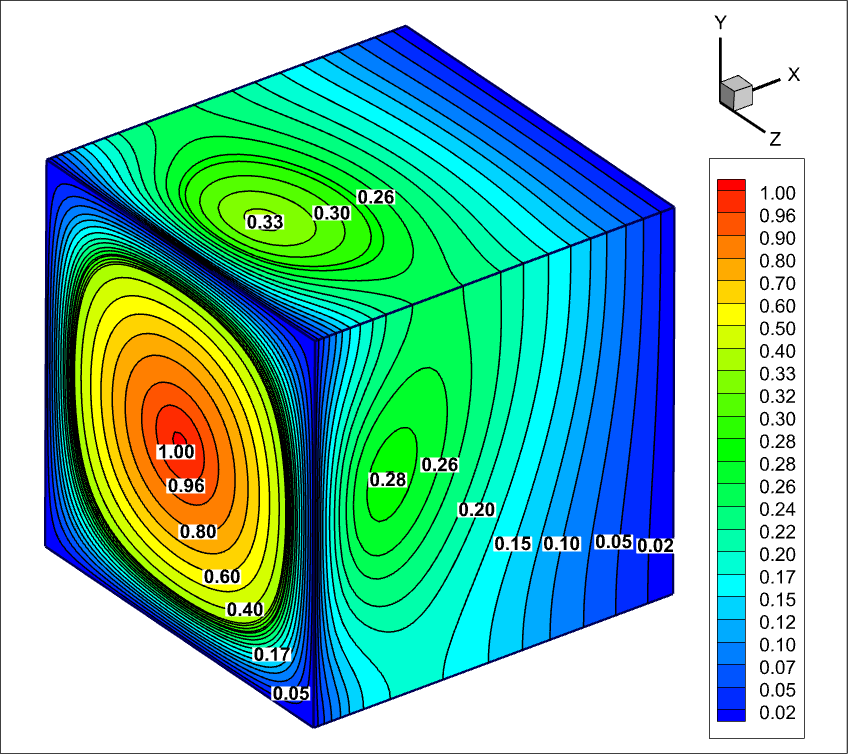}%
    \captionsetup{skip=2pt}%
    \caption{(b) $ Ha=25, Ra=10^3$}
    \label{fig:Ra_10^4_Ha_25_P3}
  \end{subfigure}
   \begin{subfigure}{0.33\textwidth}        
   \centering
    \includegraphics[width=\textwidth]{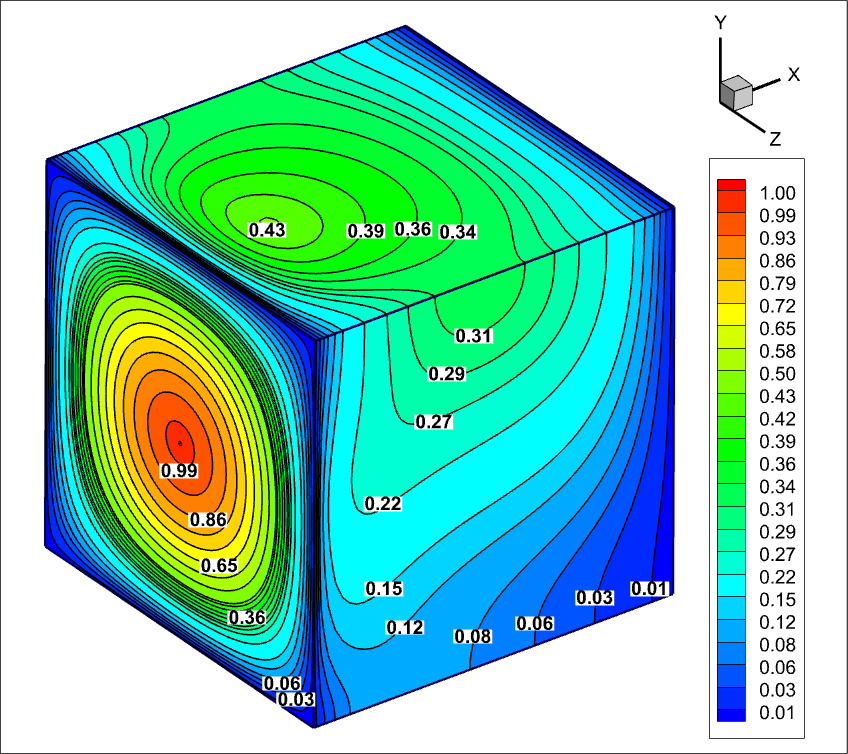}%
    \captionsetup{skip=2pt}%
    \caption{(c)  $Ha=25, Ra=10^5$}
    \label{fig:Ra_10^5_Ha_25_P3}
  \end{subfigure}%
  \hspace*{\fill}

  \vspace*{8pt}%
  \hspace*{\fill}%
  \begin{subfigure}{0.33\textwidth}     
    \centering
    \includegraphics[width=\textwidth]{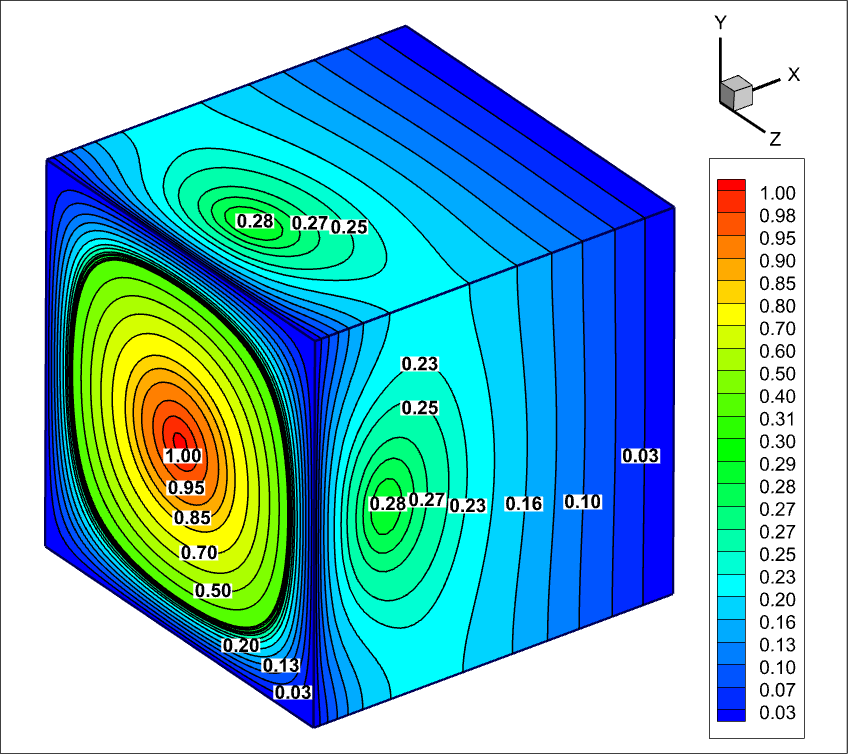}%
    \captionsetup{skip=2pt}%
    \caption{(d)  $Ha=50, Ra=10^3$}
    \label{fig:Ra_10^3_Ha_50_P3}
  \end{subfigure}%
 \begin{subfigure}{0.33\textwidth}        
   \centering
    \includegraphics[width=\textwidth]{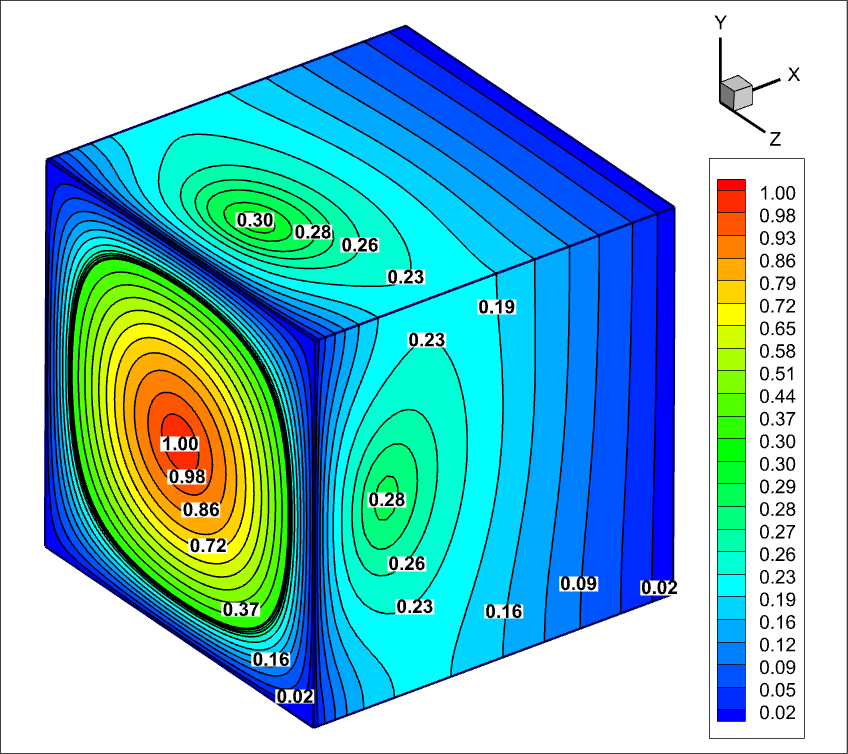}%
    \captionsetup{skip=2pt}%
    \caption{(e) $ Ha=50, Ra=10^4$}
    \label{fig:Ra_10^4_Ha_50_P3}
  \end{subfigure}
   \begin{subfigure}{0.33\textwidth}        
   \centering
    \includegraphics[width=\textwidth]{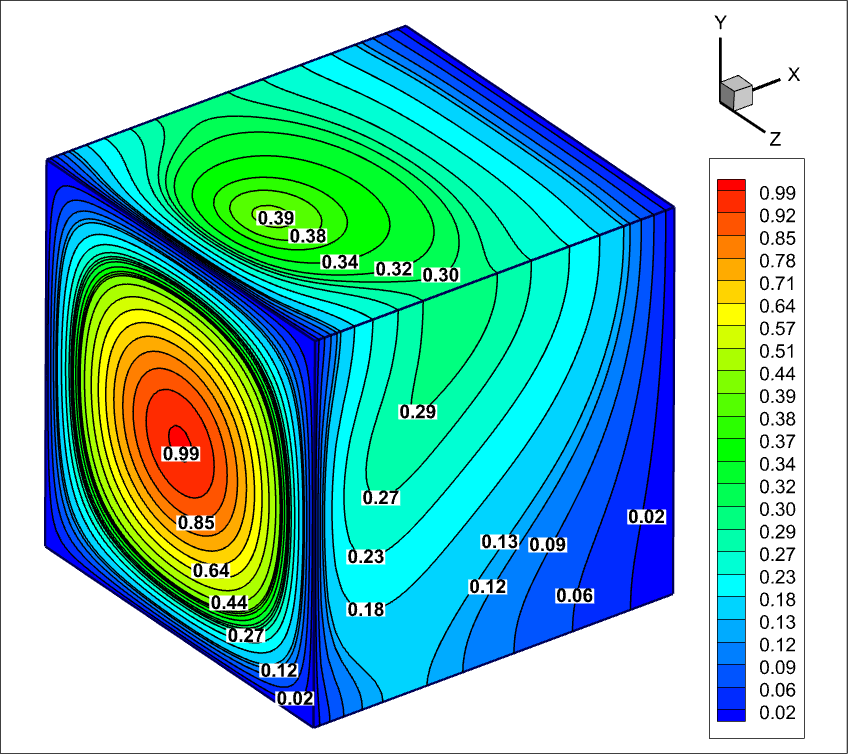}%
    \captionsetup{skip=2pt}%
    \caption{(f) $ Ha=50, Ra=10^5$}
    \label{fig:Ra_10^5_Ha_50_P3.png}
  \end{subfigure}%
  \hspace*{\fill}

  \vspace*{8pt}%
  \hspace*{\fill}%
  \begin{subfigure}{0.33\textwidth}     
    \centering
    \includegraphics[width=\textwidth]{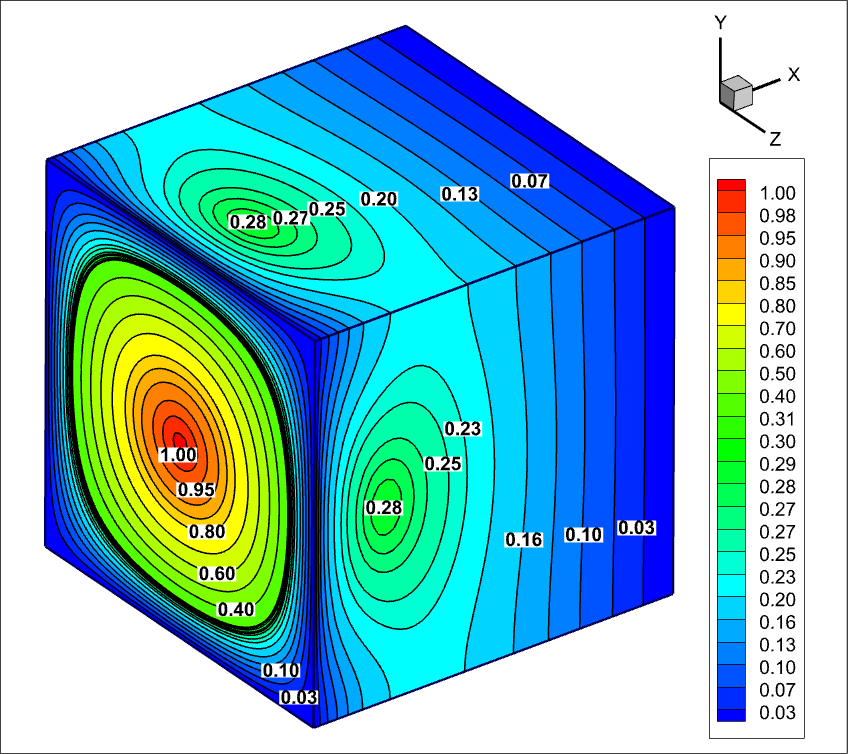}%
    \captionsetup{skip=2pt}%
    \caption{(g) $ Ha=100, Ra=10^3$}
    \label{fig:Ra_10^3_Ha_100_P3.png}
  \end{subfigure}%
 \begin{subfigure}{0.33\textwidth}        
   \centering
    \includegraphics[width=\textwidth]{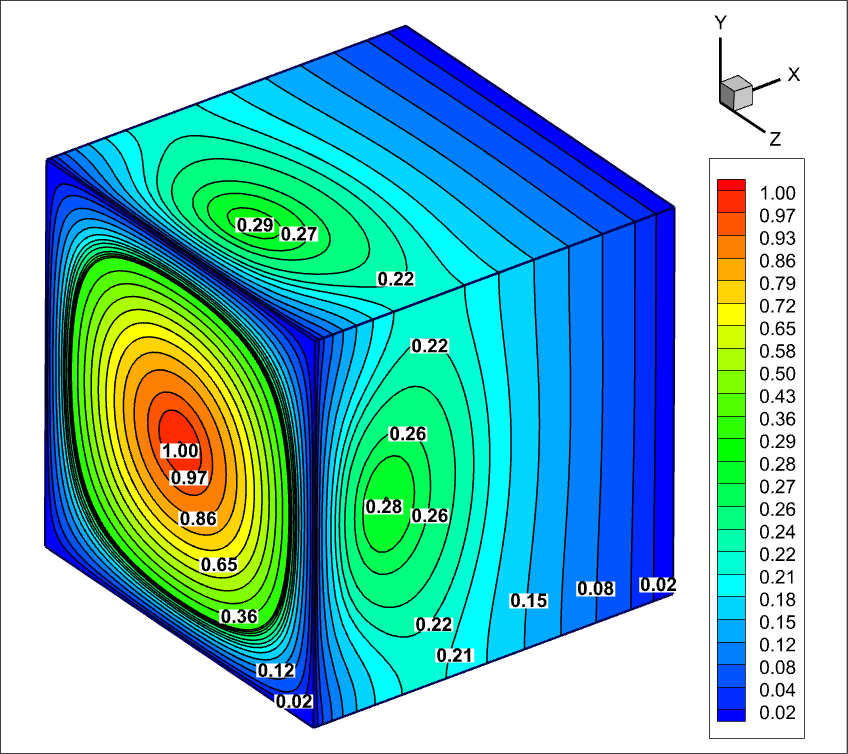}%
    \captionsetup{skip=2pt}%
    \caption{(h) $ Ha=100, Ra=10^4$}
    \label{fig:Ra_10^4_Ha_100_P3.png}
  \end{subfigure}
   \begin{subfigure}{0.33\textwidth}        
   \centering
    \includegraphics[width=\textwidth]{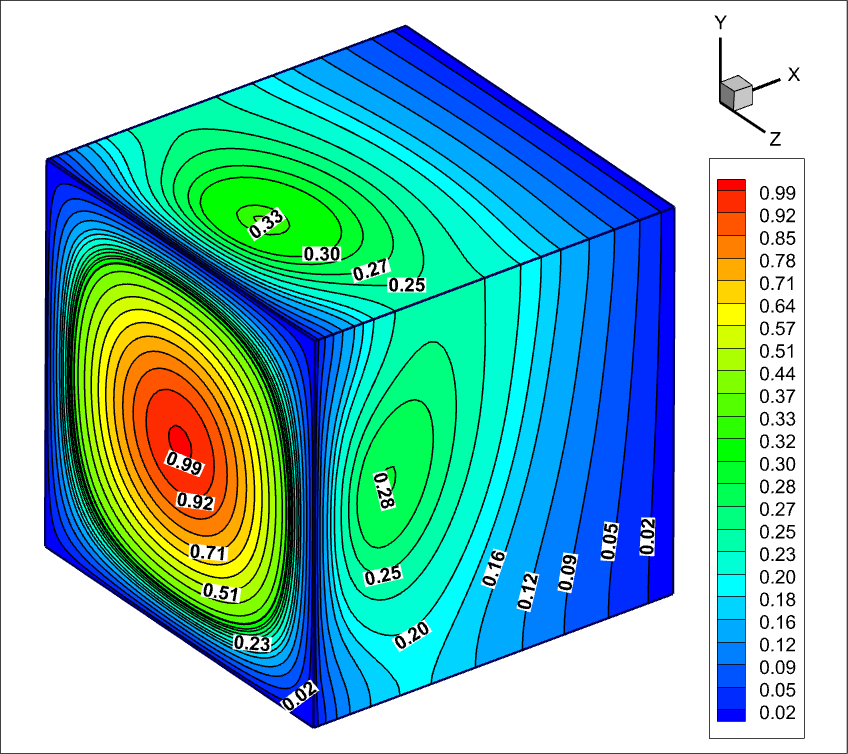}%
    \captionsetup{skip=2pt}%
    \caption{(i) $Ha=100, Ra=10^5$}
    \label{fig:Ra_10^5_Ha_100_P3.png}
  \end{subfigure}%
  \hspace*{\fill}

  \vspace*{8pt}%
  \hspace*{\fill}%
  \begin{subfigure}{0.33\textwidth}     
    \centering
    \includegraphics[width=\textwidth]{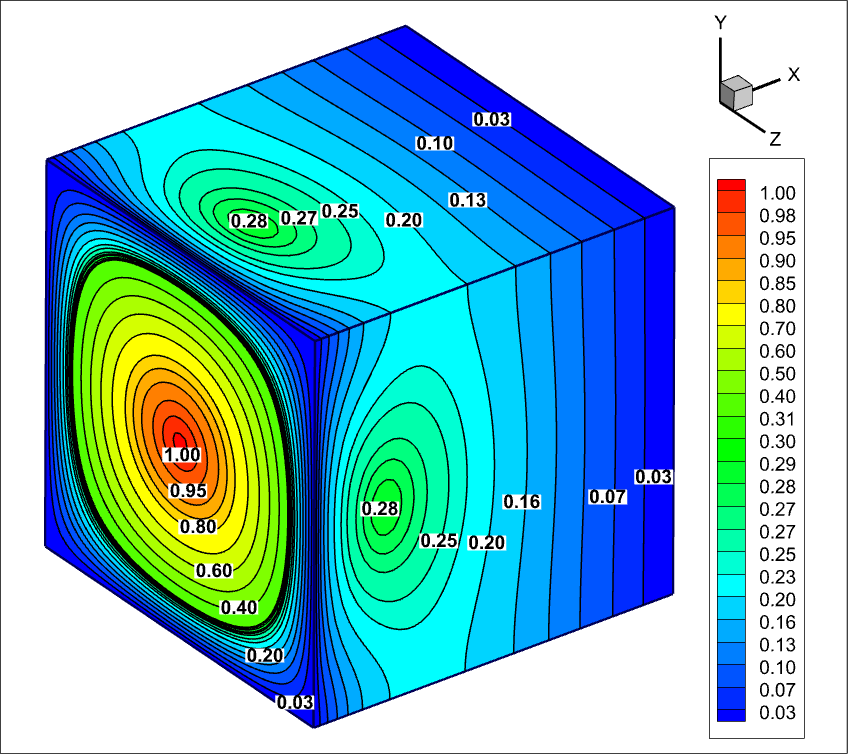}%
    \captionsetup{skip=2pt}%
    \caption{(j) $ Ha=150, Ra=10^3$}
    \label{fig:Ra_10^3_Ha_150_P3.png}
  \end{subfigure}%
 \begin{subfigure}{0.33\textwidth}        
   \centering
    \includegraphics[width=\textwidth]{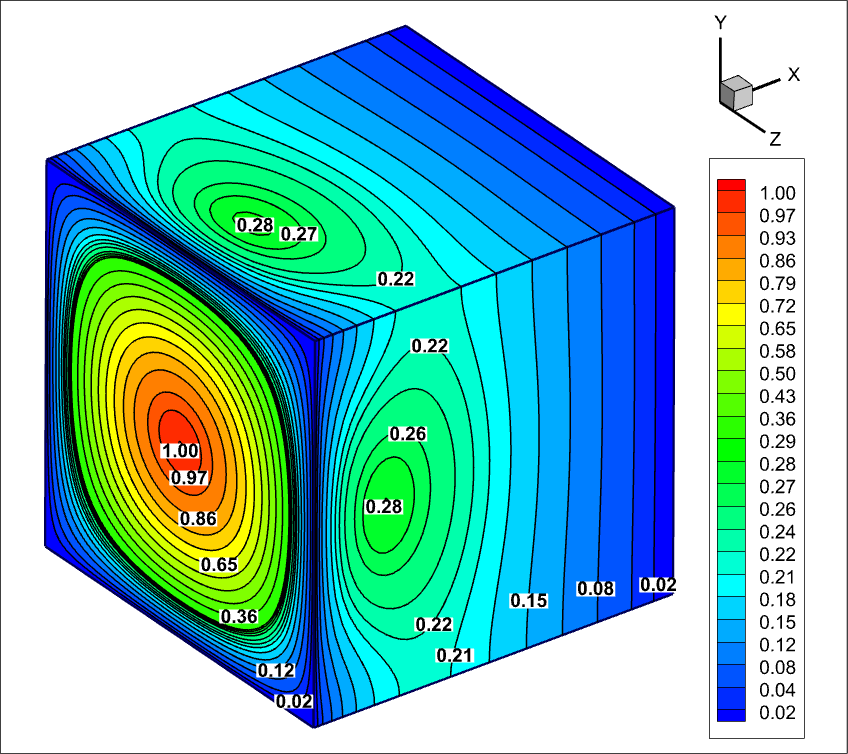}%
    \captionsetup{skip=2pt}%
    \caption{(k) $Ha=150, Ra=10^4$}
    \label{fig:Ra_10^4_Ha_150_P3.png}
  \end{subfigure}
   \begin{subfigure}{0.33\textwidth}        
   \centering
    \includegraphics[width=\textwidth]{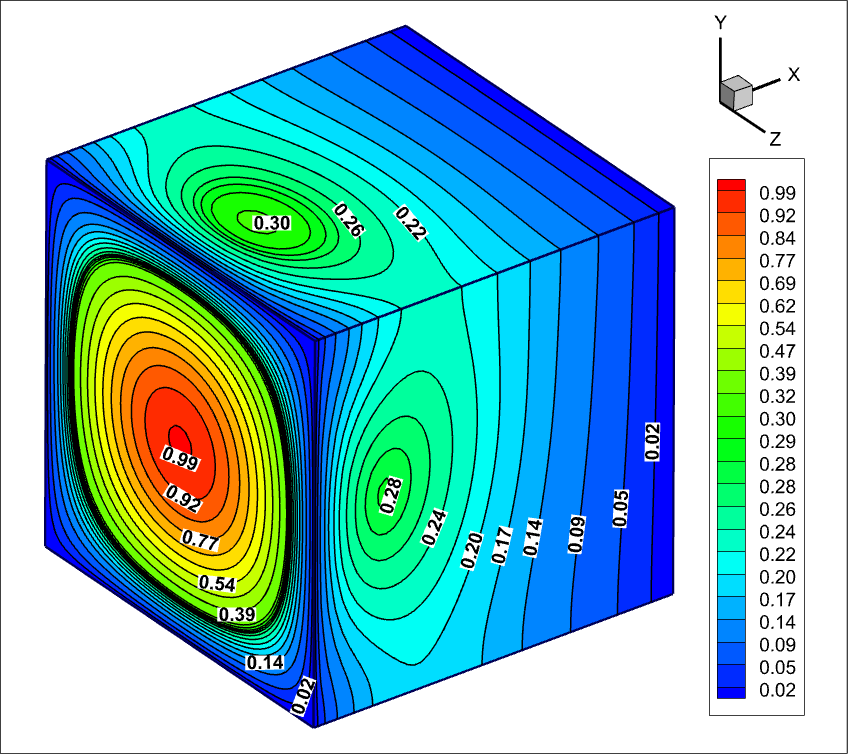}%
    \captionsetup{skip=2pt}%
    \caption{(l) $ Ha=150, Ra=10^5,$}
    \label{fig:Ra_10^5_Ha_150_P3.png}
  \end{subfigure}%
  \hspace*{\fill}
  \vspace*{2pt}%
  \hspace*{\fill}%
  \caption{Case 3. Effect of different $Ha$ (Column-wise) and $Ra$ (Row-wise) on isotherm contours at a fixed $Pr = 0.065$}
  \label{fig:case-3_Isotherm_Contours}
\end{figure}

\begin{figure}[htbp]
 \centering
 \vspace*{0pt}%
 \hspace*{\fill}%
\begin{subfigure}{0.33\textwidth}     
    \centering
    \includegraphics[width=\textwidth]{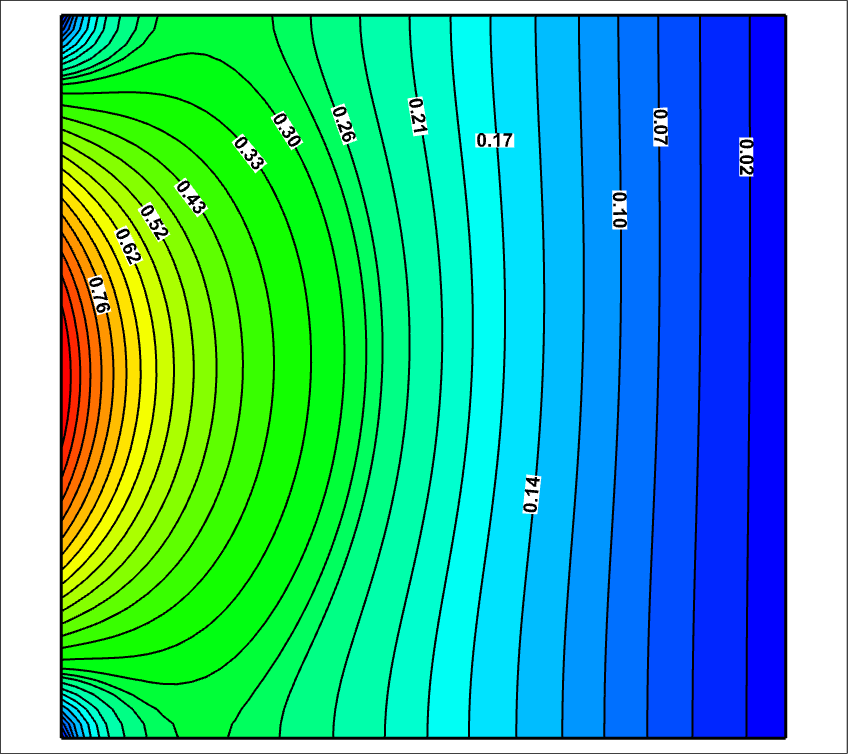}%
    \captionsetup{skip=2pt}%
    \caption{(a) $Ha=25, Ra=10^3$}
    \label{fig:2d_isotherm_Ra_10^3_Ha_25_P3}
  \end{subfigure}%
 \begin{subfigure}{0.33\textwidth}        
   \centering
    \includegraphics[width=\textwidth]{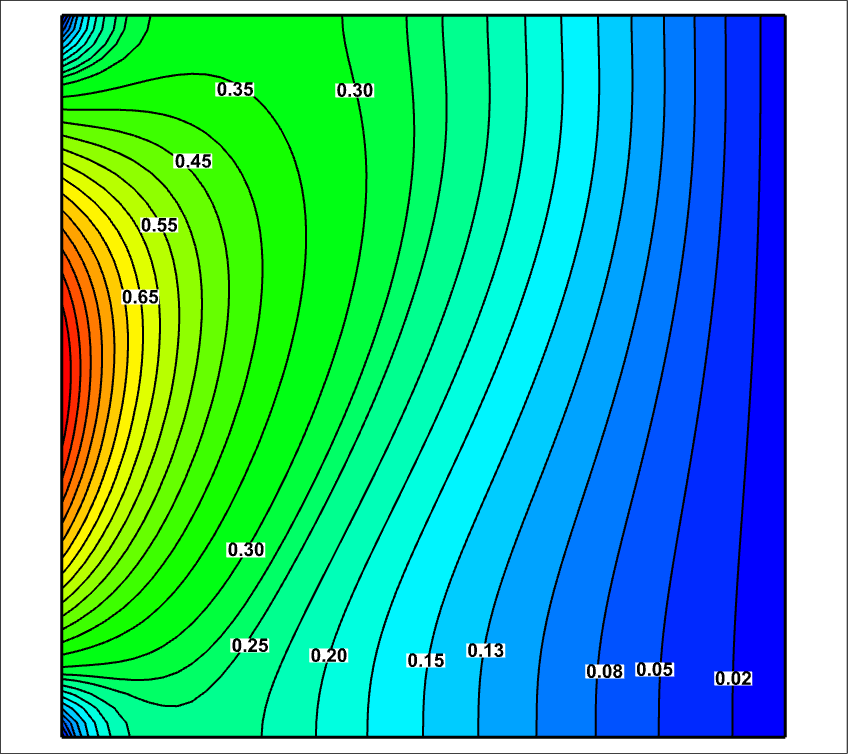}%
    \captionsetup{skip=2pt}%
    \caption{(b) $ Ha=25, Ra=10^4$}
    \label{fig:2d_isotherm_Ra_10^4_Ha_25_P3}
  \end{subfigure}
   \begin{subfigure}{0.33\textwidth}        
   \centering
    \includegraphics[width=\textwidth]{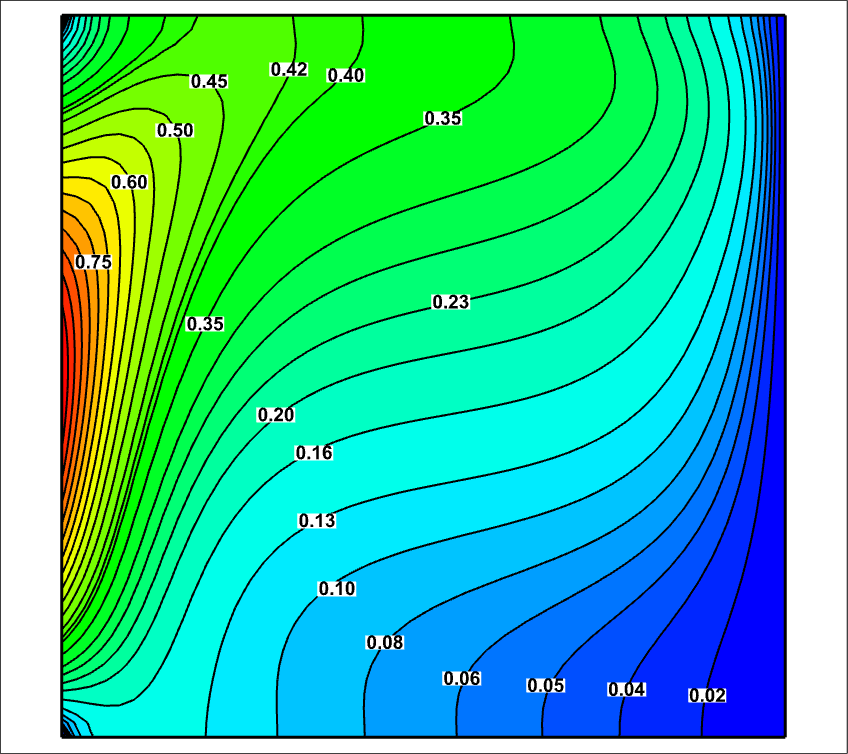}%
    \captionsetup{skip=2pt}%
    \caption{(c) $ Ha=25, Ra=10^5$}
    \label{fig:2d_isotherm_Ra_10^5_Ha_25_P3}
  \end{subfigure}%
  \hspace*{\fill}

  \vspace*{8pt}%
  \hspace*{\fill}%
  \begin{subfigure}{0.33\textwidth}     
    \centering
    \includegraphics[width=\textwidth]{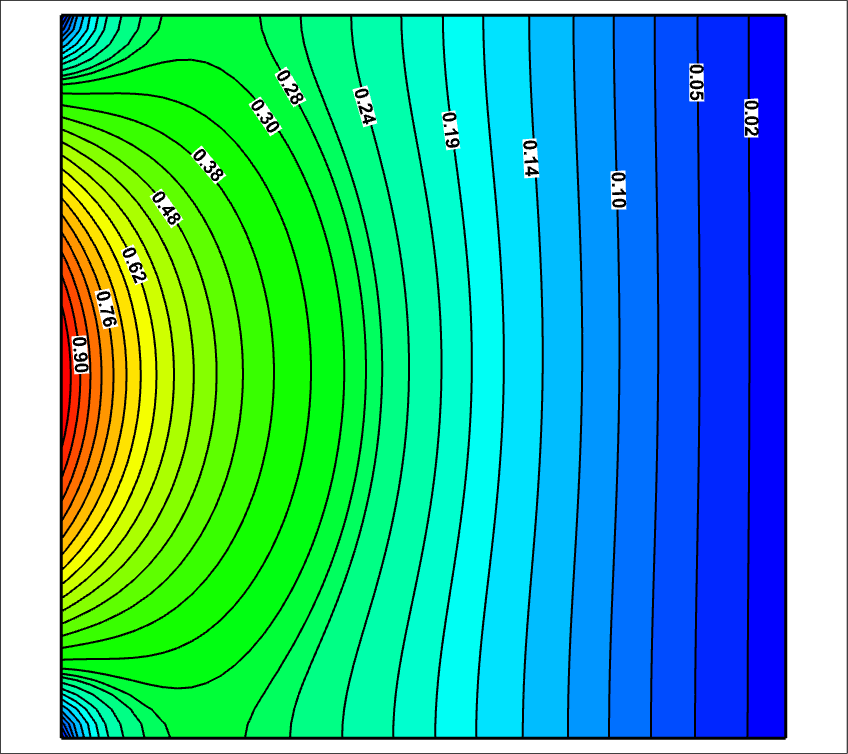}%
    \captionsetup{skip=2pt}%
    \caption{(d) $ Ha=50, Ra=10^3$}
    \label{fig:2d_isotherm_Ra_10^3_Ha_50_P3}
  \end{subfigure}%
 \begin{subfigure}{0.33\textwidth}        
   \centering
    \includegraphics[width=\textwidth]{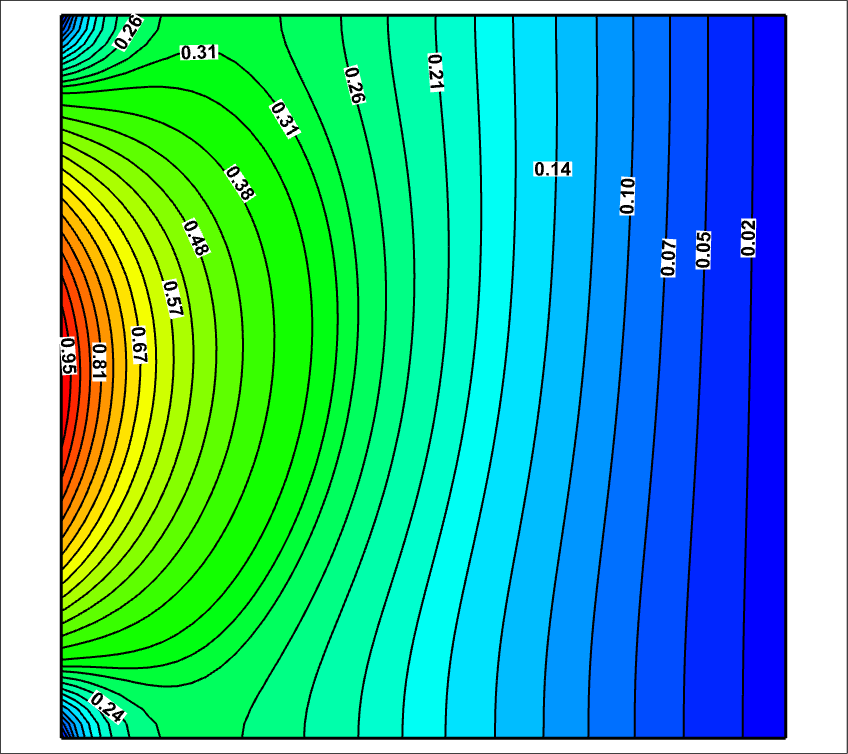}%
    \captionsetup{skip=2pt}%
    \caption{(e) $ Ha=50, Ra=10^4$}
    \label{fig:2d_isotherm_Ra_10^4_Ha_50_P3}
  \end{subfigure}
   \begin{subfigure}{0.33\textwidth}        
   \centering
    \includegraphics[width=\textwidth]{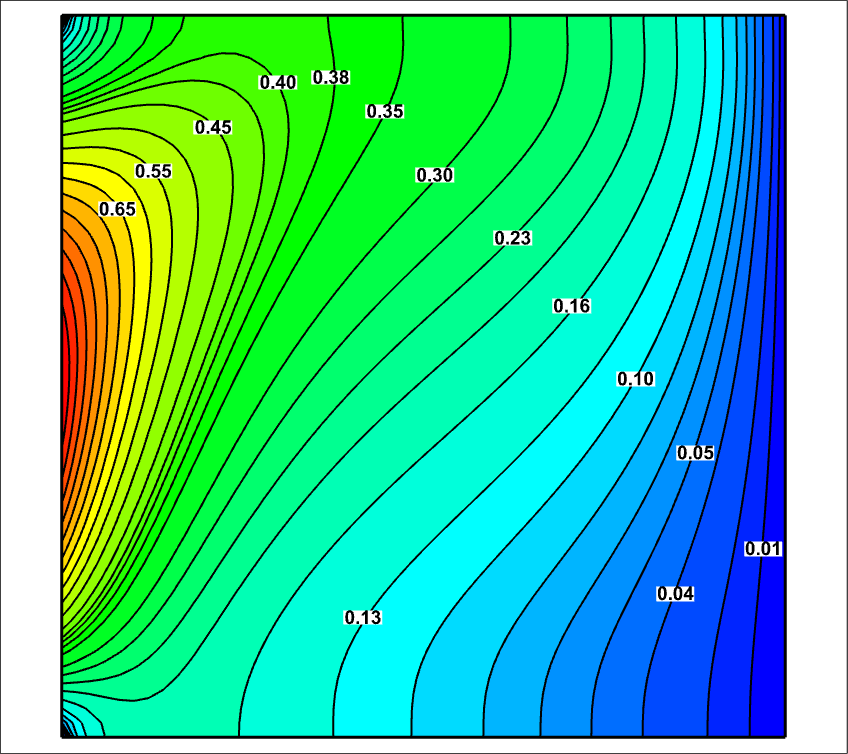}%
    \captionsetup{skip=2pt}%
    \caption{(f) $ Ha=50, Ra=10^5$}
    \label{fig:2d_isotherm_Ra_10^5_Ha_50_P3.png}
  \end{subfigure}%
  \hspace*{\fill}

  \vspace*{8pt}%
  \hspace*{\fill}%
  \begin{subfigure}{0.33\textwidth}     
    \centering
    \includegraphics[width=\textwidth]{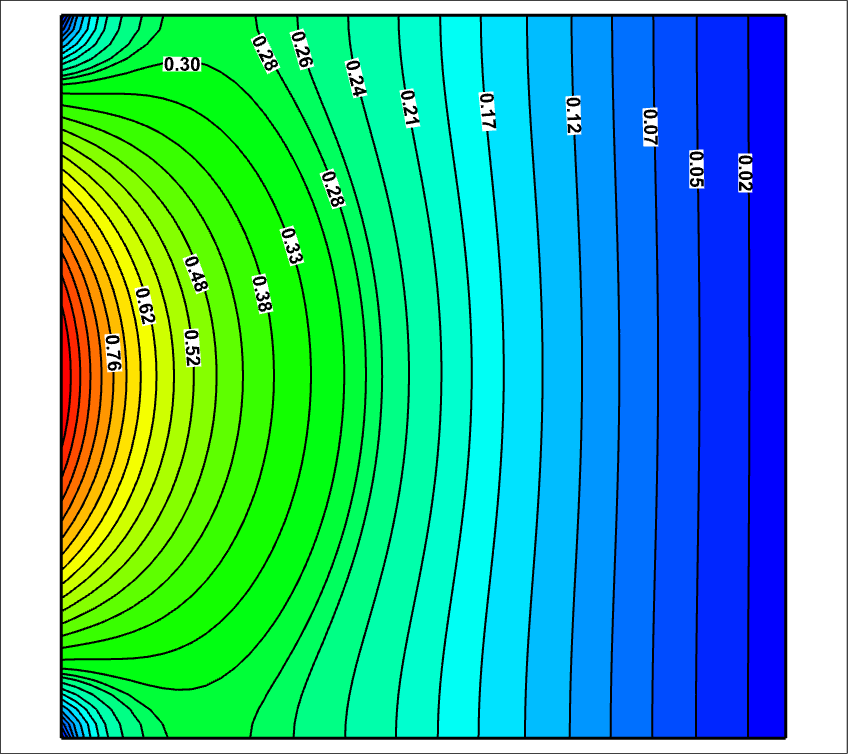}%
    \captionsetup{skip=2pt}%
    \caption{(g) $ Ha=100, Ra=10^3$}
    \label{fig:2d_isotherm_Ra_10^3_Ha_100_P3.png}
  \end{subfigure}%
 \begin{subfigure}{0.33\textwidth}        
   \centering
    \includegraphics[width=\textwidth]{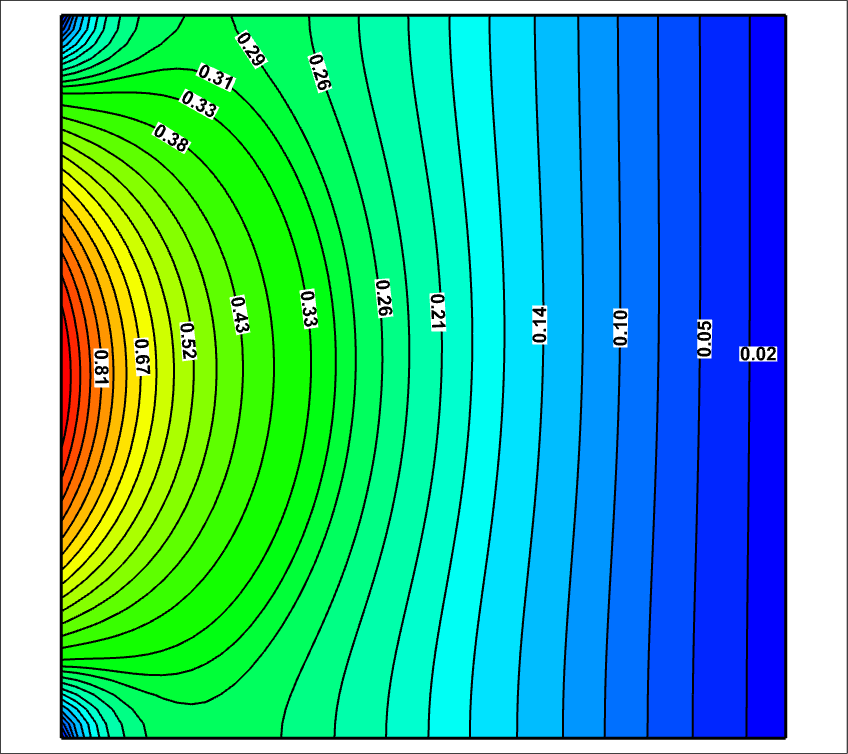}%
    \captionsetup{skip=2pt}%
    \caption{(h) $ Ha=100, Ra=10^4$}
    \label{fig:2d_isotherm_Ra_10^4_Ha_100_P3.png}
  \end{subfigure}
   \begin{subfigure}{0.33\textwidth}        
   \centering
    \includegraphics[width=\textwidth]{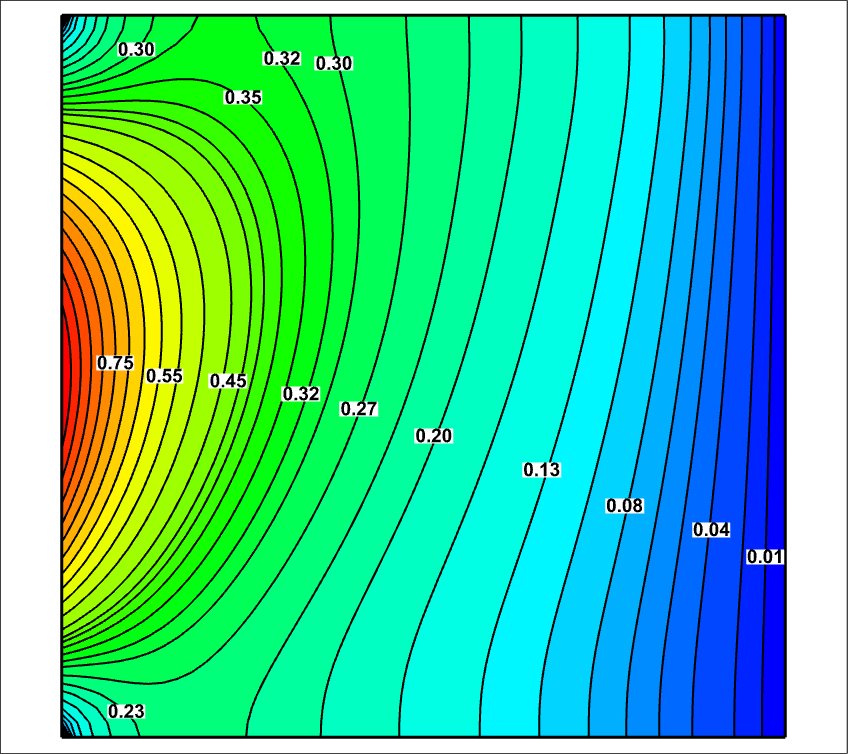}%
    \captionsetup{skip=2pt}%
    \caption{(i) $ Ha=100, Ra=10^5$}
    \label{fig:2d_isotherm_Ra_10^5_Ha_100_P3.png}
  \end{subfigure}%
  \hspace*{\fill}

  \vspace*{8pt}%
  \hspace*{\fill}%
  \begin{subfigure}{0.33\textwidth}     
    \centering
    \includegraphics[width=\textwidth]{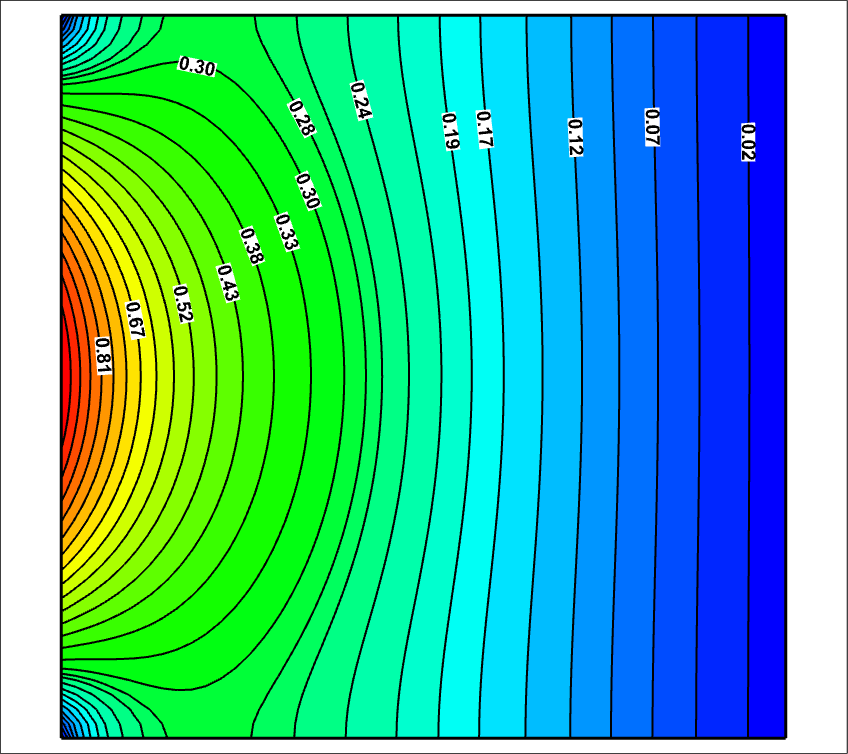}%
    \captionsetup{skip=2pt}%
    \caption{(j) $ Ha=150, Ra=10^3$}
    \label{fig:2d_isotherm_Ra_10^3_Ha_150_P3.png}
  \end{subfigure}%
 \begin{subfigure}{0.33\textwidth}        
   \centering
    \includegraphics[width=\textwidth]{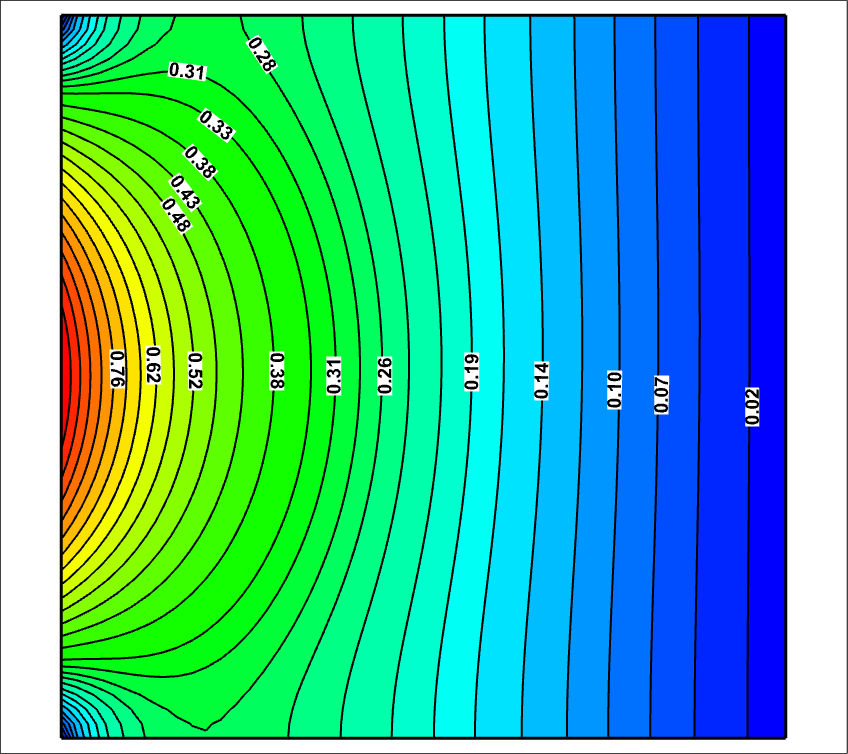}%
    \captionsetup{skip=2pt}%
    \caption{(k) $ Ha=150, Ra=10^4$}
    \label{fig:2d_isotherm_Ra_10^4_Ha_150_P3.png}
  \end{subfigure}
   \begin{subfigure}{0.33\textwidth}        
   \centering
    \includegraphics[width=\textwidth]{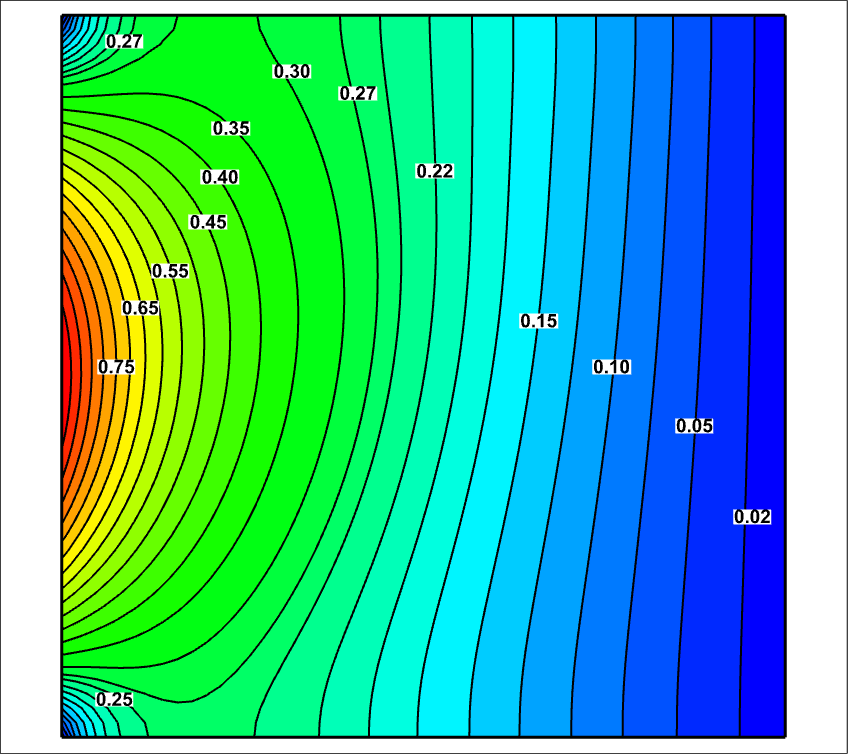}%
    \captionsetup{skip=2pt}%
    \caption{(l) $ Ha=150, Ra=10^5$}
    \label{fig:2d_isotherm_Ra_10^5_Ha_150_P3}
  \end{subfigure}%
  \hspace*{\fill}
  \vspace*{2pt}%
  \hspace*{\fill}%
  \caption{Case 3. Effect of different $Ha$ (Column-wise) and $Ra$ (Row-wise) on isotherm contours (at $z=0.5$ plane) at a fixed $Pr = 0.065$}
  \label{fig:case-3_Isotherm_Contours_2D}
\end{figure}

The schematic representation of Case 3 is depicted in Figure \ref{fig:Sche_diag}(c). In this configuration, the left wall ($x=0$) undergoes non-uniform heating with $T_h=\sin(\pi y)\sin(\pi z)$. In this case, the isotherm contours are circular in shape on the left wall boundary with maximum temperature at the core region. The corresponding surface plot of the isotherms is presented as the small subfigure in Figure \ref{fig:Sche_diag}(c). Here the non-uniform temperature boundary condition has almost an inverted bowl shape with the maximum temperature at the peak. The streamlines at three different planes ($z=0.05, 0.5, 0.95$) are presented in Figure \ref{fig:case-3_Streamlines_Contours}, showcasing the effect of various $Ha$ and $Ra$ values on the fluid flow. At $Ha=25$, the primary vortex exhibits an almost circular shape for $Ra=10^3$, but appears diagonally stretched for $Ra=10^4$ and $Ra=10^5$ (Figure \ref{fig:case-3_Streamlines_Contours}(a) - (c)). However, the centers of the primary vortex vary its position across each plane. One can also observe the formation of a pair of secondary vortices on the $z=0.05$ plane, which is extended up to the middle plane ($z = 0.5$). Whereas, on $z=0.95$, only one secondary vortex grows on the top left corner. For higher $Ha$ ($Ha=$ 50 - 150), the magnetic field's influence is prominent, resulting in a more stretched primary vortex compared to $Ha = 25$ (Figure \ref{fig:case-3_Streamlines_Contours}(d) - (l)). As we increment the values of $Ha$, the center of the primary vortex shifts towards the center of the cavity because of the reduction in heat transfer rate. One can also observe that the pair of secondary vortices appear on all planes and their sizes increase with increase $Ha$. Furthermore, with an increase in $Ra$, the primary and secondary vortices become more stretched due to the rise in fluid velocity. When compare with the previous two cases, we notice that no secondary vortex appears in Case 1, and a small pair of secondary vortices appear in Case 2 only for high $Ha$ values. Figure \ref{fig:case-3_Isotherm_Contours} and \ref{fig:case-3_Isotherm_Contours_2D} present the isotherm contours for Case 3, offering insights into the impact of $Ha$ and $Ra$. The isotherm lines transition from straight to curved as $Ra$ increases and from curved to straight as $Ha$ rises. 
This observation signifies a decrease in heat transfer with increasing $Ha$ and an opposite trend with rising $Ra$. Furthermore, the temperature contours undergo substantial variations at $Ra=10^5$ with increasing $Ha$.
With an increase in $Ha$, isothermal lines tend to be closer to the heated walls. However, the change is not as pronounced at lower $Ra$ values ($Ra=10^3, Ra=10^4$) because of the low buoyancy effect, indicating a low influence of the magnetic field for lower $Ra$ values.\\
We have thoroughly examined the influence of $Ha$ and $Ra$ on the streamlines and isotherms in all three cases, yielding interesting findings. A consistent observation across all cases is the inverse relationship between $Ha$ and heat transfer rate, where an increase in $Ha$ results in a decrease in heat transfer. Conversely, an increase in $Ra$ leads to an increase in the heat transfer rate.
However, the distinctive features in the shapes of streamlines and isotherms, as well as the magnitude of heat transfer rate, differentiate each case at specific $Ha$ and $Ra$ values. For instance, at a fixed 
$Ra=10^5$ and $Ha=25$, Case 1 exhibits diagonally symmetric isotherms due to the uniformly heated left wall. In contrast, Cases 2 and 3, with non-uniformly heated left walls, display non-symmetric isotherms. Additionally, the shape of streamlines varies; Case 1 exhibits vertically stretched streamlines at low $Ra$ values, a pattern not observed in Cases 2 and 3. The shape and size of secondary vortices near the heated wall vary across each case, as discussed earlier. These nuanced differences emphasize the significant impact of different heating conditions in determining the thermal behavior of the system. The subsequent section delves deeper into this aspect by studying the Nusselt number and entropy generation.

\newpage
\subsection{Heat Transfer characteristics: Nusselt Number Analysis}
The Nusselt number is one of the most significant dimensionless number in heat transfer and fluid dynamics. This parameter is critical for measuring the convective heat transfer between the solid and the fluid surface. The descriptions for these numbers are given below:
$$Nu_{\text {A }}(z)=\int_0^1 N u_{\text {L}}(y, z) \mathrm{d} y$$ 
where, 
$$Nu_{\text {L }}(y, z)= \frac{\partial \theta(y, z)}{\partial x} $$ at $x = 1$ or 
$x = 0.$\\ Here, $Nu_{\text{A}}$ represents the average Nusselt number, and $Nu_{\text{L}}$ represents the local Nusselt number. The definition of the total Nusselt number ($Nu_{\text {Total}}$) is as follows: 
$$N u_{\text {Total}}=\int_0^1 N u_{\text {A}}(z) \mathrm{d} z .
$$
\begin{figure}[htbp]
 \centering
 \vspace*{0pt}%
 \hspace*{\fill}%
\begin{subfigure}{0.33\textwidth}     
    \centering
    \includegraphics[width=\textwidth]{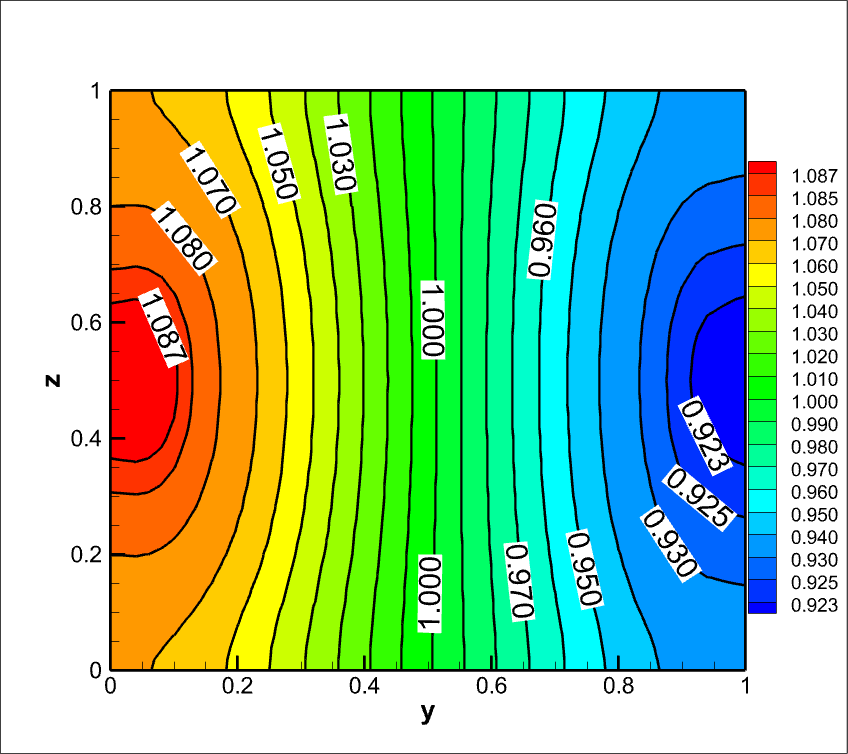}%
    \captionsetup{skip=2pt}%
    \caption{(a) $Ra=10^3, Ha=25$}
    \label{fig:LC_RA_10^3_Ha_25.png}
  \end{subfigure}%
 \begin{subfigure}{0.33\textwidth}        
   \centering
    \includegraphics[width=\textwidth]{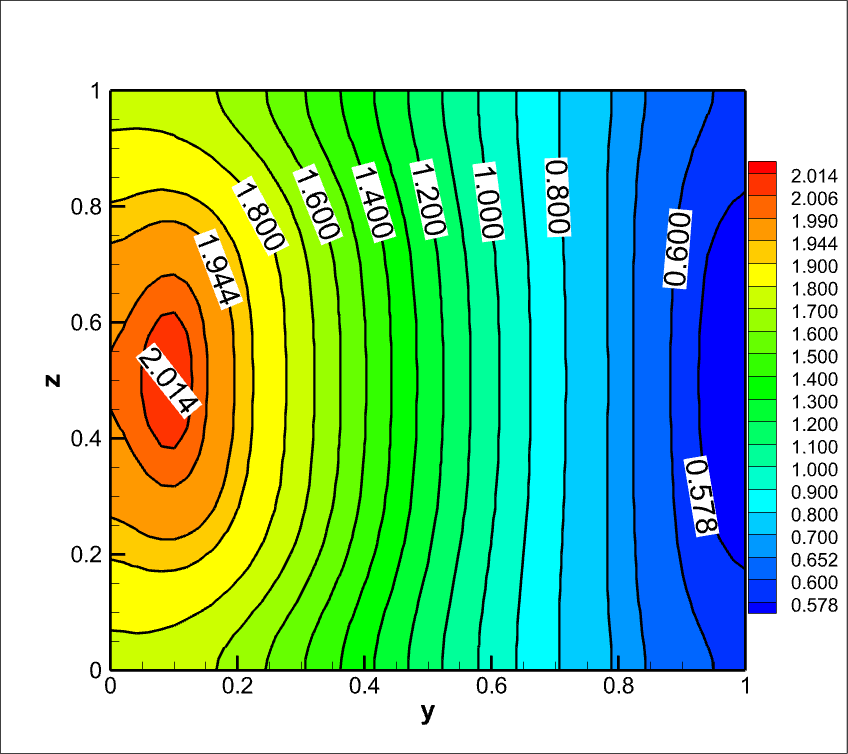}%
    \captionsetup{skip=2pt}%
    \caption{(b) $Ra=10^4, Ha=25$}
    \label{fig:LC_RA_10^4_Ha_25.png}
  \end{subfigure}
   \begin{subfigure}{0.33\textwidth}        
   \centering
    \includegraphics[width=\textwidth]{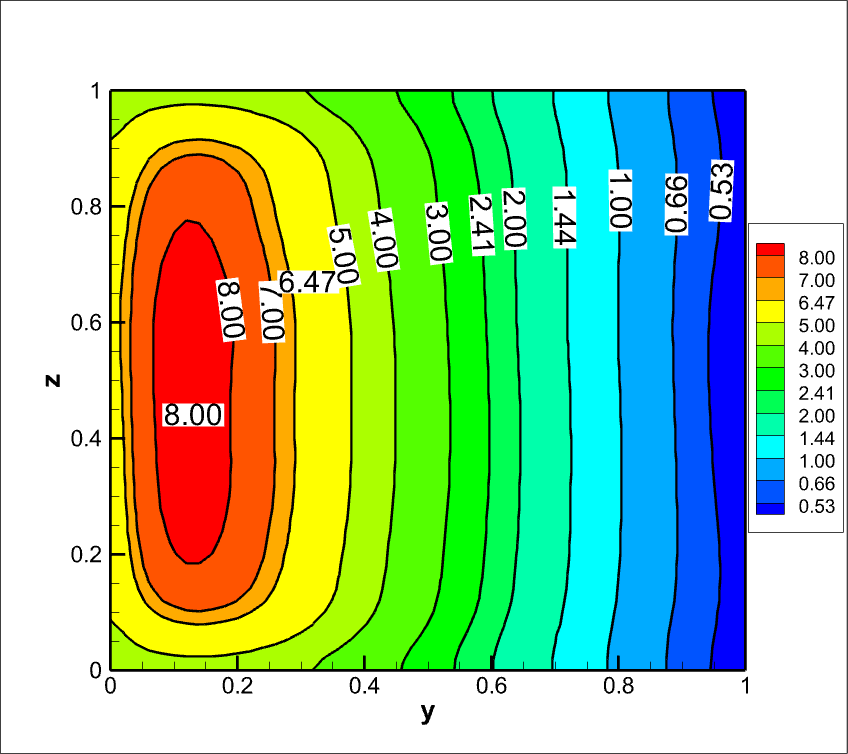}%
    \captionsetup{skip=2pt}%
    \caption{(c) $Ra=10^5, Ha=25$}
    \label{fig:LC_RA_10^5_Ha_25.png}
  \end{subfigure}%
  \hspace*{\fill}

  \vspace*{8pt}%
  \hspace*{\fill}%
  \begin{subfigure}{0.33\textwidth}     
    \centering
    \includegraphics[width=\textwidth]{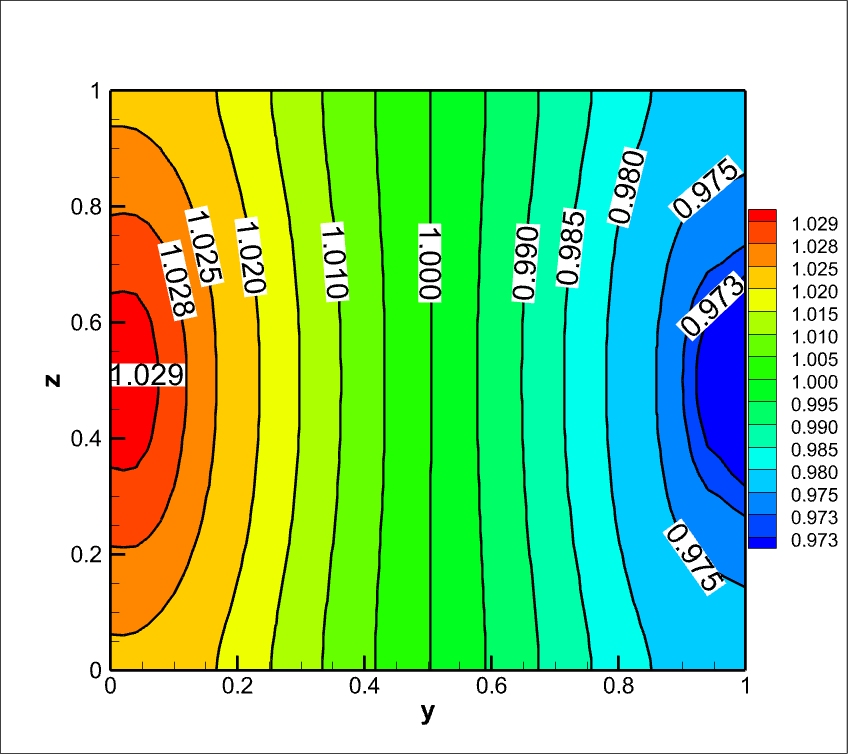}%
    \captionsetup{skip=2pt}%
    \caption{(d) $Ra=10^3, Ha=50$}
    \label{fig:LC_RA_10^3_Ha_50.png}
  \end{subfigure}%
 \begin{subfigure}{0.33\textwidth}        
   \centering
    \includegraphics[width=\textwidth]{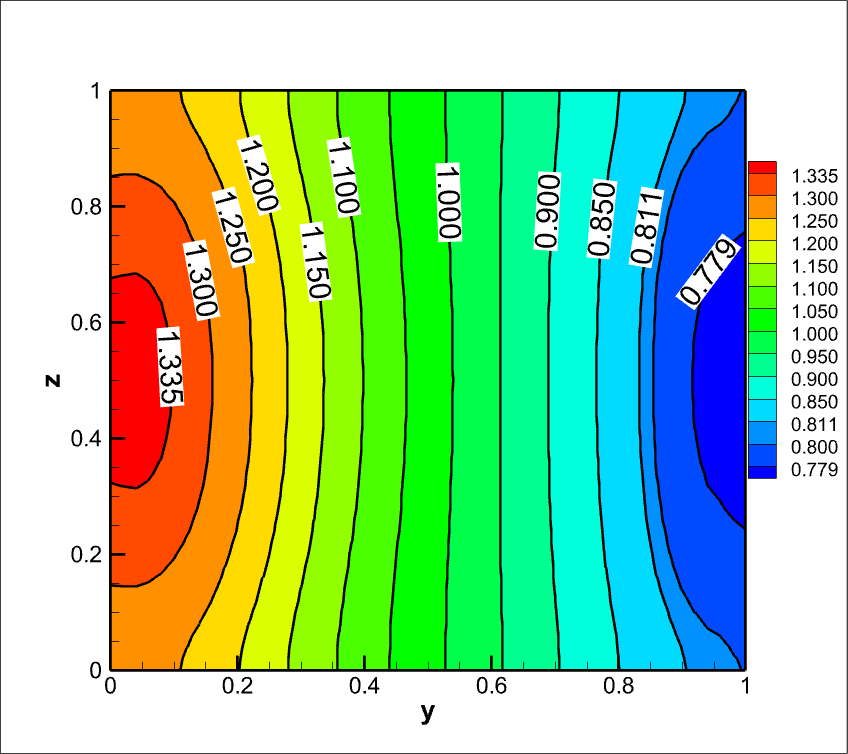}%
    \captionsetup{skip=2pt}%
    \caption{(e) $Ra=10^4, Ha=50$}
    \label{fig:LC_RA_10^4_Ha_50.png}
  \end{subfigure}
   \begin{subfigure}{0.33\textwidth}        
   \centering
    \includegraphics[width=\textwidth]{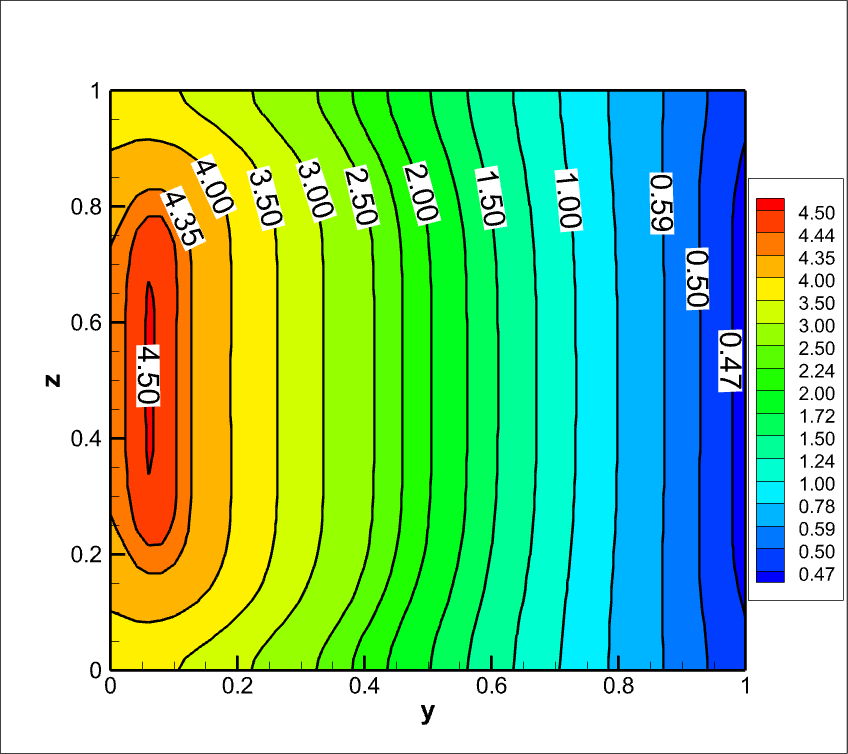}%
    \captionsetup{skip=2pt}%
    \caption{(f) $Ra=10^5, Ha=50$}
    \label{fig:LC_RA_10^5_Ha_50.png}
  \end{subfigure}%
  \hspace*{\fill}

  \vspace*{8pt}%
  \hspace*{\fill}%
  \begin{subfigure}{0.33\textwidth}     
    \centering
    \includegraphics[width=\textwidth]{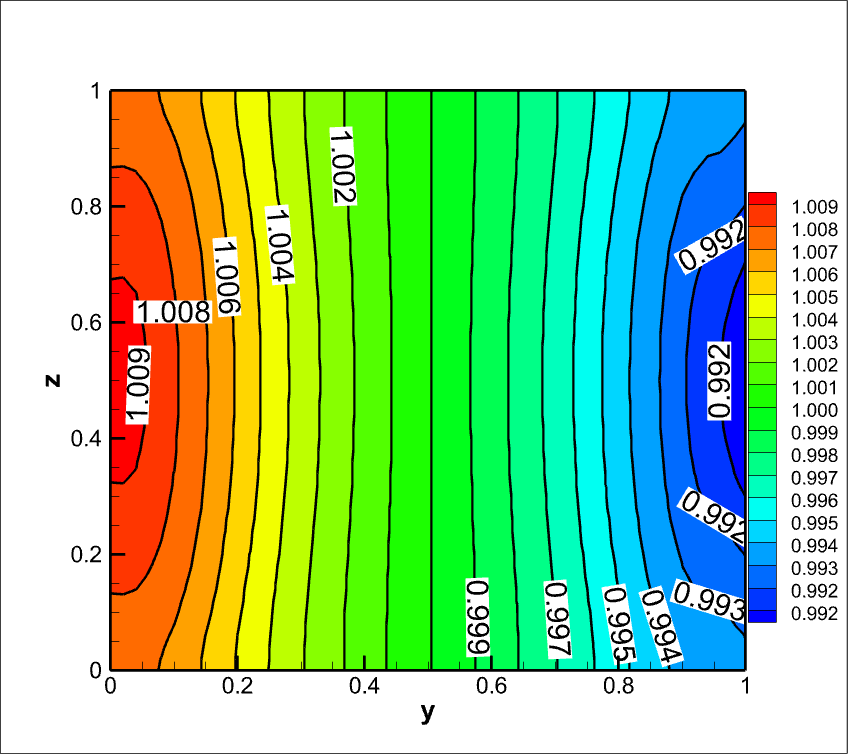}%
    \captionsetup{skip=2pt}%
    \caption{(g) $Ra=10^3, Ha=100$}
    \label{fig:LC_RA_10^3_Ha_100.png}
  \end{subfigure}%
 \begin{subfigure}{0.33\textwidth}        
   \centering
    \includegraphics[width=\textwidth]{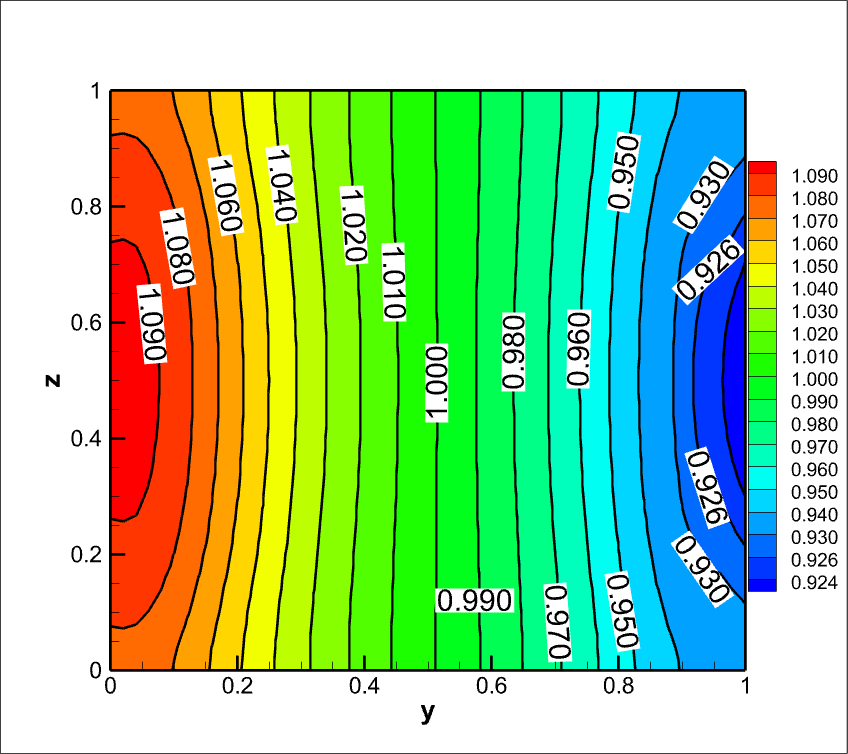}%
    \captionsetup{skip=2pt}%
    \caption{(h) $Ra=10^4, Ha=100$}
    \label{fig:LC_RA_10^4_Ha_100.png}
  \end{subfigure}
   \begin{subfigure}{0.33\textwidth}        
   \centering
    \includegraphics[width=\textwidth]{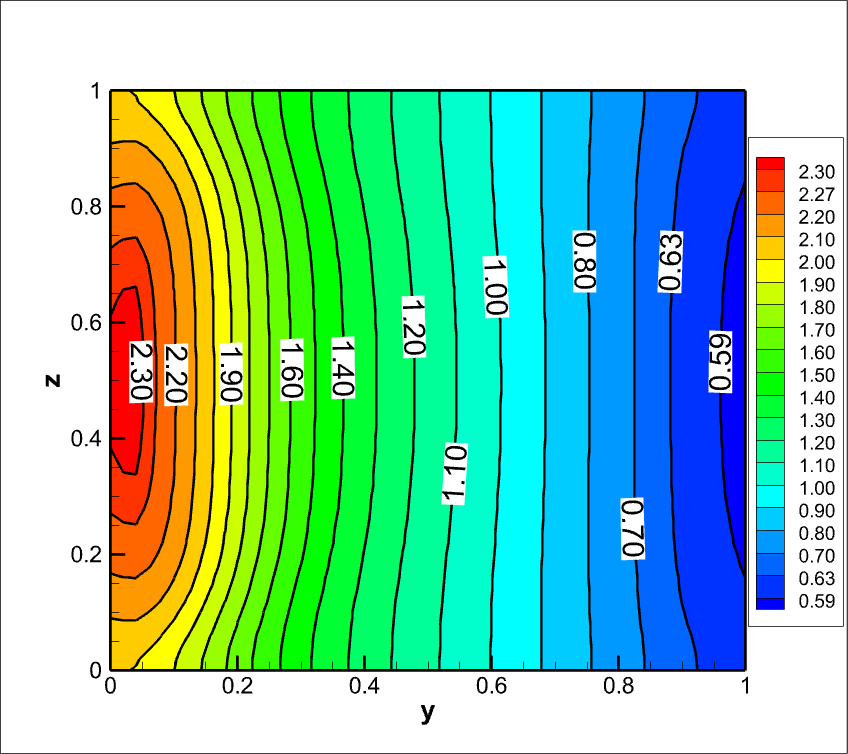}%
    \captionsetup{skip=2pt}%
    \caption{(i) $Ra=10^5, Ha=100$}
    \label{fig:LC_RA_10^5_Ha_100.png}
  \end{subfigure}%
  \hspace*{\fill}

  \vspace*{8pt}%
  \hspace*{\fill}%
  \begin{subfigure}{0.33\textwidth}     
    \centering
    \includegraphics[width=\textwidth]{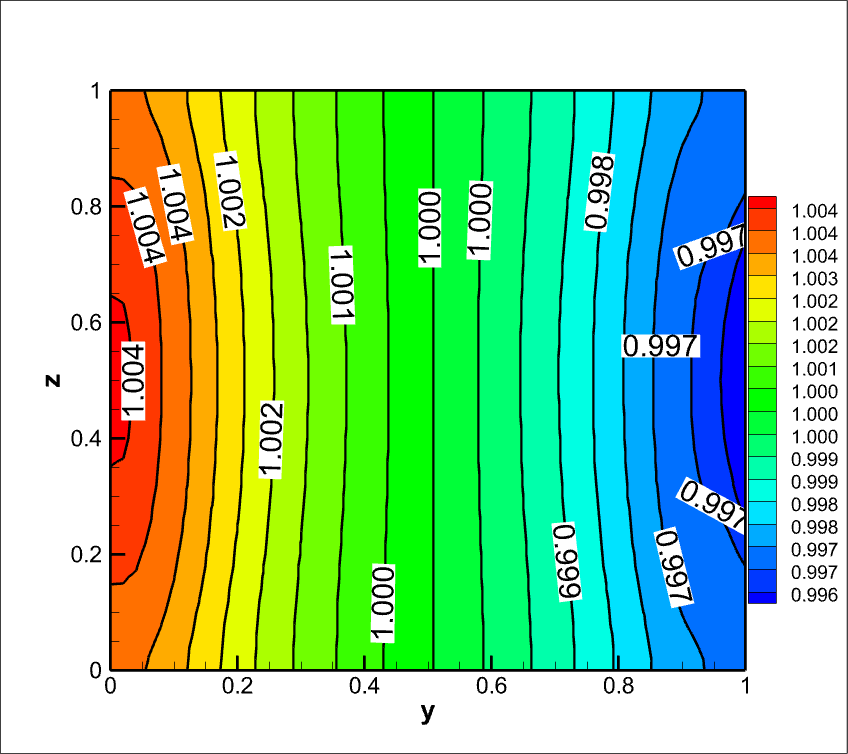}%
    \captionsetup{skip=2pt}%
    \caption{(j) $Ra=10^3, Ha=150$}
    \label{fig:LC_RA_10^3_Ha_150.png}
  \end{subfigure}%
 \begin{subfigure}{0.33\textwidth}        
   \centering
    \includegraphics[width=\textwidth]{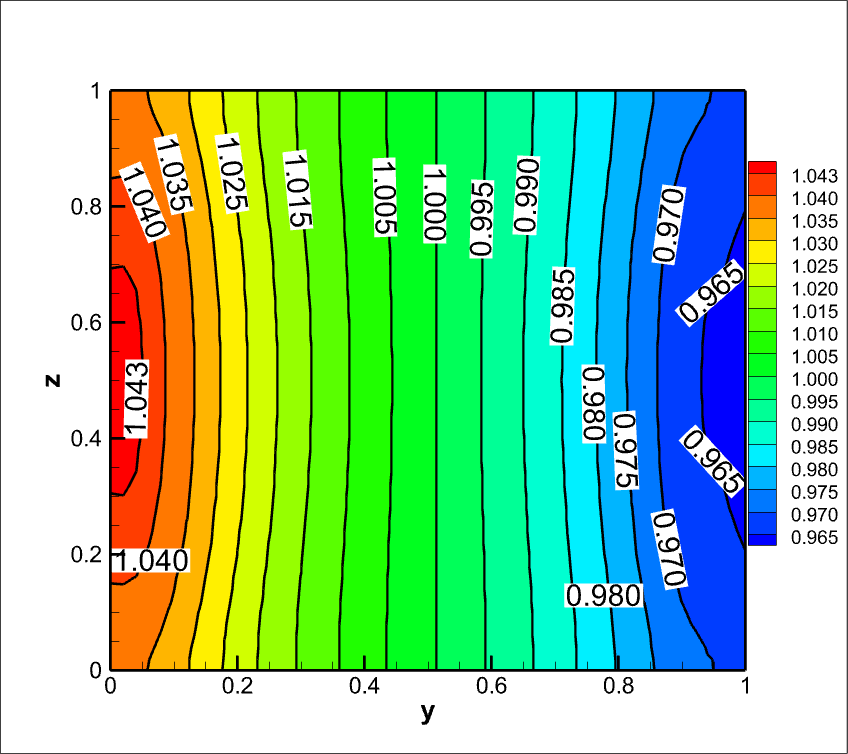}%
    \captionsetup{skip=2pt}%
    \caption{(k) $Ra=10^4, Ha=150$ }
    \label{fig:LC_RA_10^4_Ha_150.png}
  \end{subfigure}
   \begin{subfigure}{0.33\textwidth}        
   \centering
    \includegraphics[width=\textwidth]{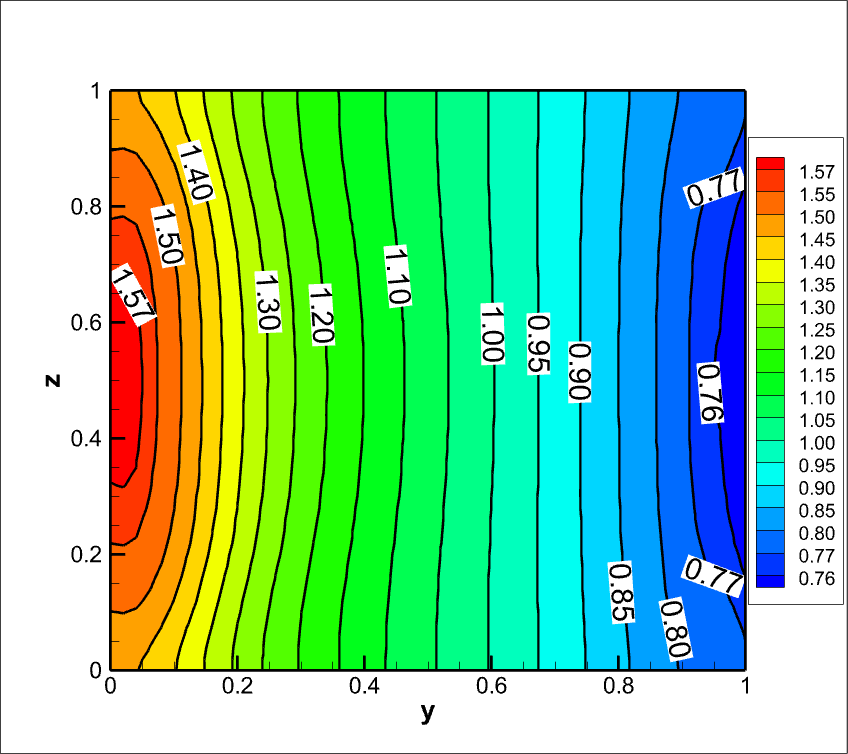}%
    \captionsetup{skip=2pt}%
    \caption{(l) $Ra=10^5, Ha=150$}
    \label{fig:LC_RA_10^5_Ha_150}
  \end{subfigure}%
  \hspace*{\fill}
  \vspace*{2pt}%
  \hspace*{\fill}%
  \caption{Case 1. $Nu_{\text{L}}$ contours on the left ($x=0$) heated wall of the cubic cavity}
  \label{fig:case-1_Local_Nusselt}
\end{figure}


\begin{figure}[htbp]
 \centering
 \vspace*{0pt}%
 \hspace*{\fill}%
\begin{subfigure}{0.33\textwidth}     
    \centering
    \includegraphics[width=\textwidth]{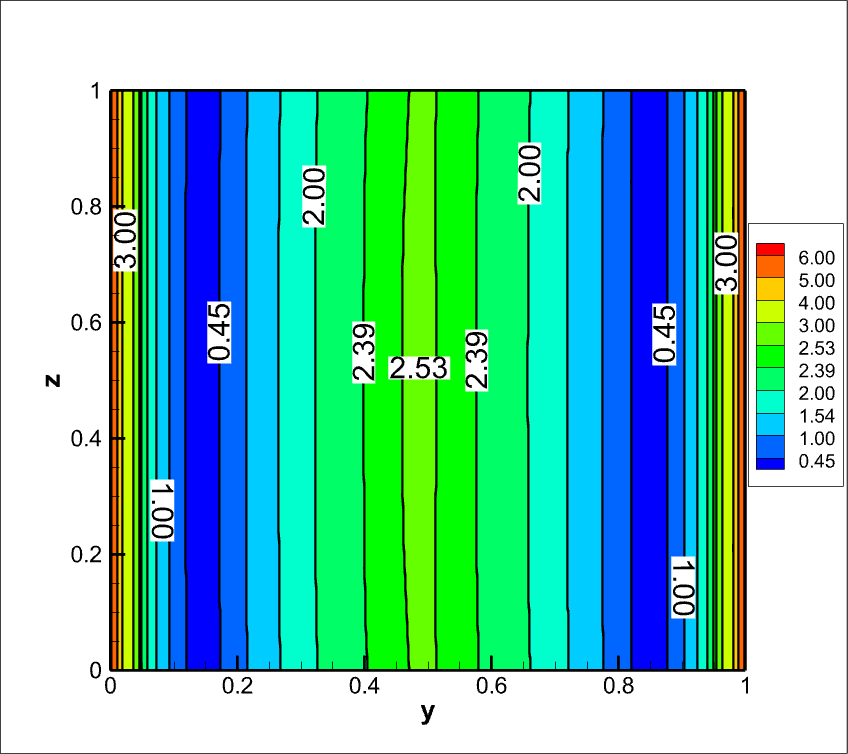}%
    \captionsetup{skip=2pt}%
    \caption{(a) $Ra=10^3, Ha=25$}
    \label{fig:P2_LC_RA_10^3_Ha_25.png}
  \end{subfigure}%
 \begin{subfigure}{0.33\textwidth}        
   \centering
    \includegraphics[width=\textwidth]{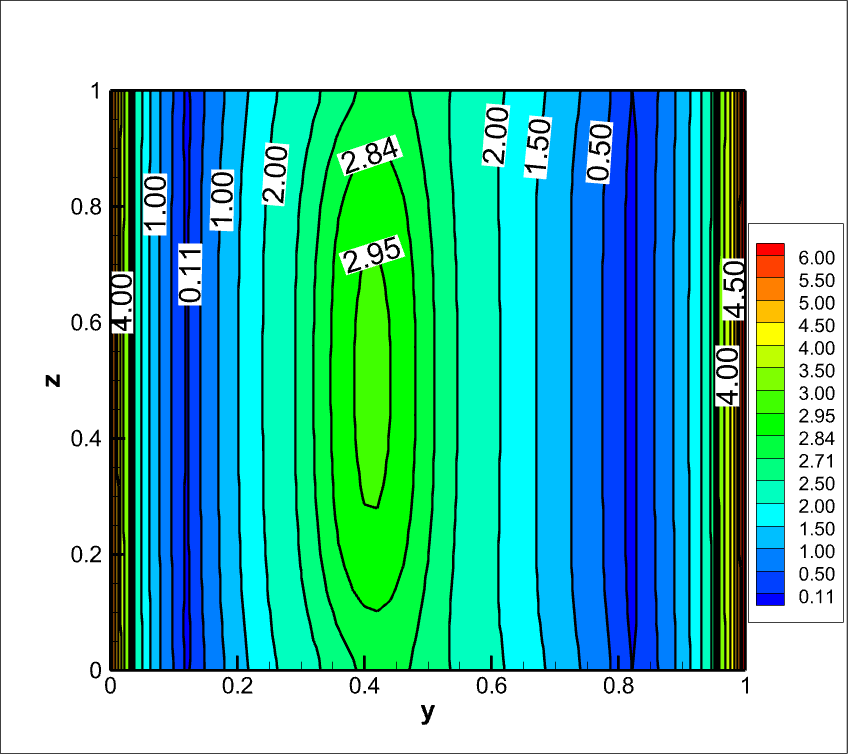}%
    \captionsetup{skip=2pt}%
    \caption{(b) $Ra=10^4, Ha=25$}
    \label{fig:P2_LC_RA_10^4_Ha_25.png}
  \end{subfigure}
   \begin{subfigure}{0.33\textwidth}        
   \centering
    \includegraphics[width=\textwidth]{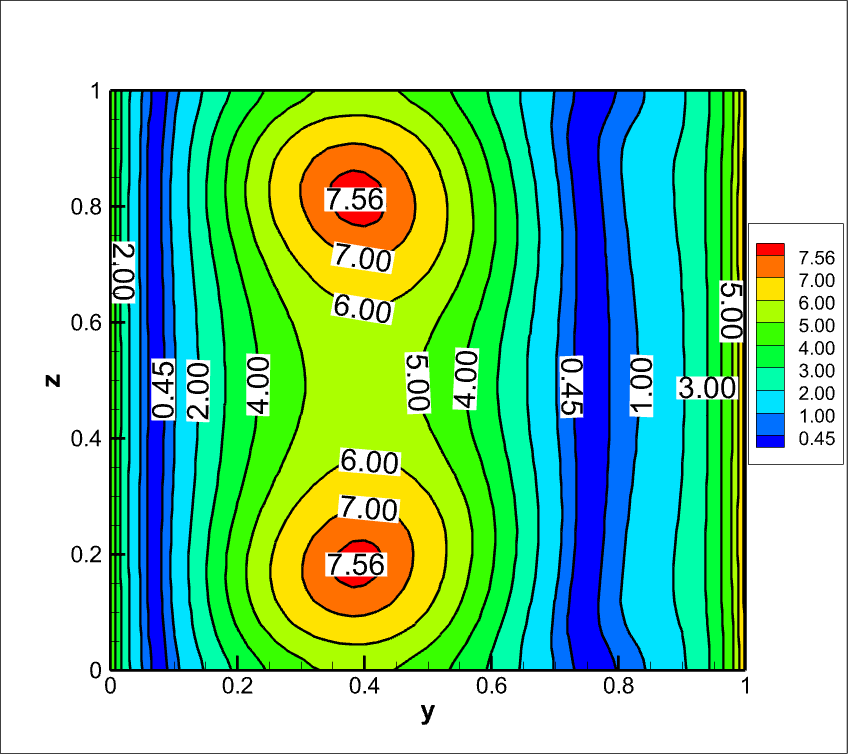}%
    \captionsetup{skip=2pt}%
    \caption{(c) $Ra=10^5, Ha=25$}
    \label{fig:P2_LC_RA_10^5_Ha_25.png}
  \end{subfigure}%
  \hspace*{\fill}

  \vspace*{8pt}%
  \hspace*{\fill}%
  \begin{subfigure}{0.33\textwidth}     
    \centering
    \includegraphics[width=\textwidth]{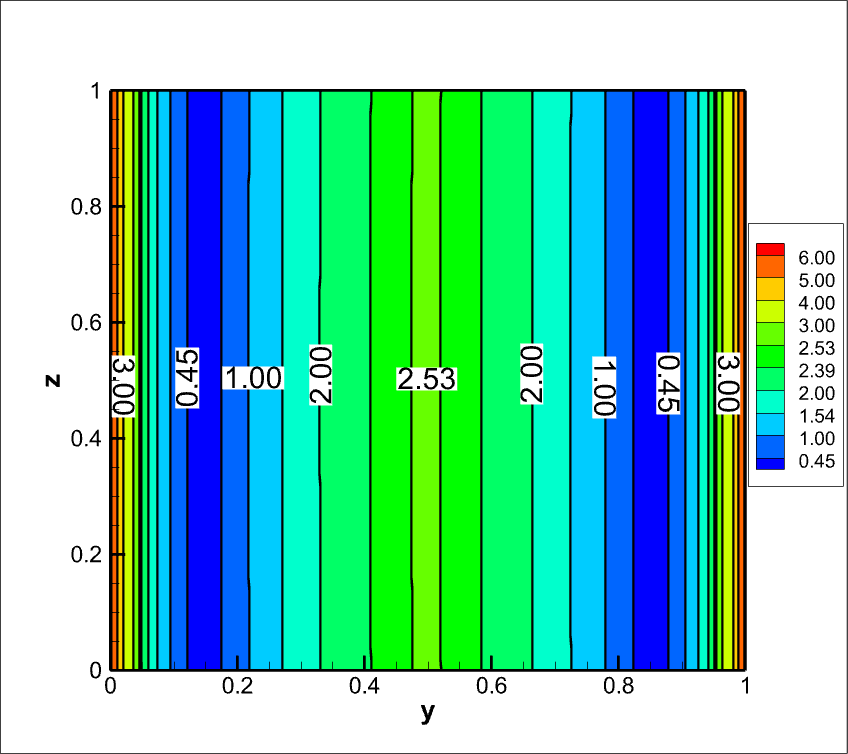}%
    \captionsetup{skip=2pt}%
    \caption{(d) $Ra=10^3, Ha=50$}
    \label{fig:P2_LC_RA_10^3_Ha_50.png}
  \end{subfigure}%
 \begin{subfigure}{0.33\textwidth}        
   \centering
    \includegraphics[width=\textwidth]{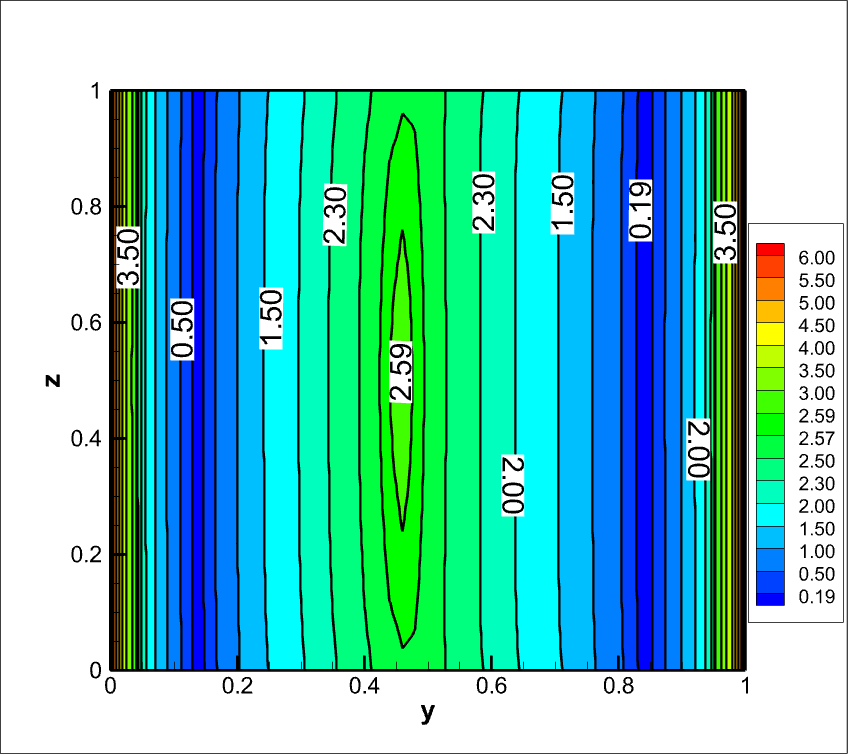}%
    \captionsetup{skip=2pt}%
    \caption{(e) $Ra=10^4, Ha=50$}
    \label{fig:P2_LC_RA_10^4_Ha_50.png}
  \end{subfigure}
   \begin{subfigure}{0.33\textwidth}        
   \centering
    \includegraphics[width=\textwidth]{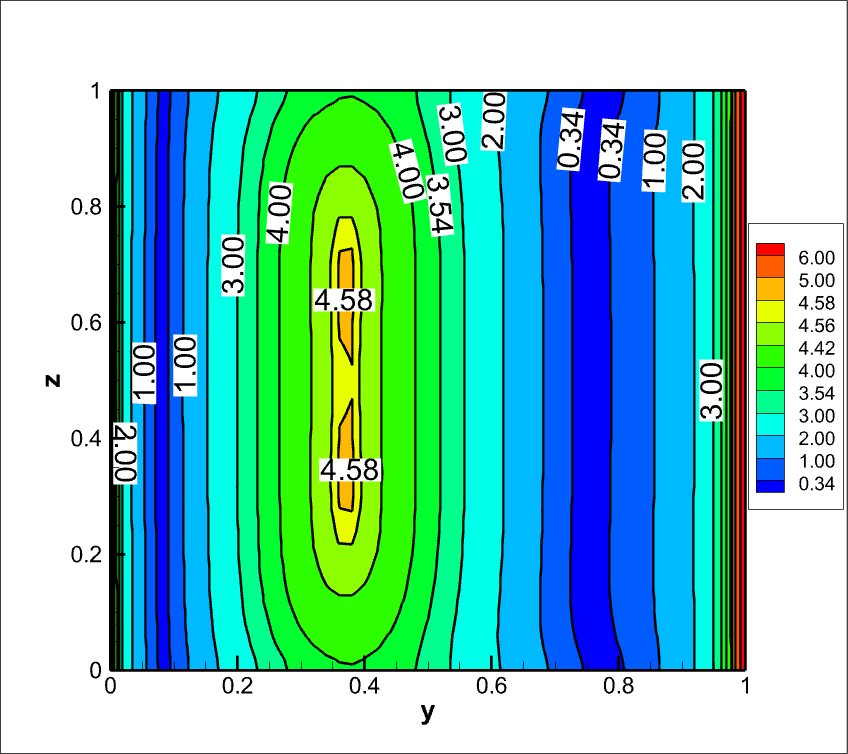}%
    \captionsetup{skip=2pt}%
    \caption{(f) $Ra=10^5, Ha=50$}
    \label{fig:P2_LC_RA_10^5_Ha_50.png}
  \end{subfigure}%
  \hspace*{\fill}

  \vspace*{8pt}%
  \hspace*{\fill}%
  \begin{subfigure}{0.33\textwidth}     
    \centering
    \includegraphics[width=\textwidth]{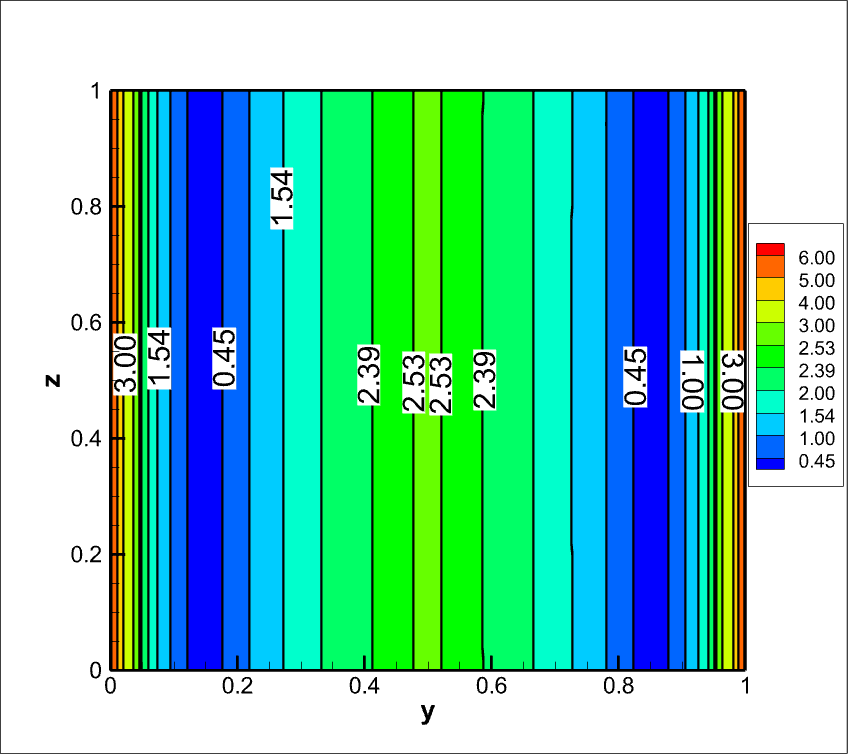}%
    \captionsetup{skip=2pt}%
    \caption{(g) $Ra=10^3, Ha=100$}
    \label{fig:P2_LC_RA_10^3_Ha_100.png}
  \end{subfigure}%
 \begin{subfigure}{0.33\textwidth}        
   \centering
    \includegraphics[width=\textwidth]{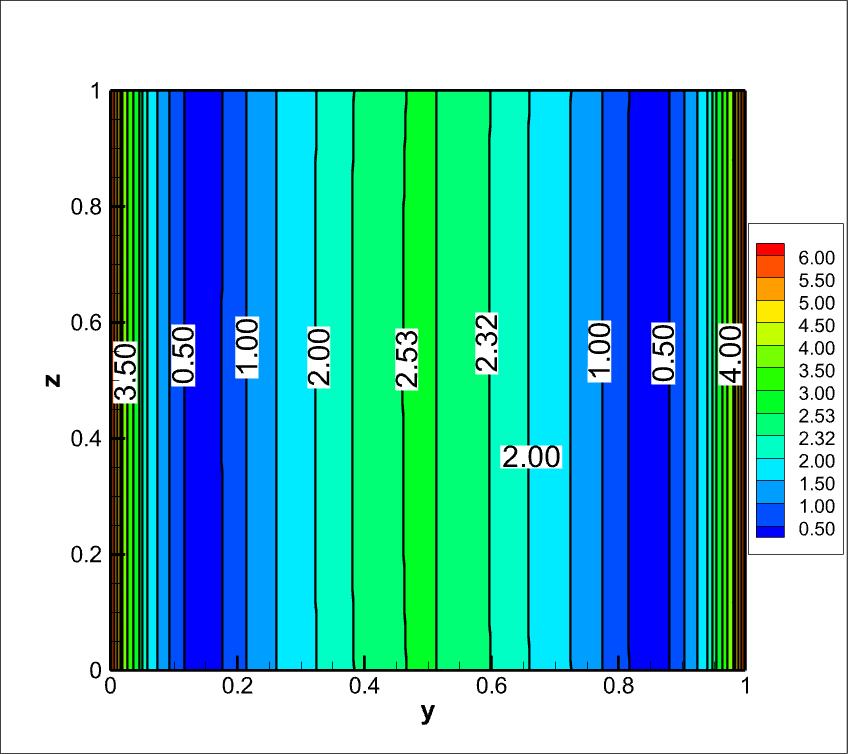}%
    \captionsetup{skip=2pt}%
    \caption{(h) $Ra=10^4, Ha=100$}
    \label{fig:P2_LC_RA_10^4_Ha_100.png}
  \end{subfigure}
   \begin{subfigure}{0.33\textwidth}        
   \centering
    \includegraphics[width=\textwidth]{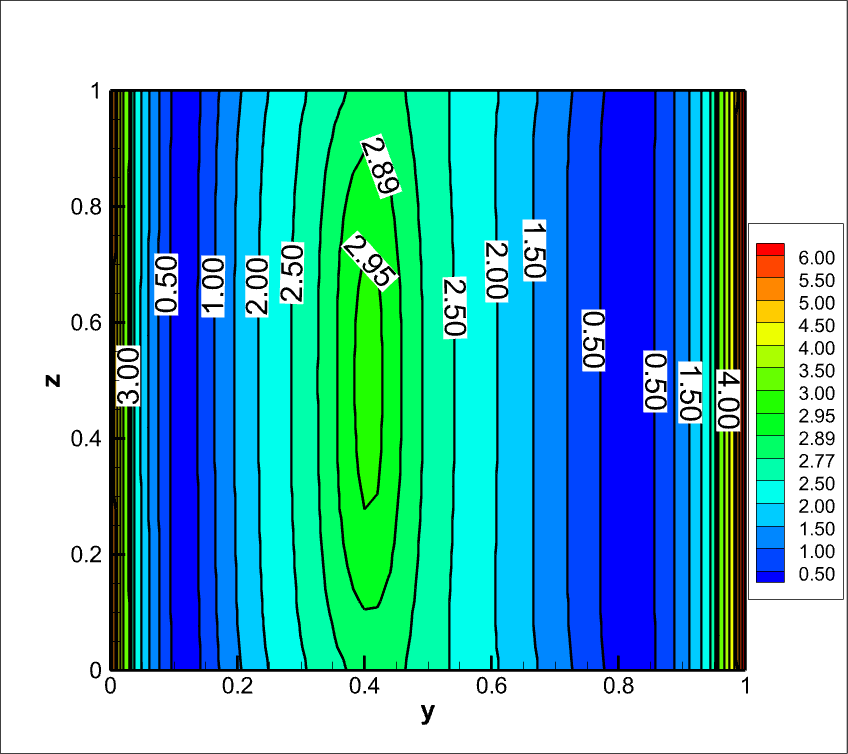}%
    \captionsetup{skip=2pt}%
    \caption{(i) $Ra=10^5, Ha=100$}
    \label{fig:P2_LC_RA_10^5_Ha_100.png}
  \end{subfigure}%
  \hspace*{\fill}

  \vspace*{8pt}%
  \hspace*{\fill}%
  \begin{subfigure}{0.33\textwidth}     
    \centering
    \includegraphics[width=\textwidth]{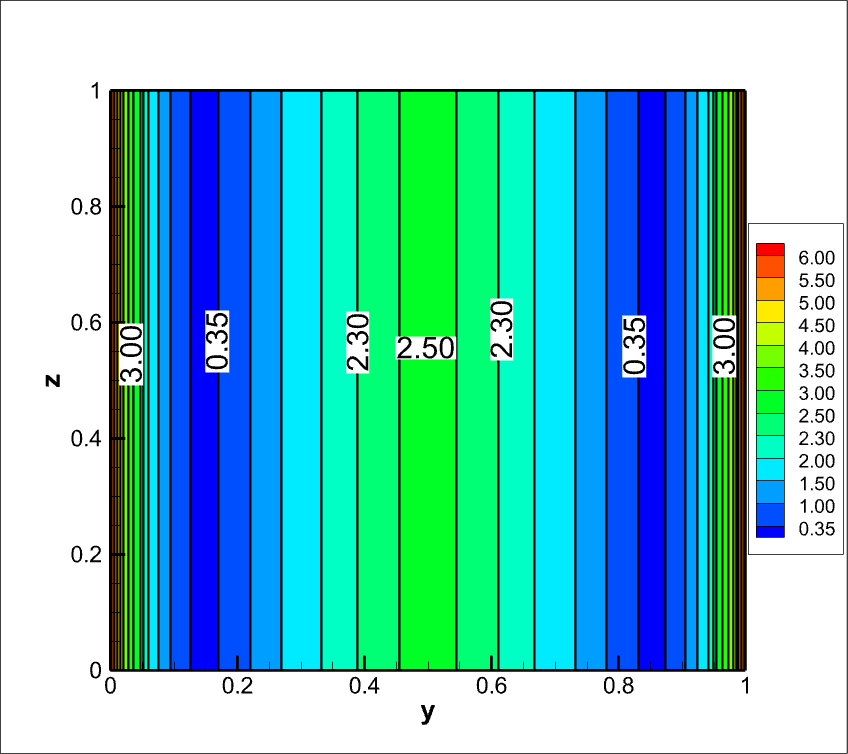}%
    \captionsetup{skip=2pt}%
    \caption{(j) $Ra=10^3, Ha=150$}
    \label{fig:P2_LC_RA_10^3_Ha_150.png}
  \end{subfigure}%
 \begin{subfigure}{0.33\textwidth}        
   \centering
    \includegraphics[width=\textwidth]{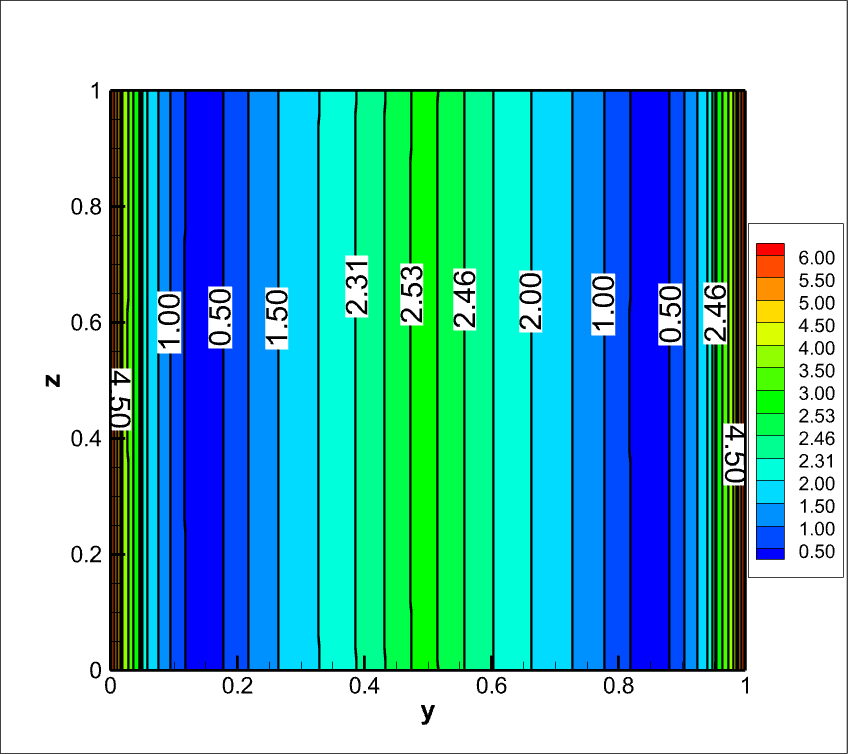}%
    \captionsetup{skip=2pt}%
    \caption{(k) $Ra=10^4, Ha=150$}
    \label{fig:P2_LC_RA_10^4_Ha_150.png}
  \end{subfigure}
   \begin{subfigure}{0.33\textwidth}        
   \centering
    \includegraphics[width=\textwidth]{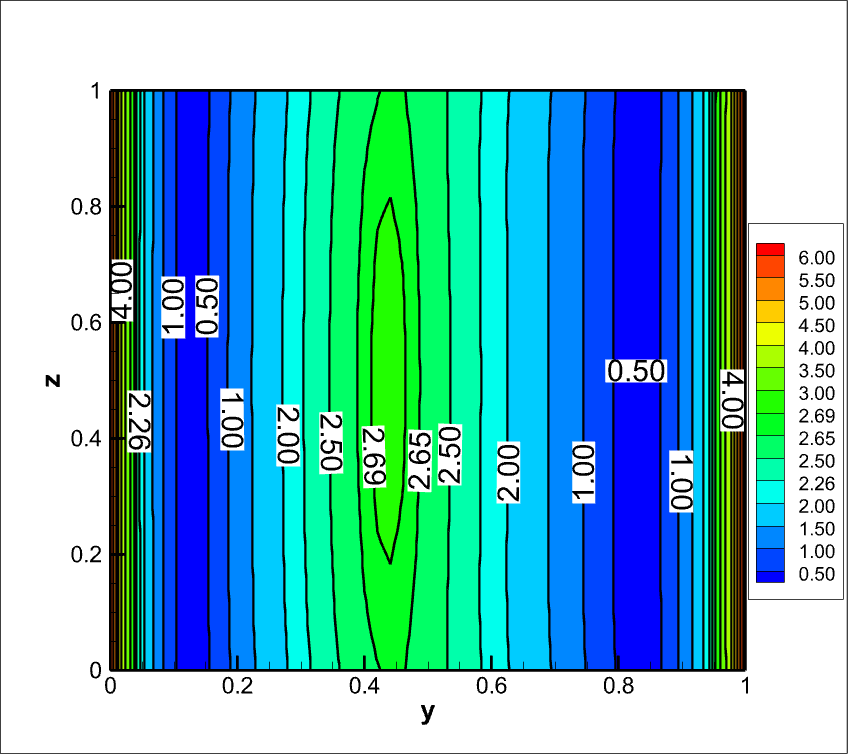}%
    \captionsetup{skip=2pt}%
    \caption{(l) $Ra=10^5, Ha=150$}
    \label{fig:P2_LC_RA_10^5_Ha_150}
  \end{subfigure}%
  \hspace*{\fill}
  \vspace*{2pt}%
  \hspace*{\fill}%
  \caption{Case 2. $Nu_{\text{L}}$ contours on the left ($x=0$) heated wall of the cubic cavity}
  \label{fig:case-2_Local_Nusselt}
\end{figure}


\begin{figure}[htbp]
 \centering
 \vspace*{0pt}%
 \hspace*{\fill}%
\begin{subfigure}{0.33\textwidth}     
    \centering
    \includegraphics[width=\textwidth]{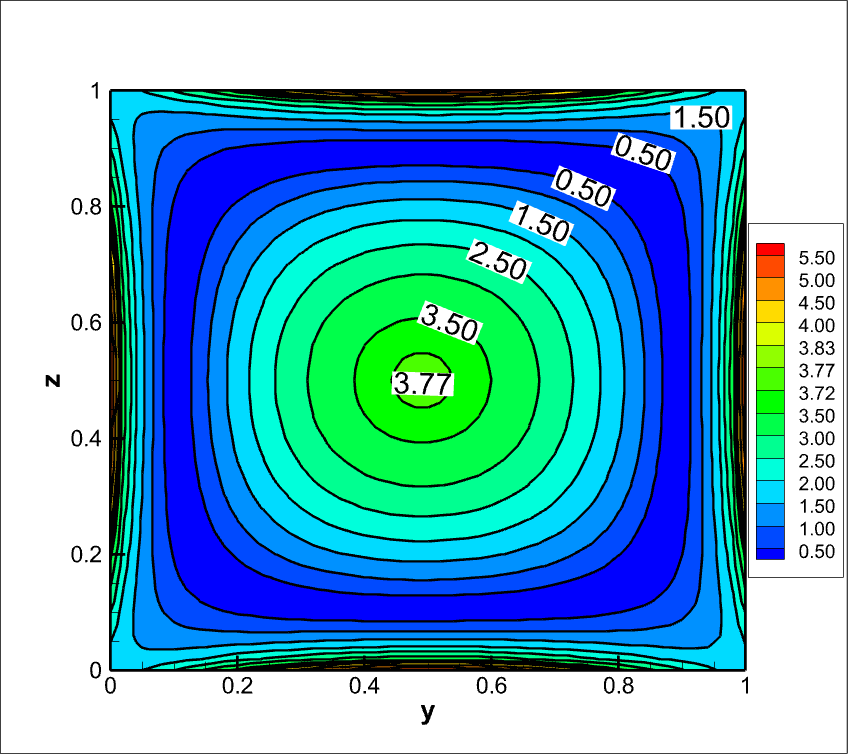}%
    \captionsetup{skip=2pt}%
    \caption{(a) $Ra=10^3, Ha=25$}
    \label{fig:P3_LC_RA_10^3_Ha_25.png}
  \end{subfigure}%
 \begin{subfigure}{0.33\textwidth}        
   \centering
    \includegraphics[width=\textwidth]{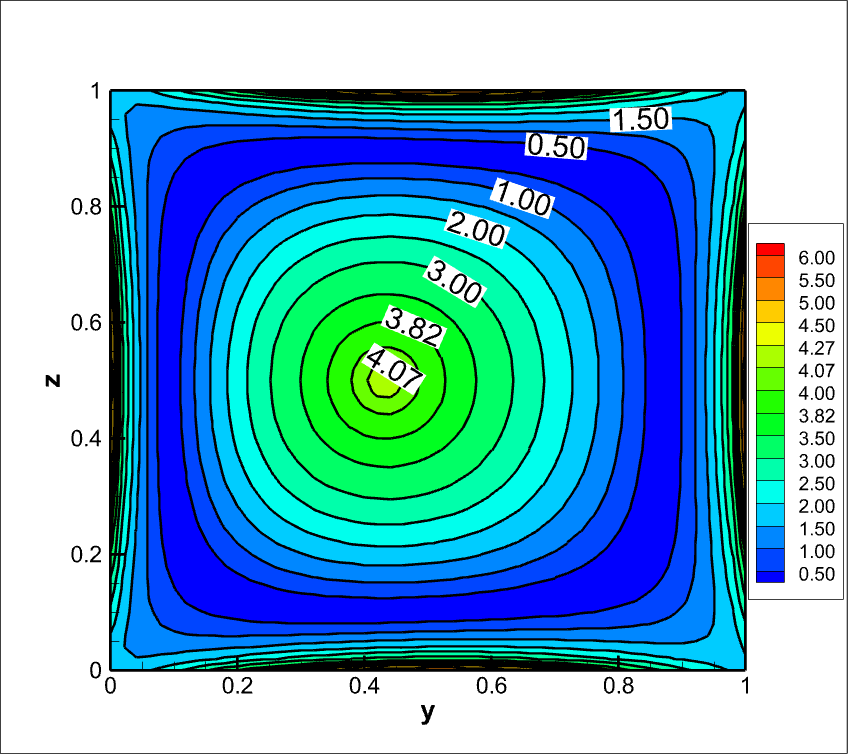}%
    \captionsetup{skip=2pt}%
    \caption{(b) $Ra=10^4, Ha=25$}
    \label{fig:P3_LC_RA_10^4_Ha_25.png}
  \end{subfigure}
   \begin{subfigure}{0.33\textwidth}        
   \centering
    \includegraphics[width=\textwidth]{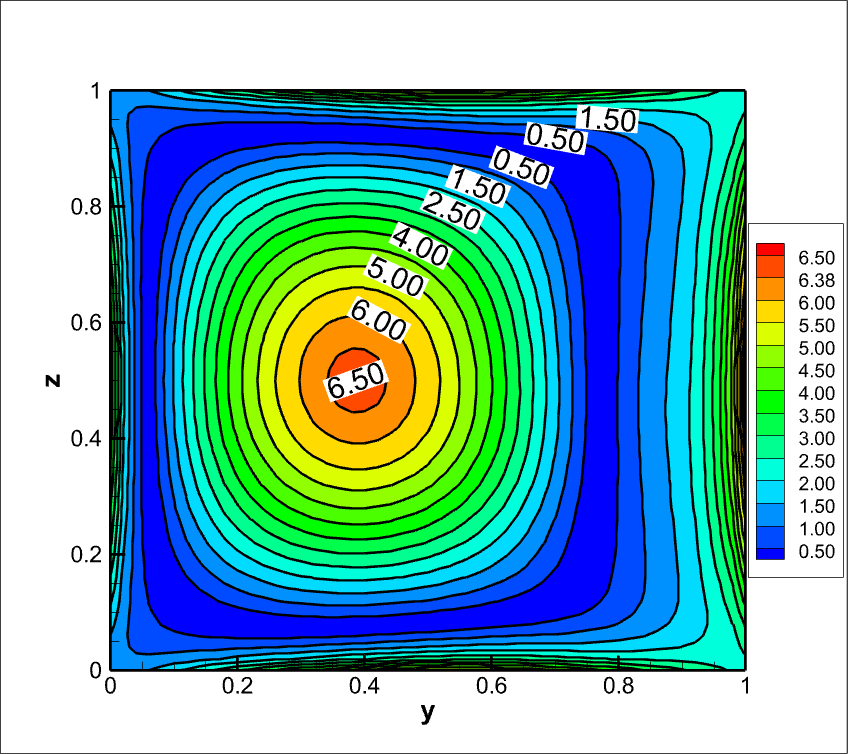}%
    \captionsetup{skip=2pt}%
    \caption{(c) $Ra=10^5, Ha=25$}
    \label{fig:P3_LC_RA_10^5_Ha_25.png}
  \end{subfigure}%
  \hspace*{\fill}

  \vspace*{8pt}%
  \hspace*{\fill}%
  \begin{subfigure}{0.33\textwidth}     
    \centering
    \includegraphics[width=\textwidth]{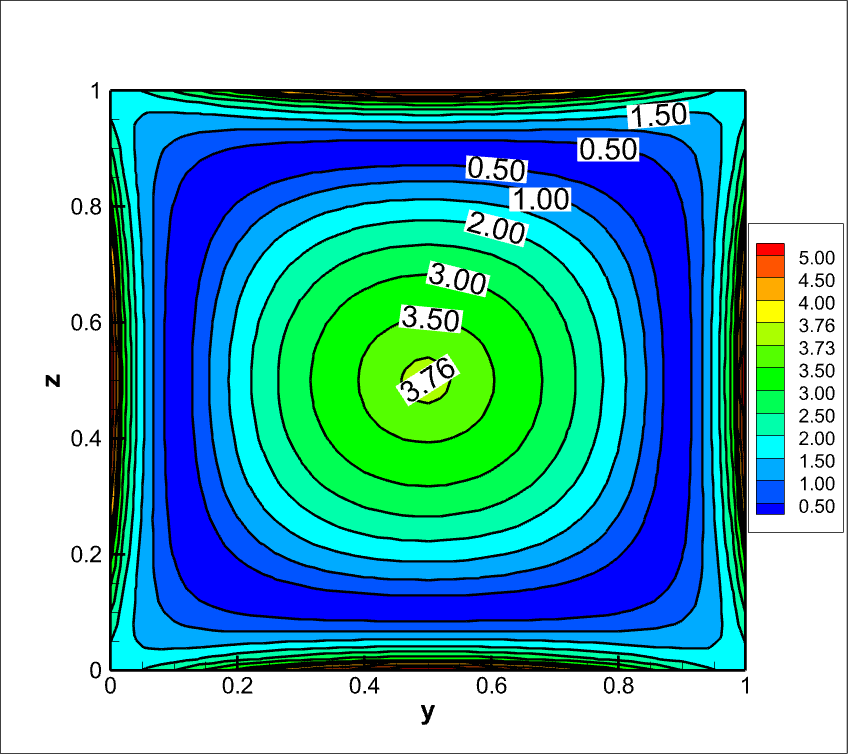}%
    \captionsetup{skip=2pt}%
    \caption{(d) $Ra=10^3, Ha=50$}
    \label{fig:P3_LC_RA_10^3_Ha_50.png}
  \end{subfigure}%
 \begin{subfigure}{0.33\textwidth}        
   \centering
    \includegraphics[width=\textwidth]{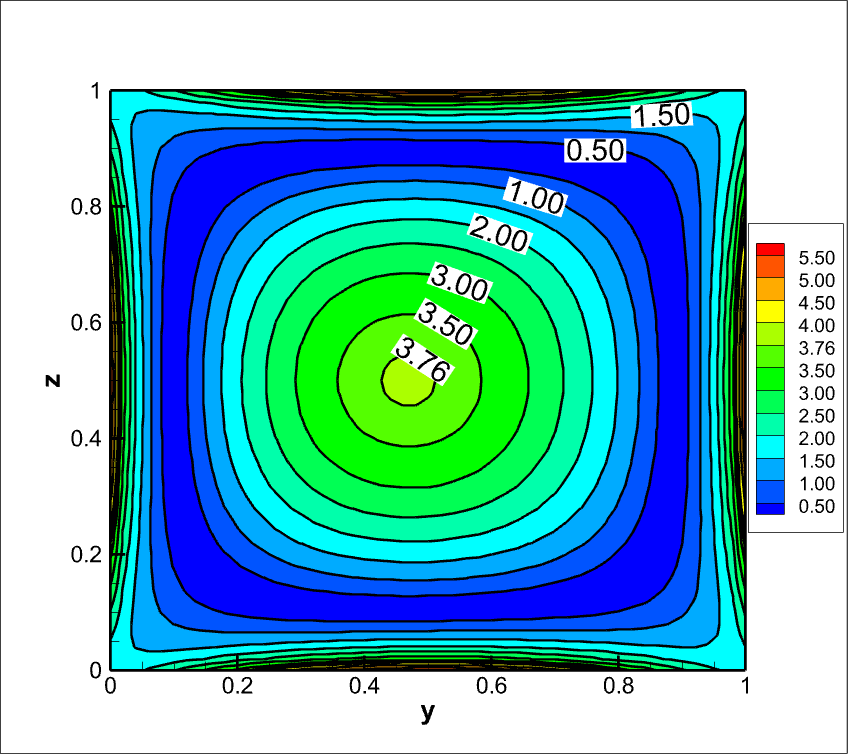}%
    \captionsetup{skip=2pt}%
    \caption{(e) $Ra=10^4, Ha=50$}
    \label{fig:P3_LC_RA_10^4_Ha_50.png}
  \end{subfigure}
   \begin{subfigure}{0.33\textwidth}        
   \centering
    \includegraphics[width=\textwidth]{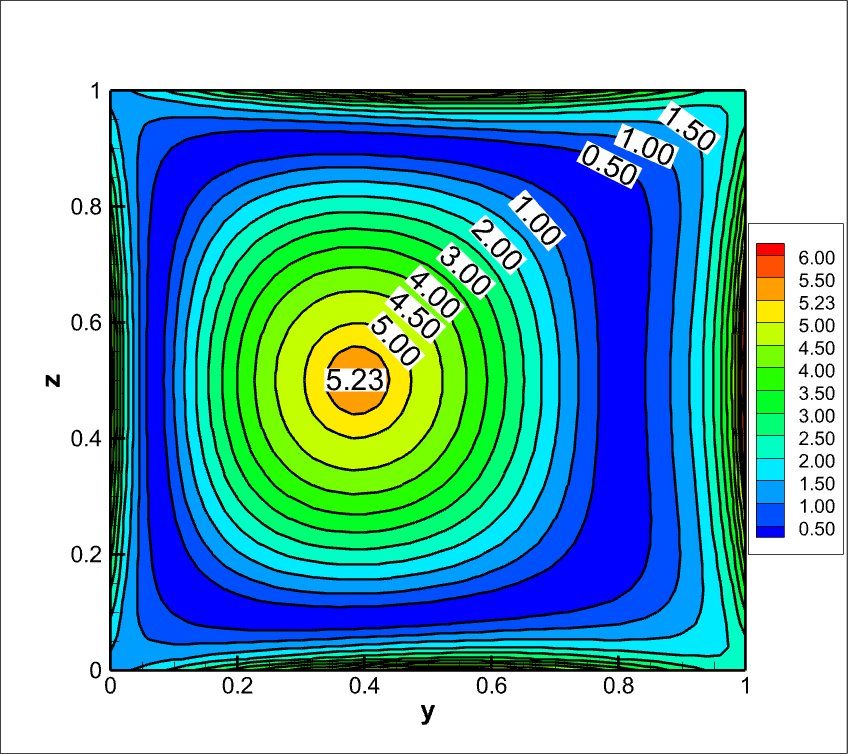}%
    \captionsetup{skip=2pt}%
    \caption{(f) $Ra=10^5, Ha=50$}
    \label{fig:P3_LC_RA_10^5_Ha_50.png}
  \end{subfigure}%
  \hspace*{\fill}

  \vspace*{8pt}%
  \hspace*{\fill}%
  \begin{subfigure}{0.33\textwidth}     
    \centering
    \includegraphics[width=\textwidth]{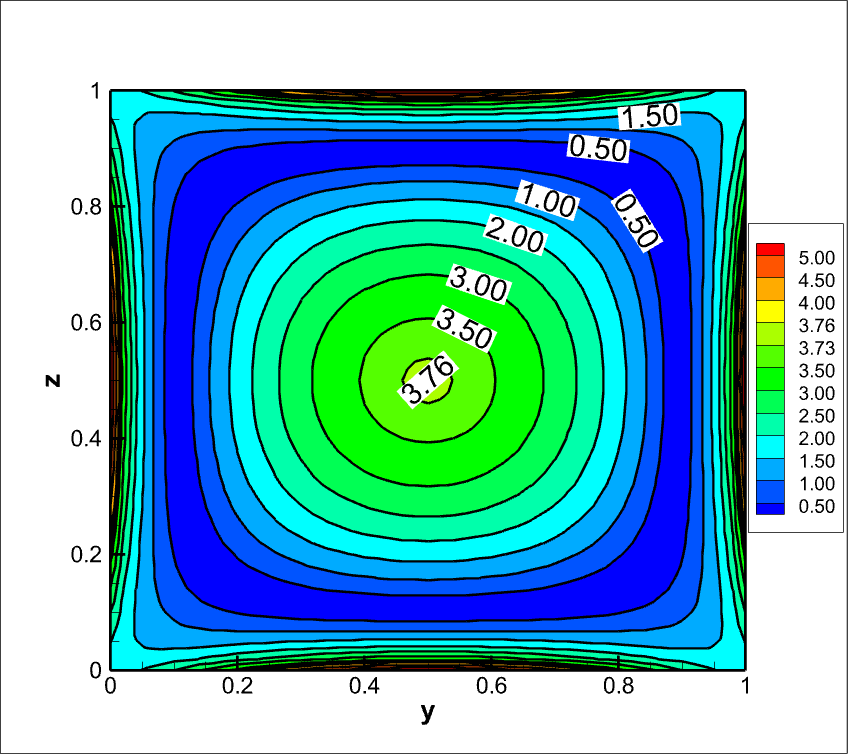}%
    \captionsetup{skip=2pt}%
    \caption{(g) $Ra=10^3, Ha=100$}
    \label{fig:P3_LC_RA_10^3_Ha_100.png}
  \end{subfigure}%
 \begin{subfigure}{0.33\textwidth}        
   \centering
    \includegraphics[width=\textwidth]{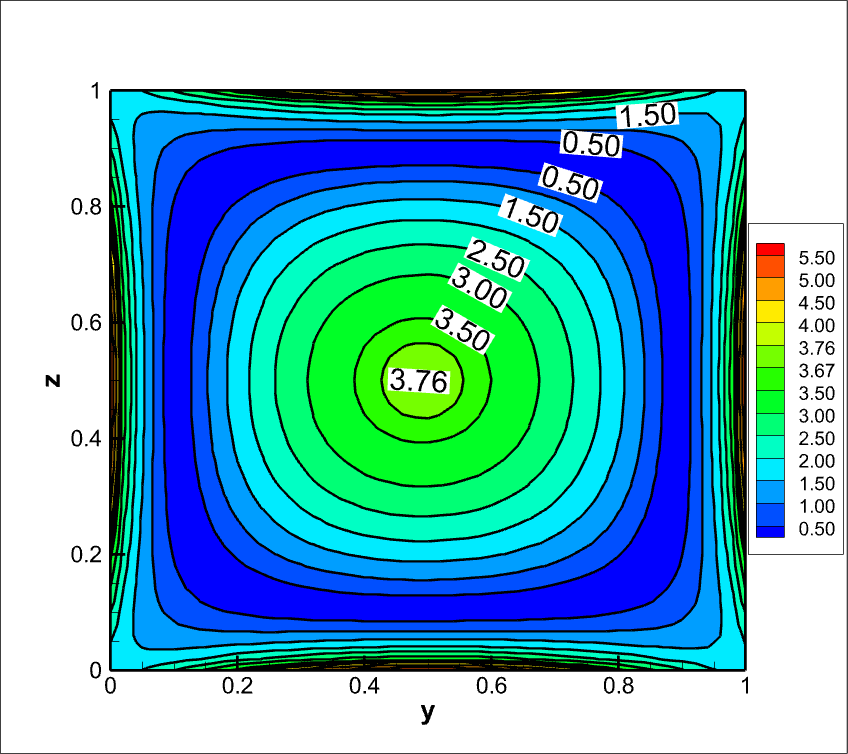}%
    \captionsetup{skip=2pt}%
    \caption{(h) $Ra=10^4, Ha=100$}
    \label{fig:P3_LC_RA_10^4_Ha_100.png}
  \end{subfigure}
   \begin{subfigure}{0.33\textwidth}        
   \centering
    \includegraphics[width=\textwidth]{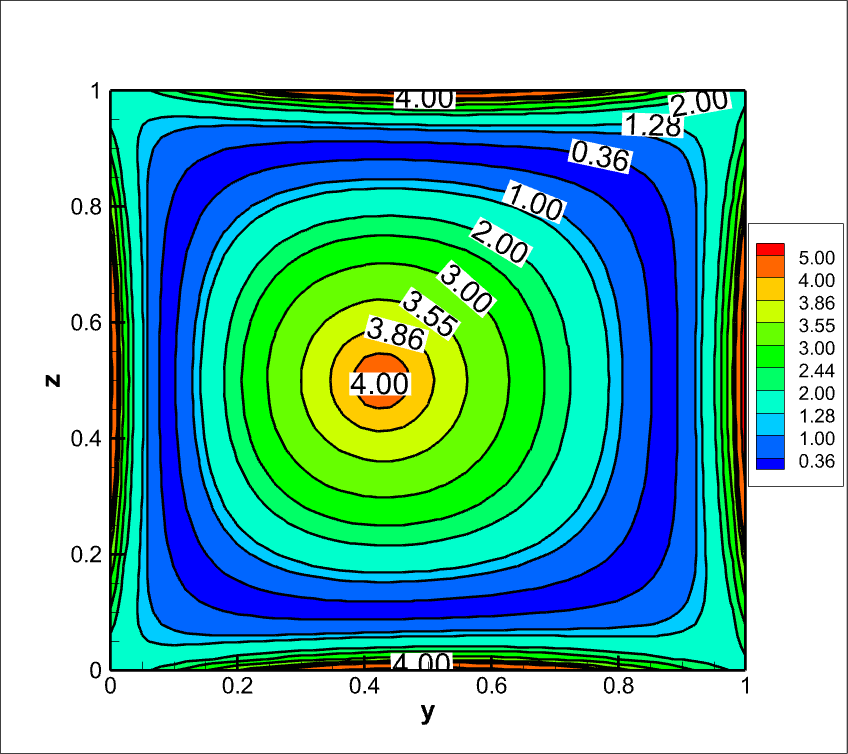}%
    \captionsetup{skip=2pt}%
    \caption{(i) $Ra=10^5, Ha=100$}
    \label{fig:P3_LC_RA_10^5_Ha_100.png}
  \end{subfigure}%
  \hspace*{\fill}

  \vspace*{8pt}%
  \hspace*{\fill}%
  \begin{subfigure}{0.33\textwidth}     
    \centering
    \includegraphics[width=\textwidth]{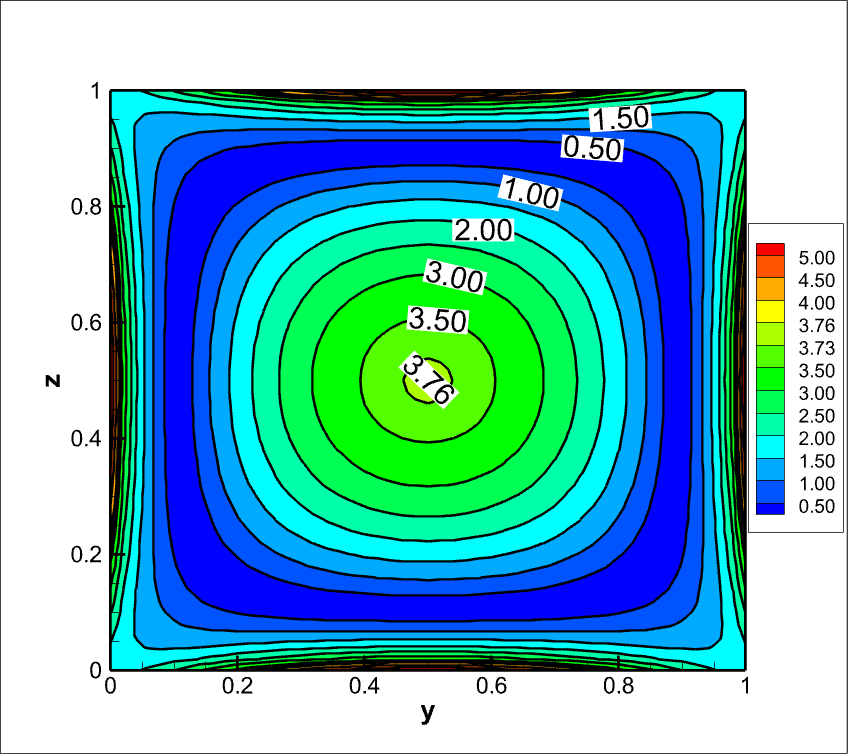}%
    \captionsetup{skip=2pt}%
    \caption{(j) $Ra=10^3, Ha=150$}
    \label{fig:P3_LC_RA_10^3_Ha_150.png}
  \end{subfigure}%
 \begin{subfigure}{0.33\textwidth}        
   \centering
    \includegraphics[width=\textwidth]{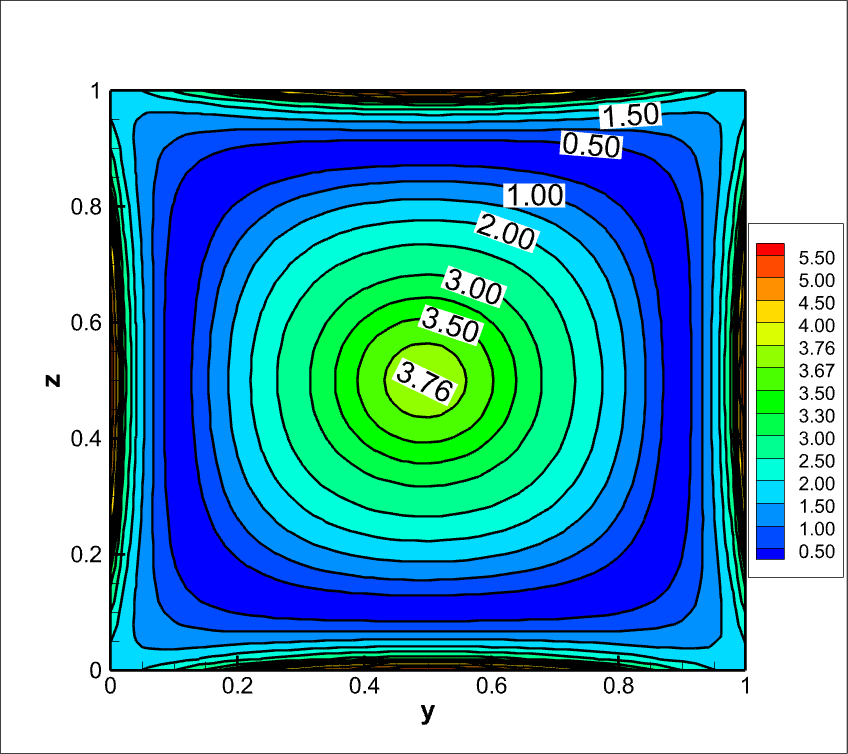}%
    \captionsetup{skip=2pt}%
    \caption{(k) $Ra=10^4, Ha=150$}
    \label{fig:P3_LC_RA_10^4_Ha_150.png}
  \end{subfigure}
   \begin{subfigure}{0.33\textwidth}        
   \centering
    \includegraphics[width=\textwidth]{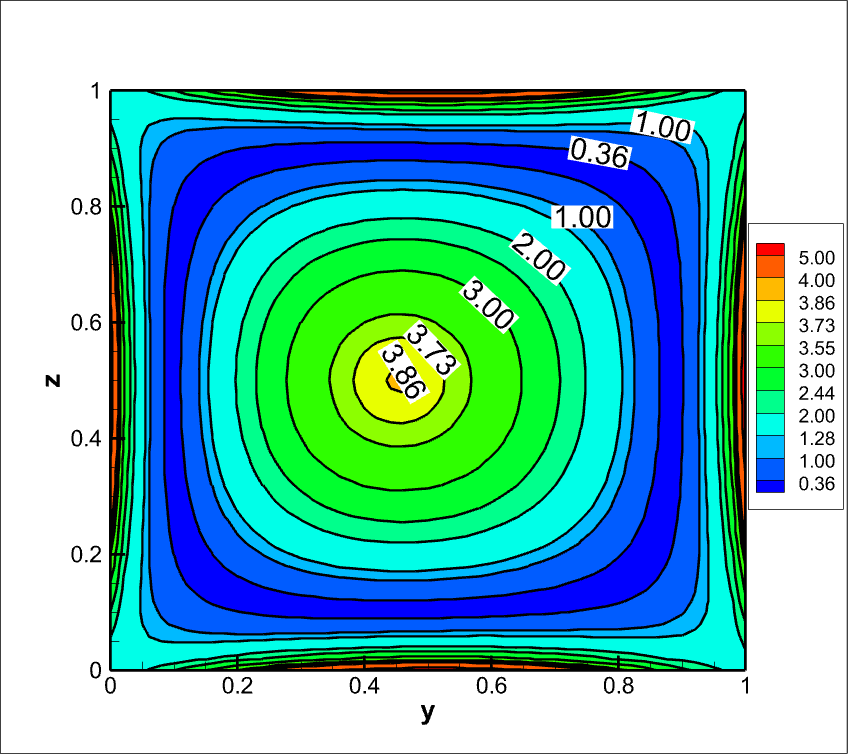}%
    \captionsetup{skip=2pt}%
    \caption{(l) $Ra=10^5, Ha=150$}
    \label{fig:P3_LC_RA_10^5_Ha_150}
  \end{subfigure}%
  \hspace*{\fill}
  \vspace*{2pt}%
  \hspace*{\fill}%
  \caption{Case 3. $Nu_{\text{L}}$ contours on the left ($x=0$) heated wall of the cubic cavity}
  \label{fig:case-3_Local_Nusselt}
\end{figure}

{\small\begin{table}[htbp] \footnotesize
\caption{\small Maximum value of $Nu_{\text{L}}$ determined at the heated wall ($x=0$) for three different cases by using $51\times51\times51$ grid size }\label{Maximum_Nusselt_Number_Comparison}
\centering
 \begin{tabular}{lccccccc}  \hline \hline
  &  &  &  &   Max. $Nu_{\text{L}}$   &   \\ \hline 
 & $Ra$
   & $Ha=25$ & $Ha=50$   & $Ha=100$  &  $Ha=150$ & Max. difference (\%) \\
   &  &  &  & & & at fixed $Ra$ \\ \hline 
 \textbf{Case 1} & $10^3$  & 1.09 & 1.03    &  1.01 & 1.01  & -7.33  \\
 & $10^4$   &    2.01 & 1.35     &  1.09 & 1.04 & -48.25   \\
 & $10^5$   & 8.22 & 4.50  &  2.35 & 2.31 & -71.77  \\
Max. difference (\%)  & --   & +654.1 & +336.8 &  +130.3 & +128.7 & --  \\
at fixed $Ha$ & \\
\hline 
 \textbf{Case 2} & $10^3$  & 6.27 & 6.24    &  6.23 & 6.22 & -0.79  \\
& $10^4$   &    6.54 & 6.35     &  6.27 & 6.25 & -4.43   \\
 & $10^5$   & 7.41  & 6.91  &  6.60 & 6.45 & -12.95  \\
Max. difference (\%) &  --  & +18.18 & +10.73&  +5.93 & +3.69 & --  \\
at fixed $Ha$  \\
\hline 
 \textbf{Case 3} & $10^3$  & 5.56 & 5.54     &  5.52 & 5.52  & -0.71 \\
& $10^4$   &    5.83 & 5.62     &  5.56 & 5.54  & -4.97  \\
 & $10^5$   & 6.96  & 6.34  &  5.81 & 5.81  & -16.52\\
Max. difference (\%) & --   & +25.17 & +14.44 &  +5.25 & +5.25 & --  \\
at fixed $Ha$  \\
\hline\hline
 \end{tabular}
\end{table}
}

Figure \ref{fig:case-1_Local_Nusselt} presents the investigation of the local Nusselt number ($Nu_{\text {L }}$) for Case 1 on the left heated wall across various $Ha$ and $Ra$, illustrating the influence on the heat transfer process. When carefully analyzing the situation at a lower Rayleigh number ($Ra = 10^3$), a notably symmetrical pattern is observed for each $Ha$ value, specifically around $y = 0.5$. But, with a rise in the $Ra$, a notable asymmetry becomes evident, specifically around $y = 0.5$. The position of the maximum Nusselt number remains relatively unchanged with variations in both the $Ha$ and $Ra$. The maximum Nusselt number values for various cases with different $Ha$ and $Ra$ are presented in Table \ref{Maximum_Nusselt_Number_Comparison}. 
This table also presents the maximum change in  $Nu_{\text{L}}$ values (in percentage) for different $Ha$ with fixed $Ra$ (row-wise), and for different $Ra$ with fixed $Ha$ (column-wise). It is readily apparent that the maximum $Nu_{\text{L}}$ decreases with an increase in $Ha$ for a fixed $Ra$, and increases with increase $Ra$ at a specific $Ha$. For Case 1, This table highlights that with an increase in $Ra$ from $10^3$ to $10^5$ with fixed $Ha=25$, there is a substantial 654.1\% rise in the $Nu_{\text {L }}$ value, and the max. $Nu_{\text {L}}$ occur near $(y,z)=(0,0.5)$ (Figure \ref{fig:case-1_Local_Nusselt}). At $Ha=50$, $Ha=100$, and $Ha=150$, there are 336.8\%, 130.3\%, and 128.7\% increase in the $Nu_{\text {L}}$ values, respectively. 
This indicates that as $Ra$ increases, $Nu_{\text {L}}$ values also increase. Moreover, the amount of increment in the $Nu_{\text {L}}$ value reduces with an increase in the magnetic strength ($Ha$), and $Nu_{\text {L }}$ values decrease with an increase in $Ha$. The table clearly shows a maximum decrease of 71.77\% for $Ra=10^5$ as we go from $Ha=25$ to $Ha=150$. For non-uniform boundary conditions (Case 2 and Case 3), the maximum changes in the calculated values of local Nusselt number as low as 12.95\% (for different $Ha$ at $Ra=10^5$) and 18.18\% (for different $Ra$ at $Ha=25$) in Case 2, and 16.52\% (for different $Ha$ at $Ra=10^5$) and 25.17\% (for different $Ra$ at $Ha=25$) in Case 3, respectively. The corresponding $Nu_{\text {L}}$ contour plots on the non-uniformly heated left wall $(x=0.0)$, for Case 2 and Case 3, are presented in Figures \ref{fig:case-2_Local_Nusselt} \& \ref{fig:case-3_Local_Nusselt}, respectively.

At low $Ra$ ($Ra= 10^3$), for Case 2 (Figure \ref{fig:case-2_Local_Nusselt}), the profiles exhibit a nearly identical pattern, indicating that the strength of the magnetic field has minimal influence on $Nu_{\text {L}}$ at lower $Ra$. The maximum $Nu_{\text {L}}$ values occur near the boundaries $(y = 0, y = 1)$ and the middle of the plane $(y = 0.5)$. For $Ra = 10^4$, there is a slight change in the contour lines at the middle of the plane for lower $Ha$ ($Ha=25$ \& $50$), but the values of $Nu_{\text {L}}$ do not change significantly. For $Ra=10^5$, the $Nu_{\text {L}}$ contours show a vertical bi-circular pattern near $y=0.4$ for $Ha=25$. This vertical bi-circular shape gradually changes to a vertical single circular contour at the same place with increasing $Ha$ from 50 to 150. When observing the figures \ref{fig:case-2_Local_Nusselt}(a)-(c) (i.e., with fixed $Ha$), the effect of $Ra$ at fixed $Ha$ is evident. There is an approximate 18\% increment in the $Nu_{\text {L}}$ values when moving from $Ra = 10^3$ to $Ra = 10^5$ at fixed $Ha = 25$. For other $Ha$ values, the changes are compatibly small, indicating that $Ra$ significantly affects heat transfer rates only at low $Ha$ for this particular case. Overall, the influence of $Ha$ and $Ra$ is relatively very weaker in comparison to Case 1. In Case 3 (Figure \ref{fig:case-3_Local_Nusselt}), the highest Nusselt number value exists near the boundaries and the center of the cavity, and the maximum value increases with an increase in $Ra$ while decreasing with an increase in $Ha$ values. For low $Ra$ values, a circular pattern is evident, but for $Ra = 10^5$, a slightly vertically stretched pattern emerges, which diminishes further for high $Ha$. This indicates that the heat transfer rate increases with an increase in $Ra$.

\begin{figure}[htbp]
 \centering
\begin{subfigure}{0.33\textwidth}     
    \centering
    \includegraphics[width=\textwidth]{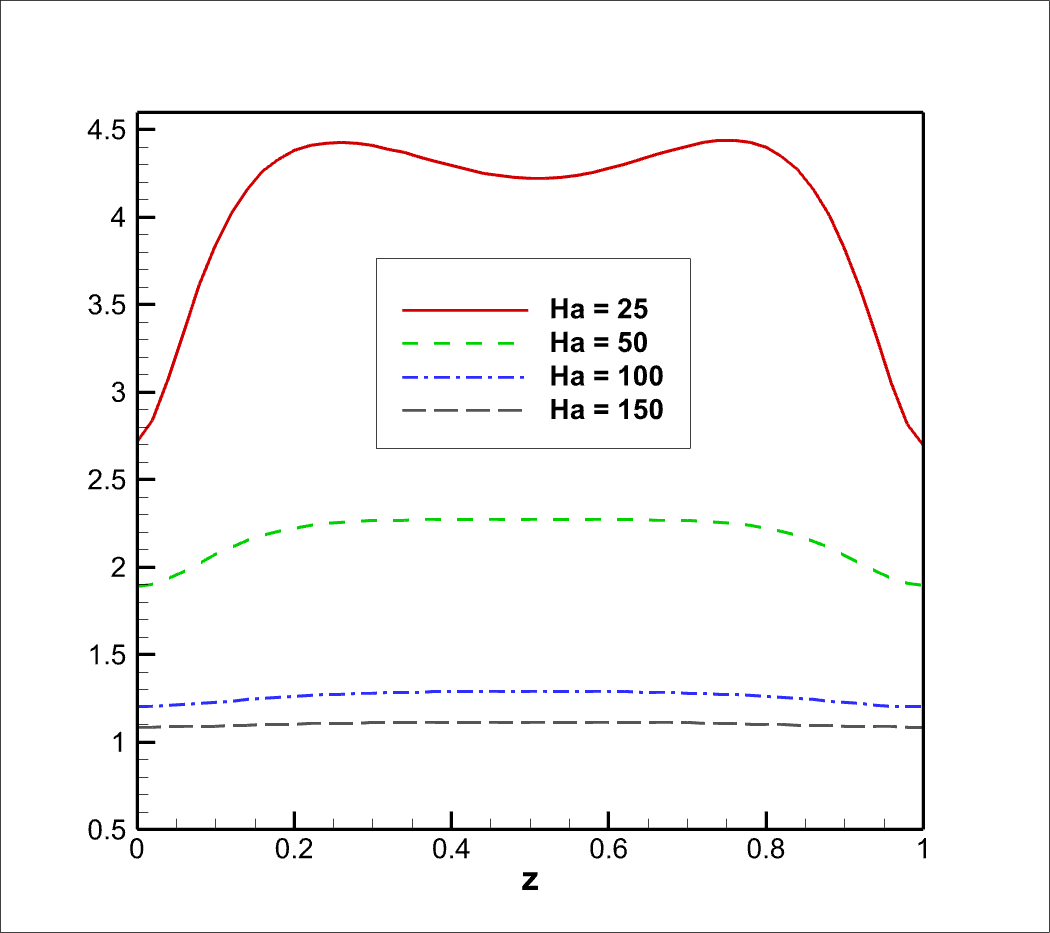}%
    \captionsetup{skip=2pt}%
    \caption{(a) Case 1}
    \label{fig:HA_Effect_P1.png}
  \end{subfigure}%
 \begin{subfigure}{0.33\textwidth}        
   \centering
    \includegraphics[width=\textwidth]{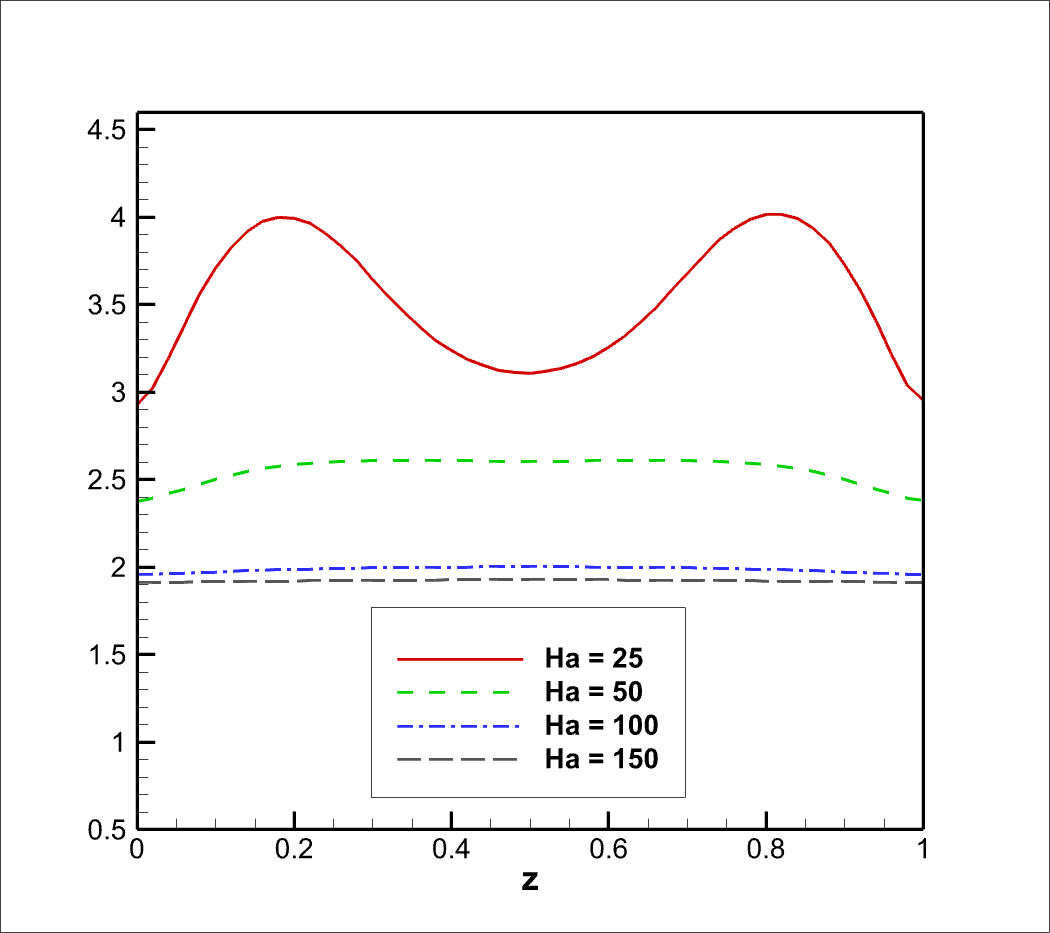}%
    \captionsetup{skip=2pt}%
    \caption{(b) Case 2}
    \label{fig:HA_Effect_P2.png}
  \end{subfigure}
   \begin{subfigure}{0.33\textwidth}        
   \centering
    \includegraphics[width=\textwidth]{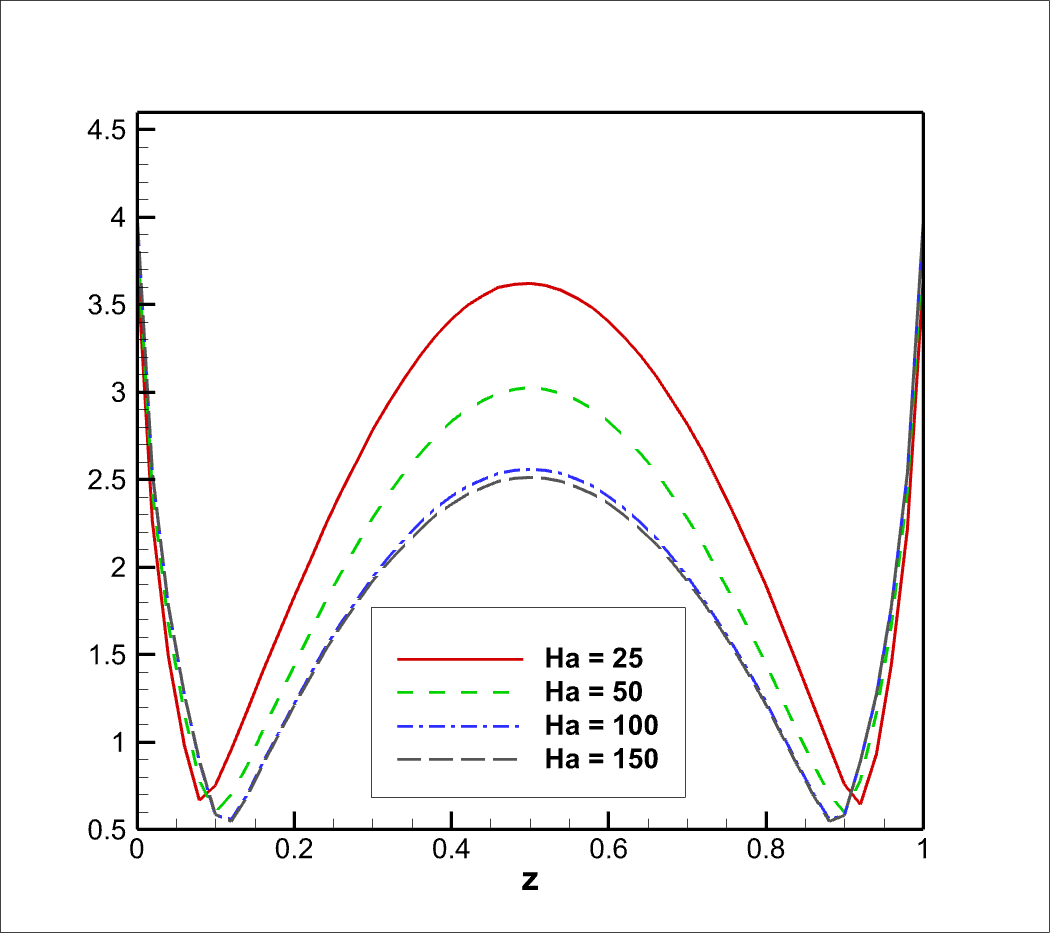}%
    \captionsetup{skip=2pt}%
    \caption{(c) Case 3}
    \label{fig:HA_Effect_P3.png}
  \end{subfigure}%
  \caption{Influence of varying $Ha$ on average Nusselt number ($Nu_{\text {A}}$) with fixed $Ra = 10^5$}
  \label{fig:Ha_Effect_On_Avg_Nusselt}
\end{figure}

Figure \ref{fig:Ha_Effect_On_Avg_Nusselt} presents the influence of varying $Ha$ on average Nusselt number with fixed $Ra$ $(Ra=10^5)$ for all three considered cases. In Figure \ref{fig:Ha_Effect_On_Avg_Nusselt}(a), results are presented to visualize the variations in the $Nu_{\text {A}}$ along the $z$-direction across the range of $Ha$ at fixed $Ra=10^5$ for the Case 1. The $Nu_{\text {A}}$ decreases with an increase in the $Ha$. The $Nu_{\text {A}}$ curve is increasing for $z<0.5$ and decreasing for $z > 0.5$ and only a single local maximum exists at $z=0.5$ for $Ha=50,100$ and $150$. However, the scenario takes a distinct turn when examining the case of $Ha = 25$. Here, in place of a single local maximum at $z = 0.5$, there are two small local maxima at $z = 0.25$ and $z = 0.75$. This interesting observation is due to the existence of extremely powerful convective flow in the $z$-direction at the lowest value of $Ha=25$. The occurrence of these dual maxima demonstrates how $Ha$ affects heat transport dynamics significantly. In this case ($Ha=25$), the convective flow is so strong that it improves heat transmission not just in the center, but also near the $z = 0.25$ and $z = 0.75$. However, Case 2 exhibits localized enhancements in heat transfer performance with increased $Nu_{\text {A}}$ value curves with respect to Case 1 (Figure \ref{fig:Ha_Effect_On_Avg_Nusselt}(b)). Whereas, Case 3 shows an entirely different behavior of $Nu_{\text {A}}$ along $z$-direction for a fixed $Ha$ with $Ra=10^5$ (Figure \ref{fig:Ha_Effect_On_Avg_Nusselt}(c)). For a fixed $Ha$, one local maximum is observed at the middle ($z = 0.5$), and two local minima are noted at $z = 0.12$ and $z = 0.88$. 
These results confirm the influence of non-uniform temperature boundary conditions on the average Nusselt number. Local maxima in the average Nusselt number curve suggest regions where convective heat transfer is particularly more efficient and local minima in the average Nusselt number curve represent regions where convective heat transfer efficiency is comparatively lower.

\begin{figure}[htbp]
 \centering
\begin{subfigure}{0.33\textwidth}     
    \centering
    \includegraphics[width=\textwidth]{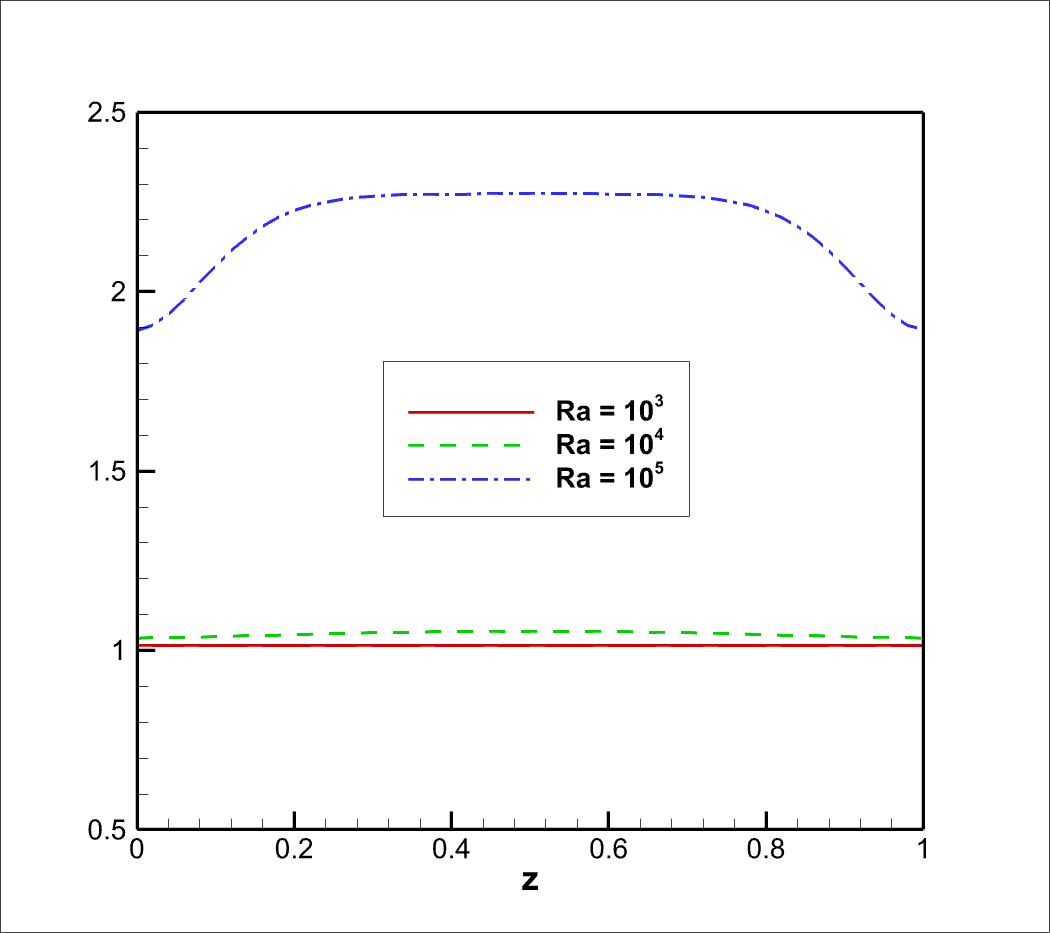}%
    \captionsetup{skip=2pt}%
    \caption{(a) Case 1}
    \label{fig:Ra_Effect_P1.png}
  \end{subfigure}%
 \begin{subfigure}{0.33\textwidth}        
   \centering
    \includegraphics[width=\textwidth]{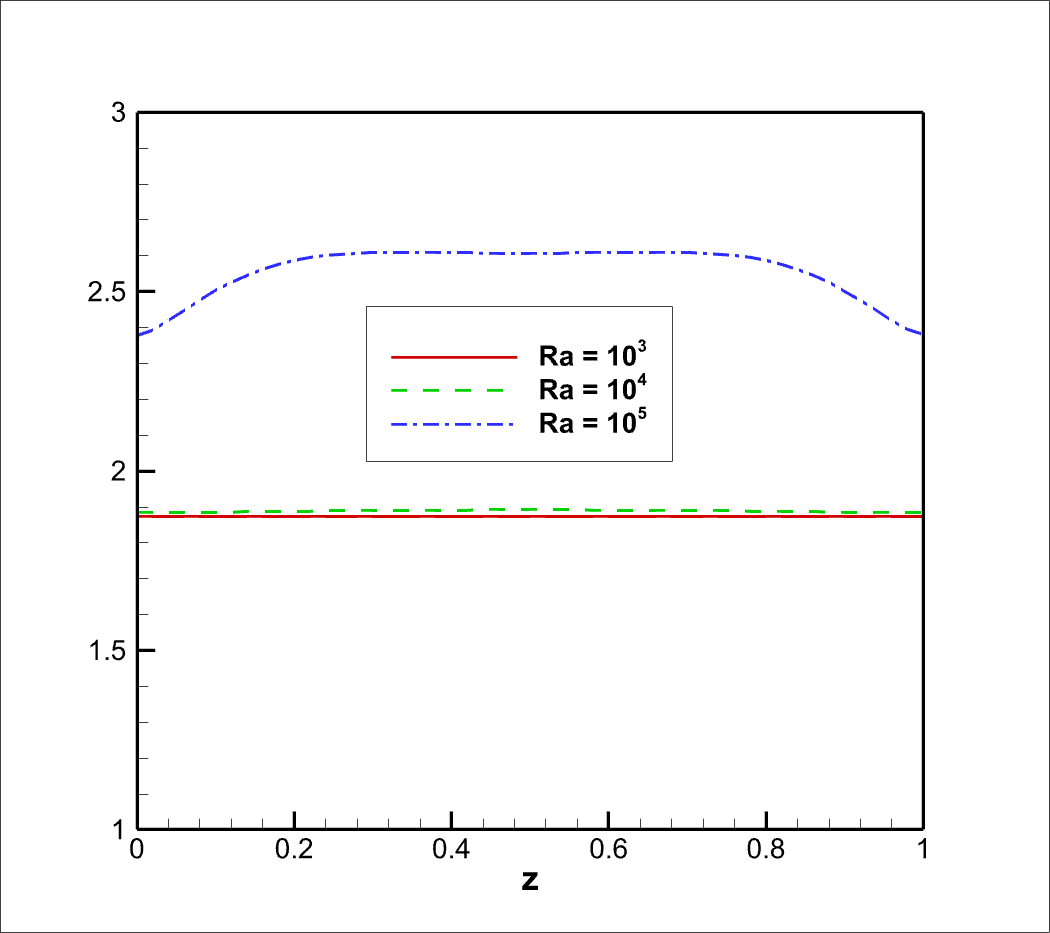}%
    \captionsetup{skip=2pt}%
    \caption{(b) Case 2 }
    \label{fig:Ra_Effect_P2.png}
  \end{subfigure}
   \begin{subfigure}{0.33\textwidth}        
   \centering
    \includegraphics[width=\textwidth]{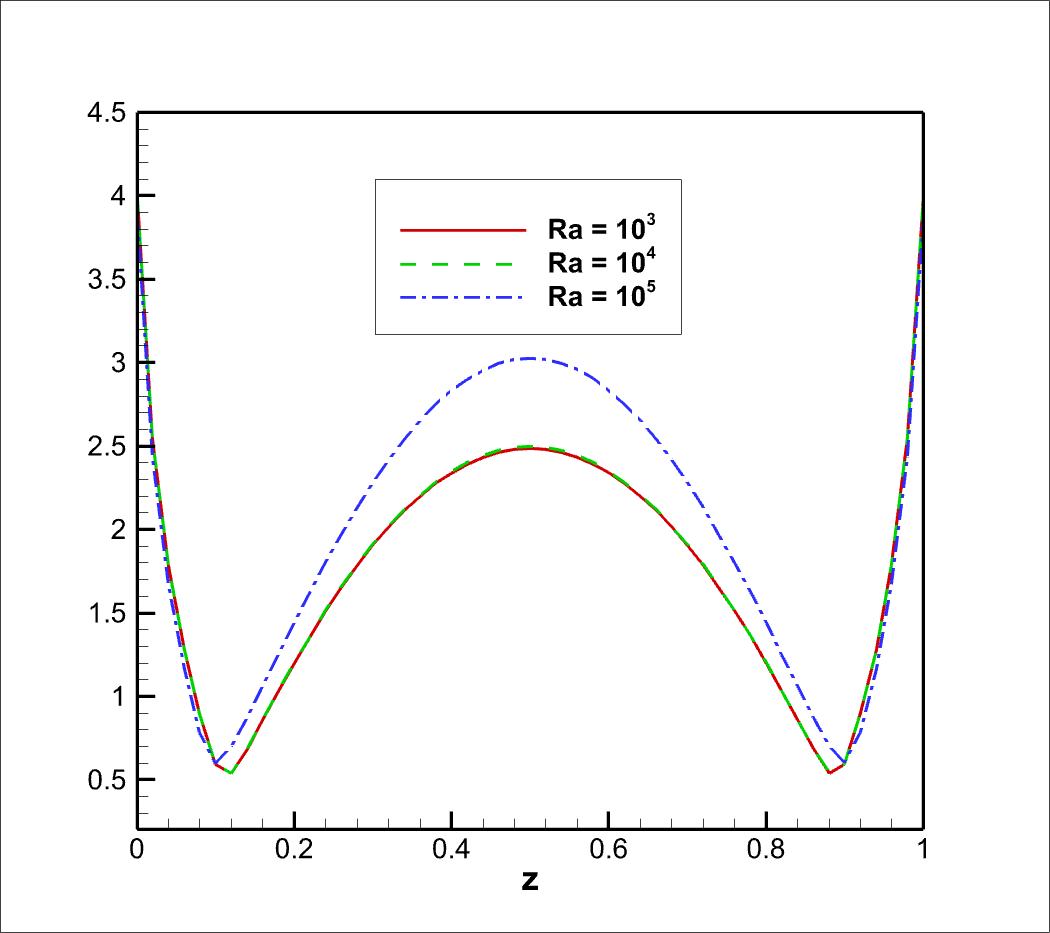}%
    \captionsetup{skip=2pt}%
    \caption{(c) Case 3}
    \label{fig:Ra_Effect_P3.png}
  \end{subfigure}%
  \caption{Influence of varying $Ra$ on average Nusselt number ($Nu_{\text {A}}$) with fixed $Ha = 50$}
  \label{fig:Ra_Effect_On_Avg_Nusselt}
\end{figure}

The impact of varying $Ra$ on $Nu_{\text{A}}$ along $z$-direction for all three cases are presented in Figure \ref{fig:Ra_Effect_On_Avg_Nusselt}. It is noticeable that as the $Ra$ increases, the average curve value of $Nu_{\text {A}}$ increases, but it is most significant for Case 1 (the uniform temperature boundary condition).

\section{Entropy Generation}
\label{sec:Entropy Generation}
The presence of a heat gradient between the cavity's vertical walls causes a non-equilibrium scenario in the fluid, which generates entropy inside the system. Entropy generation is very important for understanding and analyzing thermodynamic systems. 
It measures the energy degeneration or loss within a system. Given the changes in heat transfer phenomena, we expect entropy to vary for different temperature boundary conditions. Analyzing the entropy generation in thermodynamic systems can help to determine the best design parameters for systems with MHD natural convection.
The dimensionless local entropy generation $(E_{\mathrm{LE}})$ 
 can be expressed as follows \cite{Magherbi_2003, Rashed_2017, Tayebi_2019}:
\begin{equation}\label{eq_3_Entropy}
E_{\mathrm{LE}}=E_{\mathrm{LE,F}}+E_{\mathrm{LE,H}}+E_{\mathrm{LE,M}}
\end{equation}
where, $E_{\mathrm{LE,F}}$ represents the local entropy
generation caused by the fluid friction, $E_{\mathrm{LE,H}}$ represents the local entropy generated by heat transport, and $E_{\mathrm{LE,M}}$, represents the entropy generated by the magnetic field. These three terms can be further expressed as:
\begin{equation}\label{eq_4_Entropy}
E_{\mathrm{LE,H}}=\left[\left(\frac{\partial \theta}{\partial X}\right)^2+\left(\frac{\partial \theta}{\partial Y}\right)^2 +\left(\frac{\partial \theta}{\partial Z}\right)^2 \right]
\end{equation}

\begin{equation}
\begin{aligned}
& E_{\mathrm{LE,F}} =\varphi^{*}\left[2\left(\frac{\partial U}{\partial X}\right)^2+2\left(\frac{\partial V}{\partial Y}\right)^2+2\left(\frac{\partial W}{\partial Z}\right)^2\right]\\
&+\varphi^{*}\left[\left(\frac{\partial U}{\partial Y}+\frac{\partial V}{\partial X}\right)^2 +\left(\frac{\partial V}{\partial Z}+\frac{\partial W}{\partial Y}\right)^2+\left(\frac{\partial U}{\partial Z}+\frac{\partial W}{\partial X}\right)^2\right]
\end{aligned}\label{eq_5_Entropy}
\end{equation}
\begin{equation}\label{eq_M_Entropy}
\begin{aligned}
E_{\mathrm{LE,M}}= \varphi^{*} Ha^2V^2 
\end{aligned}
\end{equation}
$$\varphi^{*}=\frac{\mu T_0}{k}\left(\frac{a}{L(\Delta T)}\right)^2$$
Here, the value of $\varphi^{*}$, i.e., the irreversibility distribution ratio, is fixed at $10^{-4}$ for all cases. Through the integration of the dimensionless $E_{\mathrm{LE}}$ across the entire volume of the system, we can determine the total entropy generation ($E_{TE}$).
$$
E_{TE}=\int_{\vartheta} E_{\mathrm{LE}} \mathrm{d} \vartheta
$$
In thermodynamics, the Bejan number ($Be$) is defined as the ratio of entropy associated with heat transfer to the total entropy arising from the magnetic field, fluid friction, and heat transfer and is expressed as follows in the dimensionless form:
$$
Be=\frac{E_{\mathrm{LE,H}}}{E_{\mathrm{LE}}}
$$
A high $Be$ ($> 0.5$) indicates that a significant portion of the entropy generation is attributed to heat transfer, highlighting its dominance in the overall entropy generation and a low Be ($<0.5$) suggests that entropy generation due to fluid friction and magnetic field effects dominates over that caused by heat transfer. \\

\begin{figure}[htbp]
 \centering
 \vspace*{0pt}%
 \hspace*{\fill}%
\begin{subfigure}{0.24\textwidth}     
    \centering
    \includegraphics[width=\textwidth]{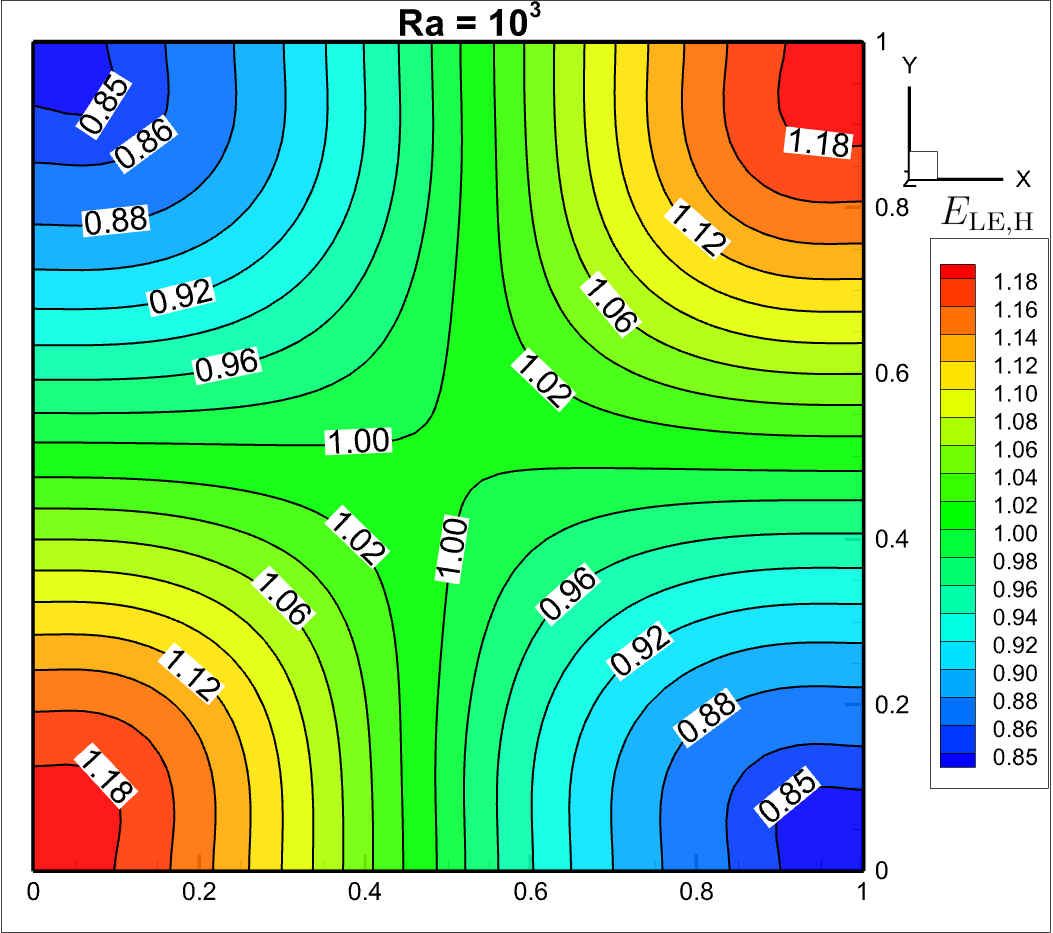}%
    \captionsetup{skip=2pt}%
    \caption{(a) $E_{\mathrm{LE,H}}$  }
    \label{fig:Ra_10^3_Ha_25_local_entropy_generation_by_HT.png}
  \end{subfigure}%
 \begin{subfigure}{0.24\textwidth}        
   \centering
    \includegraphics[width=\textwidth]{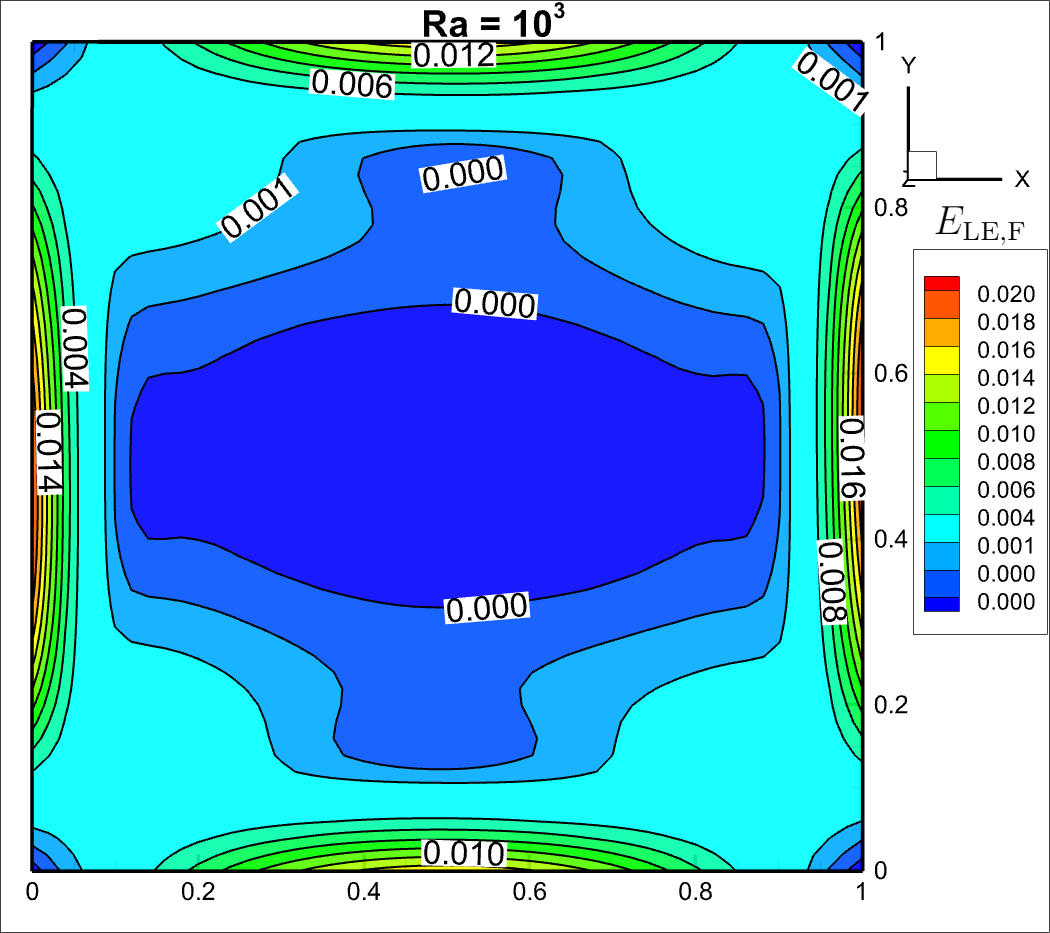}%
    \captionsetup{skip=2pt}%
    \caption{(b) $E_{\mathrm{LE,F}}$}
    \label{fig:Ra_10^3_Ha_25_local_entropy_generation_by_FF.png}
  \end{subfigure}
  \begin{subfigure}{0.24\textwidth}        
   \centering
    \includegraphics[width=\textwidth]{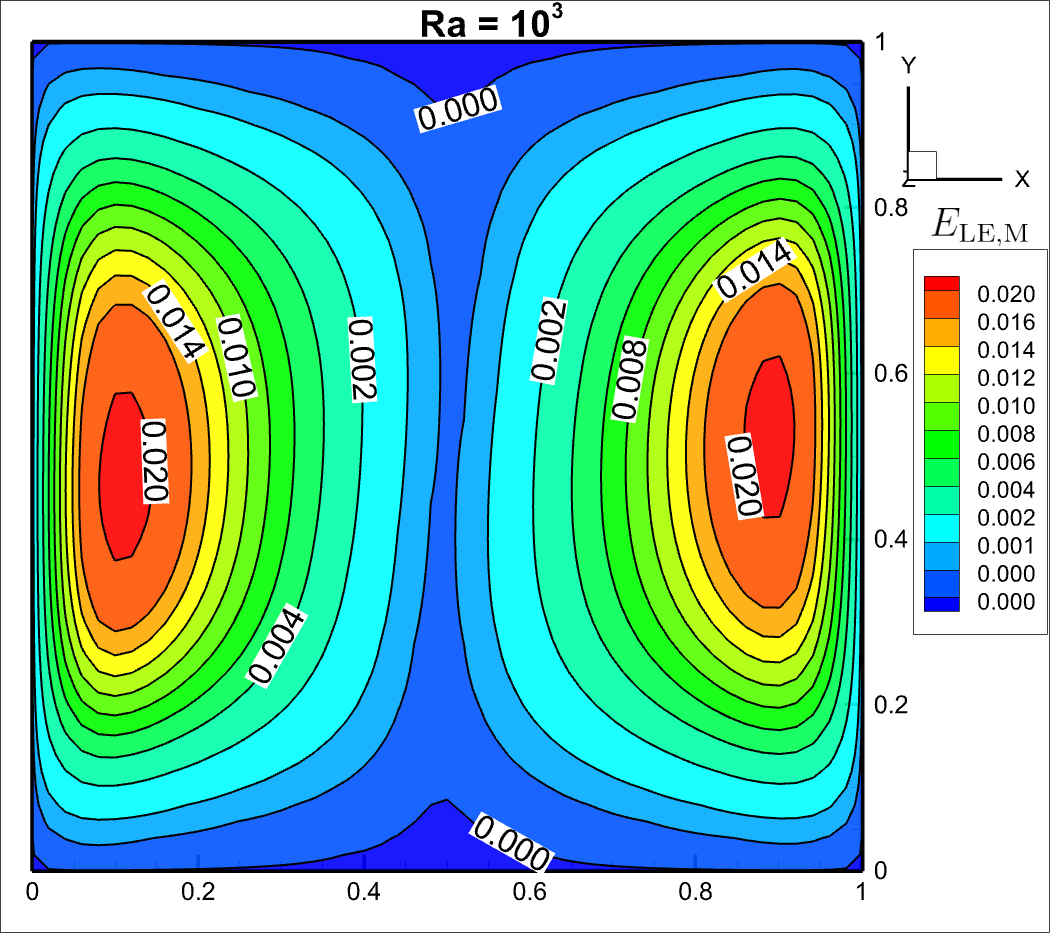}%
    \captionsetup{skip=2pt}%
    \caption{(c) $E_{\mathrm{LE,M}}$}
    \label{fig:Ra_10^3_Ha_25_local_entropy_generation_by_MF.png}
  \end{subfigure}
   \begin{subfigure}{0.24\textwidth}        
   \centering
    \includegraphics[width=\textwidth]{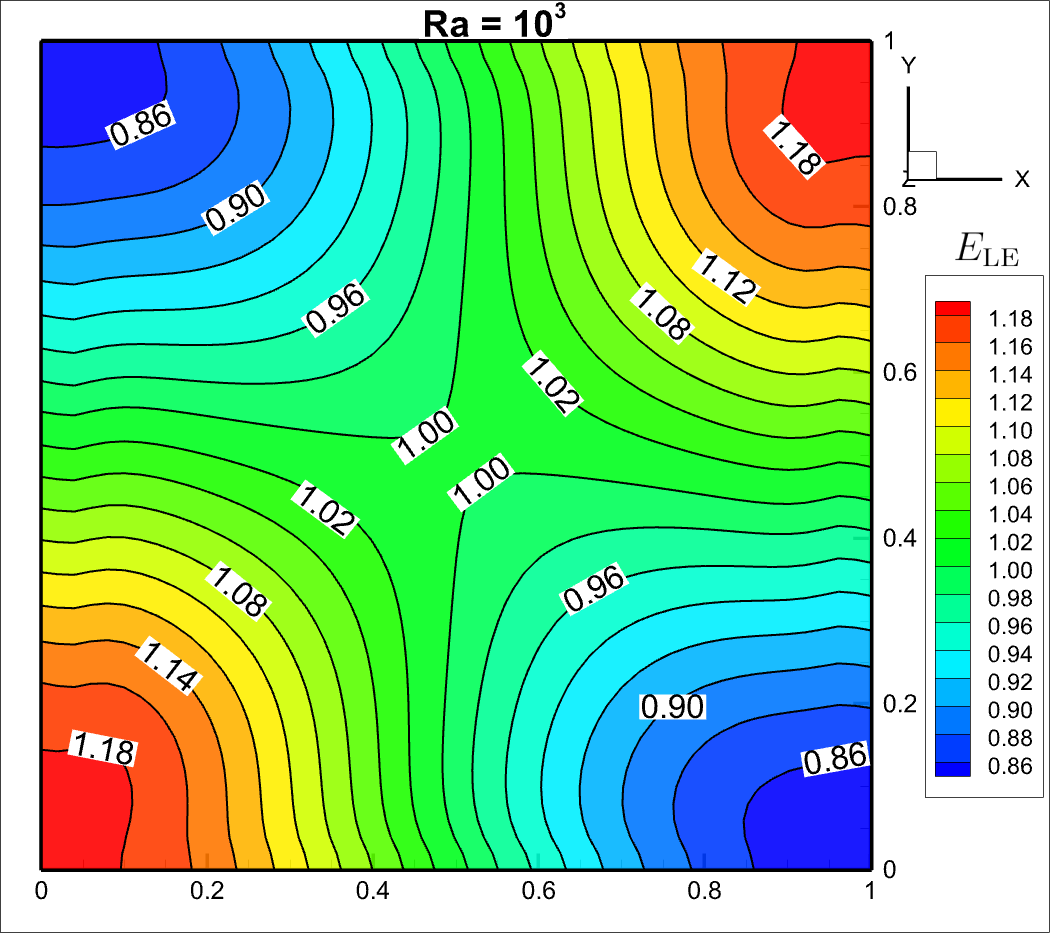}%
    \captionsetup{skip=2pt}%
    \caption{(d) $E_{\mathrm{LE}}$}
    \label{fig:Ra_10^3_Ha_25_local_entropy_generation.png}
  \end{subfigure}%
  \hspace*{\fill}

  \vspace*{8pt}%
  \hspace*{\fill}%
  \begin{subfigure}{0.24\textwidth}     
    \centering
    \includegraphics[width=\textwidth]{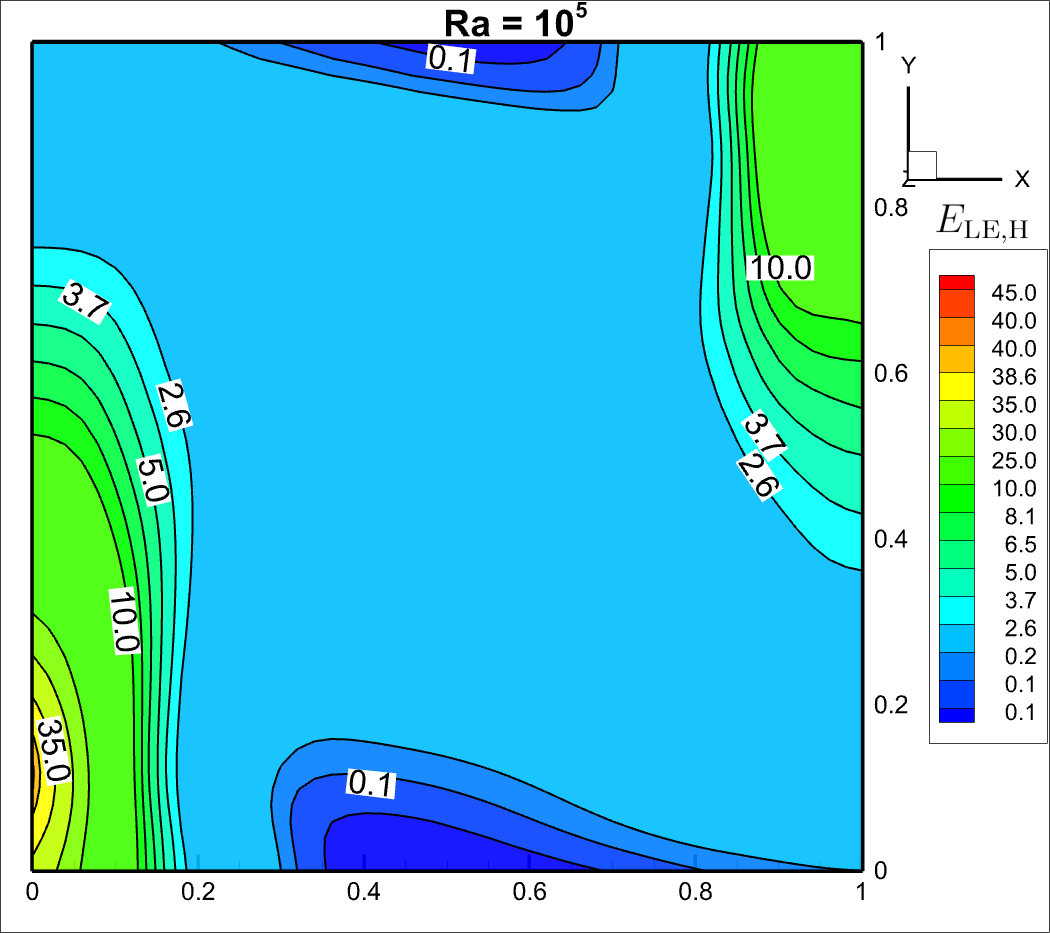}%
    \captionsetup{skip=2pt}%
    \caption{(e) $E_{\mathrm{LE,H}}$}
    \label{fig:Ra_10^5_Ha_25_local_entropy_generation_by_HT.png}
  \end{subfigure}%
 \begin{subfigure}{0.24\textwidth}        
   \centering
    \includegraphics[width=\textwidth]{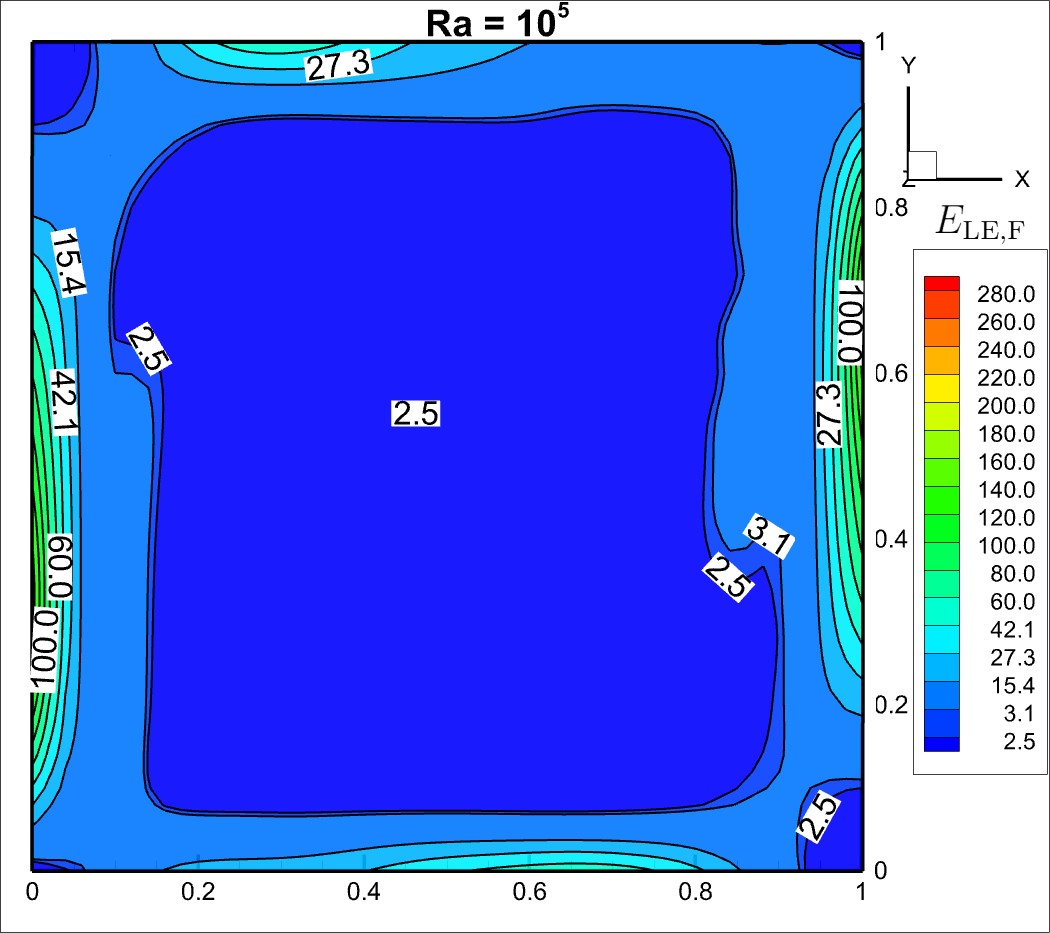}%
    \captionsetup{skip=2pt}%
    \caption{(f) $E_{\mathrm{LE,F}}$}
    \label{fig:Ra_10^5_Ha_25_local_entropy_generation_by_FF.png}
  \end{subfigure}
   \begin{subfigure}{0.24\textwidth}        
   \centering
    \includegraphics[width=\textwidth]{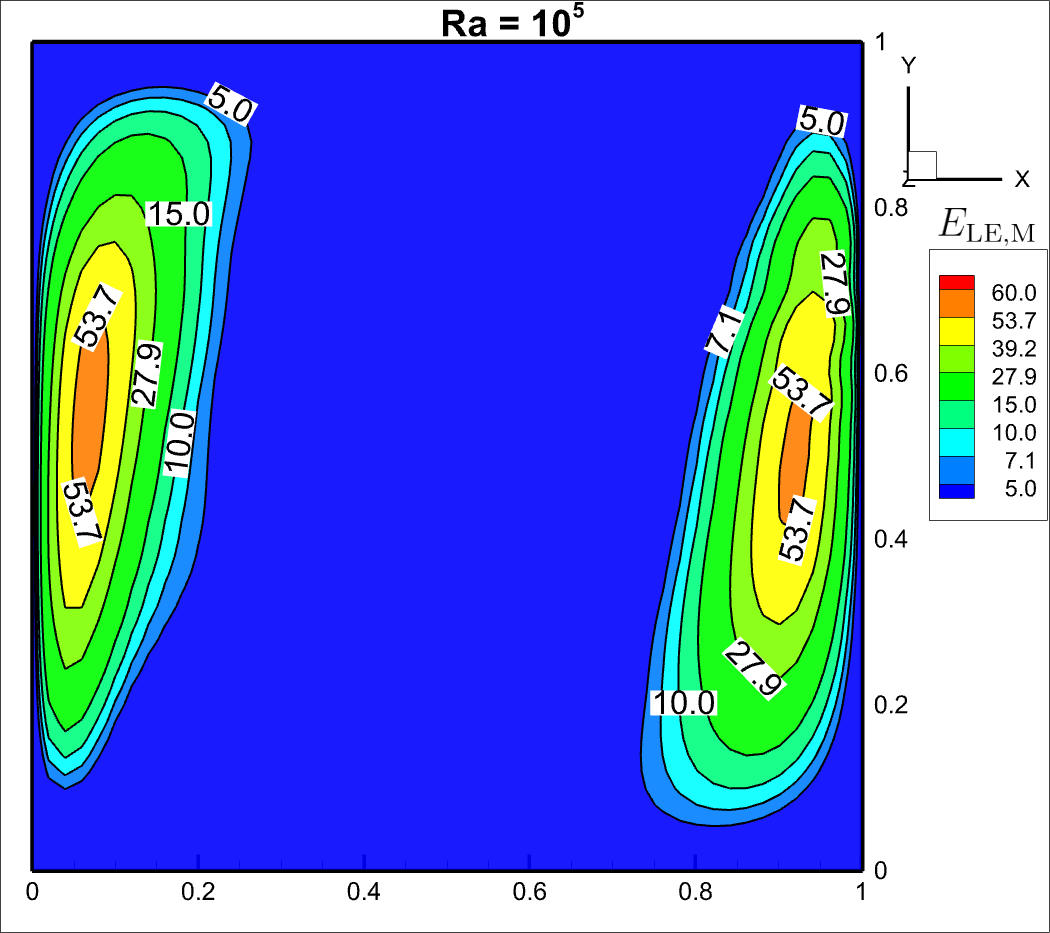}%
    \captionsetup{skip=2pt}%
    \caption{(g) $E_{\mathrm{LE,M}}$}
    \label{fig:Ra_10^5_Ha_25_local_entropy_generation_by_MF.png}
  \end{subfigure}
   \begin{subfigure}{0.24\textwidth}        
   \centering
    \includegraphics[width=\textwidth]{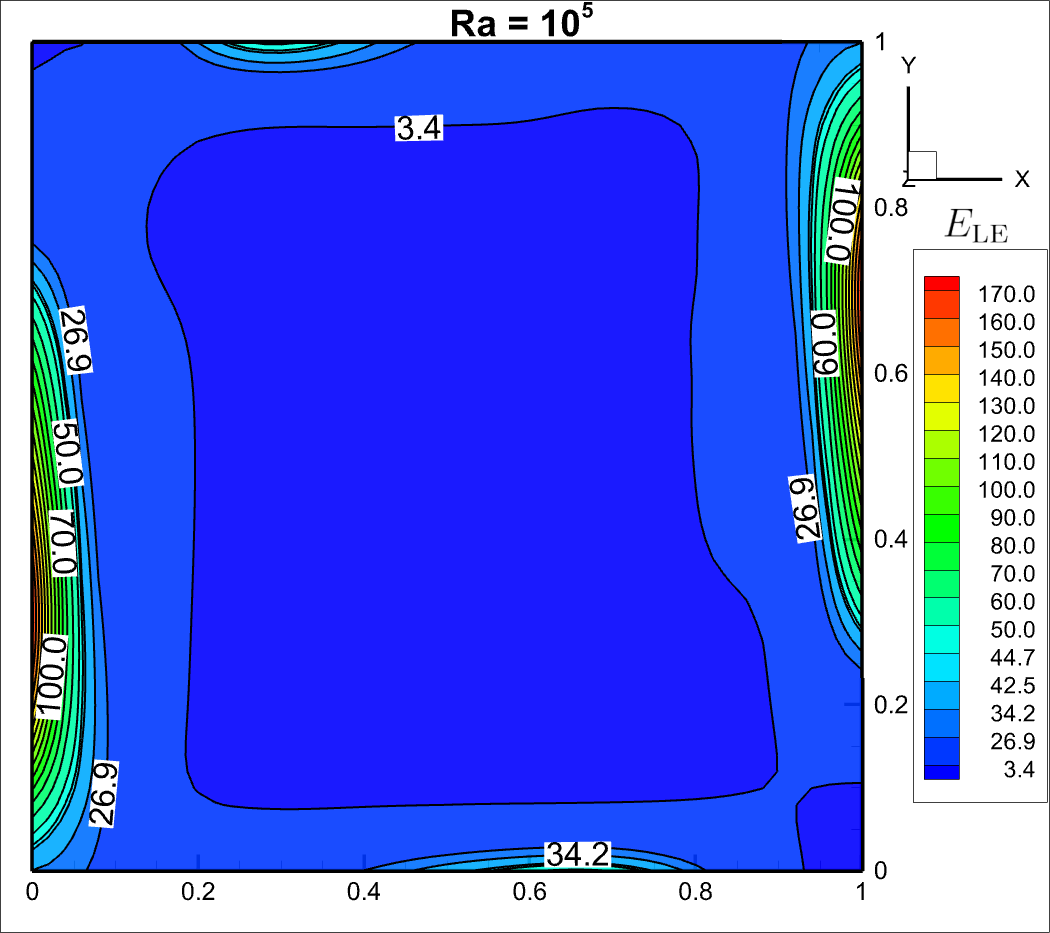}%
    \captionsetup{skip=2pt}%
    \caption{(h) $E_{\mathrm{LE}}$}
    \label{fig:Ra_10^5_Ha_25_local_entropy_generation.png}
  \end{subfigure}%
  \hspace*{\fill}
  \vspace*{2pt}%
  \hspace*{\fill}%
  \caption{Case 1. Influence of $Ra$ ((a-d) $Ra=10^3$, (e-h) $Ra=10^5$) on entropy generation contours with fixed $Ha=25$}
  \label{fig:Case_1_Effect_of_Ra_on_Entropy}
\end{figure}


\begin{figure}[htbp]
 \centering
 \vspace*{0pt}%
 \hspace*{\fill}%
\begin{subfigure}{0.24\textwidth}     
    \centering
    \includegraphics[width=\textwidth]{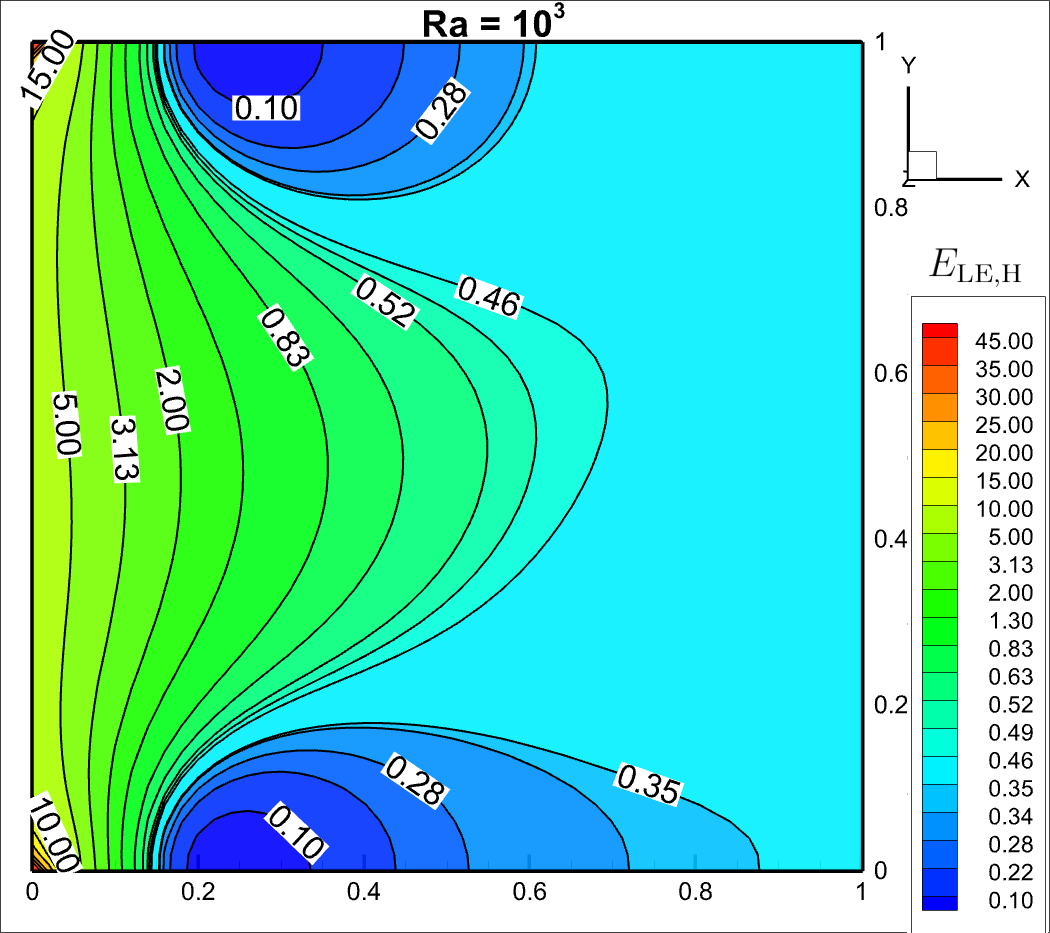}%
    \captionsetup{skip=2pt}%
    \caption{(a) $E_{\mathrm{LE,H}}$}
    \label{fig:Ra_10^3_Ha_25_local_entropy_generation_by_HT_P2.png}
  \end{subfigure}%
 \begin{subfigure}{0.24\textwidth}        
   \centering
    \includegraphics[width=\textwidth]{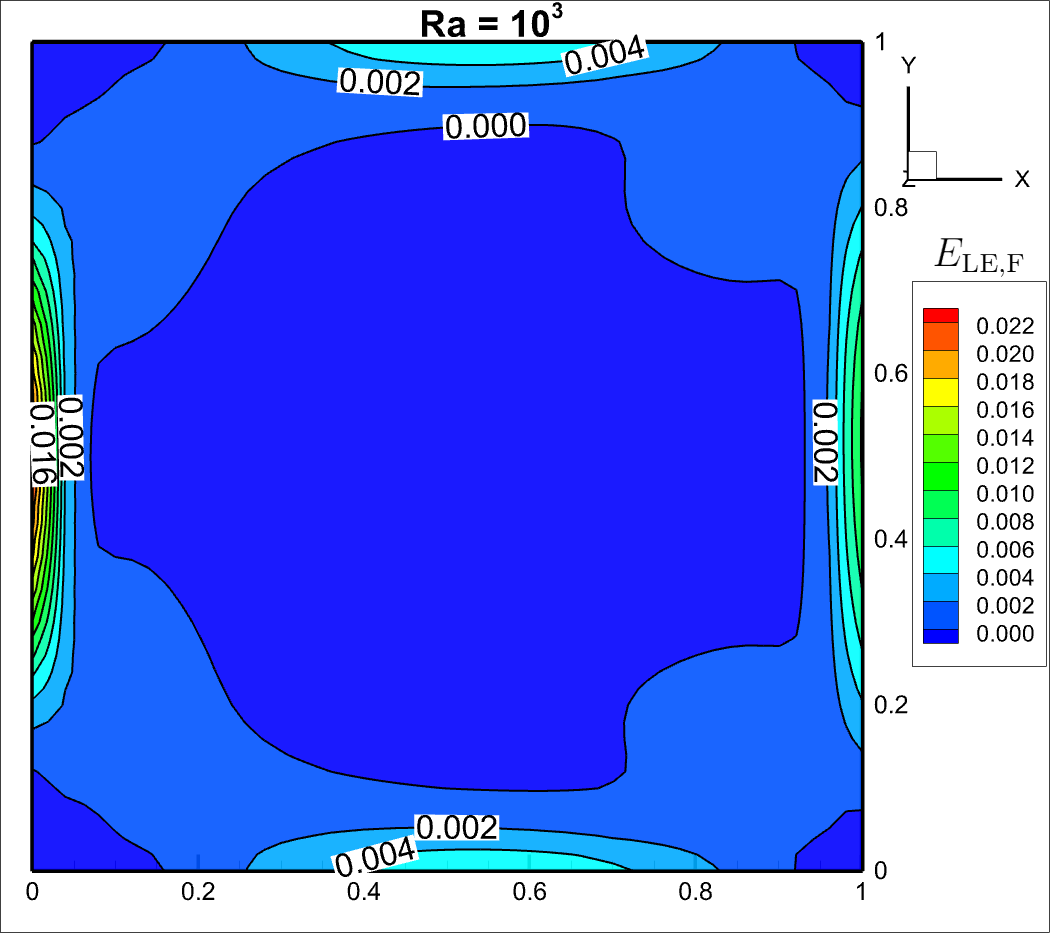}%
    \captionsetup{skip=2pt}%
    \caption{(b) $E_{\mathrm{LE,F}}$}
    \label{fig:Ra_10^3_Ha_25_local_entropy_generation_by_FF_P2.png}
  \end{subfigure}
  \begin{subfigure}{0.24\textwidth}        
   \centering
    \includegraphics[width=\textwidth]{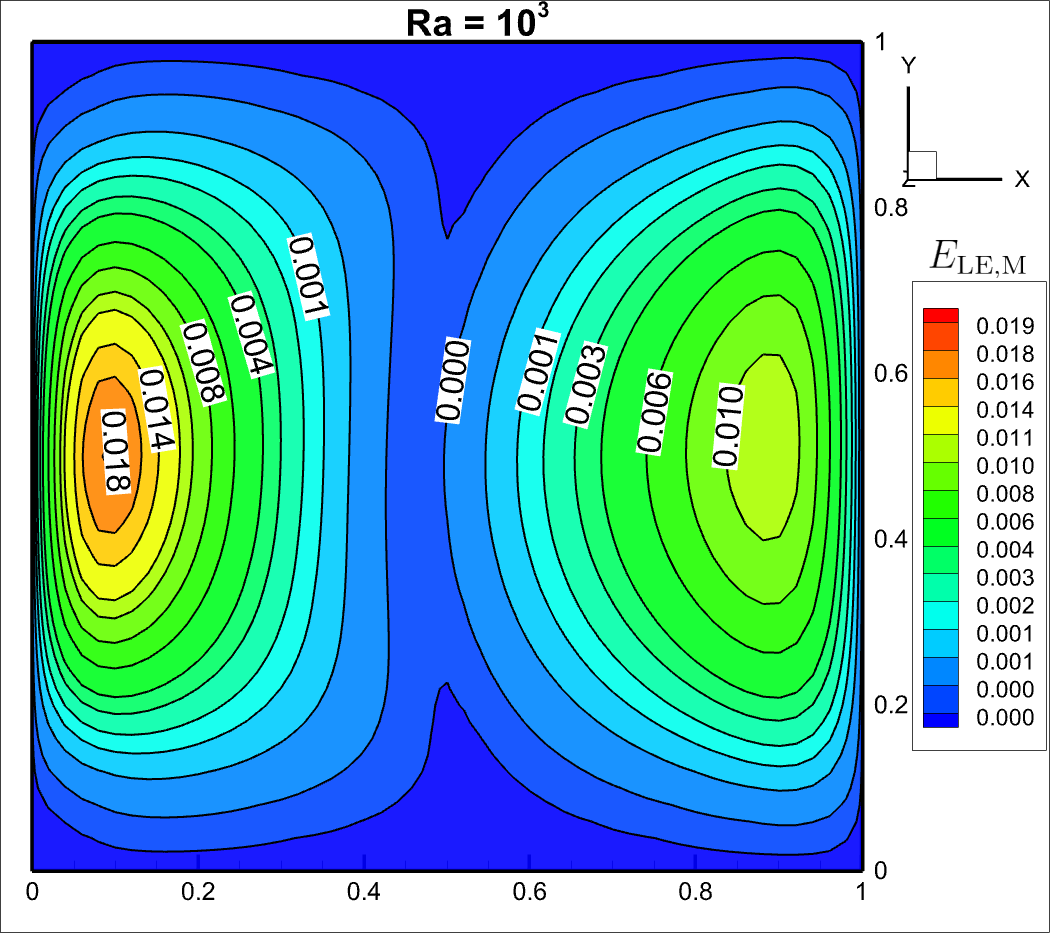}%
    \captionsetup{skip=2pt}%
    \caption{(c) $E_{\mathrm{LE,FM}}$}
    \label{fig:Ra_10^3_Ha_25_local_entropy_generation_by_MF_P2.png}
  \end{subfigure}
   \begin{subfigure}{0.24\textwidth}        
   \centering
    \includegraphics[width=\textwidth]{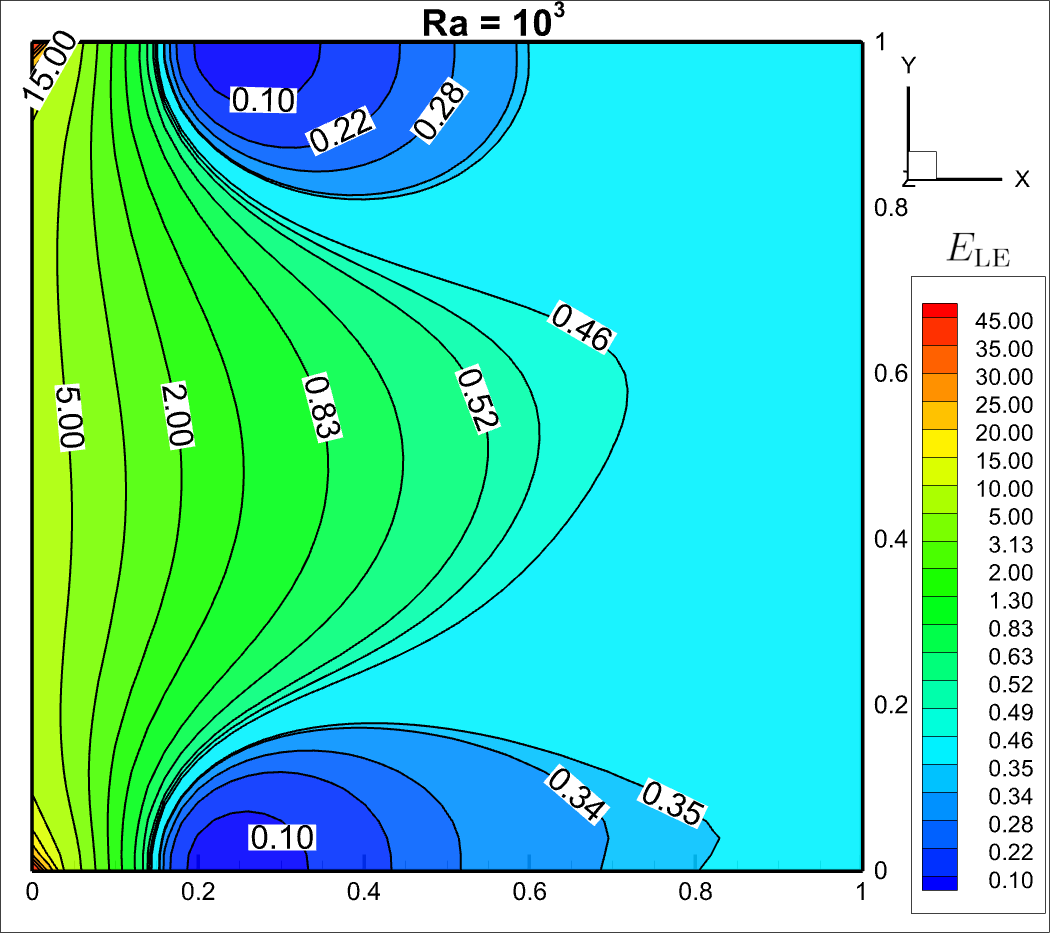}%
    \captionsetup{skip=2pt}%
    \caption{(d) $E_{\mathrm{LE}}$}
    \label{fig:Ra_10^3_Ha_25_local_entropy_generation_P2.png}
  \end{subfigure}%
  \hspace*{\fill}

  \vspace*{8pt}%
  \hspace*{\fill}%
  \begin{subfigure}{0.24\textwidth}     
    \centering
    \includegraphics[width=\textwidth]{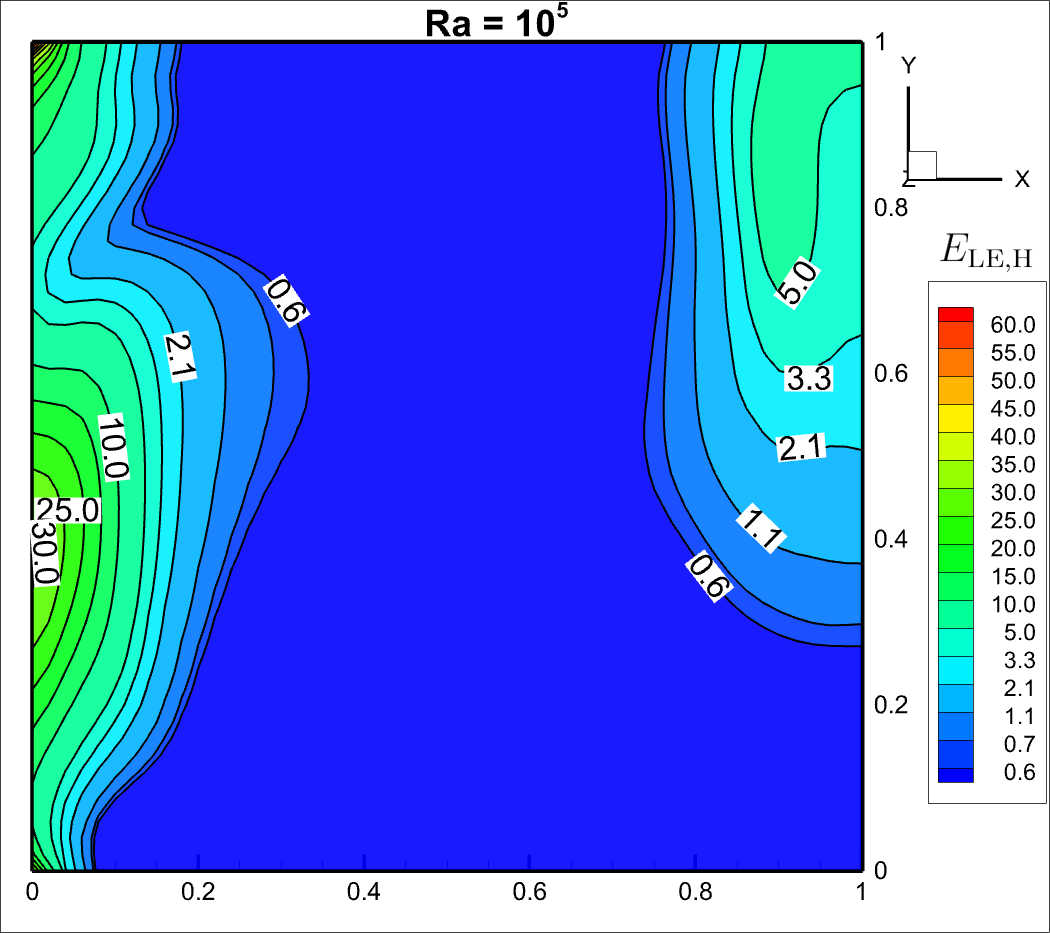}%
    \captionsetup{skip=2pt}%
    \caption{(e) $E_{\mathrm{LE,H}}$}
    \label{fig:Ra_10^5_Ha_25_local_entropy_generation_by_HT_P2.png}
  \end{subfigure}%
 \begin{subfigure}{0.24\textwidth}        
   \centering
    \includegraphics[width=\textwidth]{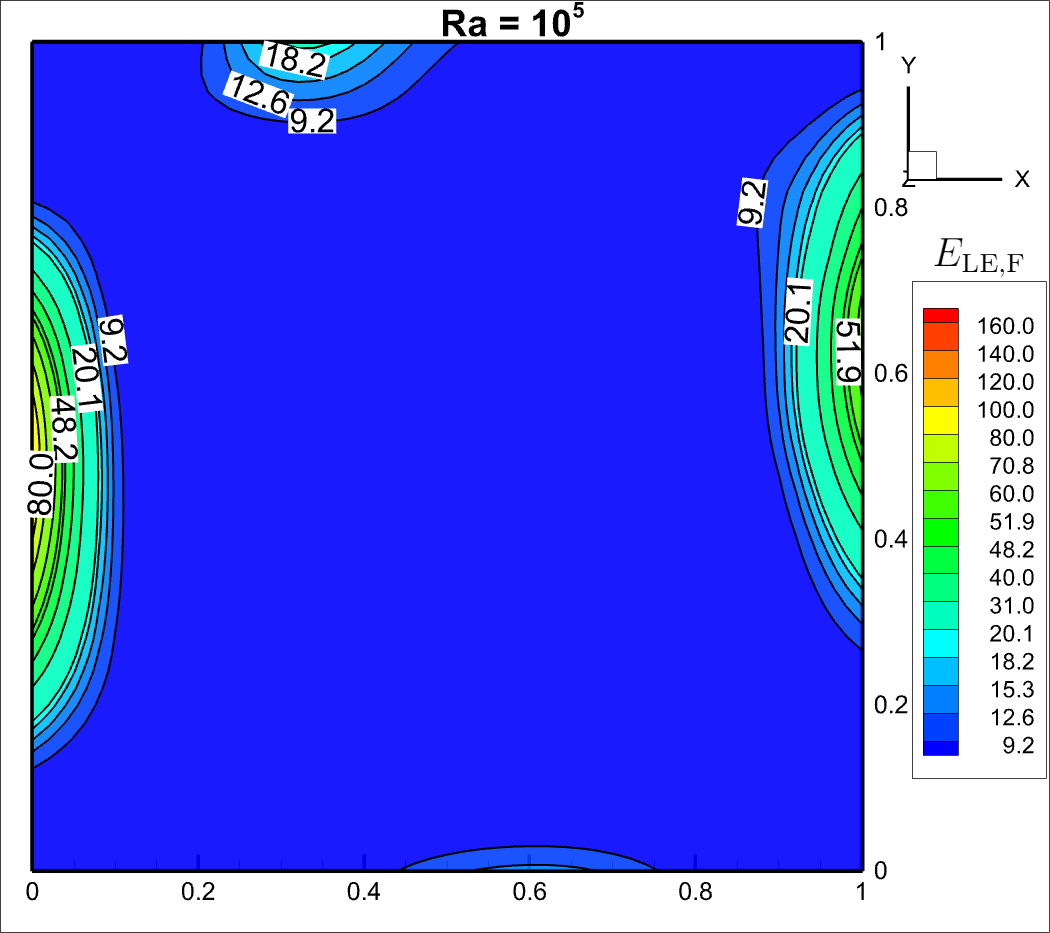}%
    \captionsetup{skip=2pt}%
    \caption{(f) $E_{\mathrm{LE,F}}$}
    \label{fig:Ra_10^5_Ha_25_local_entropy_generation_by_FF_P2.png}
  \end{subfigure}
   \begin{subfigure}{0.24\textwidth}        
   \centering
    \includegraphics[width=\textwidth]{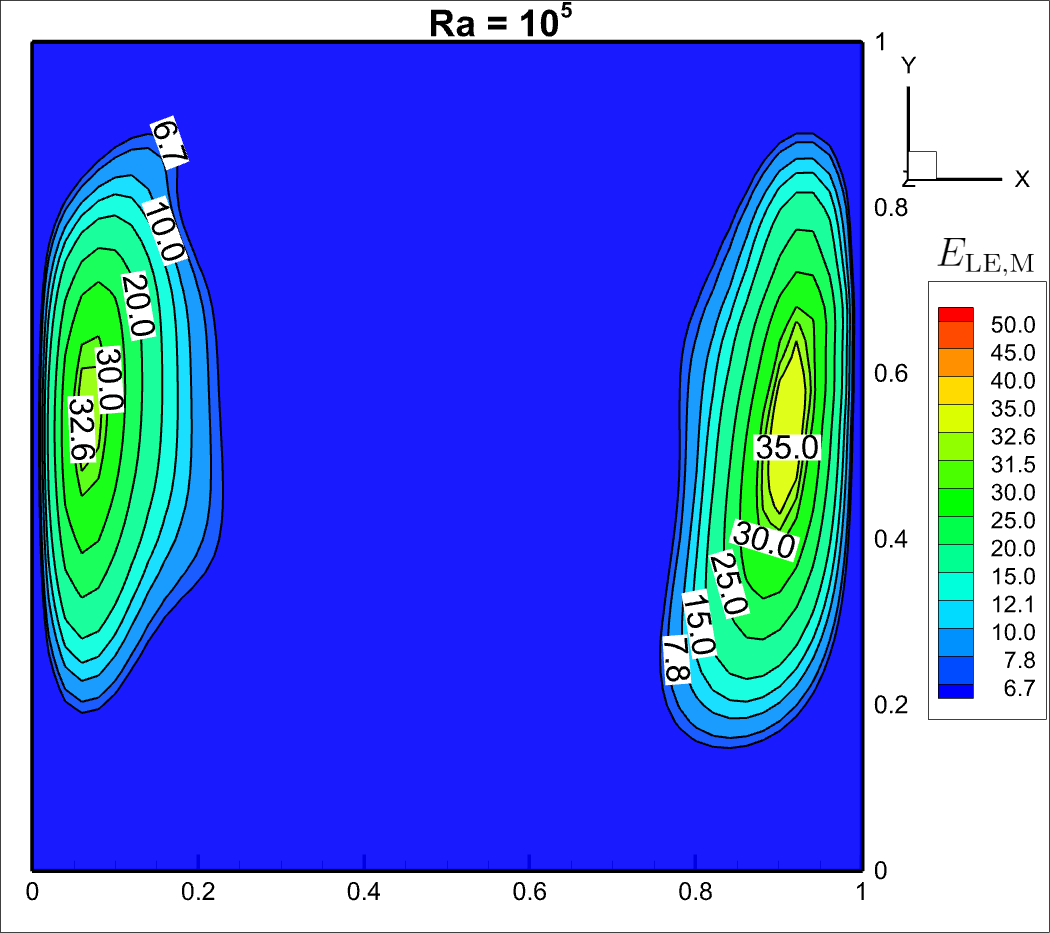}%
    \captionsetup{skip=2pt}%
    \caption{(g) $E_{\mathrm{LE,M}}$}
    \label{fig:Ra_10^5_Ha_25_local_entropy_generation_by_MF_P2.png}
  \end{subfigure}
   \begin{subfigure}{0.24\textwidth}        
   \centering
    \includegraphics[width=\textwidth]{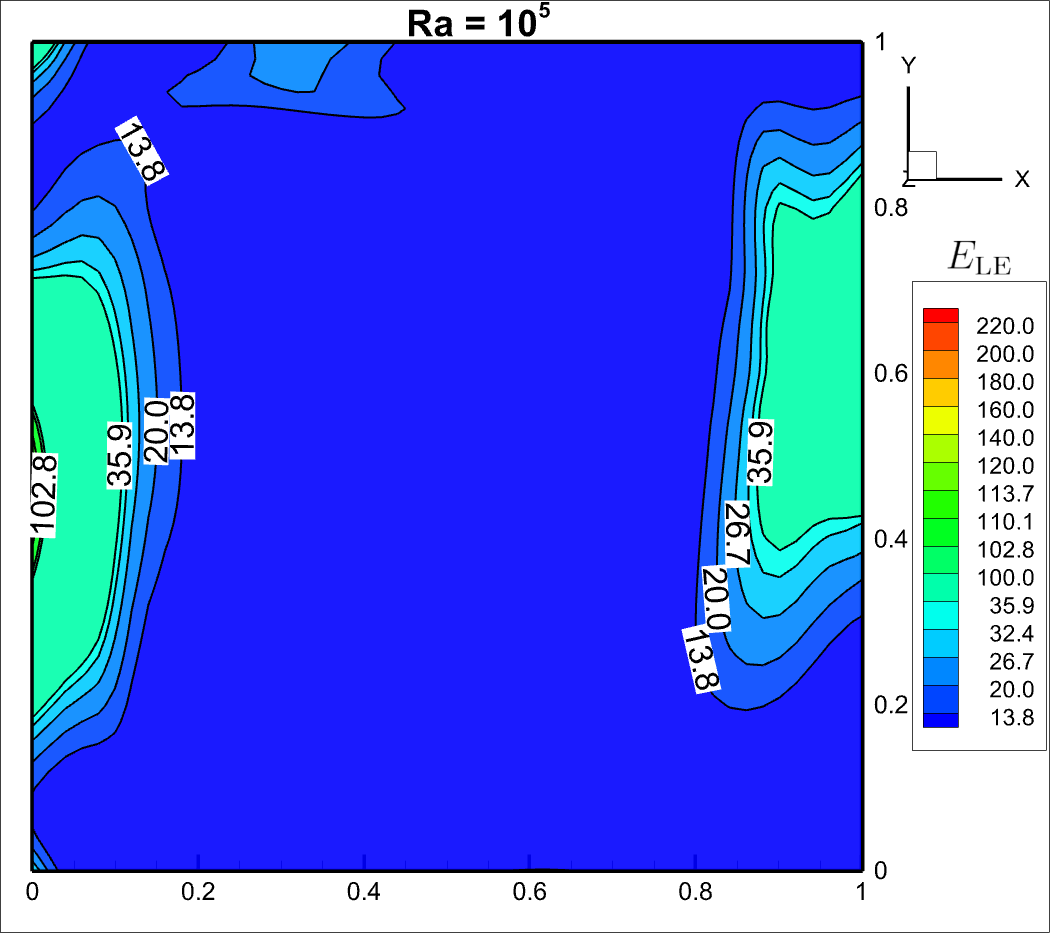}%
    \captionsetup{skip=2pt}%
    \caption{(h) $E_{\mathrm{LE}}$}
    \label{fig:Ra_10^5_Ha_25_local_entropy_generation_P2.png}
  \end{subfigure}%
  \hspace*{\fill}
  \vspace*{2pt}%
  \hspace*{\fill}%
  \caption{Case 2. Influence of $Ra$ ((a-d) $Ra=10^3$, (e-h) $Ra=10^5$) on entropy generation contours with fixed $Ha=25$}
  \label{fig:Case_2_Effect_of_Ra_on_Entropy}
\end{figure}



\begin{figure}[htbp]
 \centering
 \vspace*{0pt}%
 \hspace*{\fill}%
\begin{subfigure}{0.24\textwidth}     
    \centering
    \includegraphics[width=\textwidth]{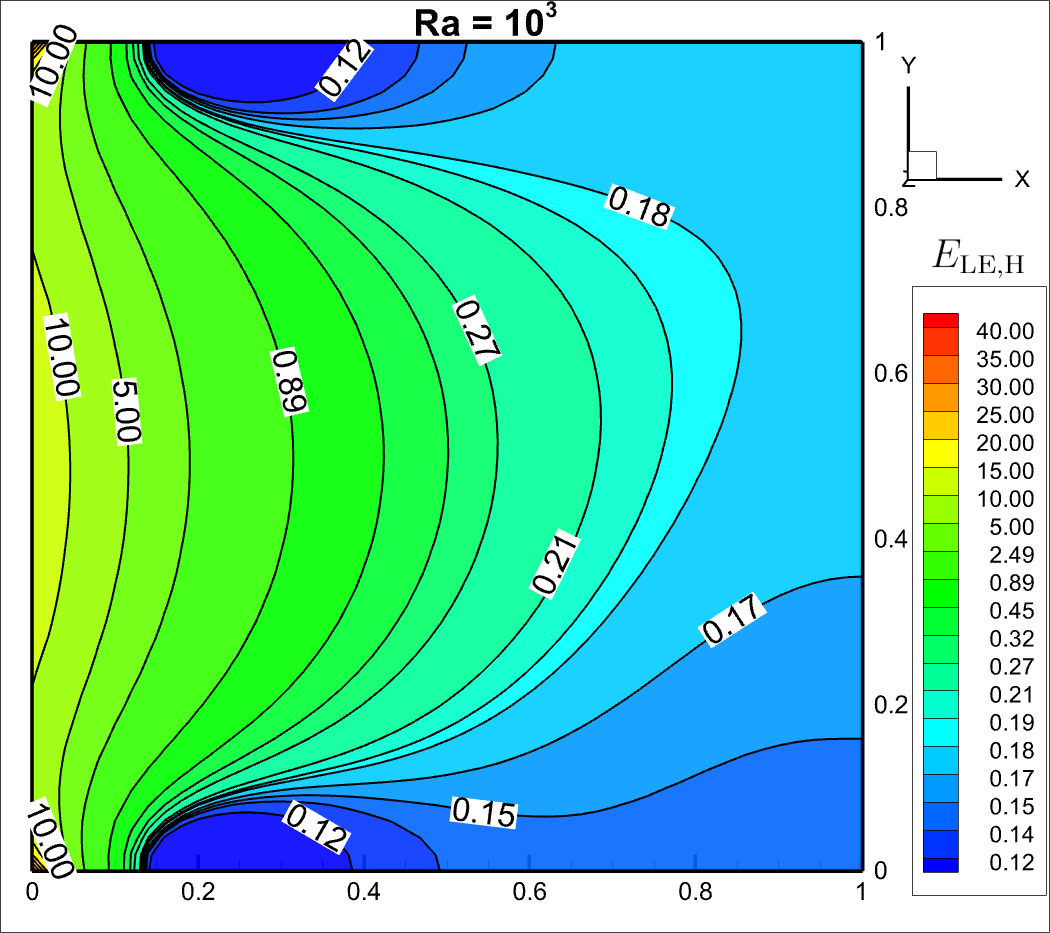}%
    \captionsetup{skip=2pt}%
    \caption{(a) $E_{\mathrm{LE,H}}$}
    \label{fig:P3_Ra_10^3_Ha_25_local_entropy_generation_by_HT.png}
  \end{subfigure}%
 \begin{subfigure}{0.24\textwidth}        
   \centering
    \includegraphics[width=\textwidth]{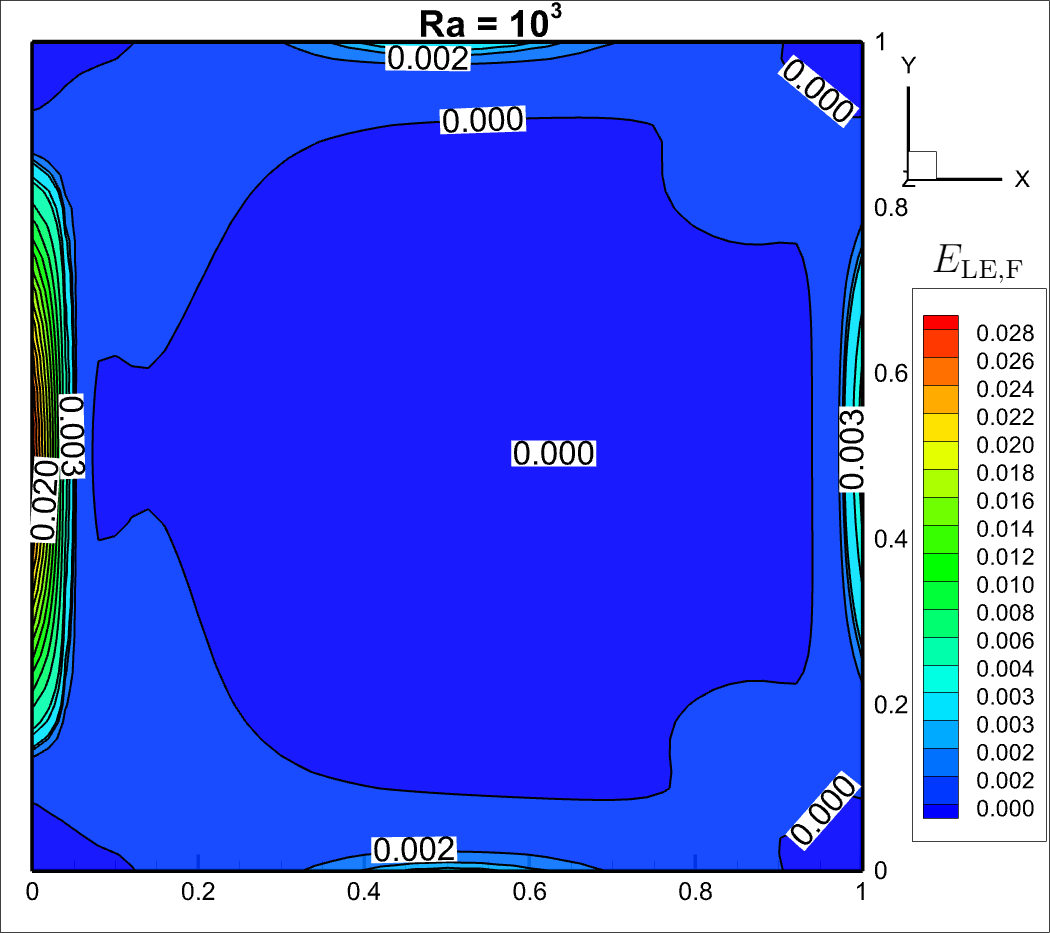}%
    \captionsetup{skip=2pt}%
    \caption{(b) $E_{\mathrm{LE,F}}$}
    \label{fig:P3_Ra_10^3_Ha_25_local_entropy_generation_by_FF.png}
  \end{subfigure}
  \begin{subfigure}{0.24\textwidth}        
   \centering
    \includegraphics[width=\textwidth]{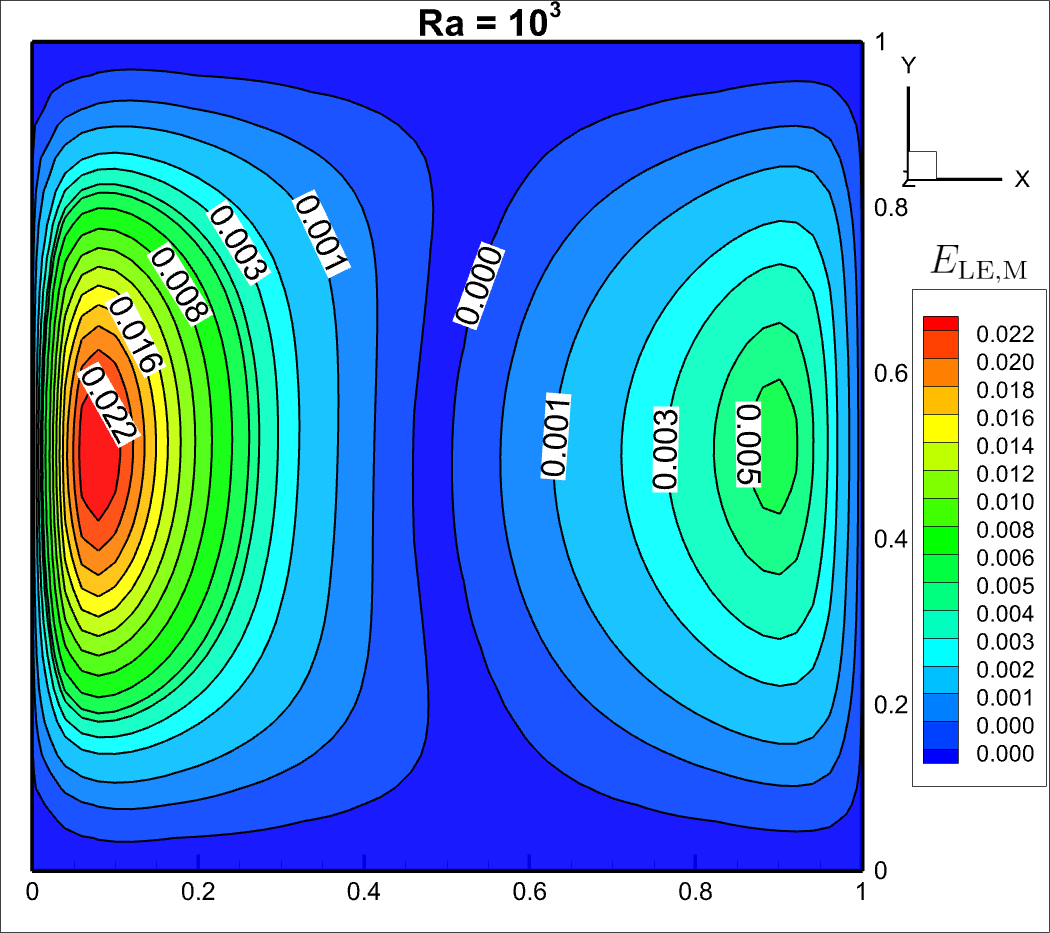}%
    \captionsetup{skip=2pt}%
    \caption{(c) $E_{\mathrm{LE,M}}$}
    \label{fig:P3_Ra_10^3_Ha_25_local_entropy_generation_by_MF.png}
  \end{subfigure}
   \begin{subfigure}{0.24\textwidth}        
   \centering
    \includegraphics[width=\textwidth]{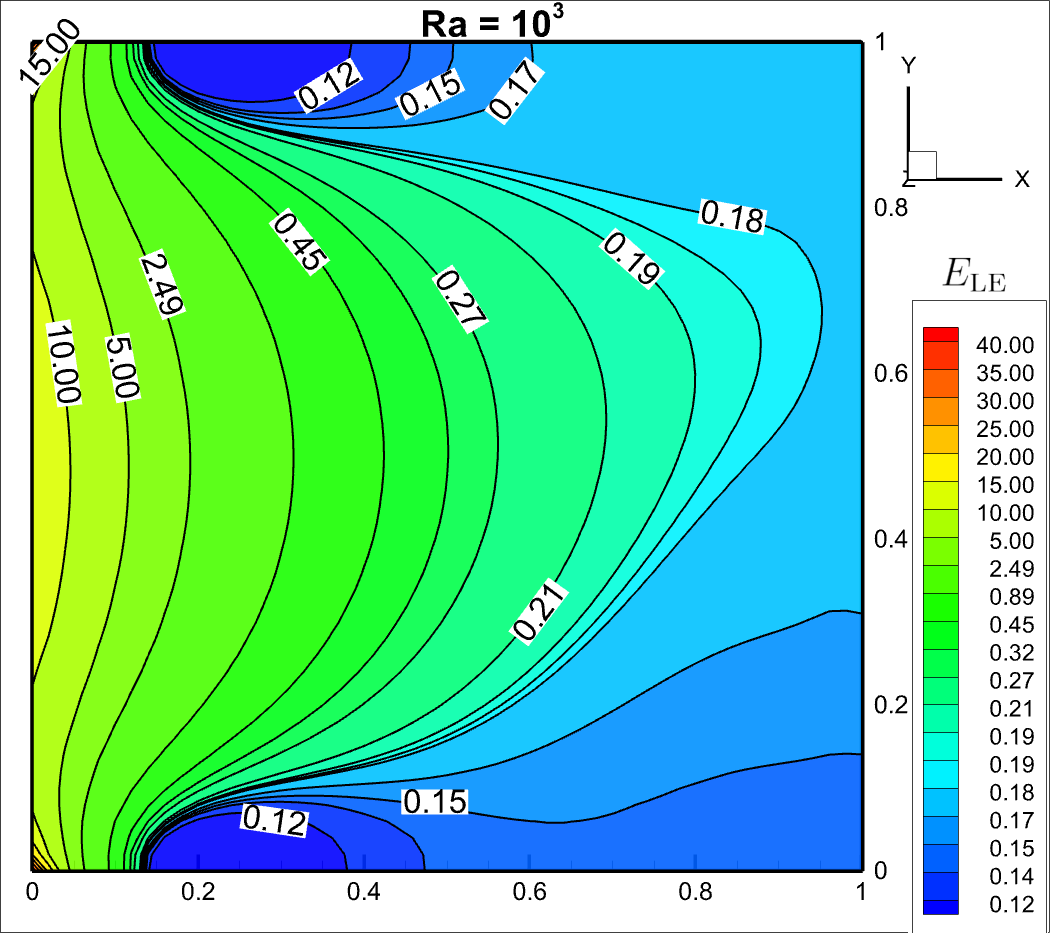}%
    \captionsetup{skip=2pt}%
    \caption{(d) $E_{\mathrm{LE}}$}
    \label{fig:P3_Ra_10^3_Ha_25_local_entropy_generation.png}
  \end{subfigure}%
  \hspace*{\fill}

  \vspace*{8pt}%
  \hspace*{\fill}%
  \begin{subfigure}{0.24\textwidth}     
    \centering
    \includegraphics[width=\textwidth]{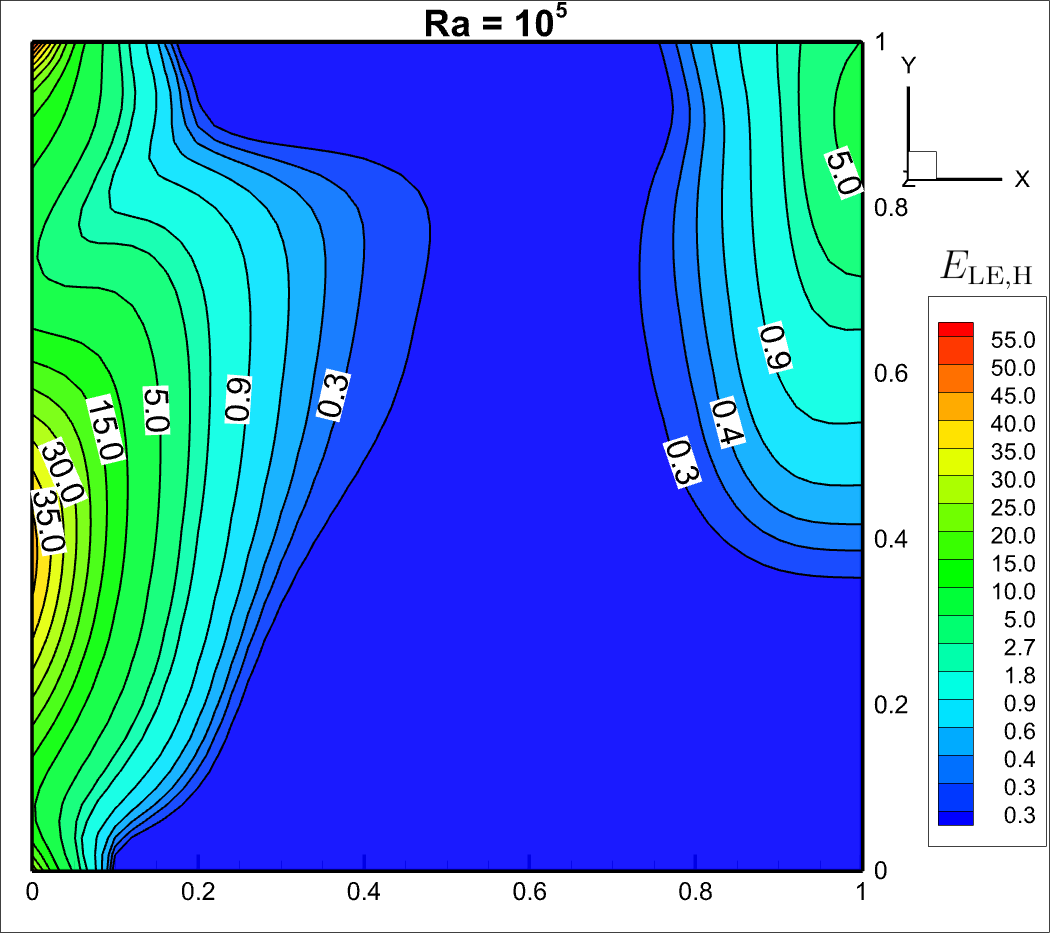}%
    \captionsetup{skip=2pt}%
    \caption{(e) $E_{\mathrm{LE,H}}$}
    \label{fig:P3_Ra_10^5_Ha_25_local_entropy_generation_by_HT.png}
  \end{subfigure}%
 \begin{subfigure}{0.24\textwidth}        
   \centering
    \includegraphics[width=\textwidth]{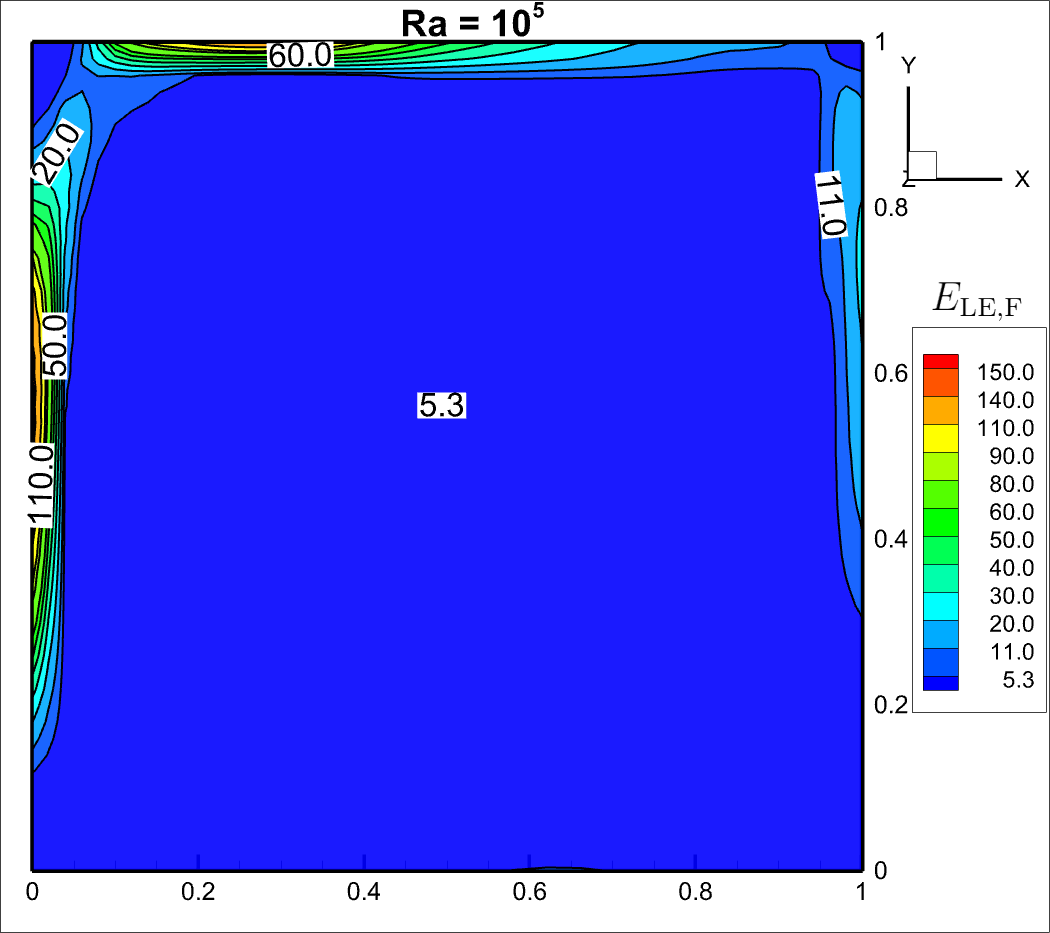}%
    \captionsetup{skip=2pt}%
    \caption{(f) $E_{\mathrm{LE,F}}$}
    \label{fig:P3_Ra_10^5_Ha_25_local_entropy_generation_by_FF.png}
  \end{subfigure}
   \begin{subfigure}{0.24\textwidth}        
   \centering
    \includegraphics[width=\textwidth]{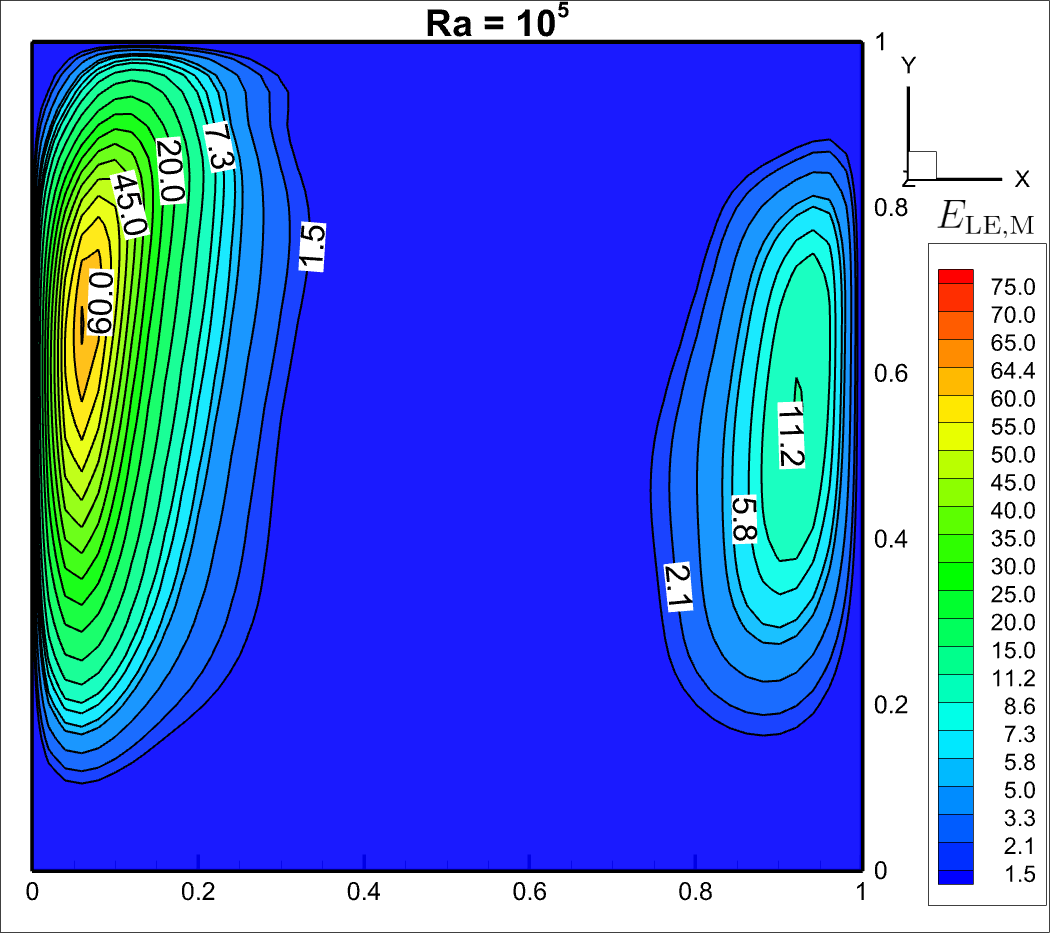}%
    \captionsetup{skip=2pt}%
    \caption{(g) $E_{\mathrm{LE,M}}$}
    \label{fig:P3_Ra_10^5_Ha_25_local_entropy_generation_by_MF.png}
  \end{subfigure}
   \begin{subfigure}{0.24\textwidth}        
   \centering
    \includegraphics[width=\textwidth]{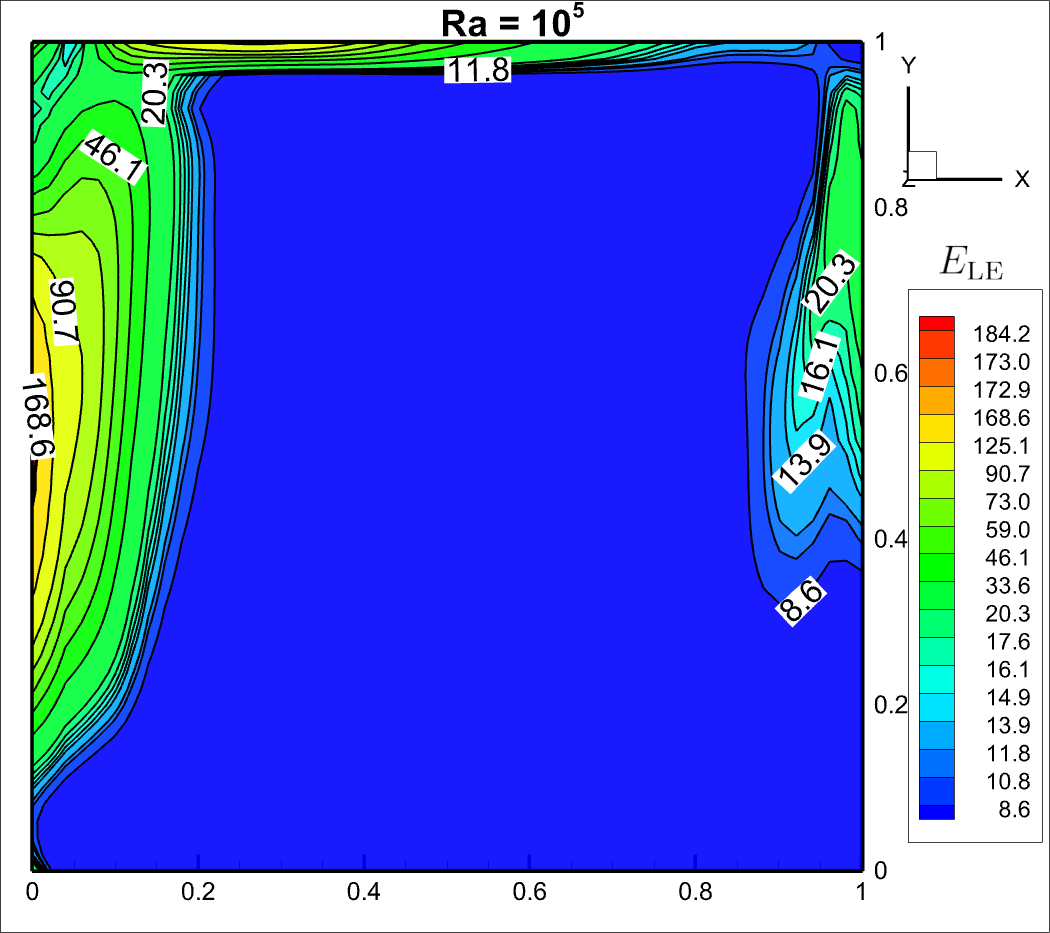}%
    \captionsetup{skip=2pt}%
    \caption{(h) $E_{\mathrm{LE}}$}
    \label{fig:P3_Ra_10^5_Ha_25_local_entropy_generation.png}
  \end{subfigure}%
  \hspace*{\fill}
  \vspace*{2pt}%
  \hspace*{\fill}%
  \caption{Case 3. Influence of $Ra$ ((a-d) $Ra=10^3$, (e-h) $Ra=10^5$) on entropy generation contours with fixed $Ha=25$}
  \label{fig:Case_3_Effect_of_Ra_on_Entropy}
\end{figure}

Figures \ref{fig:Case_1_Effect_of_Ra_on_Entropy}, \ref{fig:Case_2_Effect_of_Ra_on_Entropy} and \ref{fig:Case_3_Effect_of_Ra_on_Entropy} illustrates the local entropy generation (at $z=0.5$) for $Ra=10^3$ and $Ra=10^5$ at a constant $Ha=25$ for Case 1, Case 2 and Case 3, respectively.
The first column of these figures delineates entropy generation due to heat transfer ($E_{\mathrm{LE,H}}$), second column illustrates entropy generation attributed to fluid friction ($E_{\mathrm{LE,F}}$), third column depicts entropy generation resulting from magnetic fields ($E_{\mathrm{LE,M}}$), and the final column provides total local entropy generation ($E_{\mathrm{LE}}$). The visualization provides insights into the impact of Rayleigh number ($Ra$) on local entropy generation contours for various cases. Figure \ref{fig:Case_1_Effect_of_Ra_on_Entropy} shows the contours of local entropy generation rates for Case 1. In the depicted figure, heightened frictional (Figure \ref{fig:Case_1_Effect_of_Ra_on_Entropy}(b,f)) and thermal entropy generation (Figure \ref{fig:Case_1_Effect_of_Ra_on_Entropy}(a,e)) rates are evident near the walls and regions with high-temperature gradients, respectively, for different $Ra$ values. Examining the figure, it becomes apparent that the Lorentz force entropy generation rate reaches its maximum at the central regions of the left and right walls for $Ra =10^3$ (Figure \ref{fig:Case_1_Effect_of_Ra_on_Entropy}(c)), undergoing a change in its distribution to a more stretched form as $Ra$ is increased to $10^5$ (Figure \ref{fig:Case_1_Effect_of_Ra_on_Entropy}(g)). A comparative analysis of entropy generation rates reveals that, for $Ra=10^3$, the contributions from fluid friction and the magnetic field are relatively negligible compared to the entropy generation due to heat transfer. However, for $Ra=10^5$, a notable shift occurs, with the entropy generation from fluid friction and the magnetic field surpassing that from heat transfer. This shift indicates a convection-dominant behavior as $Ra$ rises to $10^5$. An overarching observation is that the overall local entropy generation increases with the escalation of $Ra$. 
The impact of $Ra$ on entropy generation contours for Case 2 is visually represented in Figure \ref{fig:Case_2_Effect_of_Ra_on_Entropy}.
In this case, the maximum entropy generation due to heat transfer is evident near the left wall of the cavity for both $Ra=10^3$ and $Ra=10^5$. Specifically, at $Ra=10^3$, heat transfer is primarily responsible for entropy generation. Conversely, at $Ra=10^5$, the higher entropy generation is associated with fluid friction ($E_{\mathrm{LE,F}}$) and the magnetic field ($E_{\mathrm{LE,M}}$). Notably, at low $Ra=10^3$, the patterns of $E_{\mathrm{LE}}$ and $E_{\mathrm{LE,H}}$ closely resemble each other, suggesting the dominant influence of $E_{\mathrm{LE,H}}$ on $E_{\mathrm{LE}}$. This visual analysis provides insights into the shifting mechanisms influencing entropy generation, highlighting the transition from heat transfer dominance at lower $Ra$ to the prominence of fluid friction and magnetic field effects at higher $Ra$ values. The entropy generation pattern is significantly different from that of Case 1. In this case, the pattern of entropy generation due to heat transfer is observed to be high near the left wall only, whereas in Case 1, it was near the top of the cold wall and the bottom of the hot wall. In Figure \ref{fig:Case_3_Effect_of_Ra_on_Entropy}, the impact of $Ra$ on entropy generation contours is illustrated for Case 3, with a fixed $Ha=25$. A notable observation is that akin to previous cases, an increase in $Ra$ corresponds to an escalation in the local entropy generation rate. The observed pattern differs from Case 1 but shares similarities with Case 2, attributed to the non-uniform thermal conditions on the heated wall. 
For $Ra=10^3$, there is a high concentration of local entropy around the middle of the left wall (Figure \ref{fig:Case_3_Effect_of_Ra_on_Entropy}(d)). However, with an increase in $Ra=10^5$, this concentration shifts towards the left upper corner of the cavity ((Figure \ref{fig:Case_3_Effect_of_Ra_on_Entropy}(h)). This observation suggests that local entropy generation becomes more localized around corners as $Ra$ increases. At $Ra=10^3$, the entropy generation due to Lorentz force and fluid friction appears significantly lower, almost negligible, compared to entropy generation by heat transfer. However, as $Ra$ increases to $10^5$, entropy generation due to heat transfer becomes lower than both entropy generation from fluid friction and Lorentz force. This observation suggests a growing dominance of irreversibility attributed to fluid friction and Lorentz force with increasing $Ra$.\\

\begin{figure}[htbp]
 \centering
 \vspace*{0pt}%
 \hspace*{\fill}%
\begin{subfigure}{0.24\textwidth}     
    \centering
    \includegraphics[width=\textwidth]{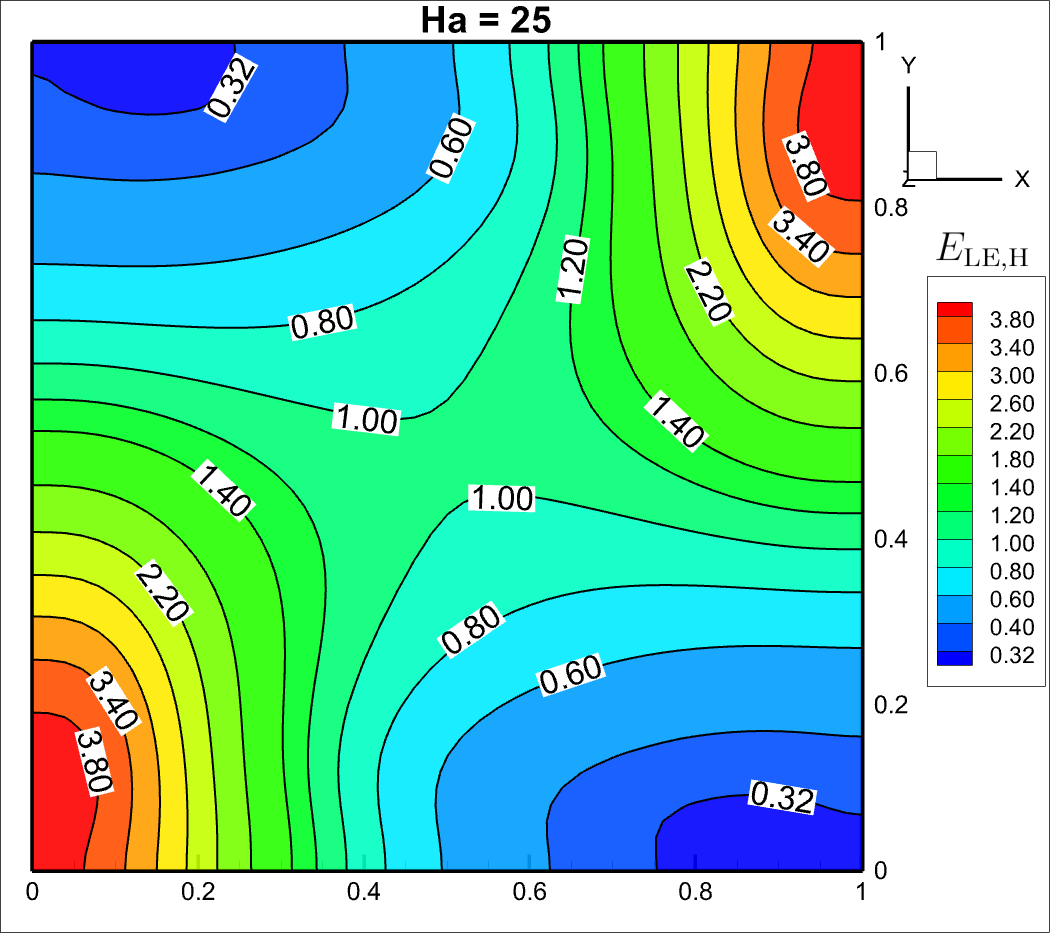}%
    \captionsetup{skip=2pt}%
    \caption{(a) $E_{\mathrm{LE,H}}$}
    \label{fig:Ra_10^4_Ha_25_local_entropy_generation_by_HT.png}
  \end{subfigure}%
 \begin{subfigure}{0.24\textwidth}        
   \centering
    \includegraphics[width=\textwidth]{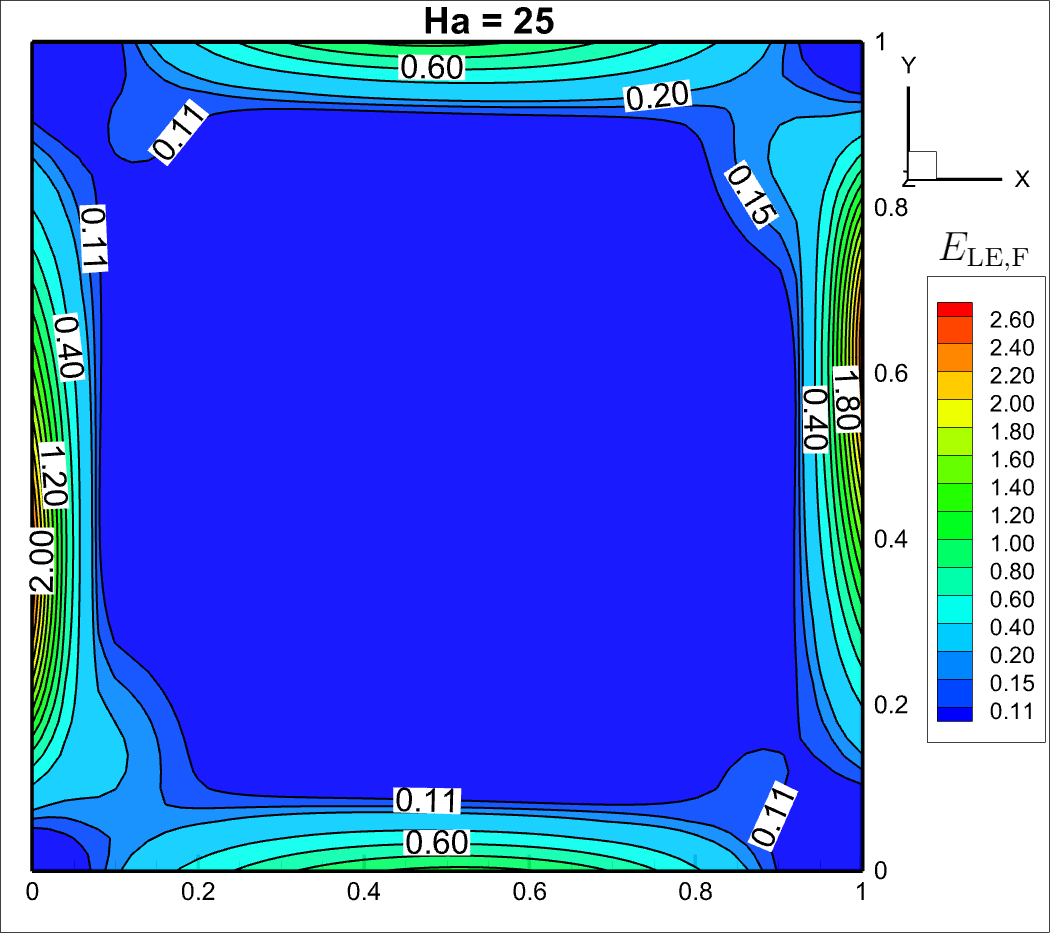}%
    \captionsetup{skip=2pt}%
    \caption{(b) $E_{\mathrm{LE,F}}$}
    \label{fig:Ra_10^4_Ha_25_local_entropy_generation_by_FF.png}
  \end{subfigure}
  \begin{subfigure}{0.24\textwidth}        
   \centering
    \includegraphics[width=\textwidth]{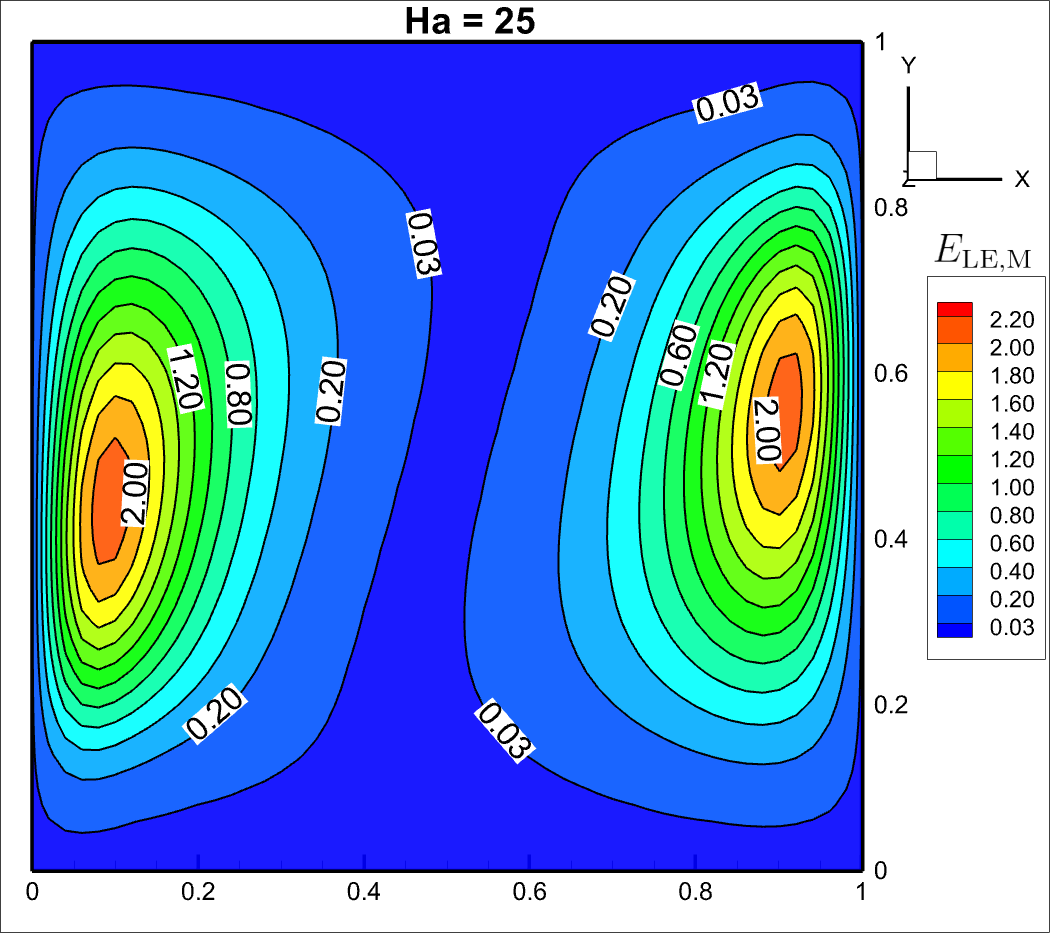}%
    \captionsetup{skip=2pt}%
    \caption{(c) $E_{\mathrm{LE,M}}$}
    \label{fig:Ra_10^4_Ha_25_local_entropy_generation_by_MF.png}
  \end{subfigure}
   \begin{subfigure}{0.24\textwidth}        
   \centering
    \includegraphics[width=\textwidth]{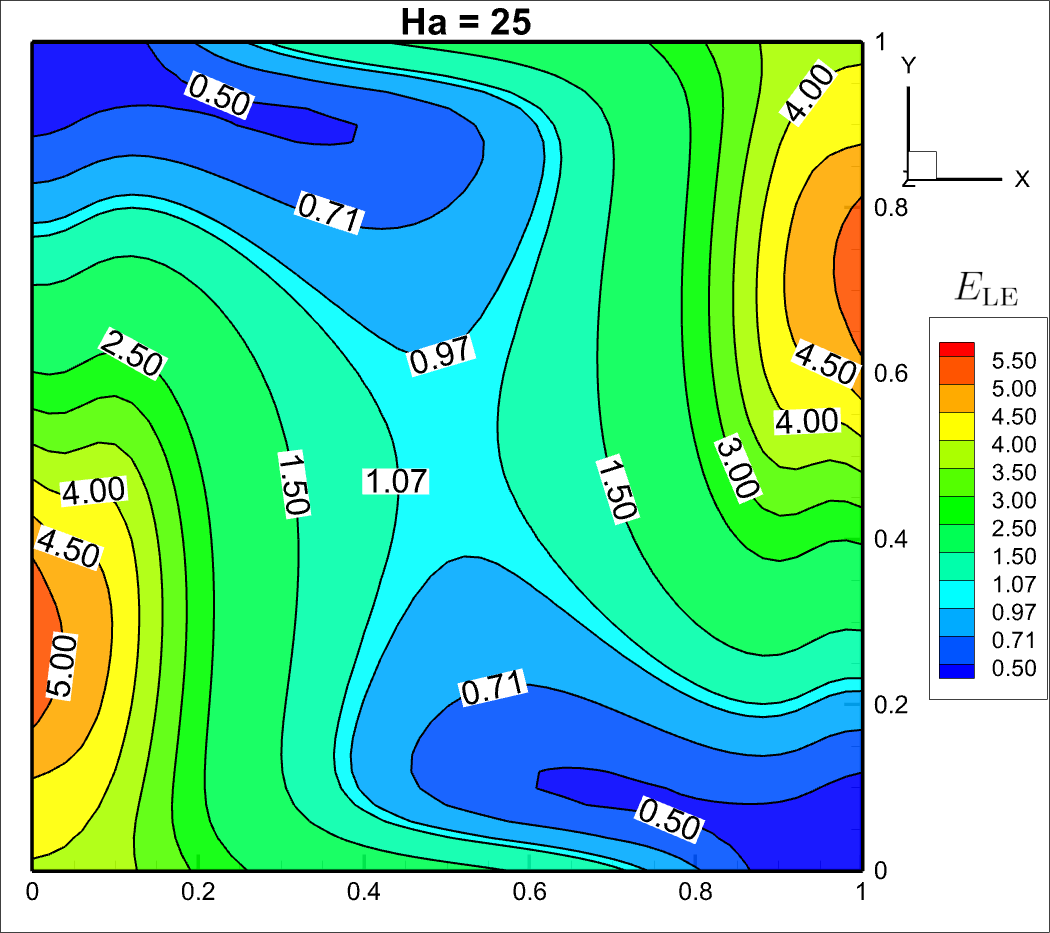}%
    \captionsetup{skip=2pt}%
    \caption{(d) $E_{\mathrm{LE}}$}
    \label{fig:Ra_10^4_Ha_25_local_entropy_generation_P1.png}
  \end{subfigure}%
  \hspace*{\fill}

  \vspace*{8pt}%
  \hspace*{\fill}%
  \begin{subfigure}{0.24\textwidth}     
    \centering
    \includegraphics[width=\textwidth]{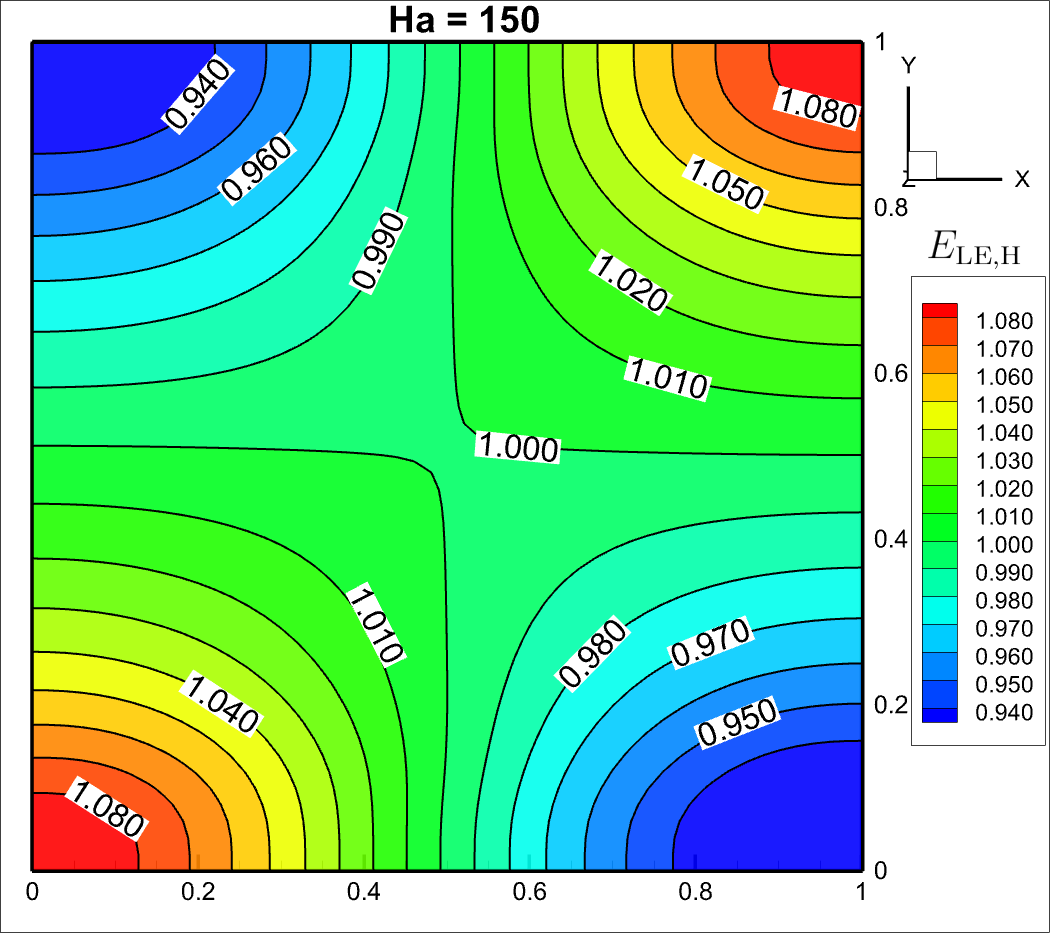}%
    \captionsetup{skip=2pt}%
    \caption{(e) $E_{\mathrm{LE,H}}$}
    \label{fig:Ra_10^4_Ha_150_local_entropy_generation_by_HT.png}
  \end{subfigure}%
 \begin{subfigure}{0.24\textwidth}        
   \centering
    \includegraphics[width=\textwidth]{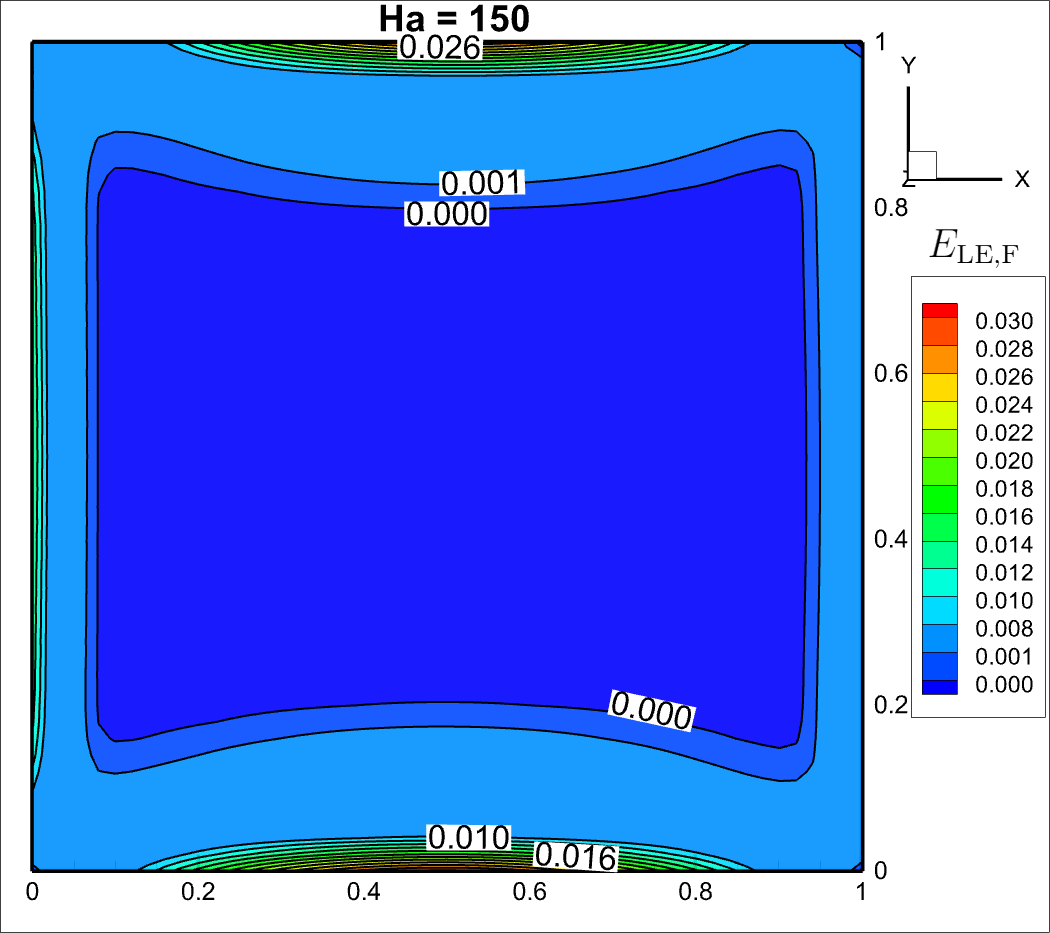}%
    \captionsetup{skip=2pt}%
    \caption{(f) $E_{\mathrm{LE,F}}$}
    \label{fig:Ra_10^4_Ha_150_local_entropy_generation_by_FF.png}
  \end{subfigure}
   \begin{subfigure}{0.24\textwidth}        
   \centering
    \includegraphics[width=\textwidth]{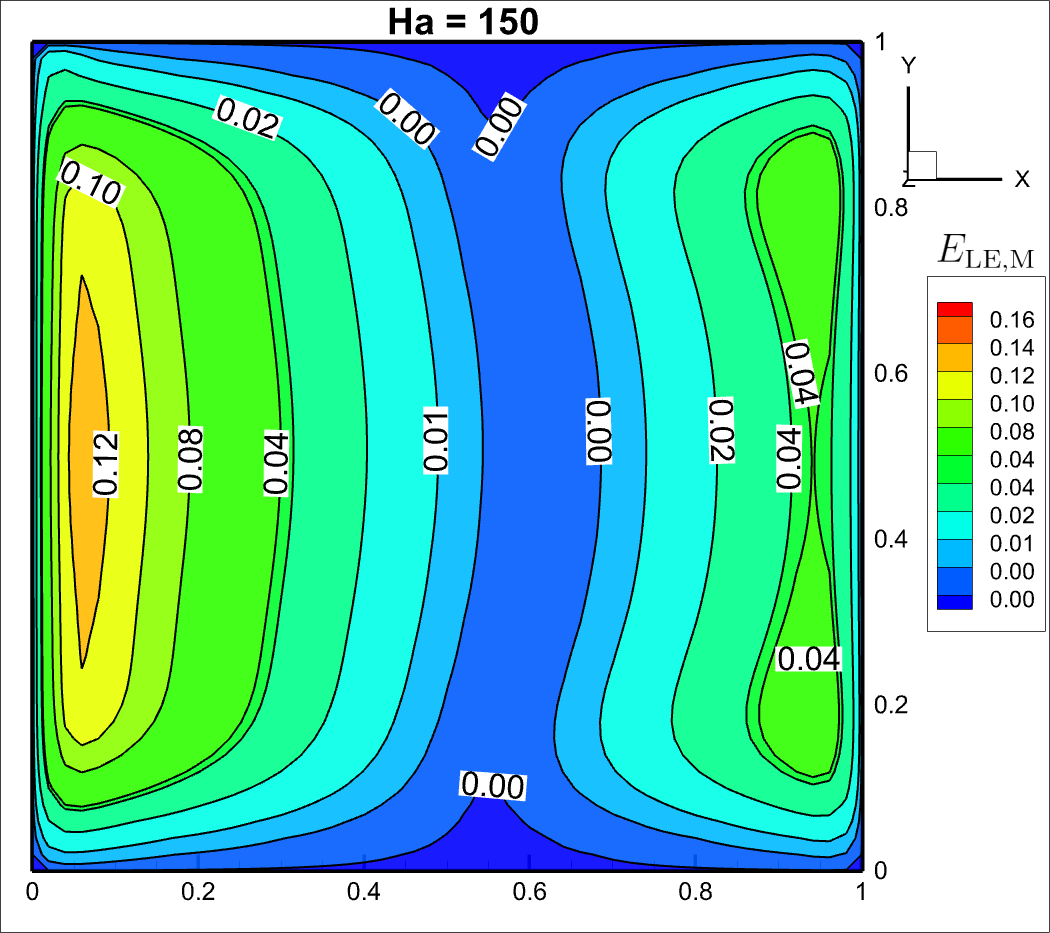}%
    \captionsetup{skip=2pt}%
    \caption{(g) $E_{\mathrm{LE,M}}$}
    \label{fig:Ra_10^4_Ha_150_local_entropy_generation_by_MF.png}
  \end{subfigure}
   \begin{subfigure}{0.24\textwidth}        
   \centering
    \includegraphics[width=\textwidth]{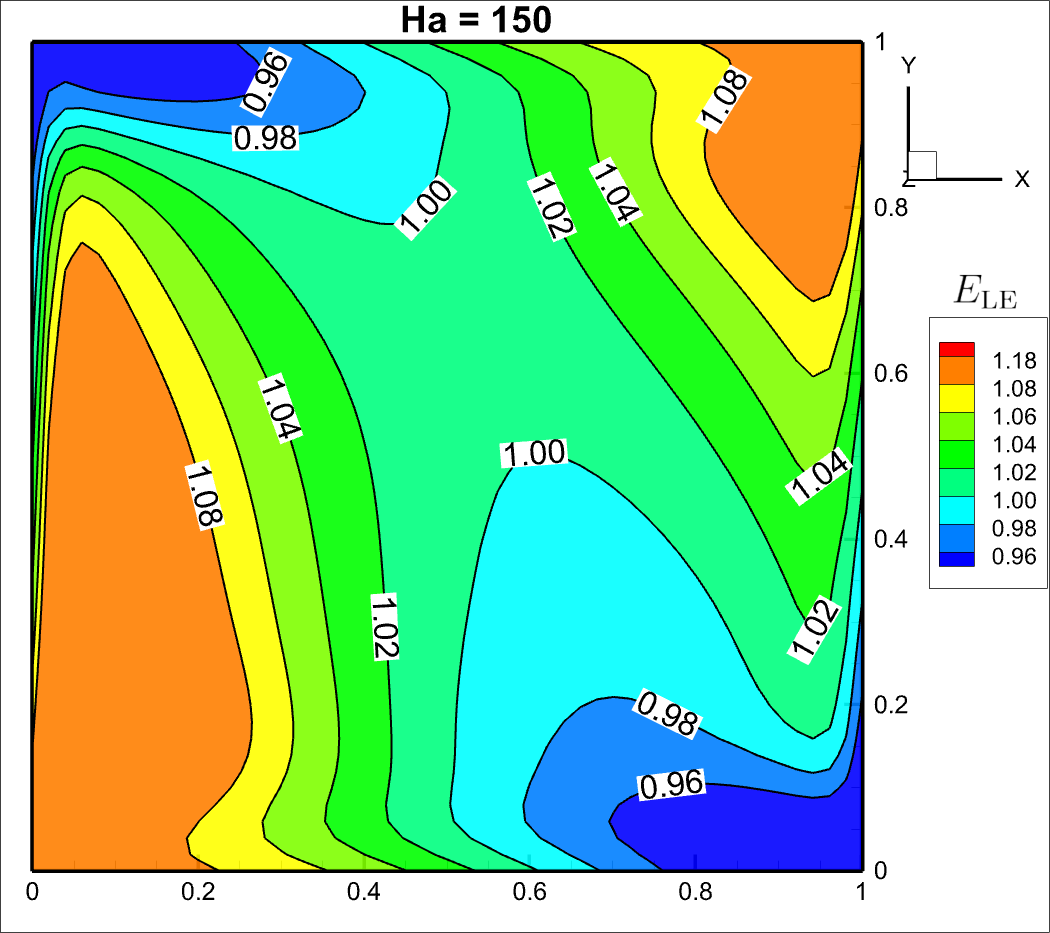}%
    \captionsetup{skip=2pt}%
    \caption{(h) $E_{\mathrm{LE}}$}
    \label{fig:Ra_10^4_Ha_150_local_entropy_generation_P1.png}
  \end{subfigure}%
  \hspace*{\fill}
  \vspace*{2pt}%
  \hspace*{\fill}%
  \caption{Case 1. Influence of $Ha$ ((a-d) $Ha=25$, (e-h) $Ha=150$) on entropy generation contours with fixed $Ra=10^4$}
  \label{fig:Case_1_Effect_of_ha_on_Entropy}
\end{figure}

\begin{figure}[htbp]
 \centering
 \vspace*{0pt}%
 \hspace*{\fill}%
\begin{subfigure}{0.24\textwidth}     
    \centering
    \includegraphics[width=\textwidth]{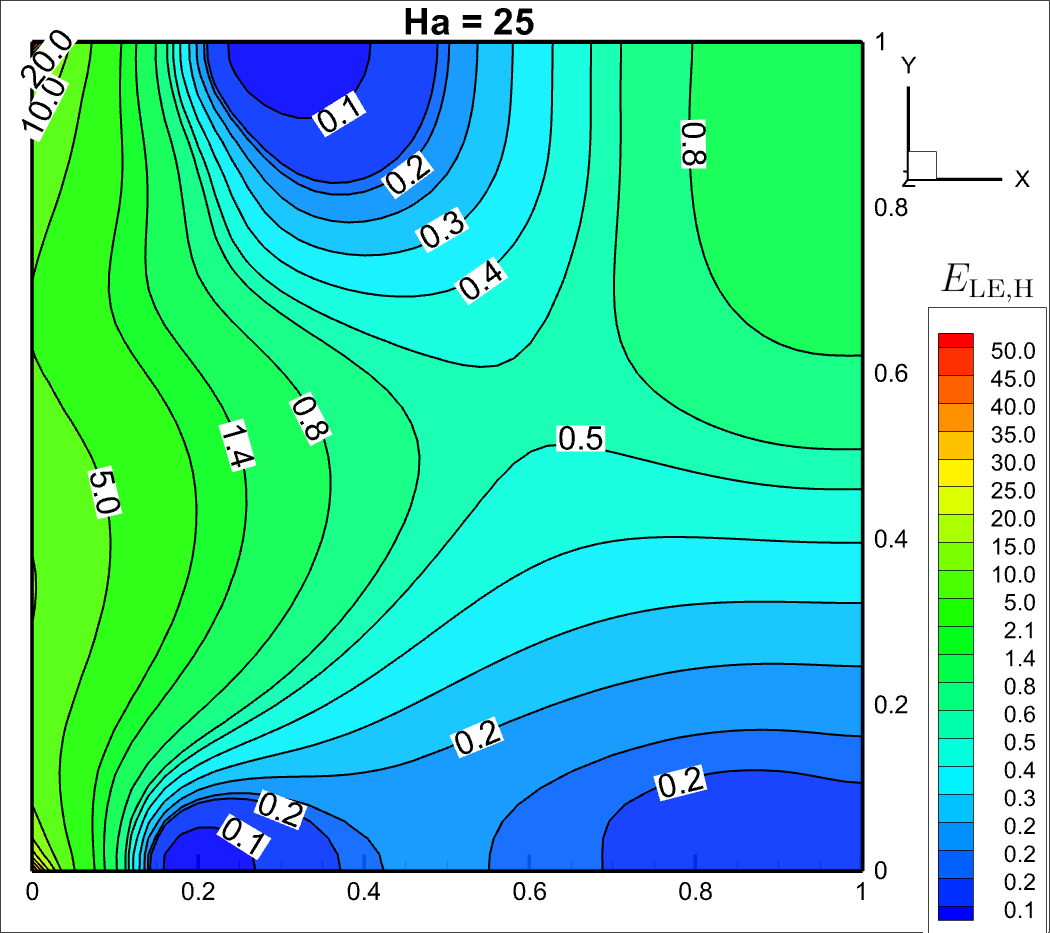}%
    \captionsetup{skip=2pt}%
    \caption{(a) $E_{\mathrm{LE,H}}$}
    \label{fig:Ra_10^4_Ha_25_local_entropy_generation_by_HT_P2.png}
  \end{subfigure}%
 \begin{subfigure}{0.24\textwidth}        
   \centering
    \includegraphics[width=\textwidth]{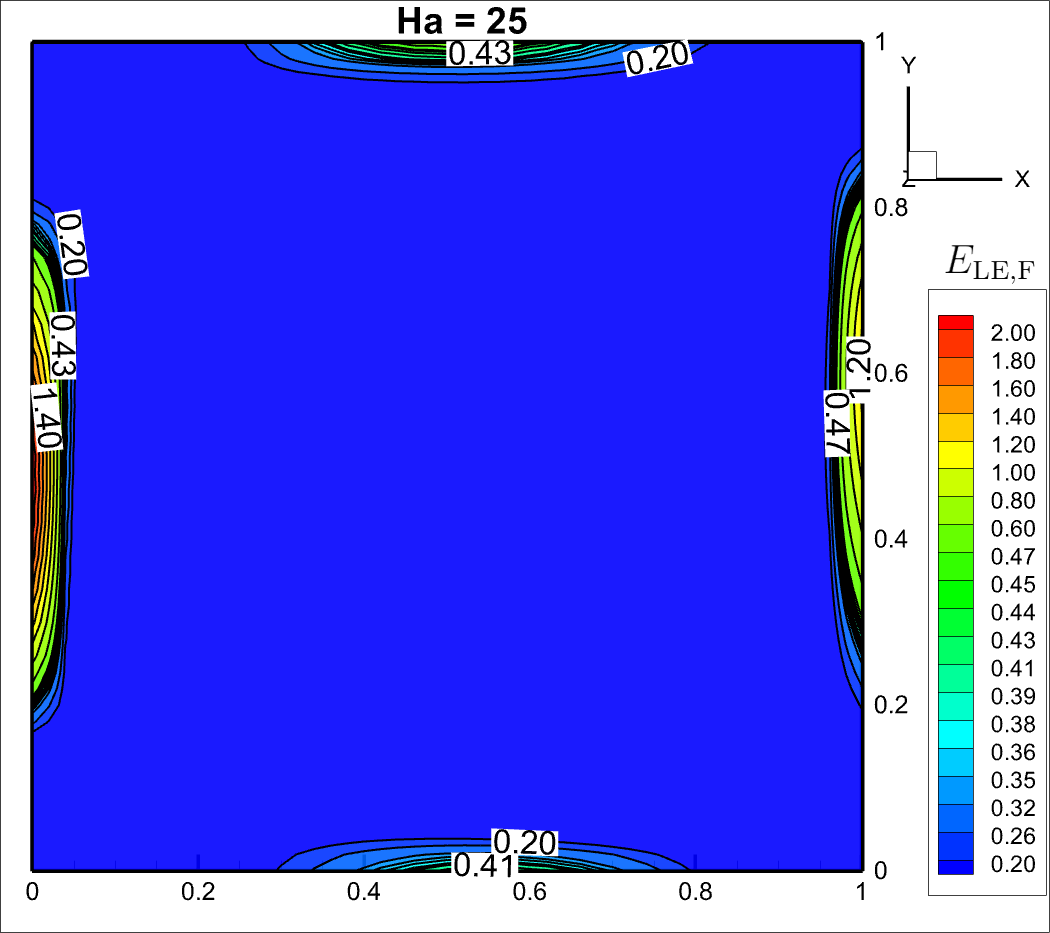}%
    \captionsetup{skip=2pt}%
    \caption{(b) $E_{\mathrm{LE,F}}$}
    \label{fig:Ra_10^4_Ha_25_local_entropy_generation_by_FF_P2.png}
  \end{subfigure}
  \begin{subfigure}{0.24\textwidth}        
   \centering
    \includegraphics[width=\textwidth]{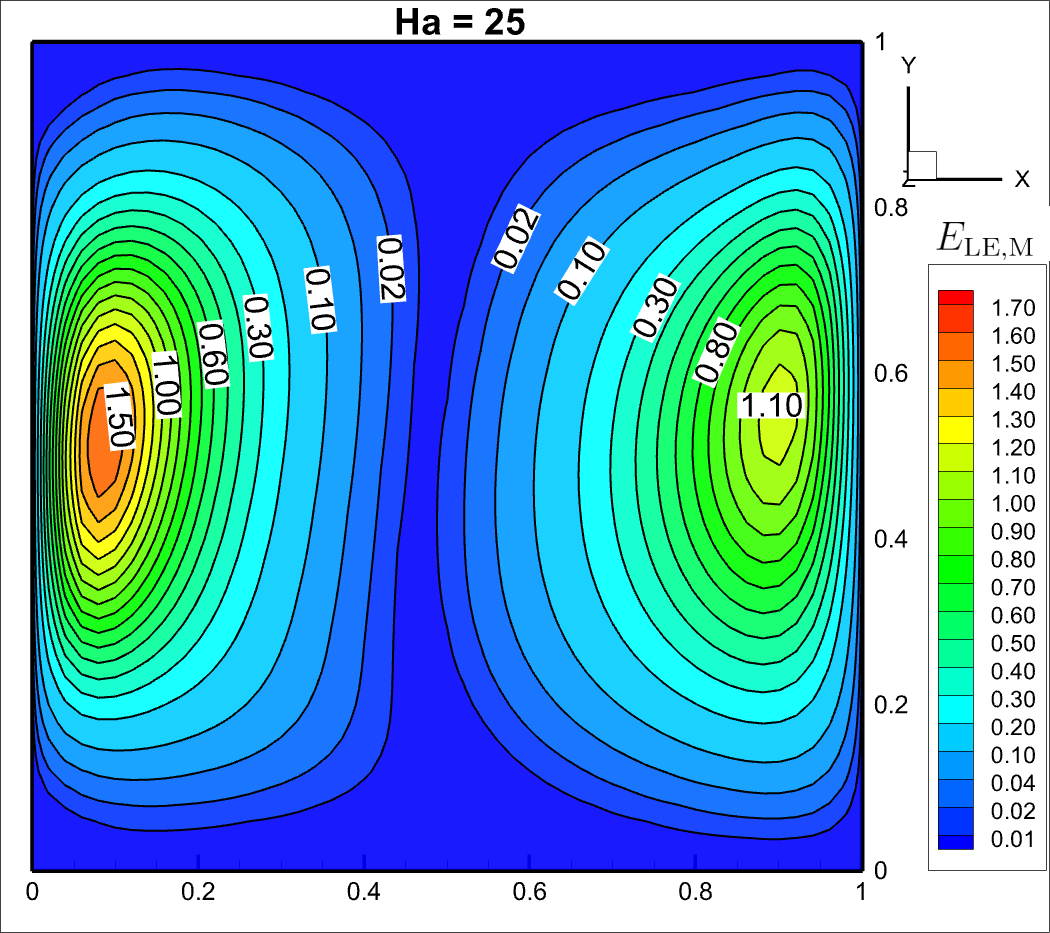}%
    \captionsetup{skip=2pt}%
    \caption{(c) $E_{\mathrm{LE,M}}$}
    \label{fig:Ra_10^4_Ha_25_local_entropy_generation_by_MF_P2.png}
  \end{subfigure}
   \begin{subfigure}{0.24\textwidth}        
   \centering
    \includegraphics[width=\textwidth]{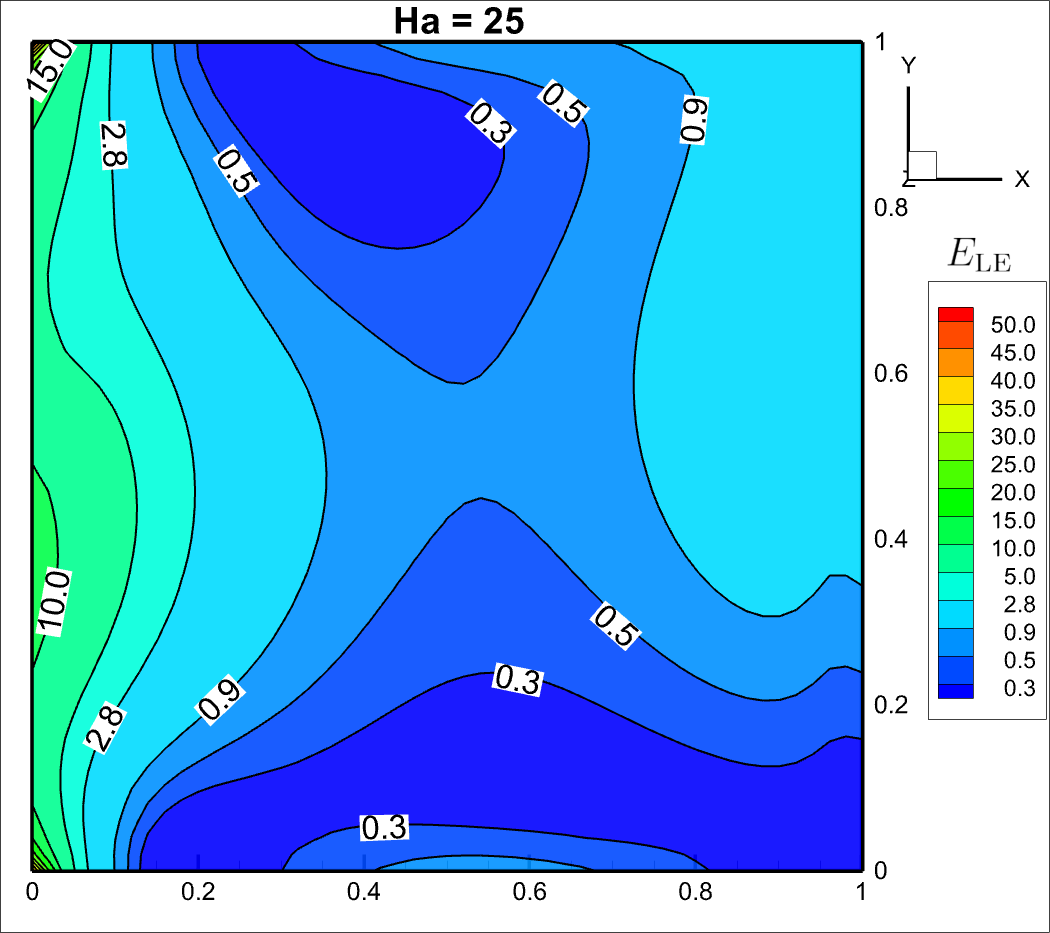}%
    \captionsetup{skip=2pt}%
    \caption{(d) $E_{\mathrm{LE}}$}
    \label{fig:Ra_10^4_Ha_25_local_entropy_generation.png}
  \end{subfigure}%
  \hspace*{\fill}

  \vspace*{8pt}%
  \hspace*{\fill}%
  \begin{subfigure}{0.24\textwidth}     
    \centering
    \includegraphics[width=\textwidth]{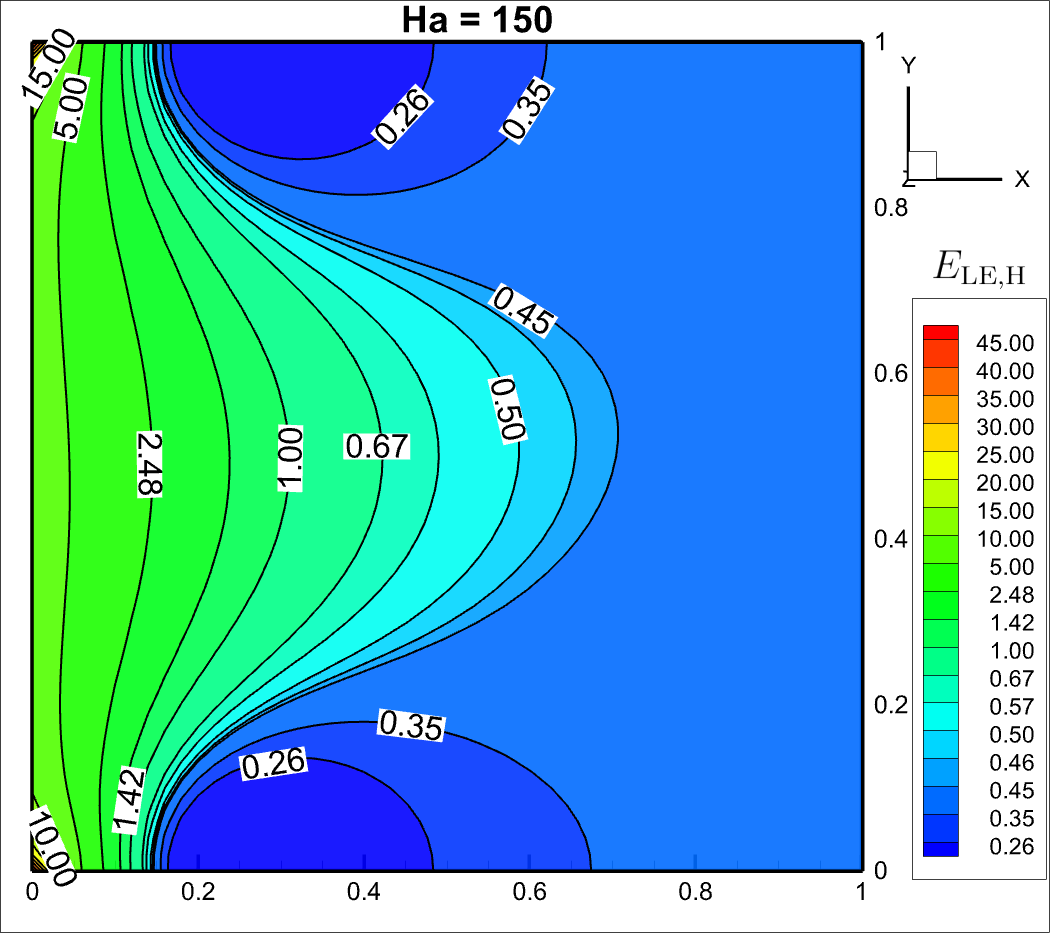}%
    \captionsetup{skip=2pt}%
    \caption{(e) $E_{\mathrm{LE,H}}$}
    \label{fig:Ra_10^4_Ha_150_local_entropy_generation_by_HT_P2.png}
  \end{subfigure}%
 \begin{subfigure}{0.24\textwidth}        
   \centering
    \includegraphics[width=\textwidth]{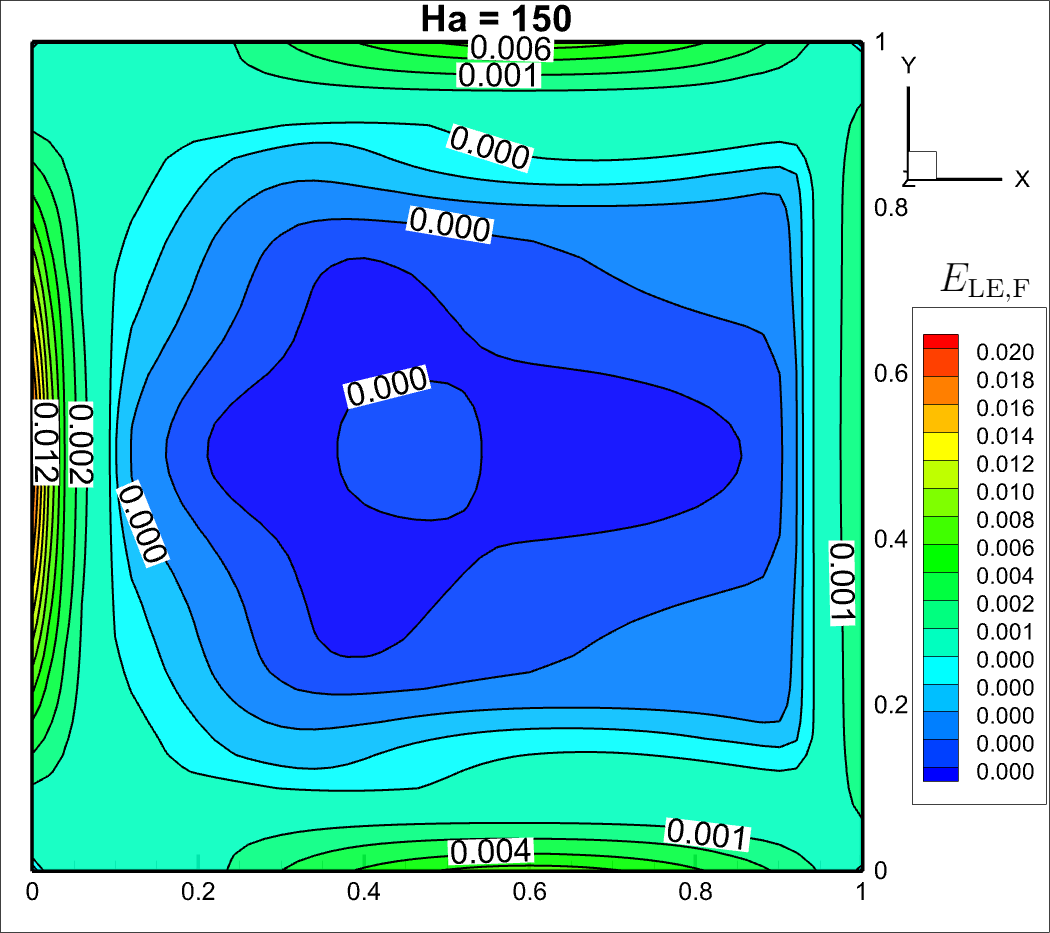}%
    \captionsetup{skip=2pt}%
    \caption{(f) $E_{\mathrm{LE,F}}$}
    \label{fig:Ra_10^4_Ha_150_local_entropy_generation_by_FF_P2.png}
  \end{subfigure}
   \begin{subfigure}{0.24\textwidth}        
   \centering
    \includegraphics[width=\textwidth]{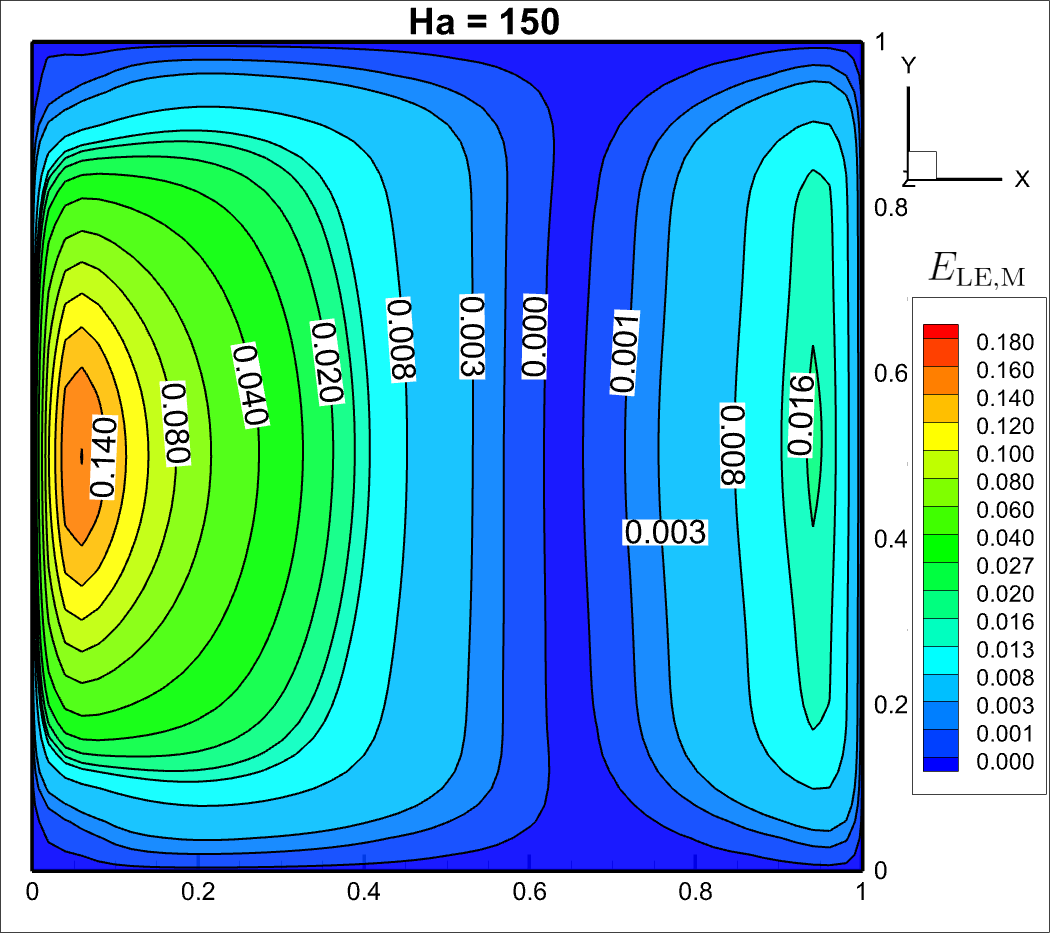}%
    \captionsetup{skip=2pt}%
    \caption{(g) $E_{\mathrm{LE,M}}$}
    \label{fig:Ra_10^4_Ha_150_local_entropy_generation_by_MF_P2.png}
  \end{subfigure}
   \begin{subfigure}{0.24\textwidth}        
   \centering
    \includegraphics[width=\textwidth]{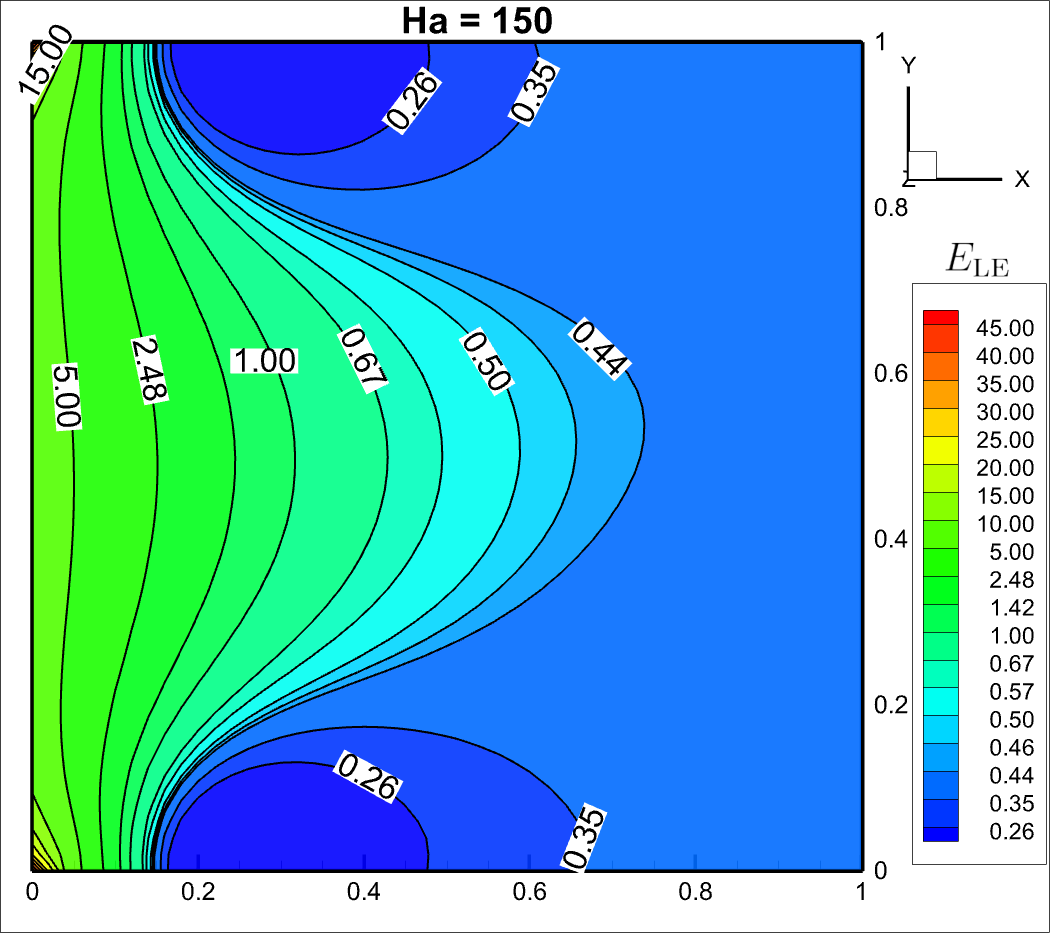}%
    \captionsetup{skip=2pt}%
    \caption{(h) $E_{\mathrm{LE}}$}
    \label{fig:Ra_10^4_Ha_150_local_entropy_generatin_P2.png}
  \end{subfigure}%
  \hspace*{\fill}
  \vspace*{2pt}%
  \hspace*{\fill}%
  \caption{Case 2. Influence of different $Ha$ ((a-d) $Ha=25$, (e-h) $Ha=150$) on entropy generation contours with fixed $Ra=10^4$}
  \label{fig:Case_2_Effect_of_Ha_on_Entropy}
\end{figure}

\begin{figure}[htbp]
 \centering
 \vspace*{0pt}%
 \hspace*{\fill}%
\begin{subfigure}{0.24\textwidth}     
    \centering
    \includegraphics[width=\textwidth]{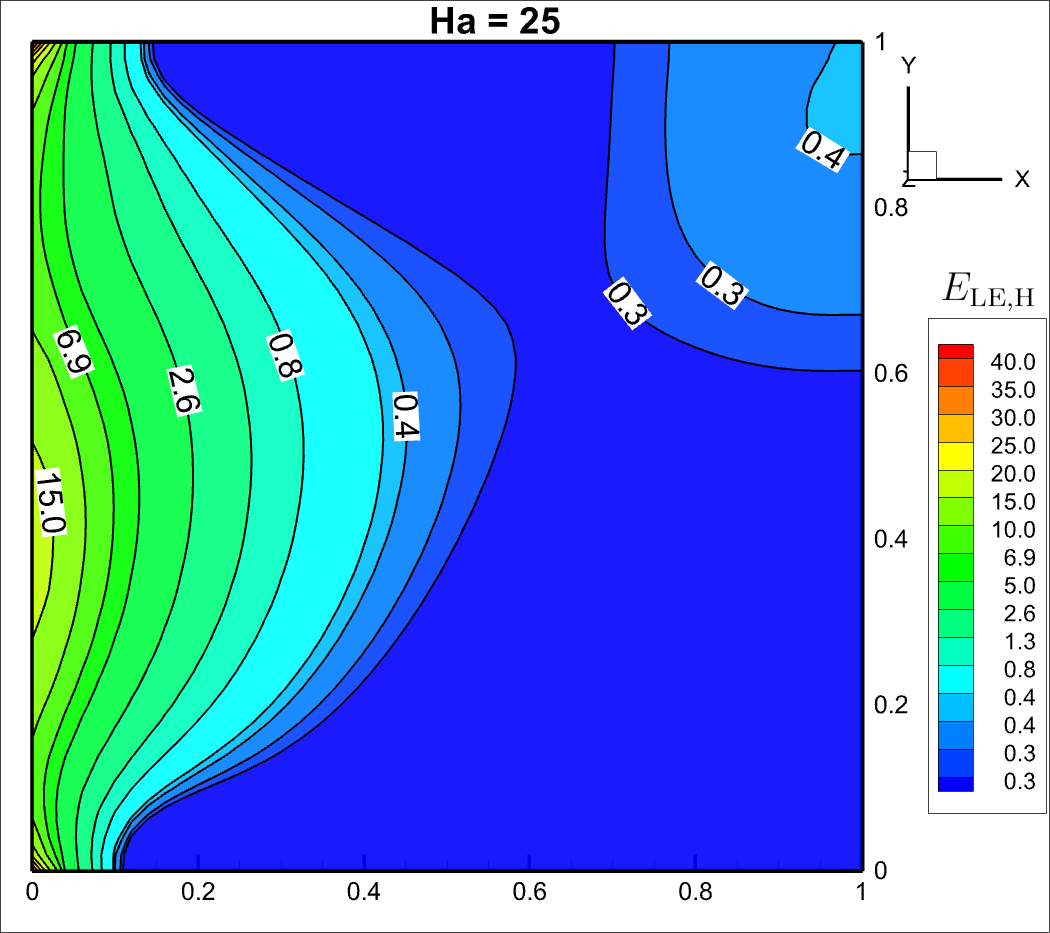}%
    \captionsetup{skip=2pt}%
    \caption{(a) $E_{\mathrm{LE,H}}$}
    \label{fig:P3_Ra_10^4_Ha_25_local_entropy_generation_by_HT.png}
  \end{subfigure}%
 \begin{subfigure}{0.24\textwidth}        
   \centering
    \includegraphics[width=\textwidth]{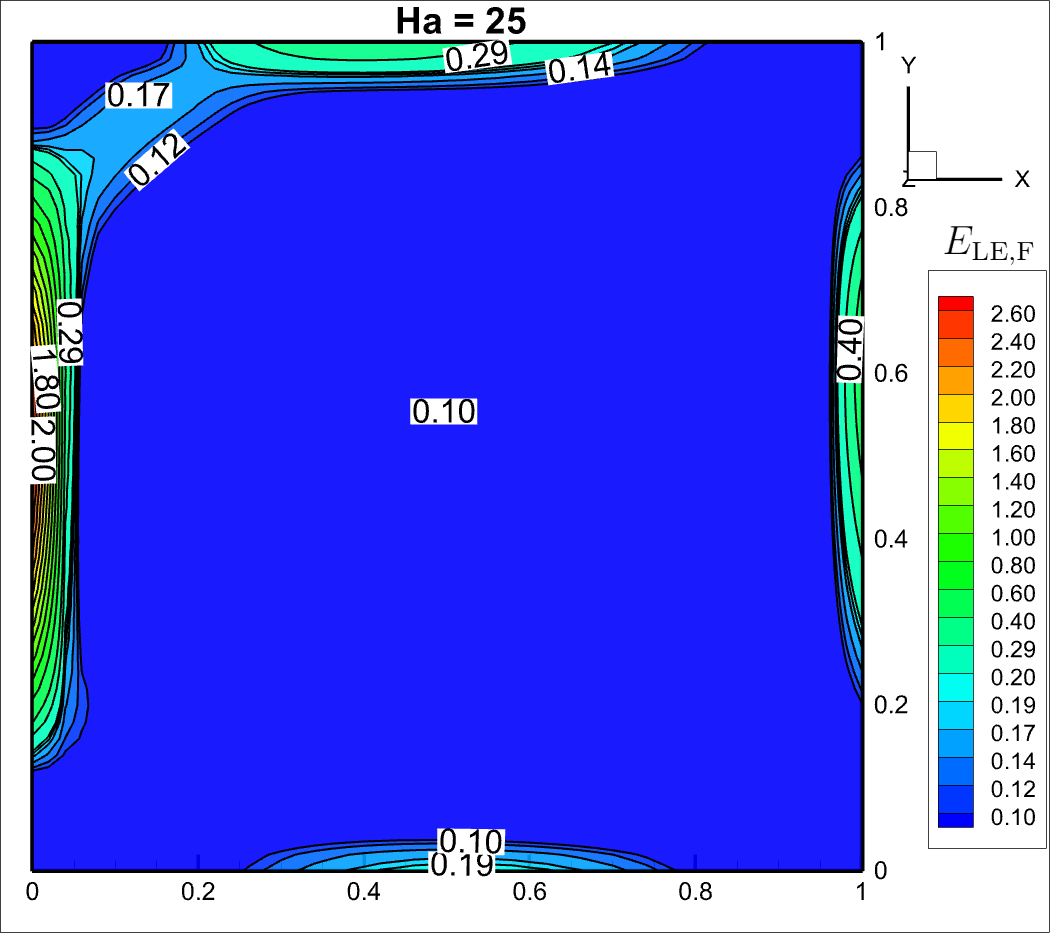}%
    \captionsetup{skip=2pt}%
    \caption{(b) $E_{\mathrm{LE,F}}$}
    \label{fig:P3_Ra_10^4_Ha_25_local_entropy_generation_by_FF.png}
  \end{subfigure}
  \begin{subfigure}{0.24\textwidth}        
   \centering
    \includegraphics[width=\textwidth]{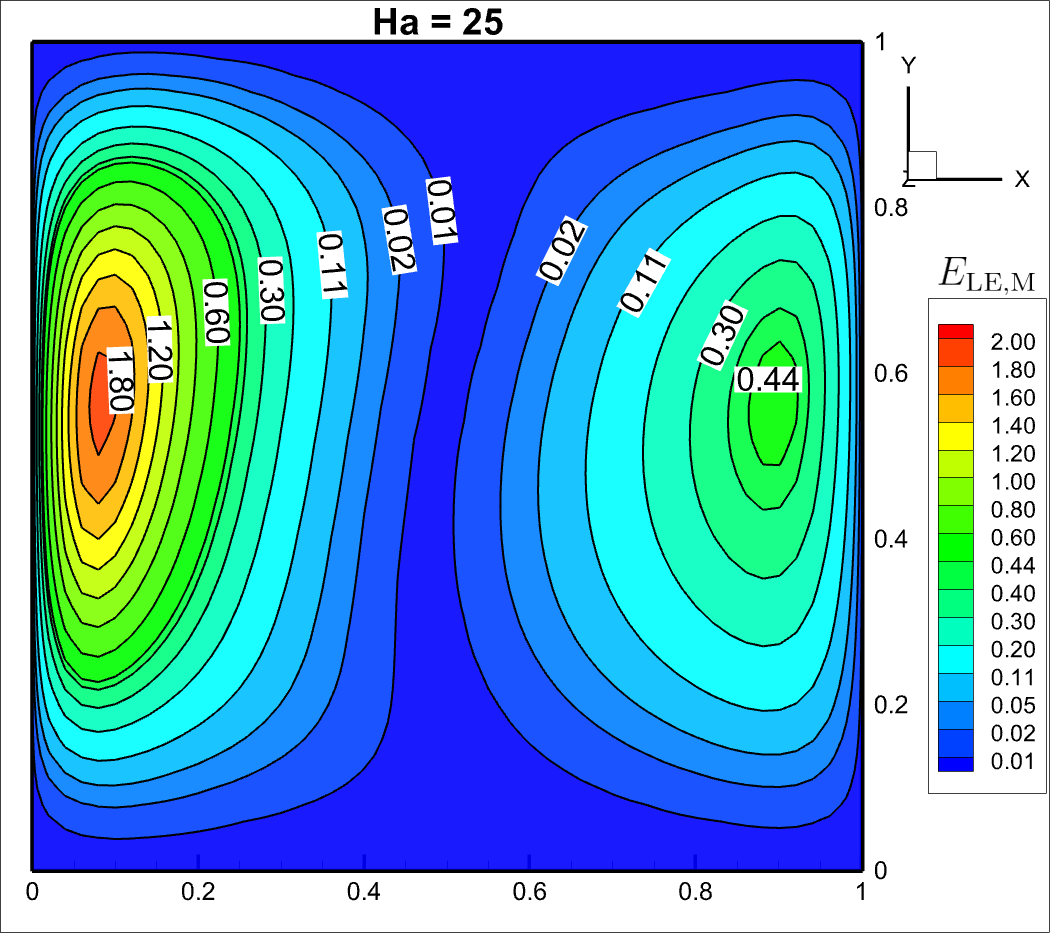}%
    \captionsetup{skip=2pt}%
    \caption{(c) $E_{\mathrm{LE,M}}$}
    \label{fig:P3_Ra_10^4_Ha_25_local_entropy_generation_by_MF.png}
  \end{subfigure}
   \begin{subfigure}{0.24\textwidth}        
   \centering
    \includegraphics[width=\textwidth]{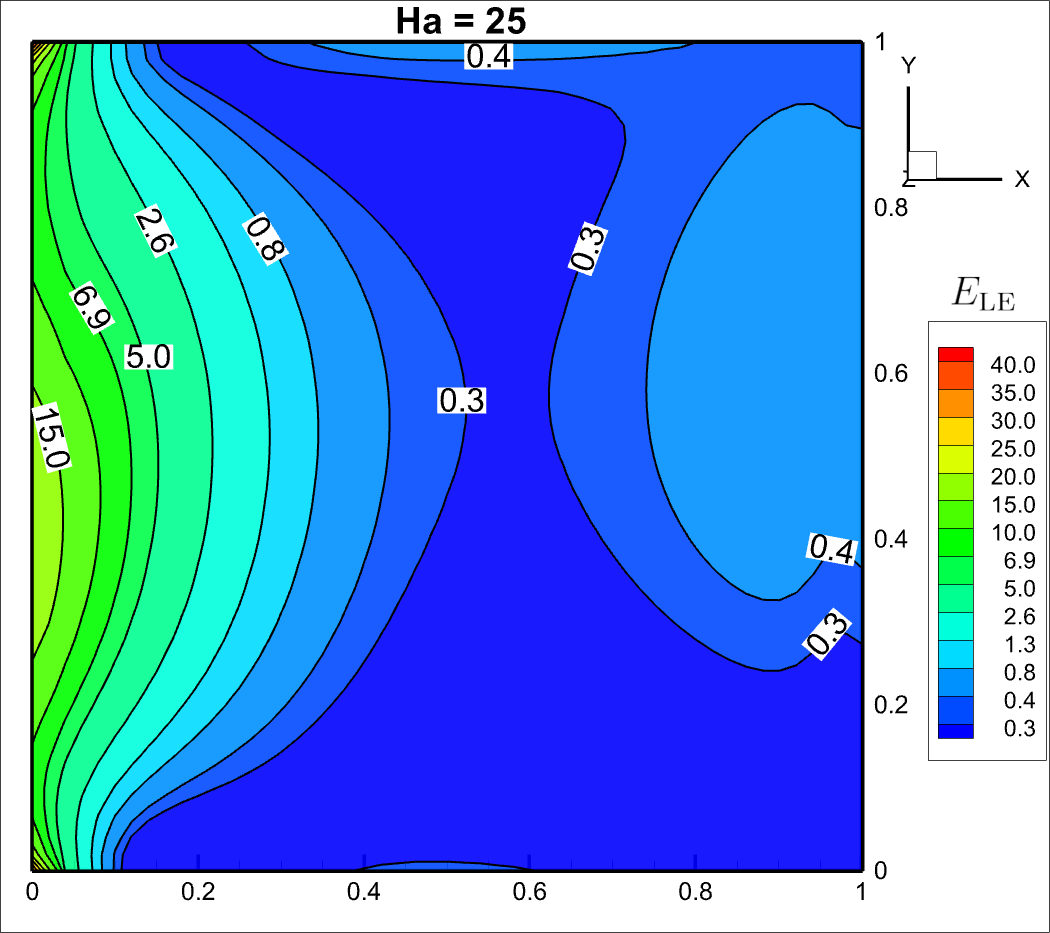}%
    \captionsetup{skip=2pt}%
    \caption{(d) $E_{\mathrm{LE}}$}
    \label{fig:P3_Ra_10^4_Ha_25_local_entropy_generation.png}
  \end{subfigure}%
  \hspace*{\fill}

  \vspace*{8pt}%
  \hspace*{\fill}%
  \begin{subfigure}{0.24\textwidth}     
    \centering
    \includegraphics[width=\textwidth]{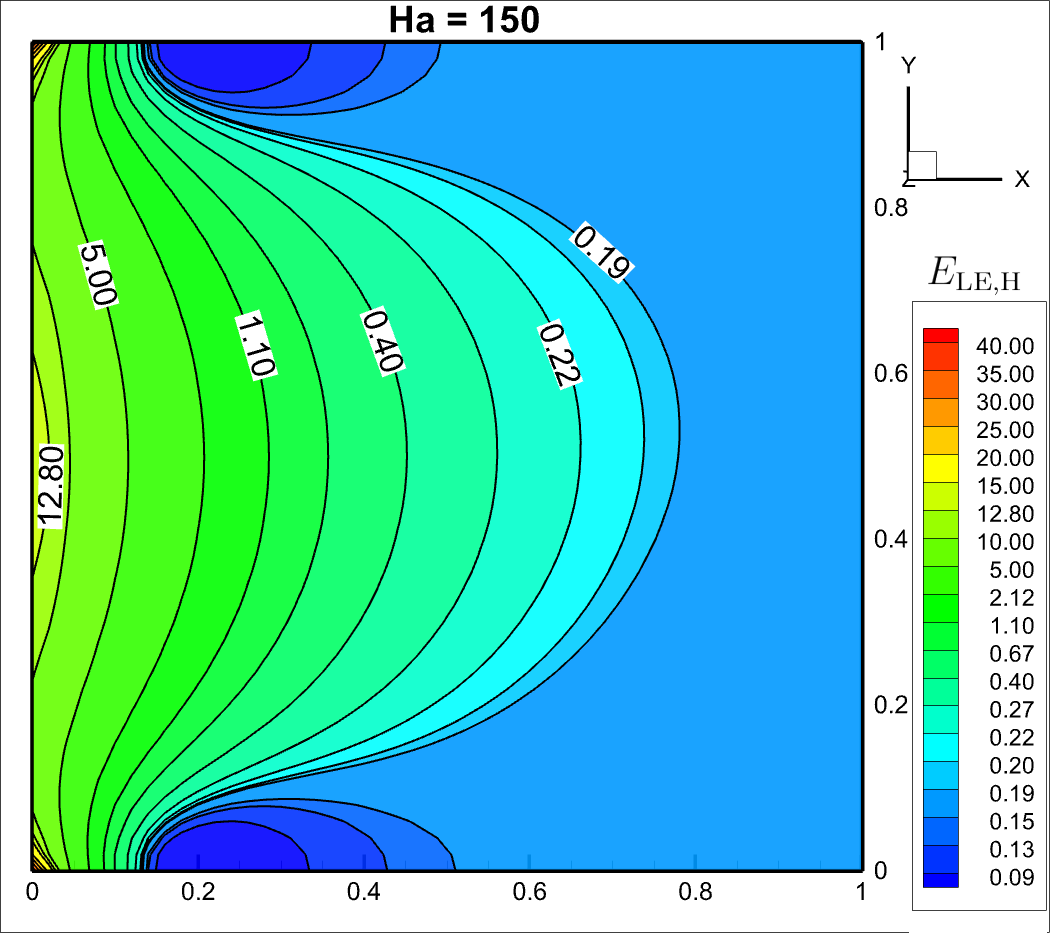}%
    \captionsetup{skip=2pt}%
    \caption{(e) $E_{\mathrm{LE,H}}$}
    \label{fig:P3_Ra_10^4_Ha_150_local_entropy_generation_by_HT.png}
  \end{subfigure}%
 \begin{subfigure}{0.24\textwidth}        
   \centering
    \includegraphics[width=\textwidth]{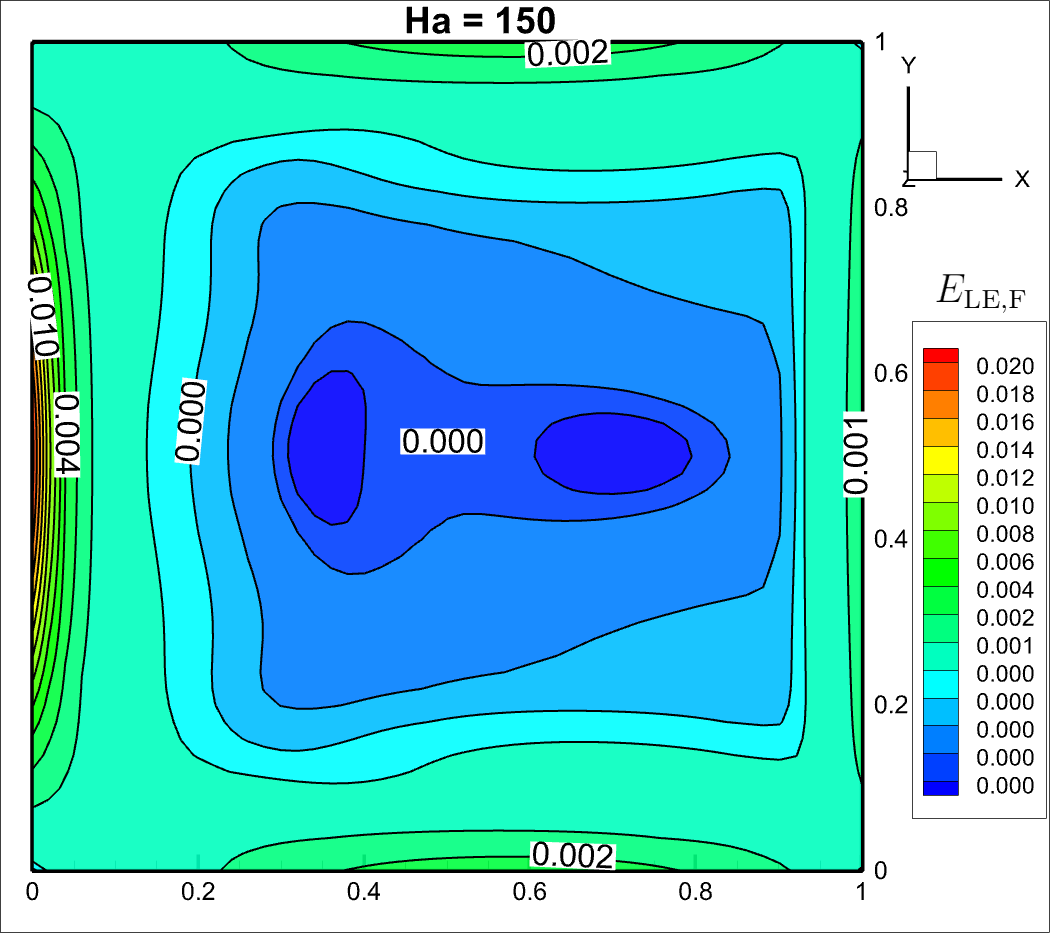}%
    \captionsetup{skip=2pt}%
    \caption{(f) $E_{\mathrm{LE,F}}$}
    \label{fig:P3_Ra_10^4_Ha_150_local_entropy_generation_by_FF.png}
  \end{subfigure}
   \begin{subfigure}{0.24\textwidth}        
   \centering
    \includegraphics[width=\textwidth]{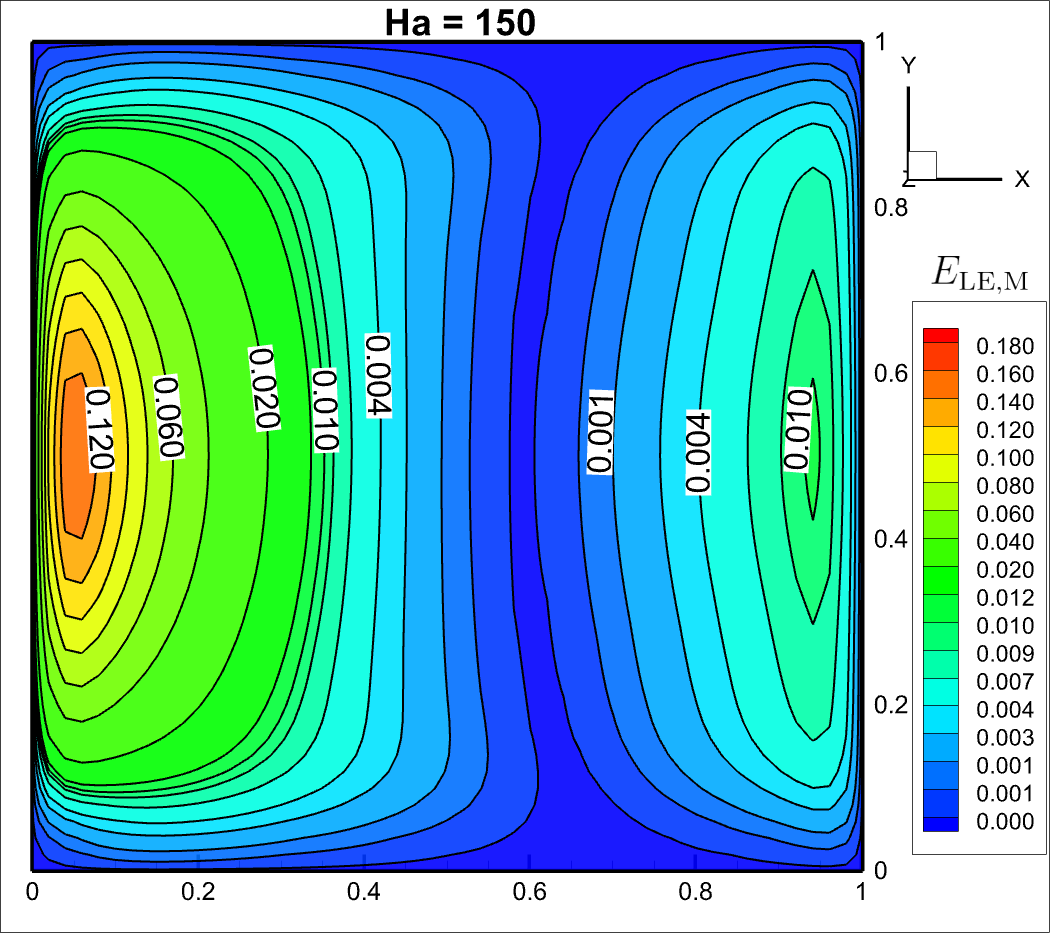}%
    \captionsetup{skip=2pt}%
    \caption{(g) $E_{\mathrm{LE,M}}$}
    \label{fig:P3_Ra_10^4_Ha_150_local_entropy_generation_by_MF.png}
  \end{subfigure}
   \begin{subfigure}{0.24\textwidth}        
   \centering
    \includegraphics[width=\textwidth]{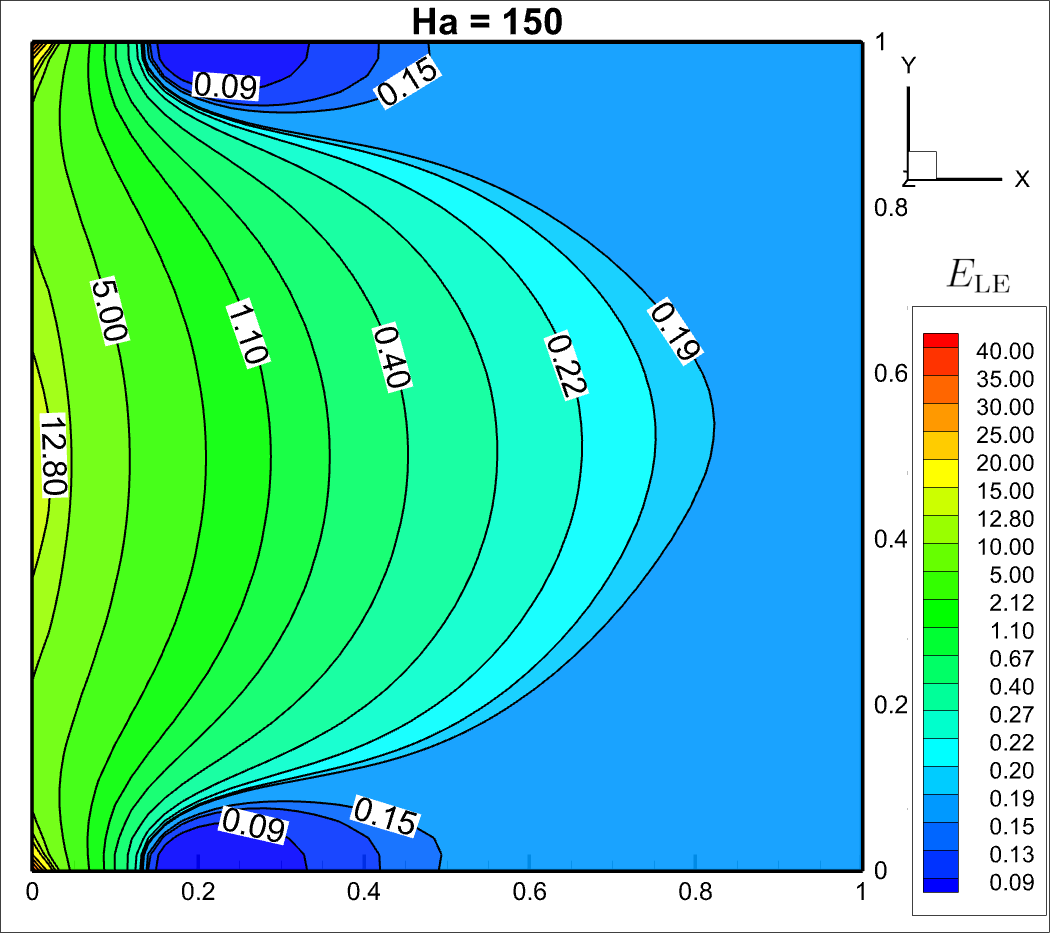}%
    \captionsetup{skip=2pt}%
    \caption{(h) $E_{\mathrm{LE}}$}
    \label{fig:P3_Ra_10^4_Ha_150_local_entropy_generation.png}
  \end{subfigure}%
  \hspace*{\fill}
  \vspace*{2pt}%
  \hspace*{\fill}%
  \caption{Case 3. Influence of different $Ha$ ((a-d) $Ha=25$, (e-h) $Ha=150$) on entropy generation contours with fixed $Ra=10^4$}
  \label{fig:Case_3_Effect_of_Ha_on_Entropy}
\end{figure}

Figures \ref{fig:Case_1_Effect_of_ha_on_Entropy}, \ref{fig:Case_2_Effect_of_Ha_on_Entropy}, and \ref{fig:Case_3_Effect_of_Ha_on_Entropy} depict the impact of $Ha$ on the spatial distribution (at $z=0.5$) of local entropy generation rates, while maintaining a constant $Ra=10^4$ as representative for Case 1, Case 2, and Case 3, respectively. In these figures, the first column displays entropy generation arising from heat transfer, second column illustrates entropy generation attributed to fluid friction, third column depicts entropy generation resulting from magnetic fields, and the final column provides the total local entropy generation. For Case 1, local entropy generation contours highlight a discernible trend: as $Ha$ increases from 25 to 150, there is a notable decrease in the local entropy generation associated with heat transfer, fluid friction, and magnetic field. This trend is also observed in Case 2 (Figure \ref{fig:Case_2_Effect_of_Ha_on_Entropy}) and Case 3 (Figure \ref{fig:Case_3_Effect_of_Ha_on_Entropy}). This observed phenomenon can be attributed to the strengthening Lorentz force induced by the elevated magnetic field associated with higher $Ha$ values. The intensified Lorentz force acts as a damping force on fluid motion, leading to the suppression of convective heat transfer. Consequently, the reduction in fluid motion leads to a decrease in entropy attributed to all contributing factors. However, the position of the maximum local entropy generation changes with different thermal boundary conditions. With uniform thermal boundary conditions (Case 1), the maximum total local entropy occurs near the left bottom and right top corners. Conversely, under non-uniform thermal boundary conditions (Case 2 and Case 3), the maximum total local entropy occurs near the left wall.   

\begin{figure}[htbp]
 \centering
   \vspace*{0pt}%
  \hspace*{\fill}%
\begin{subfigure}{0.30\textwidth}     
    \centering
    \includegraphics[width=\textwidth]{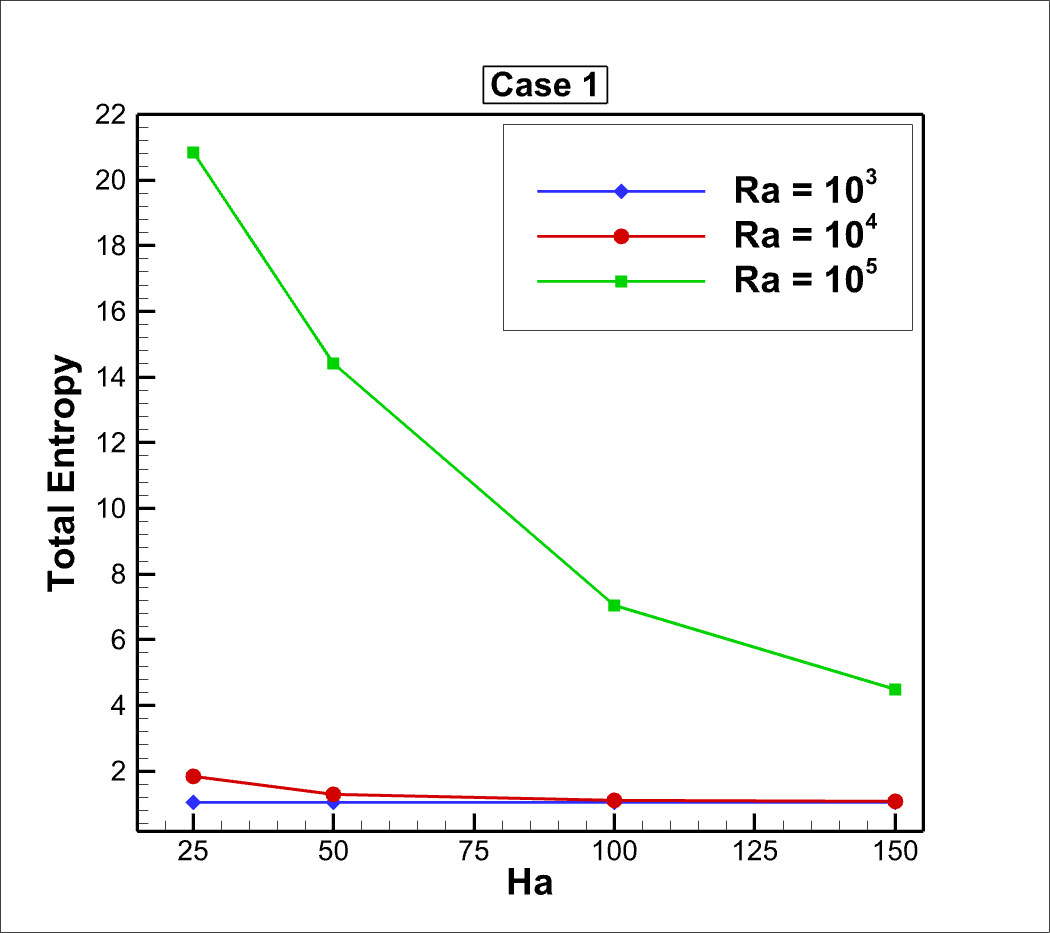}%
    \captionsetup{skip=2pt}%
    \caption{(a)}
    \label{fig:P1_Te_Effect.png}
  \end{subfigure}%
 \begin{subfigure}{0.30\textwidth}        
   \centering
    \includegraphics[width=\textwidth]{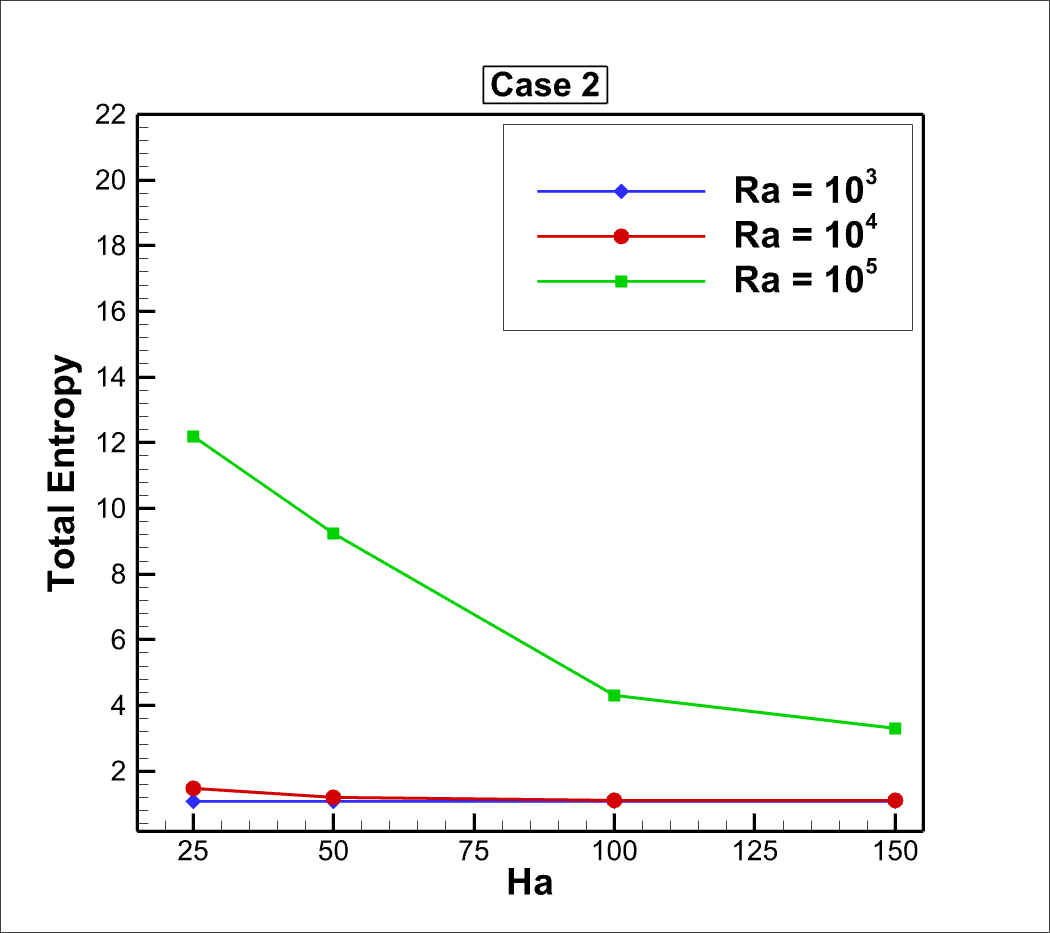}%
    \captionsetup{skip=2pt}%
    \caption{(b) }
    \label{fig:P2_Te_Effect.png}
  \end{subfigure}
   \begin{subfigure}{0.30\textwidth}        
   \centering
    \includegraphics[width=\textwidth]{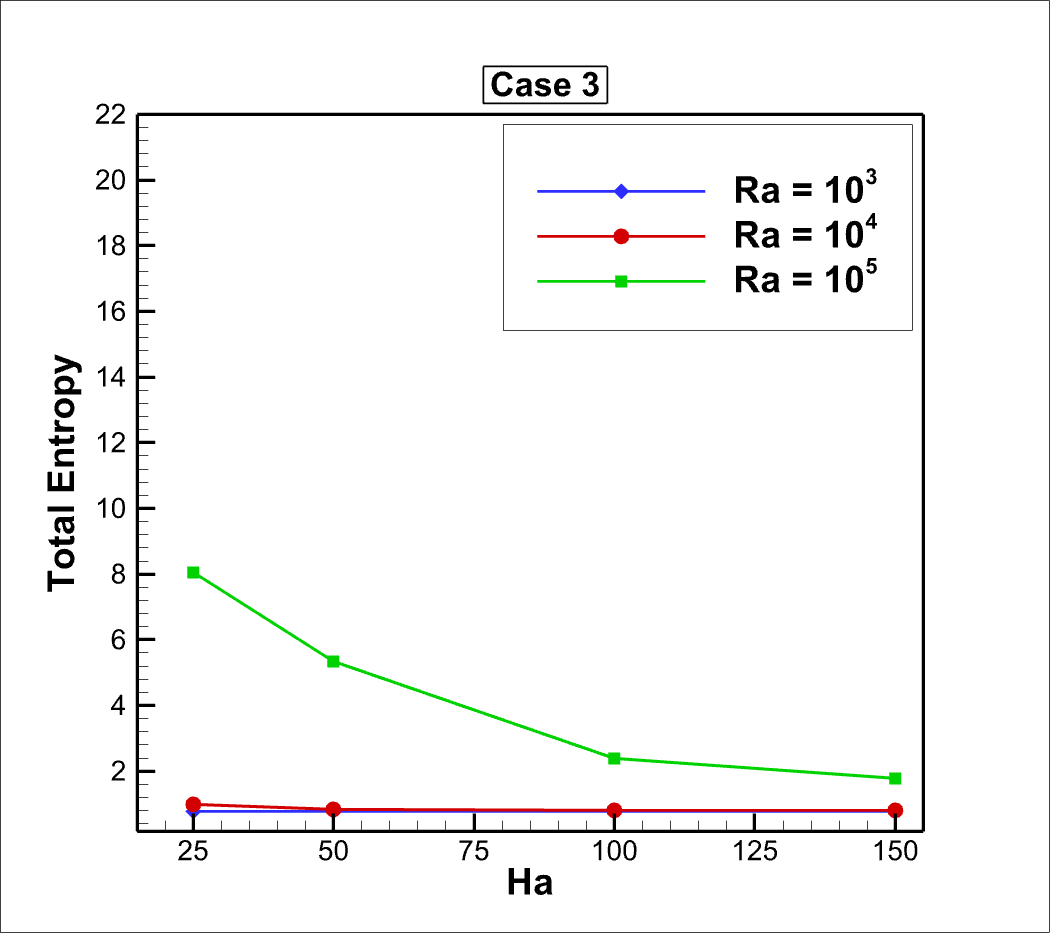}%
    \captionsetup{skip=2pt}%
    \caption{(c)}
    \label{fig:P3_Te_Effect.png}
  \end{subfigure}%
  \hspace*{\fill}
  \caption{Influence of varying $Ha$ and $Ra$ on total entropy for (a) Case 1 (b) Case 2 (c) Case 3}
  \label{fig:Ha_Ra_Effect_On_TE}
\end{figure}

\begin{figure}[htbp]
 \centering
  \vspace*{0pt}%
  \hspace*{\fill}%
  \begin{subfigure}{0.33\textwidth}     
    \centering
    \includegraphics[width=\textwidth]{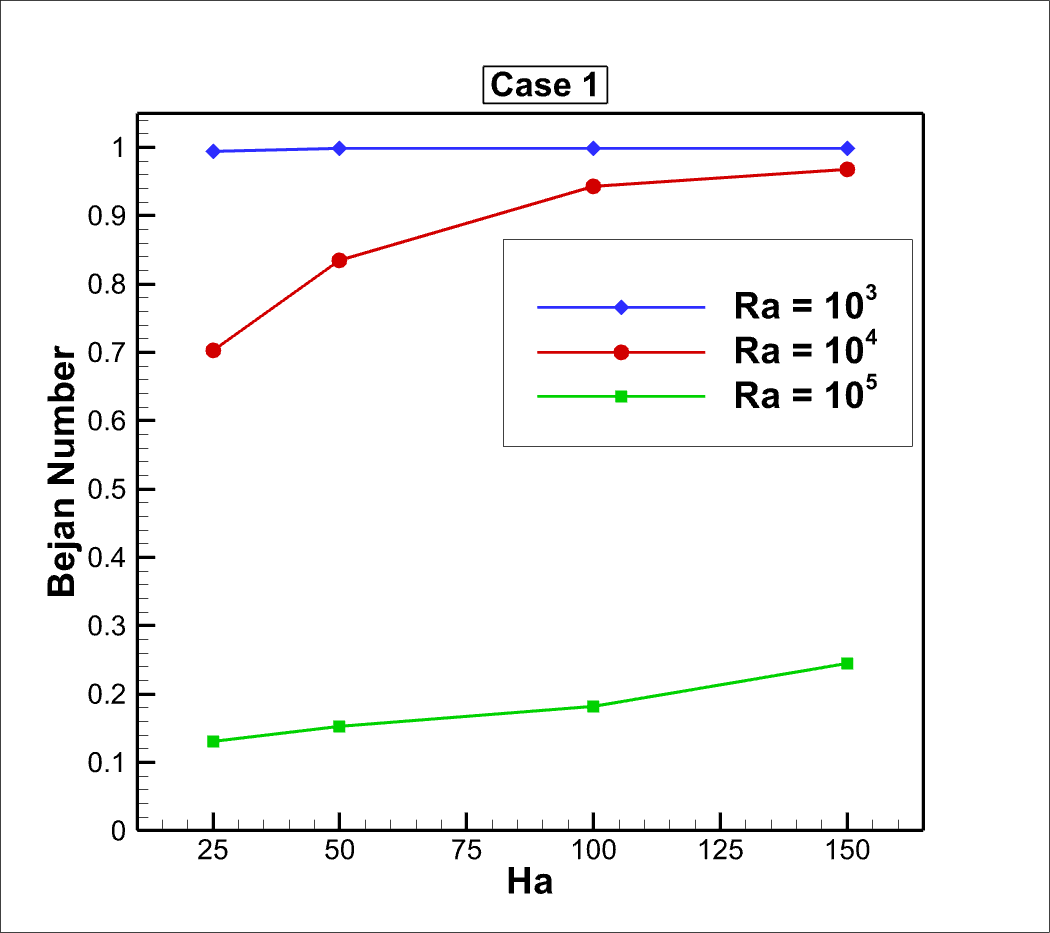}%
    \captionsetup{skip=2pt}%
    \caption{(a)}
    \label{fig:P1_Be_Effect.png}
  \end{subfigure}%
 \begin{subfigure}{0.33\textwidth}        
   \centering
    \includegraphics[width=\textwidth]{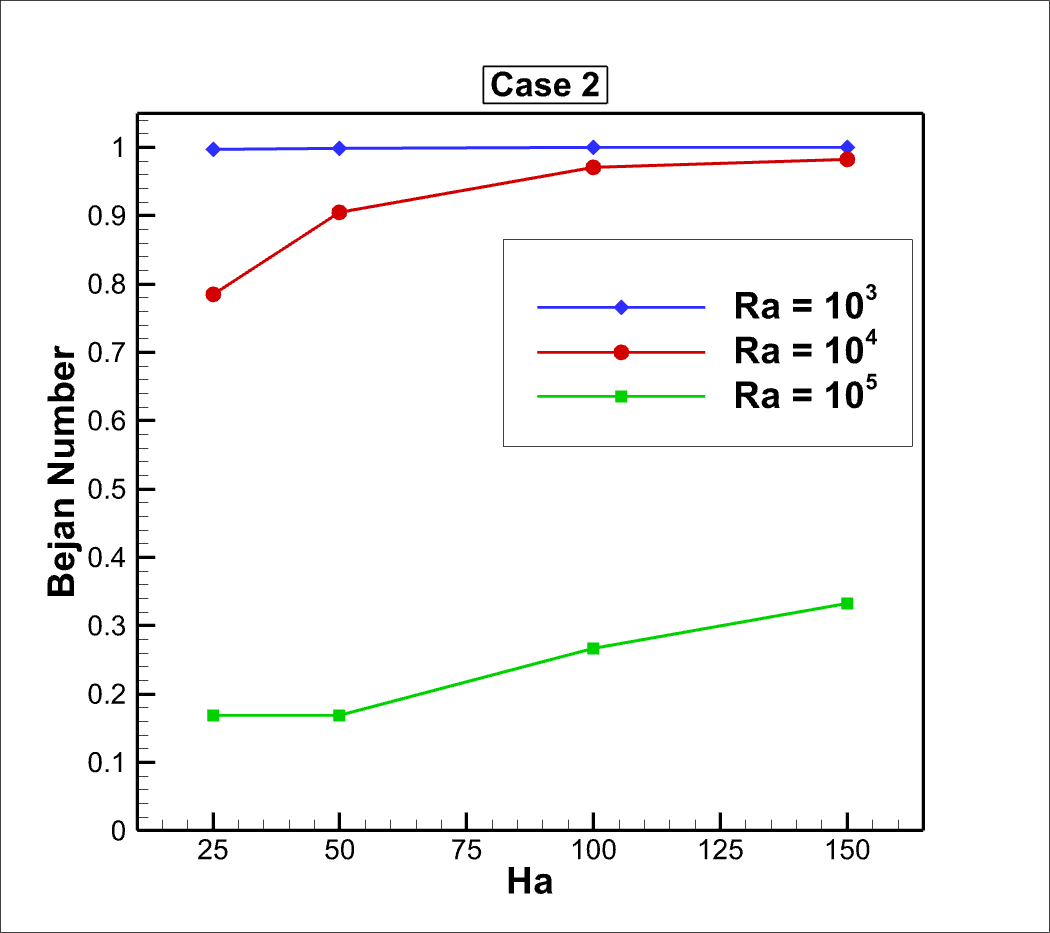}%
    \captionsetup{skip=2pt}%
    \caption{(b)}
    \label{fig:P2_Be_Effect.png}
  \end{subfigure}
   \begin{subfigure}{0.33\textwidth}        
   \centering
    \includegraphics[width=\textwidth]{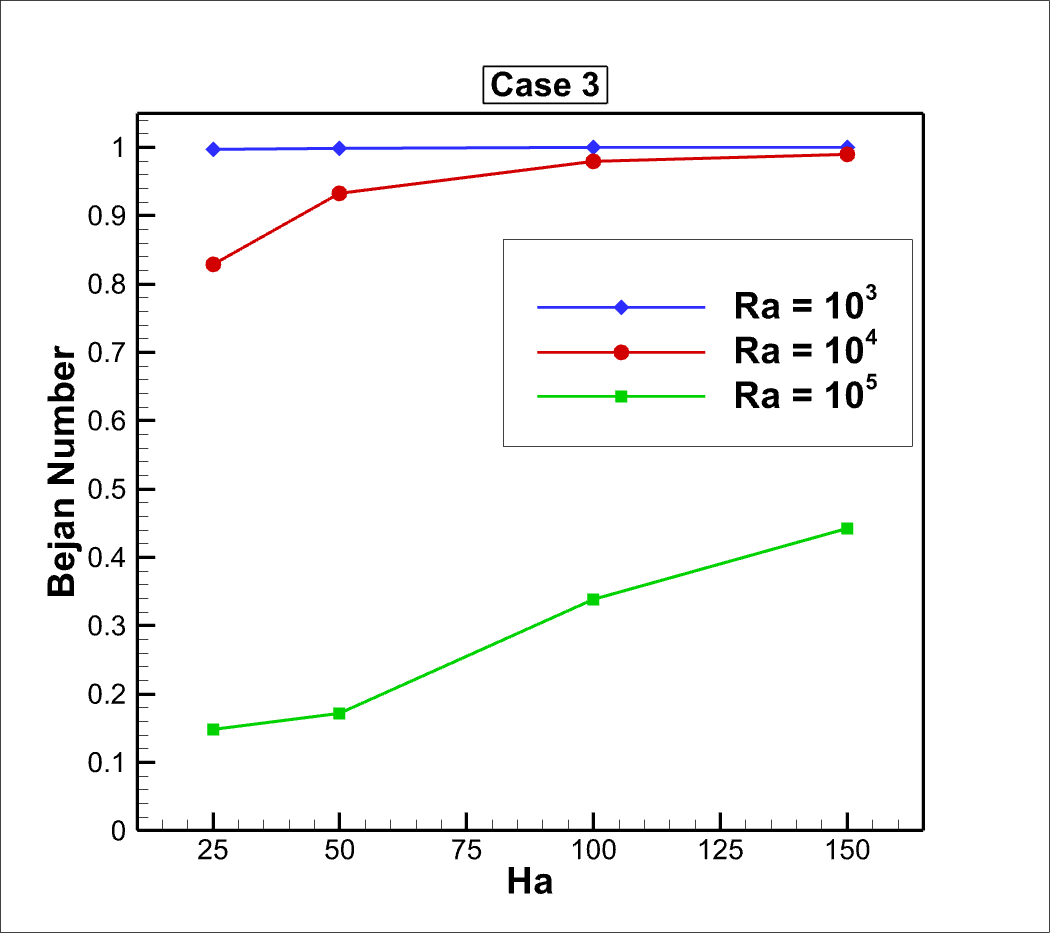}%
    \captionsetup{skip=2pt}%
    \caption{(c)}
    \label{fig:P3_Be_Effect.png}
  \end{subfigure}%
  \hspace*{\fill}
  \caption{Influence of different $Ha$ and $Ra$ on Bejan number ($Be$) for (a) Case 1 (b) Case 2 (c) Case 3 }
  \label{fig:Ha_Ra_Effect_On_BE}
\end{figure}

In Tables \ref{TE_values}, we analyze variations in total entropy (\(E_{\mathrm{TE}}\)) with changes in Hartmann number (\(Ha\)) and Rayleigh number (\(Ra\)) across all three cases. To provide a visual representation of the quantitative data in a more accessible format, we present graphical depictions in Figure \ref{fig:Ha_Ra_Effect_On_TE}. Upon observing the total entropy values for Case 1 in Table \ref{TE_values}, a discernible trend emerges, when \(Ha\) increases at a specific \(Ra\) value, the total entropy decreases. This decrement is more pronounced at higher \(Ra=10^5\). The last column in the table signifies the percentage difference in \(E_{\mathrm{TE}}\) values for various $Ha$ at specific \(Ra\). A 0.89\% decrement is observed at \(Ra=10^3\), whereas a substantial 78.44\% decrement is evident in \(E_{\mathrm{TE}}\) values at \(Ra=10^5\). This highlights the notable influence of the magnetic field on \(E_{\mathrm{TE}}\), particularly pronounced at higher \(Ra\) values. Additionally, an increase in \(Ra\) at a specific \(Ha\) results in an increment in \(E_{\mathrm{TE}}\), with a greater increase at lower \(Ha\). The last row in Table \ref{TE_values}, following Case 1, displays the percentage difference (\%) in \(E_{\mathrm{TE}}\) values for various $Ra$ at fixed $Ha$. The largest increase (1882.3\%) in \(E_{\mathrm{TE}}\) occurs when transitioning from \(Ra=10^3\) to \(Ra=10^5\) at \(Ha=25\), while the lowest increase is observed at \(Ha=150\). These observations confirm the influence of non-uniform heating condition and emphasize that the effect of \(Ra\) is more pronounced at lower \(Ha\).
When examining Cases 2 and 3, a consistent trend is observed: an increase in $Ha$ at a specific $Ra$ leads to a decrease in total entropy, while an increase in $Ra$ at a specific $Ha$ results in an increase in \(E_{\mathrm{TE}}\). This trend occurs because an elevated $Ra$ is associated with augmented temperature disparities, instigating buoyancy-driven fluid motion and intensifying convection. The heightened fluid motion, in turn, amplifies fluid friction irreversibilities, thereby fostering a general upswing in total entropy generation. Conversely, an increased $Ha$ signifies a more potent magnetic field, which exerts Lorentz forces to hinder fluid motion. The damping effect induced by the magnetic field plays an important role in diminishing total entropy. However, the rate of these changes varies across the different cases, indicating the influence of distinct thermal boundary conditions. Through comparative analysis (Table \ref{TE_values}), it becomes evident that Case 1 experiences the most significant impact on total entropy generation, followed by Case 2, and finally, Case 3 exhibits the least effect on total entropy generation. This effect can be easily observed in Figure \ref{fig:Ha_Ra_Effect_On_TE}. 
Observing Figure \ref{fig:Ha_Ra_Effect_On_TE}, it is found that the graph depicting total entropy for $Ra=10^5$ consistently remains above the graphs corresponding to lower $Ra$ values across all cases. However, it is noteworthy that the maximum value of total entropy is observed to be the highest in Case 1, followed by Case 2, and is ultimately the lowest in Case 3.\\
Table \ref{Be_values} presents quantitative data illustrating the influence of $Ha$ and $Ra$ on the Bejan number ($Be$) across all three cases. Figure \ref{fig:Ha_Ra_Effect_On_BE} provides a graphical representation of this data. 
$Be$, a dimensionless parameter, describes the amount of irreversibility caused by heat transfer in the overall irreversibility caused by heat transfer, magnetic field, and fluid friction. While the analysis of total entropy generation reveals the overall effect of $Ha$ and $Ra$ on irreversibilities in the system, the examination of $Be$ specifically highlights the role of heat transfer irreversibilities within the overall irreversibilities in a system. Notably, when $Be$ exceeds 0.5, it signifies the predominance of heat transfer irreversibility, whereas a value lower than 0.5 suggests the ascendancy of other irreversibilities over heat transfer.
When examining the $Be$ for Case 1, our analysis reveals intriguing insights. Firstly, we observe that as the $Ha$ increases, 
$Be$ demonstrates a corresponding rise at specific 
$Ra$, suggesting a growing dominance of heat transfer irreversibility with increasing $Ha$. However, upon increasing $Ra$, $Be$ experiences a decrement, indicating a shift towards other irreversibilities surpassing heat transfer irreversibility. $Be$ consistently exceeds 0.5 at $Ra=10^3$ and $Ra=10^4$, indicating a significant contribution of thermal entropy generation to the overall irreversibility. Whereas it drops below 0.5 for $Ra=10^5$ and all $Ha$ values. This observation indicates that irreversibilities arising from fluid friction and Lorentz force become dominant for $Ra=10^5$, surpassing the influence of irreversibilities due to heat transfer. The lowest $Be$ occurs at $Ra=10^5$ for all the cases. 
There is an 86.89\% decrease in the $Be$ value for Case 1, an 83.07\% decrease for Case 2, and an 82.39\% decrease for Case 3 when the $Ra$ is increased from $10^3$ to $10^5$ at a fixed $Ha=25$. This is attributed to the increase in $Ra$, resulting in the rise of the temperature difference and the buoyant forces. Consequently, there is an enhancement in convective heat transfer, contributing to the entropy generation at a faster pace. From Figure \ref{fig:Ha_Ra_Effect_On_BE}, a clearer impact of $Ha$, $Ra$, and distinct thermal boundary condition on $Be$ becomes evident. The trend observed in Case 2 and Case 3 concerning both the $Ha$ and the $Ra$ is similar to that of Case 1. However, it is noteworthy that the $Be$ value is minimum in Case 1 compared to Case 2 and 3, at specific values of $Ha$ and $Ra$.

{\small\begin{table}[htbp] \footnotesize
\caption{\small Total entropy value at different $Ha$ and $Ra$ for all three cases}\label{TE_values}
\centering
 \begin{tabular}{cccccccc}  \hline \hline
  &  &  &    &  $E_{TE}$  &   \\ \hline 
 &  $Ra$ & $Ha=25$ & $Ha=50$   & $Ha=100$  &  $Ha=150$ & Max. difference (\%) \\
   &  &  &  & & & at fixed $Ra$ \\ \hline  
 \textbf{Case 1} & $10^3$  & 1.0506 & 1.0430    &  1.0413 & 1.0412  & -0.89  \\
 & $10^4$   &    1.8438 & 1.2895     &  1.073 & 1.0743 & -41.73   \\
 & $10^5$   & 20.826 & 14.409  &  7.0438 & 4.4891 & -78.44  \\
Max. difference (\%)  & --   & +1882.3 & +1281.5 &  +576.4 & +331.1 & --  \\
at fixed $Ha$\\
\hline 
 \textbf{Case 2} & $10^3$  & 1.0829 & 1.0792    &  1.0781 & 1.0779 & -0.46  \\
& $10^4$   &    1.4854 & 1.2034     &  1.1117 & 1.0975 & -26.11   \\
 & $10^5$   & 12.205  & 9.2237  &  4.3165 & 3.3027 & -72.93  \\
Max. difference (\%)  &  --  & +1027.0 & +754.6 &  +300.3 & +206.4 & --  \\
at fixed $Ha$\\
\hline 
 \textbf{Case 3} & $10^3$  & 0.7854 & 0.7838     &  0.7833 & 0.7832  &  -0.28 \\
& $10^4$   &    0.9753 & 0.8435     &  0.7997 & 0.7915  & -18.84  \\
 & $10^5$   & 8.0617  & 5.3521  &  2.3780 & 1.7816  & -77.90 \\
Max. difference (\%)  & --   & +926.4 & +582.8 &  +203.5 & +127.4 & --  \\
at fixed $Ha$\\
\hline\hline
 \end{tabular}
\end{table}
}

{\small\begin{table}[htbp] \footnotesize
\caption{\small Bejan number at different $Ha$ and $Ra$ for all three cases}\label{Be_values}
\centering
 \begin{tabular}{cccccccc}  \hline \hline
  &  &  &    &  $Be$  &   \\ \hline 
 &  $Ra$ & $Ha=25$ & $Ha=50$   & $Ha=100$  &  $Ha=150$ & Max. difference (\%) \\
   &  &  &  & & & at fixed $Ra$ \\ \hline  
 \textbf{Case 1} & $10^3$  & 0.9938 & 0.9979    &  0.9996 & 0.9998  & +0.60  \\
 & $10^4$   &    0.7035 & 0.8349     &  0.9424 & 0.9691 & +37.75   \\
 & $10^5$   & 0.1302 & 0.1530  &  0.1816 & 0.2459 & +88.86 \\
 Max. difference (\%) & --   & -86.89 & -84.66 &  -81.83 & -75.40 & --  \\
 at fixed $Ha$\\
\hline 
 \textbf{Case 2} & $10^3$  & 0.9967 & 0.9989    &  0.9996 & 0.9998 & +0.31  \\
& $10^4$   &    0.7849 & 0.9052     &  0.9706 & 0.9824 & +25.16   \\
 & $10^5$   & 0.1687  & 0.1691  &  0.2677 & 0.3333 & +97.56  \\
Max. difference (\%)  &  --  & -83.07 & -83.07 &  -73.21 & -66.66 & --  \\
at fixed $Ha$\\
\hline 
 \textbf{Case 3} & $10^3$  & 0.9977 & 0.9992     &  0.9997 & 0.9998  & +0.21\\
& $10^4$   &    0.8290 & 0.9323     &  0.9797 & 0.9895  & +19.36  \\
 & $10^5$   & 0.1488  & 0.1759  &  0.3380 & 0.4429  & +197.6\\
Max. difference (\%)  & --   & -85.08 & -82.39 &  -66.18 & -55.70 & --  \\
at fixed $Ha$\\
\hline\hline
 \end{tabular}
\end{table}
}

\newpage
\section{Conclusion}
\label{sec:Conclusion}
Current research work deals with the impact of Magneto-Hydrodynamics (MHD) on natural convection within a 3D cavity filled with molten lithium by using a new HOSC scheme. Initially, the newly implemented scheme and code are validated against existing results, demonstrating excellent agreement. Subsequently, three distinct cases, characterized by different temperature boundary conditions, are investigated. 
The impact of varying $Ha$ and $Ra$ on fluid flow patterns and heat transfer phenomena is examined across all three cases. In summary, the findings of this study lead to the following conclusions:\\
\textbf{1.} At $Ra = 10^3$, conduction dominant effects are observed, evidenced by straight-line isothermal contours and weak clockwise circulation within the cavity for Case 1. This conduction dominant effect can be observed in the other two cases also, but with different isotherm contours. In Case 2 and Case 3, the isotherms are smoothly curved near the left wall due to the non-uniform thermal boundary condition and almost parallel near the cold wall, resulting in a reduced temperature gradient on the wall. In Case 1, a distinct transformation is evident: as Ha increases, the primary vortices change from a round shape to a stretched form in the $y$-direction while this phenomenon is absent in Case 2 and Case 3. For high $Ra=10^5$, the isotherms display a more twisted configuration for all the cases, indicating a departure from convective effects. With a rise in the $Ra$, the rate of heat transfer increases correspondingly, while an increase in $Ha$ results in a decrease in the rate of heat transfer for all considered cases.\\
\textbf{2.}  As the $Ra$ increases, the maximum values of the local Nusselt number also increase. However, the rate of increment decreases with an increase in the magnetic field ($Ha$), suggesting that higher $Ha$ values lead to a reduction in heat transfer efficiency. This effect can be observed in all the cases but the rate of change in maximum local $Nu_{\text {L }}$ values is different in each case. For example with an increase in $Ra$ from $10^3$ to $10^5$ with specific $Ha = 25$, there is a substantial 654.1\% rise in the maximum $Nu_{\text {L }}$ value while there is only 18.18\% for Case 2 and 25.17\% for Case 3.\\
\textbf{3.} The $Nu_{\text {A}}$ for the heated wall consistently increases with an increase in $Ra$ and decreases with an increase in $Ha$. Specifically, for $Ra=10^5$ and $Ha=25$, the convection dominant behavior can be seen. There are two points of local maxima and one point of local minima (at $z=0.5$) that exist in Case 1 and 2, while only one point of maxima (at $z=0.5$) and two points of minima exist in Case 3. This indicates that convection becomes more pronounced near the cavity's center for Case 3 in comparison to Case 1 and Case 2.\\
\textbf{4.} In addition, our investigation delved into entropy generation, an important measure for assessing energy dispersal or degradation within the system. We explored the correlation between total entropy and $Ra$, uncovering an increase in total entropy with the elevation of $Ra$ across all cases. Moreover, the findings reveal that an augmentation in the magnetic field power results in a reduction of the overall entropy production for all the cases. it is noteworthy that the maximum value of total entropy is observed to be the highest in Case 1, followed by Case 2, and is ultimately the lowest in Case 3.\\
\textbf{5.} Our examination of the Bejan number ($Be$) holds considerable importance, offering insights into the dominant role of heat transfer in the overall generation of entropy. We identified a definite trend: as the $Ra$ grows, the $Be$ falls, while the overall or total entropy generation increases, and with $Ha$ the inverse trend can be observed. When $Ra$ remained less than or equal to $10^4$, the $Be$ has constantly maintained above 0.5, signaling the heat transfer dominant irreversibility.
But, at $Ra = 10^5$, the $Be$ dropped below 0.5, signifying a transition towards the predominant influence of viscous and magnetic field effects on irreversibility for all the cases.\\
In conclusion, using the new HOSC scheme provided useful insights into the complex mechanics of MHD natural convection within a 3D cavity, emphasizing the pivotal significance of the Hartmann number, Rayleigh number, and different temperature boundary conditions in determining heat transfer, fluid flow, entropy generation, and the dominance of heat transfer, viscous or magnetic field effects. This comprehensive understanding provides a foundation for optimizing heat transfer processes in various applications, from thermal management systems to the manipulation of magnetic fluids. Our research not only contributes to the theoretical understanding of MHD natural convection but also offers practical implications for engineering applications. The findings of this study can help in the development and optimization of systems involving magnetic fields and fluid dynamics, with potential applications in energy transport, magnetic confinement in fusion reactors, and innovative cooling technologies.\vspace{11pt}\\
{\textbf{Author Declaration}}\\
The authors have no conflicts to disclose.
\vspace{11pt}\\
{\textbf{Data Availability}}\\
The data that support the findings of this study are available from the corresponding author upon reasonable request.
\newpage





\end{document}